\documentclass[%
	paper=A4,					
	twoside=true,				
	openright,					
	parskip=full,				
	chapterprefix=true,			
	11pt,						
	headings=normal,			
	bibliography=totoc,			
	listof=totoc,				
	titlepage=on,				
	captions=tableabove,		
	draft=false,				
]{scrreprt}

\newcommand{\thesisTitle}{Automatic Web Service Composition \\ Models, Complexity and Applications}
\newcommand{\thesisName}{Ph.D. Student \textbf{Paul Diac}}
\newcommand{\thesisSubject}{Ph.D. Thesis}
\newcommand{\thesisDate}{April 2020}

\newcommand{\thesisFirstSupervisor}{Prof. Dr. Dan Cristea}

\newcommand{\thesisFirstReviewer}{Prof. Dr. Cornelius Croitoru}
\newcommand{\thesisSecondReviewer}{Conf. Dr. Adrian Iftene}
\newcommand{\thesisThirdReviewer}{Lect. Dr. Ionu\c t Pistol}

\newcommand{\thesisUniversity}{\protect{Alexandru Ioan Cuza University of Ia\c{s}i}}
\newcommand{\thesisUniversityDepartment}{Faculty of Computer Science}

\newcommand{\thesisUniversityStreetAddress}{General Berthelot, 16, Corp C UAIC, Ia\c{s}i, Rom\^{a}nia, 700483}
\newcommand{\thesisUniversityPostalCode}{}

\usepackage[utf8]{inputenc}		
\usepackage[english]{babel} 
\usepackage{hyperref}
\usepackage[					
	figuresep=colon,%
	sansserif=false,%
	hangfigurecaption=false,%
	hangsection=true,%
	hangsubsection=true,%
	colorize=full,%
	colortheme=bluemagenta,%
	bibsys=bibtex,%
	bibfile=wsc_ref,%
	bibstyle=alphabetic,%
]{cleanthesis}

\hypersetup{					
	pdftitle={\thesisTitle},	
	pdfsubject={\thesisSubject},
	pdfauthor={\thesisName},	
	plainpages=false,			
	colorlinks=false,			
	pdfborder={0 0 0},			
	breaklinks=true,			
	bookmarksnumbered=true,		%
	bookmarksopen=true			%
}

\usepackage{listings}
\usepackage{amsmath}
\usepackage{amssymb}
\usepackage{amsthm}
\usepackage{relsize}
\usepackage{tikz}
\usetikzlibrary{
  calc, shapes, arrows, positioning,
  graphs, graphs.standard
}

\usepackage{tikz-qtree}
\usepackage{multirow}
\usepackage{booktabs}
\usepackage{makecell}
\usepackage{algorithm}
\usepackage[noend]{algpseudocode}
\usepackage{graphicx}
\usepackage{bibentry}
\usepackage[final]{pdfpages}
\usepackage{bookmark}

\usepackage[utf8]{inputenc}
\usepackage{textcomp}

\usepackage{pgfplots}
\pgfplotsset{width=7cm,compat=1.8}
\usepackage{pgfplotstable}

\definecolor{pblue}{rgb}{0.13,0.13,1}
\definecolor{pgreen}{rgb}{0,0.5,0}
\definecolor{pred}{rgb}{0.9,0,0}
\definecolor{pgrey}{rgb}{0.46,0.45,0.48}

\theoremstyle{definition}
\newtheorem{definition}{Definition}[section]
 
\newboolean{showcomments}
\usepackage{etoolbox}
\newtoggle{fullThesis}

\setboolean{showcomments}{true}
\setboolean{showcomments}{false}
\toggletrue{fullThesis}

\ifthenelse{\boolean{showcomments}}
{ \newcommand{\mynote}[3]{
   \fbox{\bfseries\sffamily\scriptsize#1}
   {\small$\blacktriangleright$\textsf{\emph{\color{#3}{#2}}}$\blacktriangleleft$}}}
{ \newcommand{\mynote}[3]{}} 
\newcommand{\comment}[1]{\mynote{comment}{#1}{blue}}

\newcommand{\algoname}[1]{\textnormal{\textsc{#1}}}

\newcommand{\HRule}[1][\medskipamount]{\par
	\vspace*{\dimexpr-\parskip-\baselineskip+#1}
	\rule{15.5em}{0.4pt}\par
	\vspace*{\dimexpr-\parskip-.5\baselineskip+#1}}

\begin{document}

\renewcaptionname{english}{\figurename}{Figure}
\renewcaptionname{english}{\tablename}{Table}

\pagenumbering{roman}			
\pagestyle{empty}				
%
\begin{titlepage}
	\pdfbookmark[0]{Cover}{Cover}
	\flushright
	\hfill
	\vfill
	{\LARGE\thesisTitle \par}
	\rule[5pt]{\textwidth}{.4pt} \par
	{\Large\thesisName}
	\vfill
	\textit{\large\thesisDate} \\
\end{titlepage}

\begin{titlepage}
	\pdfbookmark[0]{Titlepage}{Titlepage}
	\tgherosfont
	\centering

	{\Large \thesisUniversity} \\[4mm]
	\textsf{\thesisUniversityDepartment} \\

	\vfill
	{\large \thesisSubject} \\[5mm]
	{\LARGE \color{ctcolortitle}\textbf{\thesisTitle} \\[10mm]}
	{\Large \thesisName} \\

	\vfill
	
	\begin{minipage}[t]{.27\textwidth}
		\raggedleft
		\textit{Supervisor}
	\end{minipage}
	\hspace*{15pt}
	\begin{minipage}[t]{.65\textwidth}
		\Large \thesisFirstSupervisor\ 
	\end{minipage} \\[10mm]	

	\thesisDate \\

\end{titlepage}

\hfill
\vfill
{
	\small
	\textbf{\thesisName} \\
	\textit{\thesisTitle} \\
	\thesisSubject, \thesisDate \\
	Supervisor: \thesisFirstSupervisor\ \\ 
	Reviewers: \thesisFirstReviewer\ , \thesisSecondReviewer \space and \thesisThirdReviewer \\
	\textbf{\thesisUniversity} \\
	\thesisUniversityDepartment \\
	\thesisUniversityStreetAddress \\
	\thesisUniversityPostalCode\
}
\cleardoublepage

\pagestyle{plain}				
%
\pdfbookmark[0]{Abstract}{Abstract}
\addchap*{Abstract}
\label{sec:abstract}

\vspace{-1cm}

\comment{English version of the abstract. UPDATE (4 April): re-wrote all this in English and Romanian, mostly to fit one page (it does without this comment), the same information, re-structured, etc}

The automatic composition of web services refers to how services can be used in a complex and aggregate manner, to serve a specific and known functionality. Given a list of services described by the input and output parameters, and a request of a similar structure: the initially known and required parameters; a solution can be designed to automatically search for the set of web services that satisfy the request, under certain constraints.

We first propose two very efficient algorithms that solve the problem of the automatic composition of the web services as it was formulated in the competitions organized in 2005 and 2008. The algorithms obtain much better results than the rest of the participants with respect to execution time and even composition size. A test generator is also proposed, generating cases where much longer compositions are necessary, an important factor of the resulting search time. Evaluation consists of running the previous and the proposed solutions on given benchmarks. Further, we design two new models to match service's parameters, extending the semantic expressiveness of the 2008 challenge. The initial goal is to resolve some simple and practical use-cases that cannot be expressed in the previous models. We also adhere to modern service description languages, like \emph{OpenAPI} and especially \emph{schema.org}. Algorithms for the new models can solve instances of significant size, despite the increased computational complexity. Addressing a wider and more realistic perspective, we define the dynamic, online version of the composition problem. In this regard, we consider that web services and compositions requests can be added and removed in real-time, and the system must handle such operations on the fly. Despite the most common assumption in research, software systems cannot handle transient requests exclusively. It is necessary to maintain the workflows for users who actively run the compositions over time. At least, if one service becomes unavailable, a backup composition can potentially keep the request alive. As for the newly semantic model algorithms, we propose new algorithms and provide comprehensive evaluation by generating test cases that simulate all corner cases.

Thesis contributions fall under a few categories: optimizations on existing parameter matching algorithms and designing of new semantic models. Towards the end, the dynamic version of composition introduces the applications section.

\newpage
{\usekomafont{chapter}Abstract - in Romanian}
\label{sec:abstractRo}

Compunerea automată a serviciilor web studiază modalităţile prin care serviciile pot fi folosite într-un mod complex şi agregat, pentru a deservi o funcționalitate specifică și cunoscută. Mai precis, cunoscând o listă de servicii, descrise prin parametrii de intrare și ieșire, împreună cu o cerere cu structură similară: parametrii inițial cunoscuți și căutaţi; se poate proiecta un sistem care să caute automat o structură de servicii web ce satisface solicitarea, respectând anunite constrângeri.

În acest domeniu, propunem doi algoritmi eficienți care rezolvă problema compoziției automate a serviciilor web pentru versiunile sale de la competițiile organizate în 2005 și 2008. Algoritmii obțin rezultate mult mai bune decât restul competitorilor ca timp de execuție și chiar ca dimensiune a compoziției. Totodată, propunem un generator de teste care generează cazuri ce necesită compoziții mult mai lungi, un factor important pentru aproximarea timpului de căutare rezultat. Evaluarea compară rezultatele execuţiei soluțiilor anterioare și a noilor algoritmi pe testele cunoscute dar şi pe cele nou generate. Ulterior, propunem două noi modele de potrivire a parametrilor serviciilor, extinzând expresivitatea semantică a modelului prezentat în competiţia din 2008. Scopul inițial este rezolvarea unor cazuri simple și practice care nu puteau fi exprimate în modelele anterioare. În acest demers folosim limbaje moderne de descriere ale serviciilor, cum sunt \emph{OpenAPI} și în special \emph{schema.org}. Algoritmii pentru noile modele pot rezolva cazuri de dimensiuni semnificative, în ciuda complexității computaţionale crescute. Abordând o viziune mai largă și mult mai realistă, definim versiunea dinamică sau \emph{online} a problemei compoziției. În aceasta, se consideră că serviciile web și cererile utilizatorilor pot fi adăugate și şterse în timp real, iar sistemul trebuie să trateze dinamic aceste operaţii. Deşi majoritatea studiilor consideră astfel, sistemele software reale nu pot trata cererile ca fiind doar tranzitorii. Este necesară menținerea compoziţiilor pentru utilizatorii care le folosesc continuu sau de mai multe ori. Cel puțin, în cazul în care un serviciu devine indisponibil, o compoziție alternativă poate rezolva cererile afectate. Pentru noile modele, propunem şi algoritmii, împreună cu o evaluare sintetică dar cuprinzătoare, prin generarea testelor care evidenţiază toate cazurile particulare.

Contribuțiile se împart în două categorii: optimizări pentru modelele existente și proiectarea de noi modele semantice. În final, versiunea dinamică a problemei de compoziție introduce secțiunea de aplicații.

\cleardoublepage

%
\pdfbookmark[0]{Acknowledgement}{Acknowledgement}

\begingroup
\clearpage
\let\clearpage\relax
\vspace*{-1cm}
\chapter*{Acknowledgment}
\endgroup

\label{sec:acknowledgement}
\vspace*{-1cm}

All the challenges encountered during the doctoral program were overcome with the support of many helpful people: professors, students, friends. I apologize to those omitted as I am grateful to be helped by more people than I can keep in mind.
\\[10pt]
First of all, I would like to thank my supervisor \thesisFirstSupervisor, for introducing me into the fascinating domain of \emph{Natural Language Processing}, in particular to the management of workflows of services, and for all the helpful advice on my research in general and in the writing of the thesis.
\\[10pt]
I also want to thank \thesisFirstReviewer, for all the help and especially for sharing his knowledge in algorithmics, complexity analysis and related topics; moreover for directing me to \emph{automatic composition} through the service challenges. Other top-tier contributors to my Ph.D. progress are \thesisSecondReviewer \space and \thesisThirdReviewer \space who shared their experience with me all this period.
\\[10pt]
Substantial assistance was provided by the colleagues Dr. Sabin-Corneliu Buraga and Dr. Emanuel Onica, offering expert advice in the \emph{semantic web} and, respectively, the \emph{distributed systems} domains. Recent years had increased research productivity due to the collaboration with outstanding students Liana Ţucăr and Andrei Netedu, that contributed to my works with their ideas, knowledge and programming skills. Also, thanks to Dr. Pablo Rodríguez-Mier for providing the competition benchmarks.
\\[10pt]
A special thanks to Dr. \c{S}tefan Ciobâc\u{a} for both research advice and teaching support. Many other people helped me in various ways, and the following list is far from being complete: Dr. Diana Trandabăţ, Vlad Manea, Dr. Andrei Arusoaie, Radu Mereuţă, Dr. Cătălina Mărănduc.
\\[10pt]
For me, \href{https://icpciasi.bitbucket.io/}{\emph{Competitive Programming}} unfolded to a significant extracurricular activity. Therefore, my recognition also includes all participating students for their contagious enthusiasm, maintained by remarkable results. One team stands out: Valeriu Motroi, Alexandru Ioniţă and Cristian Vîntur, reaching unprecedented levels of performance that I gladely witnessed. Amplified with the challenge set by Aurelian Hreapcă, I can only await with bold curiosity for the adventure ahead!
\\[10pt]
Last and most important I want to thank my parents, my family, and my girlfriend Maria for all their support, empathy and patience.
\\[10pt]

\cleardoublepage

\setcounter{tocdepth}{2}		
\pdfbookmark{\contentsname}{}
\tableofcontents				

\pagenumbering{arabic}			
\setcounter{page}{1}			
\pagestyle{maincontentstyle} 	

%

{
\cleardoublepage
\let\clearpage\relax
\vspace*{1cm}
\cleanchapterquote{I'll tell you this...\\No eternal reward will forgive us now\\For wasting the dawn.}{James Douglas Morrison}{(singer, songwriter and poet)}
\vspace*{-0.6cm}

\chapter{Introduction}
\label{sec:intro}
}

\vspace*{-0.7cm}

\comment{The introduction chapter provides a higher-level overview of the overall domain and presents the structure of the rest of the thesis.}

Web Service Composition is a domain in Computer Science and Software Development which enables the use of multiple services in an aggregated manner, to reach a functionality of a wider range. It is also a particular area part of the Service-Oriented Architecture \cite{perrey2003service} which has been popular for many years and it will also be widely used in the future.

The core element of this domain is the Web Service. There are several definitions or agreements of what a Web Service actually is; motivated by the fact that the principle behind a service, in general, is both fundamental and widely used. However, generally and also from the thesis perspective, a Web Service is a standardized functionality or method of communication between entities in a network. Some of these entities are service developers, exposing the services interface and functionalities, and the other, the consumers use them by invoking services.  Principally, services have a precise and narrow functionality, i.e. similar to a function. We do not refer to a service as a wider application, in our view, one service is only one of the methods from a more complex WSDL file \cite{christensen2001web} for example. This is motivated by our rather theoretical view on Service Composition.

Further, the context is narrowed to the situation in which a large number of services are already known, with all the relevant interface definitions. These constitute the service repository. On top of that, there is also interest in a new functionality, usually (but not necessarily) more complex. Most often, as one can intuitively expect, there is no single service satisfying the required functionality. But there is a chance that several services from the repository can work together to reach the requirement. Web Service Composition refers to finding those services.

The first word of the title suggests more than this. Rightly so, even if the manual composition is not trivial either, automation is the key idea behind our research. Specifically, the goal is to find such a composition programmatically, ideally without any human intervention. Automation is motivated by several factors: the repositories of services can be (very) large, is less error-prone, and is cheaper and faster than manual composition; therefore it can be applied to more situations, e.g. on-the-fly switch to backup compositions when needed. But automation of the process brings two new and specific difficulties: the necessity of a model, i.e. a set of formal rules, guiding the construction of the workflow of services; and second, the algorithm which builds the construction. The two are clearly separated in the thesis by two main parts: Chapter \ref{sec:models} Models, and Chapter \ref{sec:complexity}, Complexity (wider perspective including not only algorithms but also analysis of problem complexities). Our initial interest was to implement the most efficient algorithms for known versions of composition. Afterward, we also propose some new models of the composition problem.

Finally, composition, as any research domain, should be applied in practice. This is particularly motivated by the fact that automatic composition is also a form of knowledge representation and reasoning \cite{levesque1986knowledge}. Modeling elements of software development (services, parameters), and automatically combining such pieces together, also makes automatic composition a declarative programming example. It is therefore important to develop applications, to follow how they are adopted by the community, and to consider feedback. Also, it is important to study developed applications that make direct use of the domain, or similarities of underlying concepts in others. There are also specialized domains where composition needs to be adapted to specific particularities. All of these are part of Applications (Chapter \ref{sec:applications}).

\section{Motivation}
\label{sec:intro:motivation}
\comment{Why Service Oriented Architecture is modern and important, what is a service, what is a composition in very short words, etc. About Service Oriented Architecture, very commonly used in modern distributed software development.}

Web Services provide a popular paradigm to develop a wide variety of applications. The advantages include (to some degree): ownership is distributed and well-organized, fine granularity leading to loose coupling \cite{alonso2004web}: services/developers do not interact with many other services/developers; web services are efficient, i.e. fast, provide high interoperability, and many others \cite{daigneau2012service}. Generally, Service-Oriented Architecture provides a lot of flexibility in terms of how services are designed, what standards are used to describe their interfaces, how they are published and consumed.

Service composition helps achieve more of these advantages at the same time. For example, if service granularity is too high, one service's functionality is very limited. Besides the advantages for service developers, this limits the ease of use for service consumers. Microservices \cite{newman2015building} is an emerging and recent paradigm in this direction. To enhance the usability, the composition is more obvious in this case. But composition, generally, is useful in any case in which services have to be selected from a repository and organized to resolve some specific functionality. Composition, however, requires a guiding formalism, because even manual composition is based on some rules, principles, or extra knowledge about services, other than what they do: how can they be combined?

Automating the composition process is more important when the repository is very large or when the composition conditions are complex. For a human it is difficult, if not impossible, to learn or follow the interaction between more than several services. A service repository can easily consist of thousands or even many more services. Composition search algorithms follow a set of composition rules - most often the parameter matching models. The more expressive a model is, the more it includes reasoning, the same as it happens in manual composition, but there are also at least two disadvantages. First it is more difficult to define services based on that model, and therefore it is harder to become popular among many service developers. Many service developers and in general many services are a necessity to justify automatic composition. The second issue of a complex model is that the algorithms solving the composition requests on such models are more complex themselves, not only is their design and implementation but most often in the number of operations they execute, i.e. the run time complexity. A trade-off between simplicity and expressiveness has to be made, and the models proposed in this thesis give  examples in this sense, with the analysis of advantages and disadvantages.

Further, each of the proposed model has its purpose. The \emph{relational} and the so-called \emph{object-oriented} models enhance the semantics used in parameter matching - the core formalism for automatic composition. Both of them are thesis contributions to resolve known cases where previous composition models failed to include some elementary reasoning done naturally by a human composer. So, first of all, they are motivated by relevant examples. The proposed \emph{online} version of the composition problem has its naming inspired by the meaning of \emph{online algorithms} \cite{albers2003online}. The classical view on service composition where a single request is computed on a static repository of services is not applicable in the dynamic environment of the Service-Oriented Architectures.

\section{Automatic Web Service Composition}
\label{sec:intro:whatIsWSC}
\comment{More complete, detailed definition of Composition; yet not very formal. Some history, etc.}
Until now we discussed automatic composition from a general and high-level point of view. More specifically, in the simplest version of the automatic composition, the problem is defined as in the following.

Suppose we knowing the definitions of a set of services, and the request for some new functionality, similar to a service. Also the automation is defined with the help of some functional parameter matching rules, that guide how parameters are transfered between cooperating services. A satisfying composition is a list of services from the repository, which, starting from the information given in the request, reaches the required information and respects all matching rules. More precisely, these rules work with service parameters. In the simplest form, services are defined only by a set of input and a set of output parameters. As said, the request is defined by the same structure. Generally, one service can be called or invoked if all its input parameters are known. Since the composition is computed at the abstract level, meaning that it does necessary include the execution, a parameter is known if it was conceptually known in the request or is the output of another service already added to the composition. It is easier to imagine this at execution, where a known parameter would be a parameter whose concrete value is actually known, or it is instantiated. However, our focus is on the design of the composition generally, therefore before execution. For this reason, a known parameter is just an entity, concept or an abstract value that will be effectively instantiated when the execution of the building composition would reach the current stage of the composition.

This is why the parameter matching model is fundamental to automatic composition. It is the formalism that allows the knowledge transfer from services already included in the composition to the next services that are added to the composition. The matching model can be arbitrary complex, for example, it can include: parameter concept definitions, subsumption between these concepts, other semantic properties, quality of service metrics, service state and others. 

In the simplest form, each parameter is defined by a concept from a simple set of concepts or types. A simple representation of this is through parameter \emph{names}. Each parameter has a single, textual name, that defines the information needed for input, or retrieved for the output parameters. One output parameter of a service can be matched with the input of another service if they have the same name. At some stage in the composition, a set of known parameter names can be used to call further services. To call a service, all its input parameters must be in the known set, and after the call, all output parameters are added to the set. The call is not actually a call to the service, but rather the learning of the service's output parameters and its addition to the list of processed/called services. No previously known parameter is removed: the known set is constantly growing. The goal is to add to this set, i.e. learn, all the required parameters of the request.

If successful, the result consists of the services called. It is useful to remember their order as well. Ideally, it is further useful to consider them in a partial order: some services may depend on other previously called services, but not all of them. This can improve efficiency by allowing parallel execution of services that become callable at the same time.

Throughout the thesis, services will remain stateless, meaning that they do not alter their state or any external state. This type of services is also referred to as information-providing services, like in \cite{zhao2012automatic}, highlighting that services just produce new information based on previously known information. Everything other then the \emph{knowledge} and the building composition, is stateless: the repository, the services, the semantic elements used and the request.

\section{Web Services Challenges}
\label{sec:intro:wscChallenges}
\comment{Talk about the challenge series (2005 -2010) and why it is still useful to compare results on the benchmarks. They provide one of the only synthetic tests that can be used to compare with more than just several services.}

By \emph{Web Services Challenges}, or \emph{composition challenges} or \emph{competitions}; we refer to the series of the competitions organized to centralize, evaluate and analyze different automatic composition solutions (algorithms). The competition had six editions \cite{blake2010wsc}, between 2005 and 2010, and it was part of several related conferences. The competitions take a big part of the motivation for the thesis domain, as our first research \cite{diac2017engineering} and \cite{tucar2018semantic} propose new algorithms for the competition's editions of 2005 and 2008. Also, the competition provides an excellent starting point for algorithmic improvements that which aroused interest even in recent years.

Further, two other papers closely related to the challenges, lead to our first contributions. The first is  \cite{zou2014dynamic} which reduces the composition model of the 2005 edition to Artificial Intelligence Planning instances, solved by Fast-Forward planner \cite{hoffmann2001ff}. Also, it provides a very clear problem definition that aroused interest. After the implementation of the first algorithm, the next natural step was to adapt the algorithm to more recent versions of the challenges, in particular the 2008 edition. Similarly, we compared the solution with post-challenge results, particularly to \cite{rodriguez2011automatic}. The algorithms developed for these two versions of the problems were used as a basis for other algorithms in the thesis, which solve newly proposed models of composition.

The first 2005 edition \cite{blake2005eee} of the challenge series, \emph{EEE-05 Web Services Challenge}, defines the composition problem and the parameter matching model. It also provides benchmarks and the final evaluation criteria for both composition and discovery of web services. The parameter matching model used is the simple name matching described in Section \ref{sec:models:names}. The second edition from 2006 \cite{blake2006wsc} is the first step to introduce semantics to parameter description, through a simple hierarchy of concepts, enabling subsumption. The 2007 edition \cite{blake2007wsc} adds modern semantic web standards for parameter description: the Web Ontology Language - OWL \cite{mcguinness2004owl}, and also introduces parallel execution of services. The 2008 edition \cite{bansal2008wsc} on which we have our second contribution, uses modern web standards like \emph{SOAP} \cite{gudgin2003soap} for message transmission and \emph{WS-BPEL} \cite{weerawarana2005web} to represent workflows. The 2009 challenge \cite{kona2009wsc} is the first to include non-functional properties: Quality of Service - QoS, specified by the \emph{WSLA} format \cite{ludwig2003web}. Section \ref{sec:models:qos} briefly describes the relationship between automatic composition and QoS, but more significant work in this direction is left as future work. The series ends with the 2010 edition \cite{blake2010wsc}, which extends the QoS metrics to both response time and throughput.

\section{State of the Art in Service Composition}
\label{sec:intro:stateoftheart}
\comment{Modern view on service composition, in terms of new models, language, technologies, like OpenAPI, REST; and their influence on WSC.}
\comment{reminders: category theory}

First research papers on \emph{Automatic Web Service Composition} appeared around the year 2000. Figure \ref{fig:wsc_histogram} displays the number of papers on \emph{"Automatic Web Service Composition"} from the beginning to the present time. In the first five to ten years it had continuous growth, accelerated by the same period of evolution of the wider research areas like Semantic Web, Service-Oriented Architectures, Knowledge Representation and Reasoning\footnote{Part of ACM Computing Classification System -- \url{https://dl.acm.org/ccs}}.

In the following period, there is a  stagnation or recently even decline. There are several possible reasons for this. First of all, the research domain may have reached a satisfying maturity level. However, in practice, applications using effectively automatic composition are still very limited, though there is significant interest in applying automatic composition. Secondly, the right balance between semantic expressibility and ease of use is hard to find. Much of the research proposals are not adopted in practice because they make service definitions too complex. Developers are reluctant to rigid models and specifications. Computational complexity could also have a contribution, as the number of services can be too large to allow fast computation of compositions. Finally, even the end of the web services challenge series in 2010 can be another factor. The contributions of the thesis provide some steps to mitigate the all alleged reasons for this trend.

\begin{figure}[h]
 \centering
 \caption[Growth of \textbf{Automatic Composition} Research]{Number of research papers on \emph{Automatic Web Service Composition} by the two years periods of last two decades. Data sources, as of April 2020, linked below:\\  
  \textcolor{blue}{\href{https://scholar.google.com/scholar?q=\%22Automatic+Web+Service+Composition\%22&hl=en&as_sdt=0\%2C5&as_ylo=2000&as_yhi=2001}{\underline{00-01}} \href{https://scholar.google.com/scholar?q=\%22Automatic+Web+Service+Composition\%22&hl=en&as_sdt=0\%2C5&as_ylo=2002&as_yhi=2003}{\underline{02-03}} \href{https://scholar.google.com/scholar?q=\%22Automatic+Web+Service+Composition\%22&hl=en&as_sdt=0\%2C5&as_ylo=2004&as_yhi=2005}{\underline{04-05}} \href{https://scholar.google.com/scholar?q=\%22Automatic+Web+Service+Composition\%22&hl=en&as_sdt=0\%2C5&as_ylo=2006&as_yhi=2007}{\underline{06-07}} \href{https://scholar.google.com/scholar?q=\%22Automatic+Web+Service+Composition\%22&hl=en&as_sdt=0\%2C5&as_ylo=2008&as_yhi=2009}{\underline{08-09}} \href{https://scholar.google.com/scholar?q=\%22Automatic+Web+Service+Composition\%22&hl=en&as_sdt=0\%2C5&as_ylo=2010&as_yhi=2011}{\underline{10-11}} \href{https://scholar.google.com/scholar?q=\%22Automatic+Web+Service+Composition\%22&hl=en&as_sdt=0\%2C5&as_ylo=2012&as_yhi=2013}{\underline{12-13}} \href{https://scholar.google.com/scholar?q=\%22Automatic+Web+Service+Composition\%22&hl=en&as_sdt=0\%2C5&as_ylo=2014&as_yhi=2015}{\underline{14-15}} \href{https://scholar.google.com/scholar?q=\%22Automatic+Web+Service+Composition\%22&hl=en&as_sdt=0\%2C5&as_ylo=2016&as_yhi=2017}{\underline{16-17}} \href{https://scholar.google.com/scholar?q=\%22Automatic+Web+Service+Composition\%22&hl=en&as_sdt=0\%2C5&as_ylo=2018&as_yhi=2019}{\underline{18-19}}} \textbf{G.Scholar} \\
  \textcolor{blue}{\href{https://dblp.uni-trier.de/search?q=automatic\%20web\%20service\%20composition\%20year\%3A2000\%7C2001}{\underline{00-01}} \href{https://dblp.uni-trier.de/search?q=automatic\%20web\%20service\%20composition\%20year\%3A2002\%7C2003}{\underline{02-03}} \href{https://dblp.uni-trier.de/search?q=automatic\%20web\%20service\%20composition\%20year\%3A2004\%7C2005}{\underline{04-05}} \href{https://dblp.uni-trier.de/search?q=automatic\%20web\%20service\%20composition\%20year\%3A2006\%7C2007}{\underline{06-07}} \href{https://dblp.uni-trier.de/search?q=automatic\%20web\%20service\%20composition\%20year\%3A2008\%7C2009}{\underline{08-09}} \href{https://dblp.uni-trier.de/search?q=automatic\%20web\%20service\%20composition\%20year\%3A2010\%7C2011}{\underline{10-11}} \href{https://dblp.uni-trier.de/search?q=automatic\%20web\%20service\%20composition\%20year\%3A2012\%7C2013}{\underline{12-13}} \href{https://dblp.uni-trier.de/search?q=automatic\%20web\%20service\%20composition\%20year\%3A2014\%7C2015}{\underline{14-15}} \href{https://dblp.uni-trier.de/search?q=automatic\%20web\%20service\%20composition\%20year\%3A2016\%7C2017}{\underline{16-17}} \href{https://dblp.uni-trier.de/search?q=automatic\%20web\%20service\%20composition\%20year\%3A2018\%7C2019}{\underline{18-19}}} \textbf{DBLP}
  }
 \label{fig:wsc_histogram}
 \vspace{-0.3cm}
 \begin{tikzpicture}
  \centering
  \begin{axis}[
        ybar, axis on top,
        height=8cm, width=14.5cm,
        bar width=0.4cm,
        ymajorgrids, tick align=inside,
        major grid style={draw=white},
        enlarge y limits={value=.1,upper},
        ymin=0, ymax=350,
        axis x line*=bottom,
        axis y line*=right,
        y axis line style={opacity=0},
        tickwidth=0pt,
        enlarge x limits=true,
        legend style={
            at={(0.5,-0.4)},
            anchor=north,
            legend columns=-1,
            /tikz/every even column/.append style={column sep=0.4cm}
        },
        yticklabels={,,},
        xticklabel style={rotate=90},
        symbolic x coords={
           2000-2001,2002-2003,2004-2005,2006-2007,
		2008-2009,2010-2011,2012-2013,
          2014-2015,2016-2017,2018-2019},
       xtick=data,
       nodes near coords={
        \pgfmathprintnumber[precision=0]{\pgfplotspointmeta}
       }
    ]
    \addplot [draw=none, fill=blue!30] coordinates {
      (2000-2001,6)
      (2002-2003,28) 
      (2004-2005,116)
      (2006-2007,216) 
      (2008-2009,271) 
      (2010-2011,331)
      (2012-2013,293) 
      (2014-2015,300)
      (2016-2017,229)
      (2018-2019,169) };
   \addplot [draw=none,fill=orange!30] coordinates {
      (2000-2001,0)
      (2002-2003,5) 
      (2004-2005,11)
      (2006-2007,28) 
      (2008-2009,25) 
      (2010-2011,19)
      (2012-2013,14) 
      (2014-2015,13)
      (2016-2017,8)
      (2018-2019,7) };
    \legend{Google Scholar, DBLP}
  \end{axis}
 \end{tikzpicture}
\end{figure} 

The evolution of the automated composition as a research domain and the latest advances are presented in several survey papers, like \cite{venkatachalam2016comprehensive}, \cite{syu2012survey}, or the early but popular \cite{rao2004survey}. Overcoming its incipient phase, the service composition advanced consistently with the semantic web ideas, models and standards. The need for integration and heterogeneity is obvious, and the first languages and technologies to be included in compositions studies were WSLD\footnote{WSDL -- \url{https://www.w3.org/TR/wsdl.html}} \cite{christensen2001web} for service definition, UDDI\footnote{UDDI -- \url{http://uddi.xml.org/specification}} \cite{van2000universal}, \cite{kourtesis2008combining} for service description, discovery (the stage usually preceding composition), and managing repositories, SOAP\footnote{SOAP -- \url{https://www.w3.org/TR/soap/}} \cite{gudgin2003soap} for message exchange. The compositions as workflows are described using BPEL4WS \cite{chao2004analysis}, which was also used in the services challenge, and DAML-S/OWL-S\footnote{OWL-S -- \url{https://www.w3.org/Submission/OWL-S/}} \cite{wu2003automating}. More recently, other aspects are taken into consideration, like Quality of Service, occasionally modeled with the WSLA language \cite{ludwig2003web},  \cite{nematzadeh2014qos}, enriched semantics \cite{bekkouche2017qos} the recent paradigms of Internet of Things \cite{baker2017energy}, \cite{berrani2018towards} and. Most recent of all, Micro-Services provide a favorable context for automatic composition, as the emphasis is on loosely coupled, fine-grained services \cite{ke2019implementation}, \cite{wang2018client}. Another modern approach, many publications propose composition for RESTful services \cite{richardson2008restful}, \cite{garriga2016restful}, including the OpenAPI specification\footnote{OpenAPI -- \url{https://swagger.io/specification/}} \cite{kim2016ontology}. Our paper \cite{netedu2019openapi} is also in this category. Therefore, the automatic service composition domain is always adapting to many technology trends in the web, or service-oriented computing in general. 

As a computational problem, solutions for automatic service composition use various algorithmic techniques. Some transform composition instances to artificial intelligence planning instances. Then, they can be solved with existing planners and the resulting plans are transformed back to give the compositions \cite{sirin2004htn}, \cite{omid2017context}. Similarly, Answer Set Programming has also been used for composition \cite{rainer2005web}. Another popular idea for the optimization versions is to use various heuristics to build the compositions directly. Many are based on evolutionary algorithms \cite{jatoth2015computational}, \cite{pop2010ant}, others on (hyper) graph search algorithms \cite{rodriguez2011automatic}, \cite{oh2005bf}. Even quantum inspired algorithms have approached versions that consider QoS metrics \cite{boussalia2014optimizing}. Therefore, automatic composition is a complex computational problem as well, with many versions and aspects to address.

\section{Thesis Overview}
\label{sec:intro:overview}
\comment{A short guide through the rest of the thesis, motivating its structure but also helping the reading process.}

The content of the thesis can be organized based on two different perspectives. The first would follow the chronology of our publications related to the thesis domain, which are iterated in Section \ref{sec:intro:publications}. This has the advantage of somewhat better modularity, because it generally reduces the correlation between sections or chapters, at least for nonadjacent parts. However, we opted for another perspective, which follows a higher-level prospect of the \emph{automatic service composition} problem as a whole, in broader terms, following a hierarchical view on the structure of the domain.

Thinking generally, of situations where we want to solve a higher-level, complex real-world necessity by computer science. We first have to formalize a mathematical model, expressive enough to be capable of interpreting all relevant elements. The formalism is then transformed into a computational problem for which we design algorithms, or solutions, proving that they can solve sufficiently large instances. Lastly, these results have to be applied to practice and be brought into the real world, in our case in software development.

Generally, to solve a complex real-world problem by computer science, we first formalize a model, expressive enough to include the relevant elements. Such formalism gives the computational problem for which we design algorithms, ideally showing that they can solve sufficiently large instances. Lastly, these results have to be applied to practice in our case in software development.

The thesis structure favors this unitary perspective adapted to the domain of \emph{automatic service composition}. Unfortunately, this vision also has its disadvantages: since there will be more models, they are brought into discussion each of them, being analyzed separately through the stages mentioned. This can be hard to follow at times.

The \textbf{Introduction Chapter \ref{sec:intro}} introduces the reader into the thesis, motivating the study, briefly defining the problem and presenting the state of the art of the composition domain.

The \textbf{Models Chapter \ref{sec:models}} focuses on the design-level difficulties in automatic service composition. Similarly to other chapters, it addresses the parameter matching models: initial name-matching, hierarchical, the proposed relational and object-oriented models, the QoS and stateful aspects and finally the higher-level of dynamic or online problem version. The sections here, similar to problem statements, regularly contain some introduction, motivation, one incentive example, and the formal definition.

The \textbf{Complexity Chapter \ref{sec:complexity}} first, discusses the upper bound measure of the problems associated with each model. This is done on a short description of algorithms implemented, as they are just examples for the upper bounds analysis. The run-time complexity is specified for each. The next, lower bounds section presents two complexity proofs. The third is the most consistent section. It presents implemented algorithms, with the results on various benchmarks: run-times, composition sizes, and others. We refer to this as the \emph{empirical} evaluation of the algorithms to emphasize its experimental nature, relative to the previous asymptotic computational complexity analysis. Another important contribution in this section are the four test generators. For all aspects specified here, each model is considered.

\textbf{Applications Chapter \ref{sec:applications}} connects our research to the real-world and in the first section, particularly to the software industry. Natural Language Processing (NLP in the following) specific applications are presented in the following. The last section, provides the design of a composition framework, including a data model which combines more of the proposed matching models, and briefly goes through its functionalities and open questions.

The \textbf{Conclusion Chapter \ref{sec:conclusion}} highlights the contributions of the thesis and provides many future work directions.

\section{List of Publications}
\label{sec:intro:publications}
\comment{Inspired form the thesis of Ionut Pistol. The idea is to highlight the papers published during the Ph.D., fist all that are associated with the thesis and then maybe just list the others as well (tough off-topic, they are relevant for the progress of the Ph.D. period). Here is a good place to also explain the relationship between thesis structure (previous section) and published papers. Also here present the SYNASC poster and include it as an appendix.}

The thesis is based on several research papers, published between 2014 and 2019. The chronologically first is related to the Natural Language Processing domain, while the rest, published between 2017 and 2019 belong to the more general \emph{Automatic Web Service Composition} field. They either have some algorithmic optimization focus or propose some new semantic models. The Chapter \ref{sec:models}, on models, iterates all composition versions considered and motivate, exemplifies, and formally defines the models. Their associated problems complexities are analyzed, in the same order, in the Chapter \ref{sec:complexity}, on complexity.

The list of (co)-authored published papers associated with the thesis follows.

\preto\fullcite{\AtNextCite{\defcounter{maxnames}{99}}}

\begin{itemize}
  \setlength\itemsep{-1em}
  \item \fullcite{cristea2015quo}
  \item \fullcite{diac2017engineering}
  \item \fullcite{diac2017warp}

\vspace{-0.5cm}
The poster is included in the Appendix Section \ref{sec:appendix:poster}.
  \item \fullcite{tucar2018semantic}
  \item \fullcite{diac2019relational}
  \item \fullcite{diac2019formalism}
  \item \fullcite{netedu2019openapi}
  \item \fullcite{diac2019failover}
\end{itemize}

\comment{not sure if I should add the following or how to motivate it if I do.}
\comment{TODO: these appear in the bibliography at the end, and should be removed; tough they are hard to remove without hard-coding them here.}

Several other co-authored published papers are somewhat indirectly related to the service composition domain. They belong to three general domains: a few others to Natural Language Processing which gave a good case for domain-specific interacting web services, and one paper to routing algorithms, applied to a mobile health-care scenario. Finally, some short papers describe solutions submitted to the DEBS Grand Challenge, which won the competition performance awards in 2017 and 2018\footnote{\emph{ACM Intl. Conf. on Distributed Event-Based Systems} Grand Challenge --
\url{https://debs.org/awards/}}. Just for reference, they are listed below.

\begin{itemize}
  \setlength\itemsep{-1em}
  \item \fullcite{colhon2014quovadis}
  \item \fullcite{bibiri2014statistics}
  \item \fullcite{diac2018relationships}
  \item \fullcite{pascaru2018vehicle}
  \item \fullcite{amariei2017optimized}
  \item \fullcite{rocsca2018predicting}
  \item \fullcite{amariei2018cell}
\end{itemize}

%

\chapter{Service Composition \underline{Models}}
\label{sec:models}
\comment{Describes various models used for service composition. Different 
approaches, semantic extensions and other aspects like QoS studied in previous 
research or proposed in this chapter.}

\iftoggle{fullThesis}{

Service Composition is essentially a modeling problem. Generally, in software 
development, the first step is defining a model according to required 
functionalities. Then, the resulting problem can be analyzed computationally 
(i.e. Chapter \ref{sec:complexity}) and further from a practical perspective 
(i.e. Chapter \ref{sec:applications}). Web Services - by design independent 
components - provide functionality with a narrow scope. In non-trivial 
applications, it is not expected to have a service that is useful to the end-user only by itself. A large set of services from multiple sources or providers must 
be known to allow the possibility of composition. This \emph{workflow} of 
services, named \emph{composition}, is built for the specific goal that needs to be 
reached. The ideal set of services used to reach the goal should be relatively 
small or cheap as running time and/or cost.

Automating this process significantly increases the difficulties of designing or 
choosing the model. In addition to the elements used for describing services and 
initial and goal states; the \emph{parameter matching} model is required for 
automation. The validity of a structure or workflow of service invocations is 
conditioned precisely by \emph {parameter matching} constraints. The difficulty 
introduced by automation is that parameter matching should model the reasoning 
involved in manual composition as in-depth as possible. On the other hand, the 
resulting service definition language should also be lightweight, to be easily 
adopted by service providers. Sophisticated service definition languages have a 
little chance of becoming popular. The balance between the two is delicate and 
we believe that it made Web Service Composition unpractical lately. Some of 
the models we propose in this thesis intend to solve that.

Our initial work was focused on solving some known composition models; 
prioritizing computational complexity or, empirically, execution times. This 
direction was motivated by access to comparative evaluation and public 
benchmarks of tests describing service repositories, composition request, and 
previous solutions performance. In time, it became clear that the models used in 
these benchmarks had limited expressivity and the semantics was only partially 
introduced. This motivated us for the proposal of two new models described in 
Section \ref{sec:models:hierarchical} and Section 
\ref{sec:models:object-oriented}. Both of them define elements that allow 
service parameters to relate to each other. Moreover, they introduce the 
possibility to distinguish between \emph{instances} of parameters of the same 
\emph{type}. Parameters passing from one service to another evolve through some 
service workflow as a dynamic object, as opposed to the classical model where 
they were seen just as static elements in a set or hierarchy, at most.

The composition model includes other aspects than parameter matching, that are 
not extended by this work. Some of the most important are: the 
quality of service (QoS) metrics, briefly described in Section 
\ref{sec:models:qos}, and service states as presented in Section 
\ref{sec:models:stateful}.

Finally and from a practical perspective, the use-case of solving a single 
composition request given a static repository of services is unrealistic. A 
real-life application should use the resulting composition(s) over a period of 
time - instead of only once. This motivated us to propose and study the dynamic 
version of the problem, described in Section \ref{sec:models:online}. In this 
proposal, services can become unaccessible or break. This can be seen as a 
particular case of the stateful model. Moreover, user requests for compositions 
are maintained dynamically as well: users can \emph{unsubscribe} their 
composition requests or add new requests.
}

\section{Parameter Matching by Name}
\label{sec:models:names}
\comment{Most simple problem of WSC, that was used at the first, 2005 Challenge. 
It is the starting point with little expressibility but more lightweight.}

\iftoggle{fullThesis}{

\subsection{History and Motivation}
The earliest and simplest parameter matching model was based only on 
names. By names, we refer to the \emph{strings} used for declaring parameters in 
service definitions, for example, in the WSDL language, the \emph{name} 
attribute values of \emph{part} XML nodes.
A clear description of this model, and our initial starting point resulting in 
\cite{diac2017engineering}, is presented in \cite{zou2014dynamic}. However, the 
model is much older, and it was used in the first service composition challenge 
in 2005 \cite{blake2005eee}. In this model, one output of a service can be used 
as input of another service if their names coincide.

In model described in the Section \ref{sec:models:names}, the assumption is that parameter names 
themselves represent some concepts or some atomic and independent pieces of 
information that can be completely learned from the output of a called service. 
The goal or the required functionality is also defined using parameters having 
names of the same meaning. This is the simplest model and the first that made 
automation of the composition process possible. The model can potentially be 
used in applications where service developers name the parameters following a 
specific standard, and when names are enough to reference all that is relevant 
for some concept, such as sub-domains with precise terminology. 

\subsection{Formal Definition}
The following definitions provide a more formal description of the name matching 
model, with a few other interpretations, notations, and conventions used in the 
corresponding complexity Section \ref{sec:complexity:algorithms}.

\begin{definition}{Parameter.} A parameter is, in the simplest form, an element 
of the set of all parameters, that we write as $\mathbb{P}$.\end{definition}
The definition above is enough for a formal definition but, for convenience, parameters are represented 
by strings or names. Also, $\mathbb{P}$ itself is usually not explicitly 
defined in the input of an instance but is considered to include all parameter 
names that appear in services or composition requests.

\begin{definition}{Web Service.}\label{webservicedef} A Web Service is defined 
by a pair $(I, O)$ of  input and output parameter sets ($I,O\subset \mathbb{P}$).\end{definition}
If \emph{ws} is a web service, then we write its input as \emph{ws.I} and 
output as \emph{ws.O}.
The set of all services, also named the repository, is $\mathbb{R}$. In simple 
models, and clearly in this Section \ref{sec:models:names}, \emph{I} and \emph{O} are disjoint: 
$I \cap O = \emptyset$.

\begin{definition}{Request.} A user request is defined by two sets of 
parameters: the initially known parameters and the required 
parameters.\end{definition}
It is convenient to use the structure of web services requests as well. 
Therefore, for a request \emph{req} we write the initially known parameters as 
$req.I$ and the required or goal parameters as $req.O$. In this manner, we can 
also say that the request is a web service that is needed, required by some 
user, but that is (probably) not in the repository.

\begin{definition}{Parameter Matching.} If $P$ is a set of parameters and $ws$ a 
web service, we say that the set $P$ matches the web service $ws$ if $ws.I \subseteq P$. We also define $P \oplus ws = P \cup ws.O$ as the addition of $ws.O$ to $P$ 
under the constraint of $P$ matching $ws$. \end{definition}
Parameter matching expresses when and how some known parameters can 
be used to call a new service, while the $\oplus$ operation defines the 
\emph{learning} of the output of a matched or called service.

\begin{definition}{Chained Matching.}\label{chaindef} If \emph{P} is a set of 
parameters and \emph{$\langle ws_1, ws_2, ... ws_k\rangle$} is an ordered list 
of web services, we say that \emph{$P \oplus ws_1 \oplus ws_2 \oplus ... \oplus 
ws_k$} is a chain of matching services over the set \emph{P} if: 
\newcommand\bigforall{\mbox{\Large $\mathsurround5pt\forall$}}
$$ws_1.I \subseteq P, and$$
$$ws_i.I \subseteq \Bigg( P \cup \Big(\bigcup_{j=1}^{i-1} ws_{j}.O\Big) \Bigg), 
\bigforall i = \overline{2..k}$$\end{definition}
In words, a chain of matching services is a list of services for which the input 
of each service is included in the reunion of the outputs of all previous 
services and the initial set of parameters. The result of a chain of 
matching services, \emph{$P \oplus ws_1 \oplus ... \oplus ws_k$} is $P \cup 
ws_1.O \cup ... \cup ws_k.O$, the parameters learned or known after a sequence 
of service calls. Chained matching can be extended to more complex workflows of 
services that are not limited to sequences, to conceptually allow the concurrent 
execution of services, i.e. the order between services is a \emph{partial 
order}.

\begin{definition}{Service Composition Problem.} Given a repository of web services 
$\mathbb{R}$ and a user request $req = (req.I, req.O)$, find an ordered list of 
web services $\langle req.I, ws_1, ws_2, ... ws_k, (req.O, \emptyset) \rangle$ 
with \emph{$ws_i \in R $}, $\forall$ \emph{i} = \emph{$\overline{1...k}$}; 
matching the parameter set $req.I$.\end{definition}

The pair $(req.O, \emptyset)$ is a fictive service with no output and the user 
required parameters as input. Such a notation allows a uniform way of writing 
that $req.O \in \big( req.I \cup ws_1.O \cup \dots \cup ws_k.O \big)$. Moreover, 
this can be extended to the shorter form: $\langle (\emptyset, req.I), ws_1, 
\dots, ws_k, (req.O, \emptyset) \rangle$ expressing the problem requirement as a 
single sequence of matching services without having to specify the parameter set 
separately.

Conveniently, Definition \ref{chaindef} becomes simpler if $P$ is included as a 
starting service. More precisely, we can say that a sequence of services 
$\langle ws_1, ws_2, ... ws_k \rangle$ matches iff: $ws_1.I = \emptyset \land ws_i.I \subseteq 
\big(\bigcup_{j=1}^{i-1} ws_{j}.O \big), \forall i = \overline{2...k}$.

\subsection{Name Matching Example} \label{sec:models:names:example}

We use the example presented in \cite{diac2017engineering}, with services 
related to Natural Language Processing domain. Suppose that as part of a text 
processing phase we need to replace the predicate of a sentence with a synonym 
of the verb, also the position of the predicate. The replacement should be in the 
correct conjugation. However, the service that provides synonyms takes as input 
a word sense and not a word, and there is also a word sense disambiguation 
service. More precisely, considering the web services:
\newline

$\emph{getWordSense}^{\textsc{\relsize{1}{\textsl{ .I = \big\{textualWord, 
sentence\big\}}}}}_{\textsc{\relsize{1}{\textsl{ .O = \big\{wordSense\big\}}}}} 
\hspace{1cm} \emph{getSynonim}^{\textsc{\relsize{1}{\textsl{ .I = 
\big\{wordSense\big\}}}}}_{\textsc{\relsize{1}{\textsl{ .O = 
\big\{word\big\}}}}}
\vspace{0.3cm}
\\ \emph{getPredicate}^{\textsc{\relsize{1}{\textsl{ .I = 
\big\{sentence\big\}}}}}_{\textsc{\relsize{1}{\textsl{ .O = 
\big\{textualWord\big\}}}}}$
\hspace{1cm} $\emph{getVerbProp}^{\textsc{\relsize{1}{\textsl{ .I = 
\big\{textualWord\big\}}}}}_{\textsc{\relsize{2}{\textsl{ .O = 
\Big\{$^{\textsc{\textsl{{person, tense,}}}}_{\textsc{\textsl{number, 
mood}}}$\Big\}}}}}$ \\
\vspace{0.3cm}
$\emph{conjugateVerb}^{\textsc{\relsize{2}{\textsl{ .I = 
\Big\{$^{\textsc{\textsl{word, person, 
tense,}}}_{\textsc{\textsl{\hspace{0.25cm}number, 
mood}}}$\Big\}}}}}_{\textsc{\relsize{1}{\textsl{ .O = 
\big\{conjugatedVerb\big\}}}}}$

\vspace{0.2cm}
The user request has as initial parameter \emph{sentence} and the required 
parameter is \newline \emph{conjugatedVerb}, i.e. the synonym of the verb in the 
correct form, used as a replacement in the sentence. We need to call the 
services in such an order that all input parameters are known at the time of a 
service call, as shown in Figure \ref{fig:nameExample}.

\tikzset{
    punkt/.style={
           rectangle,
           rounded corners,
           draw=black,
           text width=5.5em,
           minimum height=2em,
           text centered},
    pil/.style={
           ->, thick, shorten <=5pt, shorten >=5pt,}
}

\begin{figure}[h]
\caption[Composition example for \textbf{Name Matching} model]{Composition example: arrows show parameter matching by name.}
\label{fig:nameExample}
\begin{center}
\begin{tikzpicture}[node distance=1cm and 0.7cm]
\hspace{-1cm} \node[punkt, text width=7em] (rin) {{\smaller 
.I=$\emptyset$}\\\textbf{requestInput}\\{\smaller .O=\{sentence\}}};
 \node[punkt, text width=7em, below=0.8cm of rin] (getp) {{\smaller 
.I=\{sentence\}}\\\textbf{getPredicate}\\{\smaller .O=\{textualWord\}}};
 \path (rin) edge [pil, bend right =0] node [] 
{\textcolor{gray}{\scriptsize{sentence}}} (getp); 
 \node[below=1.5cm of getp] (dummy) {};
 \node[punkt, left=2.5cm of dummy, text width=7em] (getl) {{\smaller 
.I=\{textualWord, sentence\}}\\\textbf{getWordSense}\\{\smaller 
.O=\{wordSense\}}};
 \path (rin) edge [pil, bend right =25, sloped] node [above] 
{\textcolor{gray}{\scriptsize{sentence}}} (getl);
 \node[punkt, text width=6em, below=0.7 cm of getl] (gets) {{\smaller 
.I=\{wordSense\}}\\
\textbf{getSynonim}\\{\smaller .O=\{word\}}};
 \node[punkt, right=1cm of dummy, text width=8em] (getpp) {{\smaller 
.I=\{textualWord\}}\\\textbf{getVerbProp}\\{\smaller .O=\{person, tense, number, 
mood\}}};
 \path (getp) edge [pil, bend right =0] node [right=0.15cm, midway] 
{\textcolor{gray}{\scriptsize{textualWord}}} (getl);
 \path (getl) edge [pil, bend right =0] node [left] 
{\textcolor{gray}{\hspace{1cm}\scriptsize{wordSense}}} (gets);
 \node[punkt, below left=1.5cm and -2.7cm of getpp, text width=9em] (conj) 
{{\smaller .I=\{word, person, tense, number, 
mood\}}\\\textbf{conjugateVerb}\\{\smaller .O=\{conjugatedVerb\}}};
 \path (getp) edge [pil, bend right =0] node[right=0.1cm] 
{\textcolor{gray}{\scriptsize{textualWord}}} (getpp);
 \node[punkt, below right=3cm and 0.8cm of gets.east, text width=8em] (rout) 
{{\smaller .I=\{conjugatedVerb\}}\\\textbf{requestOutput}\\{\smaller 
.O=$\emptyset$}};
 \path (getpp) edge [pil, bend right =0] node[text width=5em, midway] 
{\hspace{-1.2cm}\textcolor{gray}{\scriptsize person, tense,\\ 
\hspace{-1.2cm}number, mood}} (conj);
 \path (gets) edge [pil, bend right =0] node[midway, below] {\hspace{0.1cm} 
\textcolor{gray}{\scriptsize{word}}} (conj);
 \path (conj) edge [pil, bend right =0] node[near start] 
{\textcolor{gray}{\hspace{-2.7cm}\vspace{1.5cm}\scriptsize{conjugatedVerb}}} 
(rout);
\end{tikzpicture}
\end{center}
\end{figure}
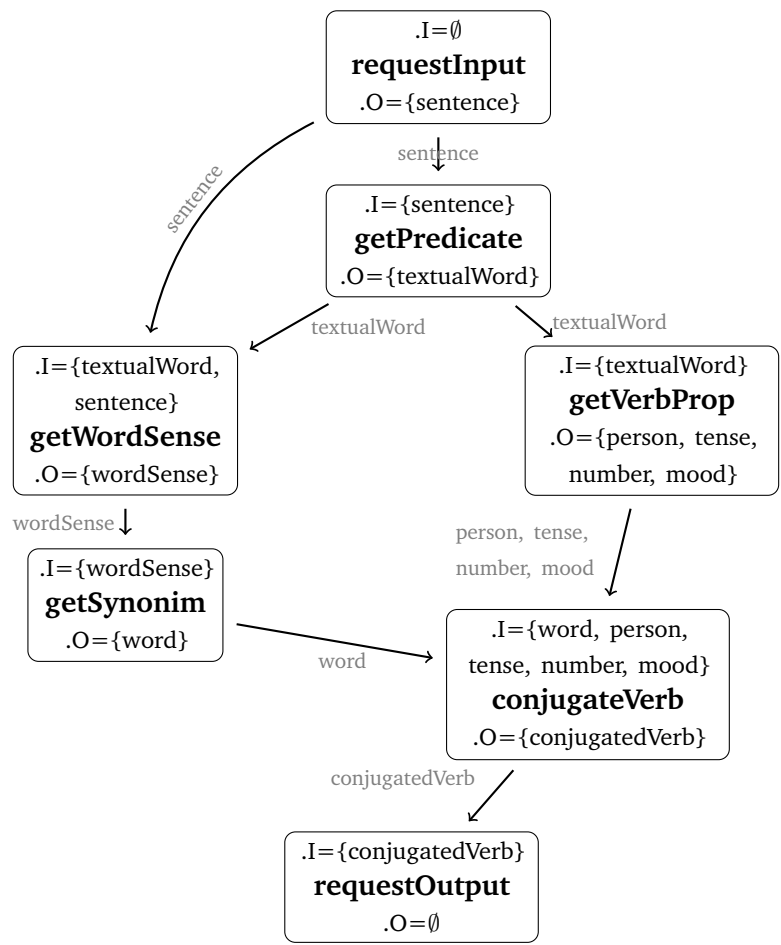

Obviously, a sequence of services resolving the example presented above can be 
$\langle$\emph{requestInput, getPredicate, getWordSense, getSynonim, 
getVerbProp, conjugateVerb, requestOutput}$\rangle$. From the graph 
representation above, it is easy to see that this sequence is not unique, as 
that the order between services is not total. The arrows displaying parameters 
passing trough the composition are represent dependencies between services.
}

\section{Hierarchical Parameter Types}
\label{sec:models:hierarchical}
\comment{Adding Semantics to composition was as first step with the simpler 
model of hierarchical parameter types. This was used in the 2008 challenge 
together with the first distinction between concepts and instances, explained 
here.}

\subsection{Introducing Semantics}

\iftoggle{fullThesis}{

It is clear that name matching model from Section \ref{sec:models:names}, used for the first 
version of automatic composition, enables only little of what is involved in 
manual matching or passing parameters between services. Needless to say, 
applications using web services can have very different use-cases from this 
point of view, because they can even have a different understanding of what a 
web service actually is. However, a vast majority of cases consider service 
parameters as some unitary information that is required at input and learned at 
output. But frequently the information contained within a parameter can be used 
in more than one way. More precisely, that information can play different roles 
in different cases. Generally, services developers write interface/definition 
of services in some particular contexts and do not know about each other. 
Parameter names usually express very local interpretations. For example, 
if a service computes the square of a number it might look like this:
$$\emph{squareService}^{\textsc{\relsize{1}{\textsl{ .I = \big\{number 
\big\}}}}}_{\textsc{\relsize{1}{\textsl{ .O = \big\{square \big\}}}}} $$

Obviously, the output \emph{square} is also a number. In this case, the name 
matching model does not allow the use of the service on its own result, for 
example, to compute $number^{4}$, i.e. $squareService(squareService(number))$. 
Parameters \emph{types} could be a solution in this particular case, as they are 
included in any service definition language, but add little expressivity in 
general. Also, \emph{typing} is used in conjunction with the parameter matching 
(i.e. parameters types should also match).

A more expressive parameter model, uses a \emph{hierarchy of concepts}, or a 
\emph{taxonomy}, to define service parameters. This was the first step towards 
\emph{semantic} service composition. Since this model was introduced, most 
research in service composition was related to the field of \emph{Semantic Web}. 
For now, we avoid using Semantic Web (methods or terminology), that will be 
covered more in Chapter \ref{sec:applications}. Instead, in this chapter, we will continue to present the model mostly theoretically.

\subsection{Formal Definition}
We will use the model of hierarchical parameters types as defined in the 2008 
edition \cite{bansal2008wsc} of the composition challenge. We first used this to 
develop the algorithm presented in Section \ref{sec:complexity:upper}, published in 
paper \cite{tucar2018semantic}. This model is not the minimal addition to the 
previous, as it includes \emph{concepts} organized in a hierarchy as well as 
\emph{instances}.

\begin{definition}{Taxonomy.} The taxonomy is composed of a set of concepts 
$\mathbb{C}$, a set of instances $\mathbb{I}$, the \emph{subTypeOf} relation 
between concepts: \emph{subTypeOf} $\subseteq \mathbb{C} \times \mathbb{C}$, and 
a function \emph{parent} that for any instance returns the (most specific) 
concept that the instance belongs to, \emph{parent} : $\mathbb{I} \to 
\mathbb{C}$. \end{definition}

The hierarchical model does not necessarily require a distinction between instances and concepts. It would be possible to have just concepts, 
but we adopted the exact model, which allows us to run our algorithm on the 
competition tests. Instances are not actual values of the concepts that contain them. Instances are used to model parameters in service definitions, they are more specific 
than concepts, but less specific than concrete values.

\emph{subTypeOf} is transitive, and over the set of all distinct concepts, forms 
a tree-like structure. Usually, in practice, the tree has a single root, named 
\emph{thing} or \emph{entity}, but this limitation has no relevance for the 
theoretical model. Again more important in practice, the set of all concepts and 
instances are explicitly present in the problem input; and \emph{subTypeOf} and 
\emph{parent} are also defined globally because service developers have to 
adhere to these specifications.

For two concepts \emph{$c_{1}$} and \emph{$c_{2}$} we can write \emph{$c_{1}$} 
is a subtype of \emph{$c_{2}$} as: \emph{subTypeOf($c_{1}$, $c_{2}$)} or $c_{1}$ 
\emph{subTypeOf} $c_{2}$; and if instance \emph{i} is of type concept \emph{c} 
then: \emph{parent(i) = c}.

\begin{definition}{Parameter.} A parameter is an element from the set of 
instances $\mathbb{I}$.\end{definition}

\begin{definition}{Web Service.} A Web Service is, as in Definition \ref{webservicedef}, a 
pair $(I, O)$ of  input and output parameter sets.\end{definition} Service 
definitions should not define parameters that are not in the set of known 
instances of the taxonomy.

\begin{definition}{\emph{subsumes} $\subseteq \Big( \mathbb{I} \times \mathbb{I} 
\Big)$} is a helper relation (or notation) defined over pairs of instances of 
the \emph{taxonomy}.
$$ subsumes(i_1, i_2) \iff i_1 = i_2 \emph{ } or \emph{ } \emph{subTypeOf} 
\big(parent(i_1), parent(i_2)\big)$$
\end{definition}

Remember that \emph{subTypeOf} is transitive, therefore an instance subsumes all 
instances that are of concepts more general than the instance's concept. If an 
instance subsumes another then it can replace it or match it in parameter 
exchange. The term \emph{subsumes} is used to express that the more specific instance can replace, include, or encompass the more generic one. In composition, this is applied for parameter matching.

\begin{definition}{\emph{subsumesSet} $\subseteq \Big( \mathcal{P}(\mathbb{I}) 
\times \mathcal{P}(\mathbb{I}) \Big) $} is another helper relation (or notation) 
defined over pairs of sets of instances of the taxonomy. $\mathcal{P}(\mathbb{I})$ is the set of all subsets of $\mathbb{I}$, or its powerset.
$$ subsumesSet(P_1, P_2) \iff \forall \emph{ } i_2 \in P_2, \exists \emph{ } i_1 
\in P_1 \emph{ such that } subsumes(i_1, i_2) $$
\end{definition}

In words, if \emph{$P_1$} \emph{subsumesSet} \emph{$P_2$} then all elements in 
\emph{$P_2$} can be represented by at least some element in \emph{$P_1$}, thus 
the set \emph{$P_2$} is covered by, or included in, encompassed by \emph{$P_1$}.

\begin{definition}{Parameter Matching.} If $P$ is a set of parameters (instances 
of the taxonomy) and $ws$ a web service, we say that the set $P$ matches the web 
service $ws$ if and only if \emph{subsumesSet(P, ws.O)}.
\end{definition}

We also define $P \oplus ws$ as the addition or the learning of \emph{ws}'s 
output as:
$$ P \cup \Big\{\emph{ } i \in \mathbb{I} \mid \exists \emph{ } 
i_o \in ws.O \emph{ such that } \emph{ subsumes } (i_o, i)\Big\} $$

\subsection{Hierarchical Model Example}

Consider the following example motivating the hierarchy of concepts model. Again 
in the domain of Natural Language Processing, we describe the following scenario. 
Suppose that some application needs to replace a phrase's main verb with a 
synonym of that verb if it exists. The ontology contains the concepts in Figure 
\ref{ontologyfig}, among others; with services defining their parameters as 
instances of concepts within this ontology. For simplicity, instances are not 
shown in the drawing and in the example services define their parameters using, 
by generalization, concept names.

\tikzset{>=latex}
\vspace {-0.4cm}
\begin{figure}[h]
\caption[Example of a \textbf{hierarchy} of NLP concepts]{NLP concepts organized hierarchically. Parent nodes are generalizations  concepts, substitutable by child subconcepts.}
\vspace {0.2cm}
\label{ontologyfig}
\vspace{-0.5cm}
\begin{center}
\begin{tikzpicture}[<-, every tree node/.style={draw,rectangle,rounded corners, 
top color=white,bottom color=gray!20, minimum height = 0.45cm},sibling 
distance=11pt, level distance=27pt]
\tikzset{edge from parent/.style={draw, edge from parent path= 
{(\tikzparentnode) -- (\tikzchildnode)}}}
\Tree [.string [.word [.{part of speech} verb noun adj. {...} ] synonym {...} ] 
[.phrase {...} ] [ .{string token} [ .substr ] substitute ] ]
\end{tikzpicture}
\end{center}
\end{figure}
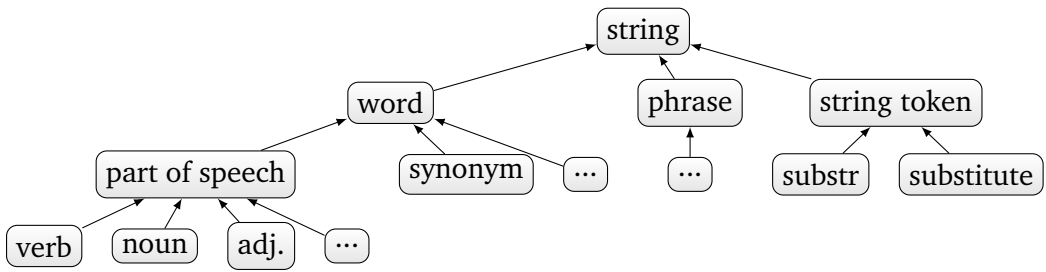

Also, suppose the repository contains the following three services:

\vspace{-0.3cm}
$\\ extractMainVerb\hspace{4pt}_{\textstyle output=\{verb\}}^{\textstyle 
input=\{phrase\}} \hspace{0.9cm} getSynonym\hspace{4pt}_{\textstyle 
output=\{synonym\}}^{\textstyle input=\{word\}} $
\vspace{12pt}
$\\ { } \hspace{2.6cm} stringReplace\hspace{4pt}_{\textstyle 
output=\{string\}}^{\textstyle input=\{string, substr, substitute\}} $
\vspace{0.5cm}

Starting with the initial \emph{phrase}  a call is made to 
\emph{extractMainVerb} service that returns the \emph{verb} in the sentence. 
Since a \emph{verb} is a \emph{part of speech} that is, more generally, a 
\emph{word}, this enables a \emph{getSynonym} service call. The final service 
call is correct if the parameters are used in the right order: 
\emph{stringReplace(phrase, verb, synonim)} that would result in the desired 
modified sentence. Note that this order is not assured by the restrictions of the 
model, as the definition of \emph{stringReplace} is designed in a more generic 
context, i.e. string replacement in general. Finally, a solution can be the 
composition: \emph{extractMainVerb}, \emph{getSynonym}, \emph{stringReplace} in 
this order. But for the reason that the \emph{verb} is a substring of \emph{phrase} and 
that the \emph{synonym} is a substitute of a word (\emph{verb}), it cannot be 
expressed in the current model.

In conclusion, the hierarchical model is a step further required to introduce 
semantics to the automatic composition of services, but by itself is still limited. 
Therefore, we proposed two new models for composition. The advantage of new 
models is that they solve a more fundamental problem of automatic composition - 
that is expressivity, and the disadvantage is that the proposed algorithms and 
solutions cannot be compared with previously implemented solutions, as the 
previous algorithms do not work on the resulting new benchmarks.
}

\section{Relational Parameters Model}
\label{sec:models:relational}
\comment{ \textcolor{red}{[contribution]}  The relational model is one of the main 
contributions of the thesis. First proposed in the paper published at the KES 
2019 conference, it adds two new elements to the composition model: binary 
relations on the parameters and \emph{instances} as composition elements. This 
sub-section does not present the algorithm which is included in the next 
chapter.}

\iftoggle{fullThesis}{

\subsection{A New and Contextual Model}

Following the conclusion from the example in the previous model defined in Section
\ref{sec:models:hierarchical}, it is clear that insufficient semantic 
information is modeled by the simple \emph{subTypeOf} relation between the 
concepts representing parameters. There are many other important types of 
relations that a human considers when manually building a workflow of services. 
For example, parameters passing between services are frequently based on the 
following kind of relations: \emph{partOf}, \emph{sameAs}, \emph{relatedTo}, 
\emph{about}, and many others. Even if not all reasoning involved in manual 
composition can be expressed by binary relations, we believe that the majority 
could. A classic example in service composition papers reveals exactly such a 
problem: \emph{latitude} and \emph{longitude} can define together as parameters an exact position on the map, but  there is no way of expressing that it is 
essential that both refer the same position, or that they are the output of the same 
service. If multiple services use these terms, it can be very easy to 
automatically mix them in a wrong way, by using, for example, the \emph{latitude} 
from the output of a service and the \emph{longitude} from another. The result 
is an irrelevant point on the map. The meanings of frequently used relations 
cannot be expressed by only one relation, like \emph{subsumption} or 
\emph{subtyping}. As in general ontologies in the Semantic Web, relations 
between concepts should be freely expressed and this is what we propose for 
Automatic Service Composition as well.

Therefore, we introduce a new model to express different types of relations. 
Moreover, by using these relations, Web Services themselves can define restrictions on the input/output parameters. Relations on output parameters are added if the service is added to the composition. This leads to a different, and more natural 
distinction between concepts and instances, or \emph{objects}. In this model, we 
refer to instances or objects as the elements passed or matched between services 
in the context of a workflow under construction, they evolve during the 
composition and are associated with some provenance information (that can be used to track their origin). Structurally, 
objects consist of a type or a concept, and the relations they have to other 
objects. The relations are dynamic, allowing updates. Another benefit, that is 
also very natural in manual composition, is that we can have multiple instances 
of the same concept or type, differentiated by their relations to other objects 
or their provenance. It is trivial to imagine an example of a workflow that uses 
two instances or objects of the same type. This was impossible to express 
in previous models, restricting these cases from automation. For this model, we 
will first present an example.

\subsection{Case Study: Foreign University Visit}

We present a nontrivial example motivating the proposed model. The following paragraphs partly reproduce a text from our paper \cite{diac2019relational}.  
\comment{the next paragraph is taken from KES 2019 paper, with some minor 
notation changes; it is useless to try to change it}

Assume a researcher is trying to schedule a meeting at some collaborating 
university. We are provided with the following information: the person's name, 
the name of the university where the researcher works, and the name of the 
university to visit. We consider the person and the two universities as 
instances of two concepts: \textbf{Person} and \textbf{University}. As we can 
already see, there is a need to distinguish between the two different instances 
of \textbf{University} and we are going to model that by the use of two 
different types of relations between our \textbf{Person} and each 
\mbox{\textbf{University}}: \textbf{isEmployeeOf} and \textbf{hasDestination}. 
Finally, we use a third relation: \textbf{isLocatedIn} and two inference rules 
that can help to expand relations. For example, if \textbf{X} is an employee of 
\textbf{Y} and \textbf{Y} is located in \textbf{Z} then \textbf{X} is located in 
\textbf{Z} (this may not be true in any situation, but it is a reasonable 
assumption that works for our example). In the composition model and in the 
further presented algorithm we can handle inference rules in the same manner as 
web services that "return" only new \emph{relations} defined on already known 
"parameters", for simplicity. If the services cost, execution time, throughput 
or other Quality of Service is considered, then the rules, modeled as services have zero cost or instant runtime, unlimited throughput.

More precisely, we have the following web services and inference rules (with 
names ending in $Rule$). A service definition contains the required 
relations between input parameters and relations generated between output and 
input; or output and output parameters.

\vspace{-0.5cm}

$\\ \textbf{\emph{getUniversityLocation}}_{\textstyle 
.O=\Bigg\{\begin{tabular}{c} $city : City,$ \\ $isLocatedIn(univ, city)$ 
\end{tabular}\Bigg\}}^{\textstyle .I=\{univ : University\}}$

$\textbf{\emph{getAirplaneTicket}}_{\textstyle .O=\{airplaneTicket : 
Ticket\}}^{\textstyle .I=\Bigg\{\begin{tabular}{c}$pers : Person; \hspace{1cm} 
souce, dest : City,$ \\ $isLocatedIn(pers, source), hasDestination(person, 
dest)$\end{tabular}\Bigg\}} $\

$\\ \textbf{\emph{locatedAtWorkRule}}_{\textstyle .O=\{isLocatedIn(X, 
Z)\}}^{\textstyle .I=\Bigg\{\begin{tabular}{c} $X, \hspace{1cm} Y, \hspace{1cm} 
Z$ \\ $isEmployeeOf(X, Y), isLocatedIn(Y, Z)$ \end{tabular}\Bigg\}} $\

$\textbf{\emph{destinationGenRule}}_{\textstyle .O=\{hasDestination(X, 
Z)\}}^{\textstyle .I=\Bigg\{\begin{tabular}{c} $X, \hspace{1cm} Y, \hspace{1cm} 
Z$ \\ $hasDestination(X, Y), isLocatedIn(Y, Z)$ \end{tabular}\Bigg\}} $

The solution for a composition is a list of services that can be called in the 
order from the list, and for which after all calls, the information required by 
the user is known. Again, we can also specify the user request using the 
structure of a service. In the composition, inference rules can and should be 
applied if possible, and we will specify them for clarity. We also specify how 
to pass the output of a service to the input of another. This was trivial on 
previous models, where if a parameter matched to more than one previous output, 
it did not matter to which it was paired. Here, as multiple instances of the 
same concept can be known, we need to distinguish between them, based on their 
\emph{relations} with other objects or some identifier based for example on 
provenance information.

\tikzset{
	>=stealth',
	box/.style={
		rectangle,
		rounded corners,
		draw=black, very thick,
		text width=20em,
		minimum height=5.5em,
		text centered},
	edgeStyle/.style={
		->,
		thick}
}

\vspace{-0.5cm}
    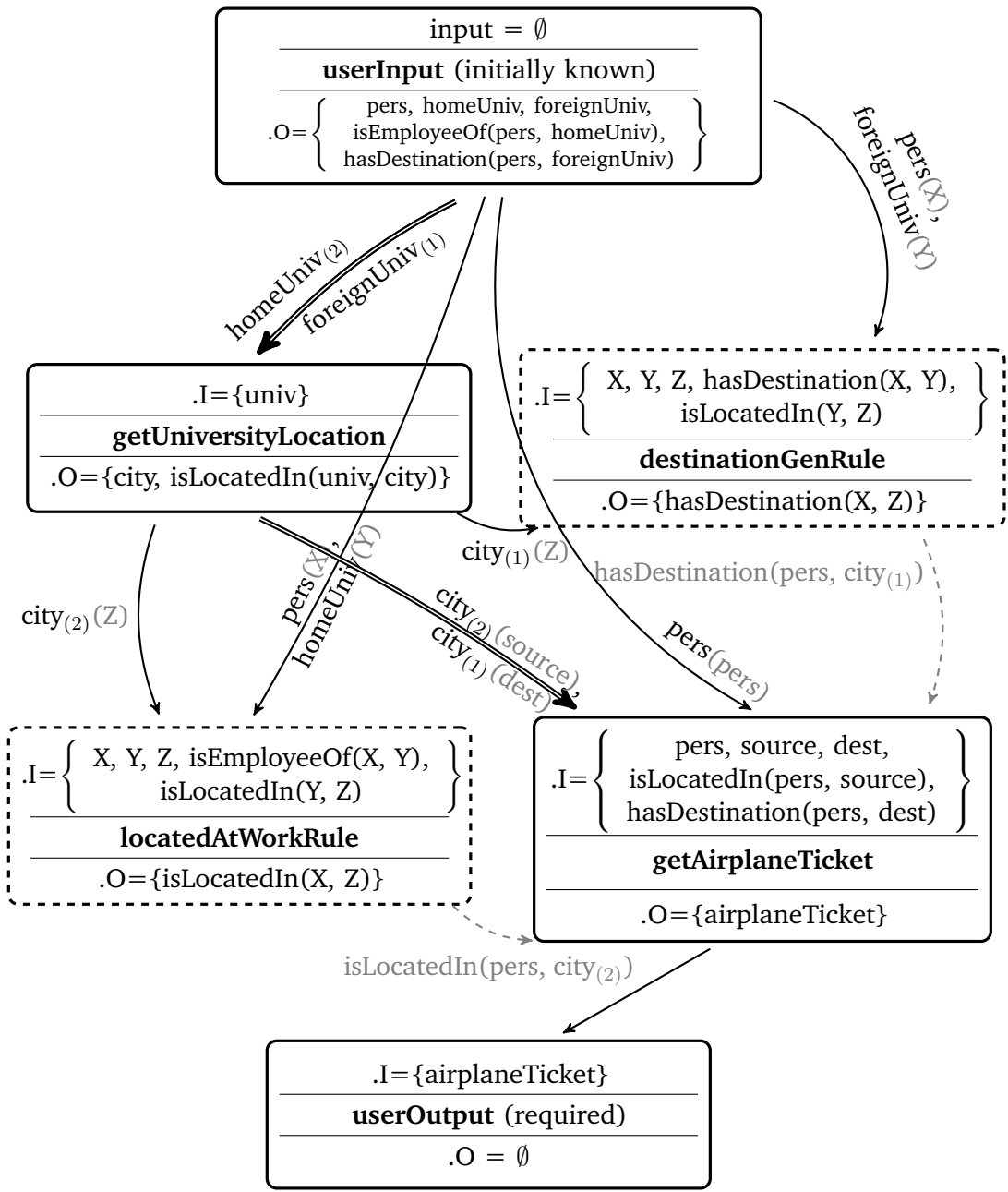
\begin{figure*}[h!]
    	\centering
	\caption[Composition example for \textbf{Relational} model]{The example motivates the need for relations and rules in the 
composition model. Service parameter types are not shown for simplicity.}
    \label{fig:largeExample}
		\vspace{1em}
		{
			\begin{tikzpicture}	[x=0.45pt,y=0.45pt,yscale=1,xscale=1]
			
			\node[box, text width=19.5em] (input) {
				input = $\emptyset$\\
				\HRule[5pt]
				\textbf{userInput} (initially known)\\
				\HRule[5pt]
				\vspace{2pt}
				\footnotesize{.O=$\left \{\begin{tabular}{c}pers, homeUniv, 
foreignUniv, \\ isEmployeeOf(pers, homeUniv), \\ hasDestination(pers, 
foreignUniv)\end{tabular}\right \}$}\\
			};
			
			\node[below=3.5cm of input] (dummy) {};
			
			\node[box, dashed, text width=16.2em, inner sep=5pt,below left =4cm 
and 0.1 cm of dummy] (locAtWorkNode) {
				.I=$\left \{\begin{tabular}{c}X, Y, Z, isEmployeeOf(X, Y), \\ 
isLocatedIn(Y, Z)\end{tabular}\right \}$ \\
				\vspace{9pt}
				\HRule[5pt]
				\textbf{locatedAtWorkRule}\\
				\HRule[5pt]
				.O=\{isLocatedIn(X, Z)\}\\
			};
			\draw (input.south) edge[edgeStyle, bend left=6, shorten <= 5pt, 
shorten >= 5pt] node[near end, 
sloped]{\begin{tabular}{c}pers\textcolor{gray}{(X)}, 
\\homeUniv\textcolor{gray}{(Y)}\end{tabular}} (locAtWorkNode.85);
			
			\node[box, text width=15.5em, inner sep=5pt,left =0.1 cm of dummy] 
(getUnivNode) {
				.I=\{univ\}\\
				\HRule[5pt]
				\textbf{getUniversityLocation}\\
				\HRule[5pt]
				.O=\{city, isLocatedIn(univ, city)\}\\
			};
			
			\node[box,  dashed, text width=17em, inner sep=5pt,right=0.3cm of 
dummy] (destGenNode) {
				.I=$\left \{\begin{tabular}{c}X, Y, Z, hasDestination(X, Y), \\ 
isLocatedIn(Y, Z)\end{tabular}\right \}$\\
				\vspace{9pt}
				\HRule[5pt]
				\textbf{destinationGenRule}\\
				\HRule[5pt]
				.O=\{hasDestination(X, Z)\}\\
			};
			\node[box, text width=16em, inner sep=5pt,below=2.7cm of 
destGenNode] (getAirplaneNode) {
				.I=$\left \{ \begin{tabular}{c}pers, source, dest, \\ 
isLocatedIn(pers, source), \\ hasDestination(pers, dest)\end{tabular} \right \} 
$\\
				\hrulefill
				\\ \textbf{getAirplaneTicket} \\
				\vspace{-5pt}
				\hrulefill
				\\.O=\{airplaneTicket\}\\
			};
			\node[box, text width=15.5em, inner sep=5pt,below =8.5cm of dummy] 
(output) {
				.I=\{airplaneTicket\}\\
				\HRule[5pt]
				\textbf{userOutput} (required)\\
				\HRule[5pt]
				.O = $\emptyset$\\
			};
			
			\draw (input.south) edge[edgeStyle, double, bend right=10,shorten <= 
15pt, shorten >= 5pt] node[sloped, below] {foreignUniv$_{(1)}$} 
(getUnivNode.north) ;
			\draw (input.south) edge[edgeStyle, double, bend right=10,shorten <= 
15pt, shorten >= 5pt] node[sloped, above, near end] {homeUniv$_{(2)}$} 
(getUnivNode.north) ;
			\draw (input.east) edge[edgeStyle, bend left=45,shorten <= 5pt, 
shorten >= 5pt] node[sloped, above] {\begin{tabular}{c} 
pers\textcolor{gray}{(X)},\\ foreignUniv\textcolor{gray}{(Y)}\end{tabular}} 
(destGenNode.40);
			\draw (getUnivNode.340) edge[edgeStyle, bend right=20] node[below, 
near end]{city$_{(1)}$\textcolor{gray}{(Z)}}  (destGenNode.202);
			\draw (getUnivNode.220) edge[edgeStyle,shorten <= 5pt, shorten >= 
5pt, bend right = 20] node[left] {city$_{(2)}$\textcolor{gray}{(Z)}}  
(locAtWorkNode.130);
			\draw (destGenNode.330) edge[edgeStyle, dashed, gray, bend left = 
20,shorten <= 5pt, shorten >= 5pt] node[left, near start] {hasDestination(pers, 
city$_{(1)}$)} (getAirplaneNode.35);
			\draw (locAtWorkNode.337) edge[dashed, gray, edgeStyle, bend 
right=20] node[below=0.2cm]{isLocatedIn(pers, city$_{(2)}$)} 
(getAirplaneNode.206);
			\draw (input.280) edge[edgeStyle, bend right=35,shorten <= 5pt, 
shorten >= 5pt] node[very near end, above, sloped] {\hspace{10pt} 
pers\textcolor{gray}{(pers)}} (getAirplaneNode.north);
			\draw (getUnivNode.south) edge[edgeStyle,double, shorten <= 5pt, 
shorten >= 5pt, bend left=5] node[near end, 
sloped]{{\begin{tabular}{c}city$_{(2)}$\textcolor{gray}{(source)},\\city$_{(1)}
$\textcolor{gray}{(dest)}\end{tabular}}} (getAirplaneNode.147);
			\draw (getAirplaneNode.250) edge[edgeStyle,shorten <= 5pt, shorten 
>= 5pt] (output.50);
			
			\end{tikzpicture}
		}
	\end{figure*} 

In Figure \ref{fig:largeExample}, solid boxes represent services and the user 
request (one for input - the initially known concepts and their relations, and 
one for the required output, as in the initial model in the Figure \ref{fig:nameExample}). Dashed boxes 
represent inference rules. Edges show information flow: solid edges - parameters 
and the dashed edges - relationships among them. Not all possible relations are 
used. Parameters are matched to rule variables (rule "parameters"), or other 
service parameters, based on the specification in gray in parenthesis. Multiple 
calls to the same service can be handled and are shown with double edges.

One composition that is a valid solution for the above instance would be the 
following list of service invocations, in order:

\vspace{-15px}
\noindent \textbf{getInput}\big($\emptyset$\big) $\Longrightarrow$ pers, 
homeUniv, foreignUniv, isEmployeeOf(pers, homeUniv), \\ hasDestination(pers, 
foreignUniv);

\vspace{-15px}
\noindent \textbf{getUniversityLocation}\big(homeUniv\big) $\Longrightarrow$ 
homeCity, \\ isLocatedIn(homeUniv, homeCity); 

\vspace{-15px}
\noindent \textbf{getUniversityLocation}\big(foreignUniv\big) $\Longrightarrow$ 
foreignCity, \\ isLocatedIn(foreignUniv, foreignCity);

The two cities: \emph{homeCity} and \emph{foreignCity} are differentiated based 
on their relations. We used some intuitive names here, but they are not relevant 
for the composition language (there is no restriction on what names they get if 
they would be automatically created; i.e. any distinct strings would work).

\vspace{-10px}
\noindent \textbf{locatedAtWorkRule}\big(pers, homeUniv, homeCity, 
isEmployeeOf(pers, homeUniv), isLocatedIn(homeUniv, homeCity)\big) 
$\Longrightarrow$ isLocatedIn(pers, homeCity);

\vspace{-15px}
\noindent \textbf{destinationGenRule}\big(pers, foreignUniv, foreignCity, 
hasDestination(pers, \\ foreignUniv), isLocatedIn(foreignUniv, foreignCity)\big) 
$\Longrightarrow$ hasDestination(pers, \\ foreignCity); 

\vspace{-15px}
\noindent \textbf{getAirplaneTicket}\big(pers, homeCity, foreignCity, 
isLocatedIn(pers, homeCity), \\ hasDestination(pers, foreignCity) 
$\Longrightarrow$ airplaneTicket;

We can immediately notice the usefulness of semantic relations as the pairs 
\textit{homeUniv} and \textit{foreignUniv}, as well as \textit{homeCity} and 
\textit{foreignCity} are essentially indistinguishable between themselves 
otherwise: if knowledge would consist only of a simple set of known types as 
before.
	
Without using semantic relations, to get this desired functionality we could 
copy services separating them for each possible instance of input parameters, 
for example, \textbf{getUniversityLocation} could be split into two different 
services: \textbf{getForeignUniversityLocation} and 
\textbf{getHomeUniversityLocation}. We can imagine cases where this workaround 
raises the number of services provided at input by a high amount for each group 
of different input parameters, and in practice, services are created 
independently by third parties ahead of the composition process.
	
For simplicity, in this example, we did not use the hierarchy of concepts. 
However, \emph{subTypeOf} can be used together with the presented additions to 
the model, as defined in the following. Even shorter, the \emph{subTypeOf} 
relation can be modeled as any other relation, therefore the relational model 
naturally includes the hierarchical model, generalizing it.

\subsection{Formal Definition}
The following formal definition of the proposed model is the same as presented in paper 
\cite{diac2019formalism}.

\begin{definition}{Concept}. A concept $c$, identified by a \emph{conceptName} 
is an element of the given set of concepts $\mathbb{C}$. Concepts are arranged 
in a hierarchy, by the use of inheritance. Specialization concepts (or 
sub-types) can replace their more generic version (or super-types). This is the 
same as in the hierarchical model from Section \ref{sec:models:hierarchical}, except for the 
definition of \emph{instances}.\end{definition}

\begin{definition}{\textbf{Object}}. An object, or an instance, is a pair of an 
\emph{id} and a \emph{type}. The set of all objects is written as $\mathbb{O} = 
\{ o = \langle id, type \rangle \}$. The $id$ is a unique identifier generated 
at object creation. The $type$ of objects are concepts: $type \in \mathbb{C}$. 
Intuitively, objects exist in the dynamic context of a set of services that have 
been called. So, they are instances of \emph{Concept}s, for which we know how 
and when they were produced and how they were used in the workflow. They are 
transformed by a series of service calls with their parameters matched by 
\emph{objects}. Clarification: in the example above we used, instead of 
\emph{id} objects, names like \emph{homeCity}. In automatic composition and 
specifically in implementation, the \emph{name} is the identifier, uniquely 
generated, and can include some provenance information: for example, the name of 
the service that created this object and by which output parameter, etc. This is 
not relevant to the theoretical model but helps to understand the meaning of 
objects in model from Section \ref{sec:models:relational}.\end{definition}

\begin{definition}{\textbf{Relation}.}\label{relationdef}
A relation $r$ is a triple consisting of the name as a unique identifier, 
relation properties, and the set of pairs of objects that are in that relation. 
The first two are static and defined at the ontology level. The latter is 
dynamic, as it can be extended through the composition process.
\vspace{-5px}
\begin{equation*}
\begin{split}\mathbb{R} = \big\{\langle \hspace{1px} name, properties, objects 
\hspace{1px} \rangle \hspace{2px} \big| \hspace{2px} properties \subseteq \{ 
transitivity, \hspace{2px} symmetry\} \hspace{3px} \\ and \hspace{3px} objects 
\subseteq \mathbb{O}^{\hspace{1px}2}\big\}
\end{split}
\end{equation*}
\end{definition}
\textbf{Relation properties}. A relation can have none, one or both of the 
properties: \emph{symmetry} and \emph{transitivity}. Symmetric relations 
\emph{r} are not oriented from the first object to the second, nor 
vice-versa, i.e. if $(o_1, o_2) \in r.objects$ then $(o_2, o_1) \in r.objects$ 
as well. Conceptually, it is similar to an edge in an undirected graph (while 
asymmetric relations are similar to labeled arcs in directed graphs). 
Transitivity implies that if $(o_1, o_2) \in r.objects$ and $(o_2, o_3) \in 
r.objects$, then $(o_1, o_3) \in r.objects$. An example of a transitive relation 
is the \emph{subTypeOf} from the hierarchical model presented in Section
\ref{sec:models:hierarchical}.
\begin{definition}\label{sec:models:relational:knowledgedef}{The 
\textbf{knowledge} $\mathbb{K}$}
 is a dynamic structure consisting of objects and relations between these 
objects. Knowledge describes what is known at a stage of the composition 
workflow, i.e. at a time when a set of services have been added to the 
composition. $\mathbb{K} = \langle \hspace{1px} \mathbb{O}, \mathbb{R} 
\hspace{1px}\rangle$. It defines what is known at especially in more complex 
models. Similar "knowledge" existed in previous models as well, but in simpler 
models, it can be omitted without much confusion. The structure of relations in 
knowledge does include \emph{properties}, as they are defined at the ontology level 
for each relation, based on the \emph{name}.
\end{definition}

\begin{definition}{\textbf{Web Services} are tuples $ws = \langle name, I, O, 
relations \rangle$ where: $I, O$ are the set of input and output parameters organized hierarchically as in the previous model, and $relations$ specify preconditions and postconditions (effects) over objects matched to the inputs and outputs. Preconditions are relations required before adding a service to the compositions, and relations specified in postconditions are generated subsequently. This model and the terms are inspired by the planning domain \cite{mcdermott1998pddl}.
Structurally, $relations$ within 
service definitions are pairs consisting of: the $name$ used to refer to an 
existing relation (the relation from $\mathbb{R}$ having the same name), and a 
binary relation over service parameters. Relations between inputs are 
preconditions, and relations between outputs are effects, i.e. they are 
generated after the call. Relations between input and output parameters are 
also effects.}
\begin{equation*}
\begin{split}
ws.relations = \big\{\langle \hspace{1px} name, parameters \hspace{1px} \rangle 
\hspace{2px} \big| \hspace{2px} name \in names \hspace{3px} from \hspace{3px} 
\mathbb{R} \hspace{3px} \\ and \hspace{3px} parameters \subseteq (ws.I \cup 
ws.O)^{2} \hspace{1px} \big\}
\end{split}
\end{equation*}
\end{definition}

\begin{definition}{\textbf{Inference Rules} are tuples of the from:  $rule = 
\langle \hspace{0.5px} name, parameters, $ \\ $preconditions, \mathit{effects} 
\rangle$ where $parameters$ are identifiers with local 
visibility (within the rule), and preconditions and effects are relations 
defined over the $parameters$ (the rule's identifiers). More precisely: 
\begin{equation*} \begin{split}
rule.preconditions, rule.\mathit{effects} \subseteq \Big\langle 
\bigcup\limits_{rel\hspace{1px}\in\hspace{1px}\mathbb{R}} rel.name, 
rule.parameters^{\hspace{1px}2} \Big\rangle
\end{split} \end{equation*}
\emph{preconditions} are separated from \emph{effects} here, because in 
\emph{rules}, unlike in \emph{services} there are no output parameters. Rules 
only produce relations, more exactly just add pairs of objects to existing 
relations. In the case of services, this distinction is based on what kind of 
parameters they take: input, output or between input and output. Another 
difference in \emph{rules} is that their \emph{parameters} do not have types, 
and as convention, we named them with capital letters in the example. \\
The set of all inference rules is written as $\mathbb{I}$. Preconditions must 
hold before applying the rule for the objects matching rule parameters, and 
relations in the effects are generated accordingly. Rules are structurally 
similar to services, but they apply automatically and, theoretically, at no 
cost. We can write properties such as \emph{transitivity} and \emph{symmetry} 
using rules. For example, the following rule expresses that $equals$ relation is 
symmetric and that $greater$ relation is transitive.}
\begin{equation*}
\begin{split} 
\overset{\mathrm{(rule \hspace{5px}name)}}{equals_{symmetric}} = \Big\langle 
\overset{\mathrm{(parameters)}}{\Big\{X, Y\Big\}}, 
\overset{\mathrm{(preconditions)}}{\Big\{\langle equals, \{ (X, Y)\} 
\rangle\Big\}},  \overset{\mathrm{(effects)}}{\Big\{\langle equals, \{ (Y, X)\} 
\rangle\Big\}}  \Big\rangle \\
\vspace{15px}
greater_{transitive} = \Big\langle \Big\{X, Y, Z\Big\}, \Big\{\langle greater, 
\{ (X, Y)\} \rangle, \langle greater, \{ (Y, Z)\} \rangle\Big\}, \\ 
\Big\{\langle greater, \{ (X, Z)\} \rangle\Big\}  \Big\rangle
\end{split}
\end{equation*}
\end{definition}

\begin{definition}{The \textbf{Ontology} $\mathbb{G}$ consists of: 
the concepts, organized hierarchically, relations with objects of these concepts 
and inference rules over the relations.}
$\mathbb{G} = \langle \hspace{3px} \mathbb{C},\mathbb{R}, \mathbb{I} 
\hspace{3px} \rangle$. At the ontological level, relations are static and defined 
only by names and properties. At the \textbf{knowledge} level (Definition \ref{sec:models:relational:knowledgedef}), relations are dynamic in what 
objects they materialize to. For simplicity, we refer to both using $\mathbb{R}$.
\end{definition}

\begin{definition}{\textbf{Parameter Matching}. In the ontology 
$\mathbb{G} = \langle \hspace{3px} \mathbb{C}, \mathbb{R}, \mathbb{I} 
\hspace{3px} \rangle$, a web service $ws$ matches (or is "callable" in) a 
knowledge state $\mathbb{K} = \langle \hspace{1px} \mathbb{O}, \mathbb{R} 
\hspace{1px}\rangle$, iff:}

\begin{equation*}
\begin{split}
\exists \hspace{3px} function \hspace{5px} f : ws.I \rightarrow \mathbb{O} 
\hspace{4px} such \hspace{3px} that: \textrm{(1)}\\
\forall \hspace{3px} i \in ws.I, \big(f(i).type,i.type\big) \in 
\mathit{subtypeOf} \hspace{5px} and \hspace{5px} \textrm{(2)}\\
\forall \hspace{3px} \hspace{3px} j \in ws.I \hspace{5px} and \hspace{5px} \forall \hspace{5px}
r_{ws} \in ws.relations, \hspace{3px} with \hspace{5px} (i,j) \in 
r_{ws}.parameters \hspace{2px} \textrm{(3)}\\
\exists \hspace{3px} r_{obj} \in \mathbb{R} \hspace{5px} with: 
\hspace{5px} r_{obj}.name=r_{ws}.name \hspace{5px} and \hspace{5px} 
\big(f(i),f(j)\big) \in r_{obj}.objects \hspace{2px}\textrm{(4)}
\end{split}
\end{equation*}
\end{definition}

The definition requires the elements above: both the ontological constructs and the current knowledge information; because it is not enough to keep a set of known objects: their relations state is also relevant.

In (1), the matching function \emph{f} associates any input of the service to 
one of the objects in the knowledge, that, as specified in (2) can be of any more 
specialized type than the input type. This association has to respect the 
service's preconditions: (3) any other input parameter \emph{j} that is in any 
precondition-relation $r_{ws}$ with \emph{i} has to be associated with an object 
from the knowledge \emph{f(j)} that is in a relation with the same name 
(identifier) with \emph{f(i)} at the knowledge level.

\begin{definition}{\textbf{Parameter learning}}. For a knowledge state 
$\mathbb{K}$ and a service \emph{ws} matching the knowledge state by the 
function \emph{f : ws.I $\rightarrow \mathbb{O}$}, we define the new knowledge 
state that includes the objects and relations learned from the \emph{ws} match 
over \emph{f} as $\mathbb{K} \oplus_{f} ws$:
\begin{equation*}
\begin{split}
\mathbb{K} \oplus_{f} ws = \Bigg\langle \mathbb{O} \cup \Big\{ c = \langle new 
\textrm{ } id, c.type \rangle \hspace{5px} \Big| \hspace{5px} c \in ws.O \Big\} \hspace{7px} \textrm{(1)},\\
\Big\{ r = \langle name, objects \cup \{ \big(f(i), f(j)\big) \in (ws.I \cup 
ws.O)^2 \hspace{2px} | \exists \hspace{5px} r_{ws} \in ws.relations, \\ 
\hspace{5px} with \hspace{5px} r_{ws}.name = r.name \land (i,j) \in 
r_{ws}.parameters \} \rangle \Big| r \in \mathbb{R} \Big\} \hspace{7px} \textrm{(2)} \\
\bigcup \Big\{ r \in ws.relations \Big| r \notin \mathbb{R} \Big\} \hspace{7px} \textrm{(3)} 
\Bigg\rangle
\end{split}
\end{equation*}
\end{definition}

In (1), the objects created by output parameters are added to the knowledge, in 
(2) new matching object pairs are added to relations that are already existing 
in knowledge and in (3) new relations are added to knowledge. All added 
relations must be defined at the ontology level, including relations added in 
(3), i.e. they are defined but no objects are yet known to be in that relation.

We skip other similar definitions of the relational model that are intuitively 
similar, such as \textbf{chained matching}, and the result $\oplus$ of a chain 
matching.

\begin{definition}{Request}. The \textbf{user request} structure is similar to a 
web service. The \emph{input} specifies what the user initially knows, that are 
objects with types and known relations between them, and the \emph{output} is 
the user's required objects with required relations between them. Like in some 
service, there can be relations between inputs and outputs, specifying 
restrictions on the user required outputs relative to his initially known 
objects.\end{definition}
	
\textbf{Relation based Web Service Composition Problem}. Given an ontology 
defining: \emph{concepts} and their hierarchy, \emph{relations} with 
\emph{properties} and \emph{inference rules},  a repository of services, and a 
user request, all defined over the ontology; find an ordered list of services 
that are validly callable, that solves the user request, starting with the 
initial information. For each service in the composition, the source of its 
input parameter must be specified: resolved or bound to an output of a specified 
previously called service, or user query. From a high-level view the definition 
is the same as in previous models, but in the relational models implies extra 
conditions: relations have to match and parameter-object matching has to be 
exactly specified (to avoid confusion). Some \textbf{further clarifications} on 
the proposed model follow.

\textbf{Services and Rules.} Services and inference rules are structurally 
similar, with the distinction that: services must output at least some new 
parameter and cannot generate only new relations. Services do not "generate" 
relations between their input ($input  \times input$) parameters. If their 
definition specifies relations between input parameters, then they represent 
restrictions, i.e. conditioning the call of the service. Rules however never 
"generate" any parameters but only new relations based on outputs of previously 
called services, and; their parameters are not restricted by type i.e. they do 
not have types.

\textbf{Parameters and Types.} Each parameter of a service has a type, but 
unlike previous models, objects of the same type can be differentiated by their 
relations. This is fundamental in manual compositions or other types of 
workflows, and we believe the impossibility to express this was the main flaw of 
previous models. This greatly increases the problem computational complexity, 
but as we will see in Chapter \ref{sec:complexity}, our proposed algorithm is 
still able to find compositions on non-trivial instances of significant size, 
that a human user could not work with.

\textbf{Objects.} Objects are defined by their types, and all relations they are 
in. An object is similar to an instance of that type but is not an actual value 
or an instantiation. Objects are still, at the abstract level, dynamic 
elements that pass through the composition stages, and we keep information about 
their semantic context, through relations. On the current model, it may be 
useless to have multiple objects of the same type with the same set of 
relations, so we keep only one for each possible context. Even with this 
reduction, the total possible number of objects is exponential compared to the 
number of relations. Moreover, when considering a relation of the current 
object, the type of the paired object on which the relation is defined, may also 
be relevant to the current object state. This is motivated by parameter matching 
definition, and without considering it the model (or algorithm) might fail to 
find valid compositions even if they would exist. Also, it may be the case that 
a service can generate an unlimited number of objects if it can match in an 
unlimited number of times. Computationally, this can be a problem but is 
a natural case in manual composition, even the simple service 
\emph{squareService} from Section \ref{sec:models:names} can generate an unlimited 
number of objects if it is called on its output successively. We can 
compute the square of the square and so on, and even on this trivial example, 
relations make sense.

The defined Relational Model is, as it can be seen from previous pages, far more 
complex than the previous models. Without increasing the complexity the 
expressiveness is drastically limited, and all the elements introduced are 
designed to solve some known issue/example, that is frequently met in 
practice, and that could not be modeled before. The relational model, in this thesis, is not 
intended to be completely rigorously defined, as more important is the 
understanding of the high-level concepts. For example, the choice of having 
typed service parameters and untyped rule parameters or variables is not 
essential. Also, depending on the implementation of such model, the object 
provenance and semantic information can be tracked to some level of depth, 
depending on the application: what distinguishes an object from another can be 
the relations to other object or can also include the type of the objects in 
relation; the service where the object was created but eventually, also other 
services that updated the respective object. In our implementation in 
Section \ref{sec:complexity:relational}, we specify how we made such choices and motivate 
them; but they might be different in other cases.

Finally, we consider the Relational Model one of the main contributions of the 
thesis, and one of the most promising as future work. This is motivated by the 
many examples where human reasoning is able to manually create a composition, 
but all previous models lack the expressivity to specify any of the essential 
constraints that make a composition meaningful or valid. Moreover, as the Semantic 
Web generally had a similar evolution in the past, the need for relational 
concepts was eventually a necessity.
}

\section{Object-Oriented Parameters}
\label{sec:models:object-oriented}
\comment{\textcolor{red}{[contribution]} Mainly inspired from schema.org data 
model, similar to the object-oriented paradigm. Each concept from a hierarchy 
has a set of properties that are inherited. Properties have types from the same 
set of concepts. For one object or instance, a subset of properties can be 
defined. This is a partially defined concept.}

\iftoggle{fullThesis}{

\subsection{Modern Web Aware Approach}

The second parameter matching model proposed is partially similar to the 
Relational Model from Section \ref{sec:models:relational}; as it is also based primarily on 
the hierarchical model, enriching its semantic expressivity. Similarly, it is 
inspired by real-world cases for which previous models failed, providing means 
for expressing the necessary restrictions. The initial intention was to integrate 
more modern Web Service definition standards into our composition methods, such as 
the OpenAPI standard \cite{sferruzza2018extending}. The most relevant standard that was integrated, and the basis of "object-oriented" 
model, is the \emph{schema.org} \cite{guha2016schema} data model
\footnote{Schema.org Data Model -- 
\url{https://meta.schema.org/docs/datamodel.html}}. \emph{Schema.org} is a 
popular ontology, created as a community initiative to structure data on the Internet. It is relatively rigorously defined; which allowed the 
possibility to express precise constraints for parameter matching and therefore the automation of composition. While not all the elements from \emph{schema.org} are introduced in our model, the most important of them are enough to solve the examples.

\subsection{Formal Definition}
\textbf{Formal Definition of the Object-Oriented Model} is described as in our paper \cite{netedu2019openapi} - presenting it more from the perspective of the semantic web standards that inspired the model, and more formally, in \cite{diac2019formalism}.

Service parameters are defined over a set of \emph{concepts}. Like in previous models, $\mathbb{C}$ is the notation for the set of all concepts that appear in a repository of services, which are all possible concepts or the problem universe.
Following the hierarchical model, the concepts are first organized by using the \textit{isA} or \textit{subsumes} relation, equivalent to \emph{subTypeOf}. This is a binary relation between concepts and, in this context, it can also be considered somewhat similar to the inheritance in object-oriented programming. If a concept $c_{a}$ \textit{isA} $c_{b}$ then $c_{a}$ can substitute when there is need of a $c_{b}$. Also, for any concept $c$ in $\mathbb{C}$, there can be only one directly, more generic concept then $c$, i.e. we do not allow multiple inheritance. Note: in \emph{schema.org}, multiple inheritance is allowed, but we avoid it, because of several reasons: first, it is rarely met among \emph{schema.org} concepts, it significantly increases the complexity of the model and the computational complexity of composition algorithms, and finally it is not useful for the examples that we wanted to solve. Obviously, \textit{isA} is transitive and, for convenience, reflexive: $c_{a}$ \textit{isA} $c_{a}$. This implies that $\mathbb{C}$ together with the \textit{isA} relation forms a tree (a taxonomy) or, more generally, a forest (a set of taxonomies or an ontology).

\smallskip
The new elements added to the problem are concept \emph{properties}. These resemble to members of classes in object-oriented programming, hence the name of our model: object-oriented.
Any concept has a set of properties, possibly empty. 

Each property $p$ is a pair $\langle name, type\rangle$. The name is just an identifier of the property, and for simplicity, it can be seen as a string (not necessarily unique: different concepts can have properties with the same name - e.g. common properties like \emph{id}, \emph{name}, etc).   The type of a property is a concept also from the same set of concepts $\mathbb{C}$ and can be considered as the range of a property.

From an ontological point of view~\cite{allemang2011semantic}, a property $p$ is defined as a relation between Web resources. From this point of view, the two models proposed in Section \ref{sec:models:relational} and Section \ref{sec:models:object-oriented} are similar, but they are not equivalent. The values of a property are instances of one or more concepts (classes) -- expressed by $range$ property. Any resource that has a given property is an instance of one or more concepts -- this is denoted by the $domain$ property.

\textbf{Properties are inherited}: if $c_{a}$ \textit{isA} $c_{b}$ and $c_{b}$ has some property $\langle name_{x}, type_{x} \rangle$ then $c_{a}$ also has the property $\langle name_{x}, type_{x}\rangle$. For example, if an \textit{apple} is a \textit{fruit}, and \textit{fruit} has property $\langle hasColor, Color\rangle$ stating that \textit{fruit} instances have a color, then \textit{apple}s must have a color as well.

It is important that property names do not repeat for any unrelated concepts. For any concepts that are in a \textit{isA} relation, all properties are passed to the specialization concepts by inheritance. This restriction is just to avoid confusion and does not reduce the expressiveness, because properties can be renamed.

Syntactically, to define a property we need to know: its name, its type, and the most general concept that the property can describe. For example, consider that the \textit{hasColor} property can describe the \textit{apple} concept, but also the more general \textit{fruit} concept. If concept \textit{fruit} \textit{isA} \textit{physicalObject}, the next more general concept than \textit{fruit} (i.e. its parent in the concepts tree), under the assumption that not all physical objects are colored, then we can say that \textit{hasColor} can most generally describe \textit{fruit}, but not any \textit{physicalObject} or any other more general concept. However, it can describe other concepts in the tree, together with all their descendants.
For simplicity, we will consider further that all the properties are defined within $\mathbb{C}$, thus the concepts, \textit{isA} relation, and properties structure are structurally in $\mathbb{C}$ -- the ontology.

A \textit{partially defined concept} is a pair \textit{(c, propSet)}, where $c$ is a concept from $\mathbb{C}$ and \textit{propSet} is a subset of the properties that $c$ has defined directly or through inheritance from more generic concepts. At some moment in time (or in some stage of a workflow), a partially defined concept describes what is currently known about a concept. It does not refer to a specific concept instance, but rather generally to the information that could potentially be found for any instance of that concept at execution time.

A Web Service \textbf{w} is defined by a pair of input and output parameters: (\textbf{w}$_{in}$, \textbf{w}$_{out}$). Both are sets of partially defined concepts. All are defined over the same structure $\mathbb{C}$ so all service providers must adhere to $\mathbb{C}$, thus adding the requirement that $\mathbb{C}$ is publicly available and defined ahead of time. To be able to validly call a service all input parameters in \textbf{w}$_{in}$ must be known together with their specified required properties. After calling the service all output parameters \textbf{w}$_{out}$ will be learned with the properties specified at output.

\textbf{Parameter matching}. Let \textbf{P} be a set of partially defined concepts, and suppose \textbf{w} = (\textbf{w}$_{in}$, \textbf{w}$_{out}$) is a Web service. The set \textbf{P} matches service \textbf{w} (or, equivalently, \textbf{w} is \textit{callable} if \textbf{P} is known) iff: for any partially defined concept \textbf{pdc} = \textit{(c, propSet)} $\in$ \textbf{w}$_{in}$,  $\exists$ \textbf{p} = \textit{($c_{spec}$, propSuperSet)} $\in$ \textbf{P} such that $c_{spec}$ \textit{isA} $c$ and $propSet \subseteq propSuperSet$. 
We further define the addition of \textbf{w}$_{out}$ to \textbf{P} as:
\begin{flushleft}
  \begin{align*}
     \begin{matrix}
     \textbf{P} \oplus  \textbf{w} \hspace{5px} ( \hspace{1px} or \hspace{3px} \textbf{P}  \cup  \textbf{w}_{out} ) \hspace{3px} = \hspace{3px}
     \end{matrix} \\
     \begin{Bmatrix}
     \Bigg(c,\bigg\{p \Big| \exists(c', propSet')\in  w_{out}  \hspace{3px} and \hspace{3px} \begin{matrix}
     c\ has \ p\\
     p \in propSet'\\
     c'\ isA\ c\\ 
     \end{matrix} \bigg\}\Bigg) \hspace{3px} \bigg| \hspace{3px} \nexists (c,propSet)\in P \\
     \end{Bmatrix} \hspace{3px} \bigcup \\
      \begin{Bmatrix}
      \Bigg(c,propSet  \cup  \bigg\{p \Big| \exists(c',propSet')\in  w_{out} \hspace{3px} and \hspace{3px} \begin{matrix}
      c\ has \ p\\
      p \in propSet'\\ 
      c'\ isA\ c\\ 
      \end{matrix} \bigg\}\Bigg) \hspace{3px} \bigg| \hspace{3px} \exists  (c,propSet)\in P \\
      \end{Bmatrix}
  \end{align*}
\end{flushleft}
\noindent
or the union of \textbf{w}$_{out}$ with \textbf{P} under the constraint of \textbf{P} matching \textbf{w} (defined as parameter matching above). Also, by $c$ $has$ $p$ we refer to the fact that property $p$ is defined for $c$ directly or by inheritance. \textbf{P} $\oplus$ \textbf{w} contains new concepts that are in \textbf{w}$_{out}$ - the first line in the equation, and concepts already in \textbf{P} possibly with new properties from \textbf{w}$_{out}$ defined for corresponding concepts or their specializations - the second line.

In words, after a call to a service, all its output parameters are selected, and for each concept together with its selected properties \textit{(c, propSet)} in \textbf{w}$_{out}$, \textit{propSet} is added to \textit{c}, \textit{c}$'$s parent in the concepts tree or the ascendants until we reach the first node that gains no new information, or the root. More precisely, for each $p$ in $propSet$ we add $p$ to our knowledge base for $c$, for the parent of $c$, and so on until $p$ is no longer defined for the node we reached. The node where this process stops can differ from one $p$ property to another $p'$ property, but once the process stops for all properties in $propSet$ there is no need to go further.

\textbf{Chained matching}. Let \textbf{P} be a set of partially defined concepts and $(w_{1}, w_{2}, \dots, w_{k})$ an ordered list of services. We say that $\textbf{P} \oplus w_{1} \oplus w_{2} \oplus \dots \oplus w_{k}$ is a chain of matching services iff $w_{i}$ matches $\textbf{P} \oplus w_{1} \oplus w_{2} \oplus \dots \oplus w_{i-1}; \forall i = 1 \dots k$. This is the rather primitive model for multiple service calls, which is a requirement for defining composition. For simplicity, we avoid for now more complex workflows that could handle parallel and sequential service execution constructs.

\textbf{Web Service Composition problem}. Given an ontology having a set of concepts $\mathbb{C}$ and a repository of web services $W = (w_{1}, w_{2}, \dots, w_{n})$, and two sets of partially defined concepts \textbf{Init} and \textbf{Goal}, all defined over $\mathbb{C}$, find a chain of matching services $(w_{c1}, w_{c2}, \dots w_{ck})$ such that $(\emptyset, $\textbf{Init}$) \oplus w_{c1} \oplus w_{c2} \oplus \dots \oplus w_{ck} \oplus ($\textbf{Goal}$, \emptyset)$.

The $(\emptyset, \textbf{Init})$ and $(\textbf{Goal},$ $\emptyset)$ are just short ways of writing the initially known and finally required parameters, by using mock services. We can also imagine (\textbf{Init}, \textbf{Goal}) as a web service, then the problem requires finding an ``implementation'' of a web service using the services in some repository.

\subsection{Case Study: Transport Agency}

To illustrate an application of our approach, we have considered the following example, according to the problem definition above. The concepts and properties used are actual, real concepts and properties from \emph{schema.org}. Technically, the service interfaces were stored in a OpenAPI compliant document and the "resources" (to follow OpenAPI terminology), i.e. service parameters, were described in an extended JSON-LD format. For our mathematical model all these standards and formats are not relevant, but the model from Section \ref{sec:models:object-oriented} itself was inspired by them. More details on the JSON-LD extension are presented in \cite{netedu2019openapi}.

The case study illustrates a car company operating via web services. Describing services using OpenAPI and JSON-LD, the approach shows how to represent complex relations between resources by combining object-oriented concepts in both structure and semantics. 
For the conducted experiment, our example contains six services, several resources, and the user request to be solved by a composition.

The scenario is the following: supposing that a customer needs a \emph{vehicle} to transport a given \emph{payload}. This person knows his/her current \emph{GeoLocation(latitude, longitude)} and a time frame \emph{Action(startTime, endTime)} in which the transport should arrive. The \textbf{Goal} is to obtain the \emph{Action(location)} where the \emph{vehicle} will arrive. 
The six services -- each of them implementing a single operation -- are specified below:

} 

\begin{lstlisting}[basicstyle=\small,escapeinside={<}{>}\ttfamily]
getCountryFromLocation
   in  = GeoLocation(lat,lon)
   out = Country(name)
    
getTransportCompany
   in  = AdministrativeArea(name)
   out = Organization(name)
    
getClosestCity
   in  = GeoLocation(lat,lon)
   out = City(name)
    
getLocalSubsidiary
   in  = Organization(name), City(name)
   out = LocalBusiness(email)

getVehicle
   in = Vehicle(payload), LocalBusiness(email)
   out= Vehicle(vehicleIdentificationNumber)

makeArrangements
   in = Vehicle(vehicleIdentificationNumber),
        Organization(name,email), Action(startTime,endTime)
   out = Action(location)
\end{lstlisting}

\iftoggle{fullThesis}{

In OpenAPI terms, a HTTP GET method is defined to obtain a JSON representation of the desired Web resource, for each \textit{getResource} operation -- i.e. using \texttt{GET /country} with \emph{GeoLocation} as input parameter and \emph{Country} as output. Without JSON-LD constructs, these parameters have regular JSON datatypes like string or number. 

As defined in \emph{schema.org}, \emph{LocalBusiness}\footnote{LocalBusiness, a particular physical business or branch of an organization -- \url{https://schema.org/LocalBusiness}} \textit{isA} \emph{Organization} -- or, equivalent: \emph{LocalBusiness} $\sqsubseteq$ \emph{Organization}. Similarly, \emph{Country} \textit{isA} \emph{AdministrativeArea}, etc.

\smallskip
A valid composition satisfying the user request can consist of the services in the following order: \textbf{Init} $\rightarrow$ \emph{getCountryFromLocation} $\rightarrow$ \emph{getTransportCompany} $\rightarrow$ \emph{getClosestCity} $\rightarrow$ \emph{getLocalSubsidiary} $\rightarrow$ \emph{getVehicle} $\rightarrow$ \emph{makeArrangements} $\rightarrow$ \textbf{Goal}. The order is relevant, but not unique in this case. This can be verified by considering all resources added by each service and also by the use of the \textit{isA} relation. For example, \emph{LocalBusiness(email)} can be used as \emph{Organization(email)}.

\begin{figure}[h!]
\caption[\textbf{Schema.org} concepts used in \textbf{Object-Oriented} model]{Concepts (on the left in the boxes) and properties (on the right) from \emph{schema.org} used in the example for object-oriented model.}
\label{oopExampleFig}
\begin{center}
\tikzset{every picture/.style={line width=0.85pt}} 

\begin{tikzpicture}[x=0.85pt,y=0.75pt,yscale=-1,xscale=1]

\draw   (59,20.97) .. controls (59,16.81) and (62.38,13.43) .. (66.55,13.43) -- (211.45,13.43) .. controls (215.62,13.43) and (219,16.81) .. (219,20.97) -- (219,43.61) .. controls (219,47.78) and (215.62,51.16) .. (211.45,51.16) -- (66.55,51.16) .. controls (62.38,51.16) and (59,47.78) .. (59,43.61) -- cycle ;
\draw    (152,13.47) -- (152,51.11) ;

\draw    (50,95.43) -- (100.47,52.72) ;
\draw [shift={(102,51.43)}, rotate = 499.76] [fill={rgb, 255:red, 0; green, 0; blue, 0 }  ][line width=0.75]  [draw opacity=0] (8.93,-4.29) -- (0,0) -- (8.93,4.29) -- cycle    ;

\draw   (9,102.43) .. controls (9,99.11) and (11.69,96.43) .. (15,96.43) -- (117,96.43) .. controls (120.31,96.43) and (123,99.11) .. (123,102.43) -- (123,120.43) .. controls (123,123.74) and (120.31,126.43) .. (117,126.43) -- (15,126.43) .. controls (11.69,126.43) and (9,123.74) .. (9,120.43) -- cycle ;
\draw    (75.26,96.46) -- (75.26,126.39) ;

\draw   (137,102.43) .. controls (137,99.11) and (139.69,96.43) .. (143,96.43) -- (237,96.43) .. controls (240.31,96.43) and (243,99.11) .. (243,102.43) -- (243,120.43) .. controls (243,123.74) and (240.31,126.43) .. (237,126.43) -- (143,126.43) .. controls (139.69,126.43) and (137,123.74) .. (137,120.43) -- cycle ;
\draw    (189.26,96.46) -- (189.26,126.39) ;

\draw    (165,96.43) -- (122.4,52.86) ;
\draw [shift={(121,51.43)}, rotate = 405.64] [fill={rgb, 255:red, 0; green, 0; blue, 0 }  ][line width=0.75]  [draw opacity=0] (8.93,-4.29) -- (0,0) -- (8.93,4.29) -- cycle    ;

\draw   (12.26,161.97) .. controls (12.26,157.81) and (15.64,154.43) .. (19.81,154.43) -- (148.45,154.43) .. controls (152.62,154.43) and (156,157.81) .. (156,161.97) -- (156,184.61) .. controls (156,188.78) and (152.62,192.16) .. (148.45,192.16) -- (19.81,192.16) .. controls (15.64,192.16) and (12.26,188.78) .. (12.26,184.61) -- cycle ;
\draw    (69.26,154.47) -- (69.26,192.11) ;

\draw    (69.26,173.29) -- (156,173.43) ;

\draw   (176,196.97) .. controls (176,192.81) and (179.38,189.43) .. (183.55,189.43) -- (321.45,189.43) .. controls (325.62,189.43) and (329,192.81) .. (329,196.97) -- (329,219.61) .. controls (329,223.78) and (325.62,227.16) .. (321.45,227.16) -- (183.55,227.16) .. controls (179.38,227.16) and (176,223.78) .. (176,219.61) -- cycle ;
\draw    (259.5,189.47) -- (259.5,227.11) ;

\draw    (259.5,208.29) -- (329,208.43) ;

\draw   (276,26.97) .. controls (276,22.81) and (279.38,19.43) .. (283.55,19.43) -- (428.45,19.43) .. controls (432.62,19.43) and (436,22.81) .. (436,26.97) -- (436,49.61) .. controls (436,53.78) and (432.62,57.16) .. (428.45,57.16) -- (283.55,57.16) .. controls (279.38,57.16) and (276,53.78) .. (276,49.61) -- cycle ;
\draw    (369,19.47) -- (369,57.11) ;

\draw   (309,110.5) .. controls (309,106.23) and (312.46,102.77) .. (316.73,102.77) -- (443.27,102.77) .. controls (447.54,102.77) and (451,106.23) .. (451,110.5) -- (451,133.7) .. controls (451,137.97) and (447.54,141.43) .. (443.27,141.43) -- (316.73,141.43) .. controls (312.46,141.43) and (309,137.97) .. (309,133.7) -- cycle ;
\draw    (402.26,102.81) -- (402.26,141.38) ;
\draw    (356,102.43) -- (335.85,59.24) ;
\draw [shift={(335,57.43)}, rotate = 424.98] [fill={rgb, 255:red, 0; green, 0; blue, 0 }  ][line width=0.75]  [draw opacity=0] (8.93,-4.29) -- (0,0) -- (8.93,4.29) -- cycle    ;

\draw    (369,38.29) -- (436,38.43) ;

\draw    (402,121.92) -- (450,122.1) ;
\draw   (357,173.83) .. controls (357,168.09) and (361.66,163.43) .. (367.4,163.43) -- (464.5,163.43) .. controls (470.24,163.43) and (474.9,168.09) .. (474.9,173.83) -- (474.9,205.03) .. controls (474.9,210.78) and (470.24,215.44) .. (464.5,215.44) -- (367.4,215.44) .. controls (361.66,215.44) and (357,210.78) .. (357,205.03) -- cycle ;
\draw    (407,163.49) -- (407,215.37) ;
\draw    (407.7,181.02) -- (474.28,181.19) ;
\draw    (407,197.99) -- (474.62,198.25) ;

\draw (105,33) node [scale=0.9] [align=center] {Administrative\\Area};
\draw (185,31) node [scale=0.9] [align=center] {name};
\draw (41.77,113) node [scale=0.9] [align=center] {Country};
\draw (98.77,113) node [scale=0.9] [align=center] {name};
\draw (163.77,113) node [scale=0.9] [align=center] {City};
\draw (214.77,111) node [scale=0.9] [align=center] {name};
\draw (40.77,174) node [scale=0.9] [align=center] {Vehicle};
\draw (112.77,164) node [scale=0.9] [align=center] {vehicleId.Nr.};
\draw (108.77,183) node [scale=0.9] [align=center] {payload};
\draw (217.77,209) node [scale=0.9] [align=center] {GeoLocation};
\draw (293.77,199) node [scale=0.9] [align=center] {latitude};
\draw (294.77,218) node [scale=0.9] [align=center] {longitude};
\draw (322,38) node [scale=0.9] [align=center] {Organization};
\draw (401,28.86) node [scale=0.9] [align=center] {name};
\draw (356.77,123.12) node [scale=0.9] [align=center] {LocalBusiness};
\draw (401,47.86) node [scale=0.9] [align=center] {email};
\draw (426,113.06) node [scale=0.9] [align=center] {name};
\draw (426,132.68) node [scale=0.9] [align=center] {email};
\draw (382.82,189.81) node [scale=0.9] [align=center] {Action};
\draw (439.88,173.4) node [scale=0.9] [align=center] {startTime};
\draw (440.88,190.45) node [scale=0.9] [align=center] {endTime};
\draw (441.52,206.92) node [scale=0.9] [align=center] {location};

\end{tikzpicture}
\end{center}
\end{figure}
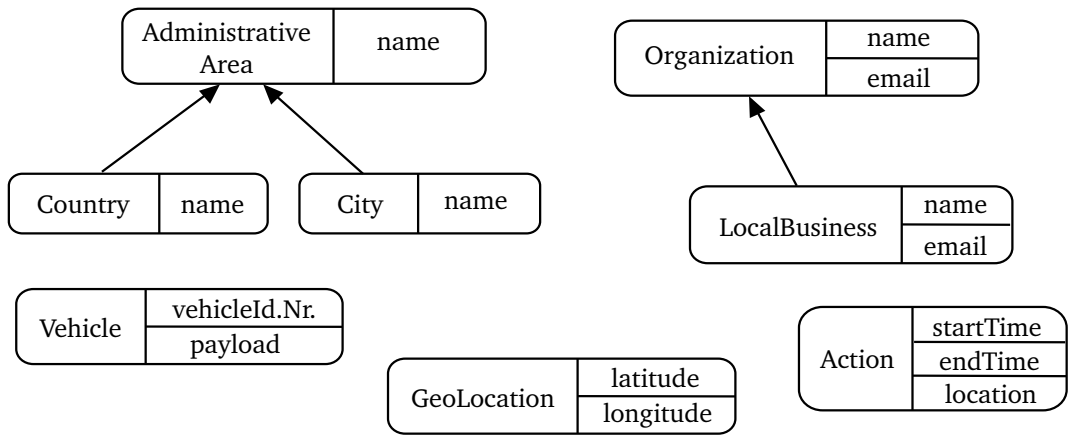

The example described was not intended to motivate the proposed theoretical model, but its compliance with the actual concepts and properties from in \emph{schema.org}. However, the model can work on other ontologies following the same constructs; and to motivate the model itself a more appropriate example can be easily found.
}

\section{QoS Aware Composition}
\label{sec:models:qos}
\comment{Considering Quality of Service in Composition. Iterate through  types of 
QoS, describe each and reason about how new cost function aggregates on serial 
or parallel service invocations.}

\iftoggle{fullThesis}{

Almost any system developed using Service Oriented Architecture (SOA) takes 
into consideration the Quality of Service (QoS) of the components involved. QoS 
refers to all different measurable aspects that analyze service behavior, 
performance, for example response time, reliability, throughput and many more. 
Fundamentally, this is especially important because services are developed by 
multiple third parties - the service developers; making services less 
trustworthy. Most often, if a system depends on the functionality of many 
services, it depends on the functionality of \textbf{all} those services or at 
least a high percentage of them. Therefore, it is not surprising that research 
about QoS-aware Service Composition started just a few years after the service 
composition concept itself was proposed. This is true for automatic composition, 
but QoS adds significant complexity even to manual composition (for example, 
see the work of \cite{berbner2006heuristics}).

Considering service metrics requires the addition of a new model, that is 
independent of the semantic model used to represent parameters (discussed in the 
previous three sections). Therefore, generally, any model used for QoS can be 
used together with any parameter model; but the complexity of the arising 
problems may not be independent between the two choices, so there are many cases to analyze.

In our previous work, we did not focus particularly on the QoS aspect of  the 
composition problem. However, we did consider some of the most simple metrics 
for QoS, like composition length (number of services used in a composition), or 
execution path (the service-length of a composition, if parallel execution of 
services is allowed). These are not proper QoS metrics for services, but for a 
composition, they are the most basic measures approximating efficiency. This is 
more relevant for  the complexity analysis of Chapter \ref{sec:complexity}. 
However, we will broadly describe the possible models of QoS metrics. As our 
research is for the most part based on the Web Services Challenge editions, it 
is worth mentioning that the challenge included the QoS aspect starting with the 
2009 edition. This was described in \cite{kona2009wsc}, where, in particular, 
the WSLA language was used to describe service metrics - the standard for 
defining Web Service Level Agreement.

To define QoS metrics for a composition of services, we can consider two types 
of service execution withing compositions. Previously, we referred to the 
composition as either an ordered list of services or a "workflow" of services, 
without defining exactly what is a workflow. Service restrictions 
(preconditions), or input parameters eventually described with more complex 
semantic elements (e.g. relations or properties in our cases), are matched with 
some output of previous services already included in the composition; or to the 
information initially provided by the user. The dependencies between such services are usually displayed by arrows or arcs in workflow graphs, showing the order in 
which services must be executed. However, after the execution of a service, 
several other services might become available at once, so more than one may be 
included in the composition as the next step. In this case, those services could 
be invoked in parallel at execution time. For specific QoS metrics, this is 
relevant, as their values are computed differently depending on the type of 
execution. Figure \ref{fig:workflowfefintion} displays the main types of service 
execution that we will consider.

\begin{center}
\begin{figure}
\caption[Aggregated \textbf{QoS} on different execution types]{Different service execution cases, resulting in different QoS metrics for the resulting composition service.}
\tikzset{every picture/.style={line width=0.75pt}} 

\begin{tikzpicture}[x=1pt,y=1pt,yscale=-1,xscale=1]
0, and has height of 244.42855834960938

\draw    (85.33,49.33) -- (100.67,49.29) ;
\draw [shift={(102.67,49.29)}, rotate = 539.8399999999999] [fill={rgb, 255:red, 
0; green, 0; blue, 0 }  ][line width=0.75]  [draw opacity=0] (8.93,-4.29) -- 
(0,0) -- (8.93,4.29) -- cycle    ;

\draw    (134.67,49.33) -- (150,49.29) ;
\draw [shift={(152,49.29)}, rotate = 539.8399999999999] [fill={rgb, 255:red, 0; 
green, 0; blue, 0 }  ][line width=0.75]  [draw opacity=0] (8.93,-4.29) -- (0,0) 
-- (8.93,4.29) -- cycle    ;

\draw    (171.33,49.33) -- (186.67,49.29) ;
\draw [shift={(188.67,49.29)}, rotate = 539.8399999999999] [fill={rgb, 255:red, 
0; green, 0; blue, 0 }  ][line width=0.75]  [draw opacity=0] (8.93,-4.29) -- 
(0,0) -- (8.93,4.29) -- cycle    ;

\draw    (219.33,50) -- (234.67,49.96) ;
\draw [shift={(236.67,49.95)}, rotate = 539.8399999999999] [fill={rgb, 255:red, 
0; green, 0; blue, 0 }  ][line width=0.75]  [draw opacity=0] (8.93,-4.29) -- 
(0,0) -- (8.93,4.29) -- cycle    ;

\draw    (35.67,50) -- (51,49.96) ;
\draw [shift={(53,49.95)}, rotate = 539.8399999999999] [fill={rgb, 255:red, 0; 
green, 0; blue, 0 }  ][line width=0.75]  [draw opacity=0] (8.93,-4.29) -- (0,0) 
-- (8.93,4.29) -- cycle    ;

\draw    (354.17,52.33) -- (384.64,71.4) ;
\draw [shift={(386.33,72.46)}, rotate = 212.04] [fill={rgb, 255:red, 0; green, 
0; blue, 0 }  ][line width=0.75]  [draw opacity=0] (8.93,-4.29) -- (0,0) -- 
(8.93,4.29) -- cycle    ;

\draw    (353.83,20.33) -- (387.61,63.88) ;
\draw [shift={(388.83,65.46)}, rotate = 232.21] [fill={rgb, 255:red, 0; green, 
0; blue, 0 }  ][line width=0.75]  [draw opacity=0] (8.93,-4.29) -- (0,0) -- 
(8.93,4.29) -- cycle    ;

\draw    (353.83,102) -- (384.76,77.7) ;
\draw [shift={(386.33,76.46)}, rotate = 501.84] [fill={rgb, 255:red, 0; green, 
0; blue, 0 }  ][line width=0.75]  [draw opacity=0] (8.93,-4.29) -- (0,0) -- 
(8.93,4.29) -- cycle    ;

\draw    (289,66.43) -- (320.85,21.07) ;
\draw [shift={(322,19.43)}, rotate = 485.07] [fill={rgb, 255:red, 0; green, 0; 
blue, 0 }  ][line width=0.75]  [draw opacity=0] (8.93,-4.29) -- (0,0) -- 
(8.93,4.29) -- cycle    ;

\draw    (290.5,72.93) -- (319.69,52.61) ;
\draw [shift={(321.33,51.46)}, rotate = 505.16] [fill={rgb, 255:red, 0; green, 
0; blue, 0 }  ][line width=0.75]  [draw opacity=0] (8.93,-4.29) -- (0,0) -- 
(8.93,4.29) -- cycle    ;

\draw    (290.5,80.93) -- (320.19,101.33) ;
\draw [shift={(321.83,102.46)}, rotate = 214.5] [fill={rgb, 255:red, 0; green, 
0; blue, 0 }  ][line width=0.75]  [draw opacity=0] (8.93,-4.29) -- (0,0) -- 
(8.93,4.29) -- cycle    ;

\draw    (83.33,188.33) -- (101.12,152.22) ;
\draw [shift={(102,150.43)}, rotate = 476.22] [fill={rgb, 255:red, 0; green, 0; 
blue, 0 }  ][line width=0.75]  [draw opacity=0] (8.93,-4.29) -- (0,0) -- 
(8.93,4.29) -- cycle    ;

\draw    (83.33,188.33) -- (148.01,195.22) ;
\draw [shift={(150,195.43)}, rotate = 186.08] [fill={rgb, 255:red, 0; green, 0; 
blue, 0 }  ][line width=0.75]  [draw opacity=0] (8.93,-4.29) -- (0,0) -- 
(8.93,4.29) -- cycle    ;

\draw    (243.33,162.33) -- (258.67,162.29) ;
\draw [shift={(260.67,162.29)}, rotate = 539.8399999999999] [fill={rgb, 255:red, 
0; green, 0; blue, 0 }  ][line width=0.75]  [draw opacity=0] (8.93,-4.29) -- 
(0,0) -- (8.93,4.29) -- cycle    ;

\draw    (133.33,173) -- (210.02,161.72) ;
\draw [shift={(212,161.43)}, rotate = 531.63] [fill={rgb, 255:red, 0; green, 0; 
blue, 0 }  ][line width=0.75]  [draw opacity=0] (8.93,-4.29) -- (0,0) -- 
(8.93,4.29) -- cycle    ;

\draw    (33.67,189) -- (49,188.96) ;
\draw [shift={(51,188.95)}, rotate = 539.8399999999999] [fill={rgb, 255:red, 0; 
green, 0; blue, 0 }  ][line width=0.75]  [draw opacity=0] (8.93,-4.29) -- (0,0) 
-- (8.93,4.29) -- cycle    ;

\draw    (83.33,188.33) -- (99.43,175.67) ;
\draw [shift={(101,174.43)}, rotate = 501.79] [fill={rgb, 255:red, 0; green, 0; 
blue, 0 }  ][line width=0.75]  [draw opacity=0] (8.93,-4.29) -- (0,0) -- 
(8.93,4.29) -- cycle    ;

\draw    (132.33,150.33) -- (152.93,182.74) ;
\draw [shift={(154,184.43)}, rotate = 237.57] [fill={rgb, 255:red, 0; green, 0; 
blue, 0 }  ][line width=0.75]  [draw opacity=0] (8.93,-4.29) -- (0,0) -- 
(8.93,4.29) -- cycle    ;

\draw    (181,193.43) -- (210.42,170.65) ;
\draw [shift={(212,169.43)}, rotate = 502.25] [fill={rgb, 255:red, 0; green, 0; 
blue, 0 }  ][line width=0.75]  [draw opacity=0] (8.93,-4.29) -- (0,0) -- 
(8.93,4.29) -- cycle    ;

\draw    (53.33,41.33) .. controls (53.33,40.23) and (54.23,39.33) .. 
(55.33,39.33) -- (83.33,39.33) .. controls (84.44,39.33) and (85.33,40.23) .. 
(85.33,41.33) -- (85.33,59.33) .. controls (85.33,60.44) and (84.44,61.33) .. 
(83.33,61.33) -- (55.33,61.33) .. controls (54.23,61.33) and (53.33,60.44) .. 
(53.33,59.33) -- cycle  ;
\draw (69.33,50.33) node  [align=left] {ws{\scriptsize 1}};
\draw    (103,41.33) .. controls (103,40.23) and (103.9,39.33) .. (105,39.33) -- 
(133,39.33) .. controls (134.1,39.33) and (135,40.23) .. (135,41.33) -- 
(135,59.33) .. controls (135,60.44) and (134.1,61.33) .. (133,61.33) -- 
(105,61.33) .. controls (103.9,61.33) and (103,60.44) .. (103,59.33) -- cycle  ;
\draw (119,50.33) node  [align=left] {ws{\scriptsize 2}};
\draw    (188.5,41.33) .. controls (188.5,40.23) and (189.4,39.33) .. 
(190.5,39.33) -- (217.5,39.33) .. controls (218.6,39.33) and (219.5,40.23) .. 
(219.5,41.33) -- (219.5,59.33) .. controls (219.5,60.44) and (218.6,61.33) .. 
(217.5,61.33) -- (190.5,61.33) .. controls (189.4,61.33) and (188.5,60.44) .. 
(188.5,59.33) -- cycle  ;
\draw (204,50.33) node  [align=left] {ws{\scriptsize k}};
\draw (26.67,50) node  [align=left] {...};
\draw (161,49.29) node  [align=left] {...};
\draw (245.67,49.95) node  [align=left] {...};
\draw (135,85.33) node  [align=left] {{\small (A) sequential execution}};
\draw    (322.33,11.33) .. controls (322.33,10.23) and (323.23,9.33) .. 
(324.33,9.33) -- (352.33,9.33) .. controls (353.44,9.33) and (354.33,10.23) .. 
(354.33,11.33) -- (354.33,29.33) .. controls (354.33,30.44) and (353.44,31.33) 
.. (352.33,31.33) -- (324.33,31.33) .. controls (323.23,31.33) and 
(322.33,30.44) .. (322.33,29.33) -- cycle  ;
\draw (338.33,20.33) node  [align=left] {ws{\scriptsize 1}};
\draw    (322,43.33) .. controls (322,42.23) and (322.9,41.33) .. (324,41.33) -- 
(352,41.33) .. controls (353.1,41.33) and (354,42.23) .. (354,43.33) -- 
(354,61.33) .. controls (354,62.44) and (353.1,63.33) .. (352,63.33) -- 
(324,63.33) .. controls (322.9,63.33) and (322,62.44) .. (322,61.33) -- cycle  ;
\draw (338,52.33) node  [align=left] {ws{\scriptsize 2}};
\draw    (322.5,93.33) .. controls (322.5,92.23) and (323.4,91.33) .. 
(324.5,91.33) -- (351.5,91.33) .. controls (352.6,91.33) and (353.5,92.23) .. 
(353.5,93.33) -- (353.5,111.33) .. controls (353.5,112.44) and (352.6,113.33) .. 
(351.5,113.33) -- (324.5,113.33) .. controls (323.4,113.33) and (322.5,112.44) 
.. (322.5,111.33) -- cycle  ;
\draw (338,102.33) node  [align=left] {ws{\scriptsize k}};
\draw (278.67,70) node  [align=left] {...};
\draw (337,76.29) node  [align=left] {...};
\draw (396.67,70.95) node  [align=left] {...};
\draw (343,125.33) node  [align=left] {{\small (B) parallel \ execution}};
\draw    (51.33,180.33) .. controls (51.33,179.23) and (52.23,178.33) .. 
(53.33,178.33) -- (81.33,178.33) .. controls (82.44,178.33) and (83.33,179.23) 
.. (83.33,180.33) -- (83.33,198.33) .. controls (83.33,199.44) and 
(82.44,200.33) .. (81.33,200.33) -- (53.33,200.33) .. controls (52.23,200.33) 
and (51.33,199.44) .. (51.33,198.33) -- cycle  ;
\draw (67.33,189.33) node  [align=left] {ws{\scriptsize 1}};
\draw    (101,130.33) .. controls (101,129.23) and (101.9,128.33) .. 
(103,128.33) -- (131,128.33) .. controls (132.1,128.33) and (133,129.23) .. 
(133,130.33) -- (133,148.33) .. controls (133,149.44) and (132.1,150.33) .. 
(131,150.33) -- (103,150.33) .. controls (101.9,150.33) and (101,149.44) .. 
(101,148.33) -- cycle  ;
\draw (117,139.33) node  [align=left] {ws{\scriptsize 2}};
\draw    (212.5,153.33) .. controls (212.5,152.23) and (213.4,151.33) .. 
(214.5,151.33) -- (241.5,151.33) .. controls (242.6,151.33) and (243.5,152.23) 
.. (243.5,153.33) -- (243.5,171.33) .. controls (243.5,172.44) and 
(242.6,173.33) .. (241.5,173.33) -- (214.5,173.33) .. controls (213.4,173.33) 
and (212.5,172.44) .. (212.5,171.33) -- cycle  ;
\draw (228,162.33) node  [align=left] {ws{\scriptsize k}};
\draw (24.67,189) node  [align=left] {...};
\draw (197,188.29) node  [align=left] {...};
\draw (132,225.33) node  [align=left] {{\small (C) general workflow}};
\draw    (101,165.33) .. controls (101,164.23) and (101.9,163.33) .. 
(103,163.33) -- (131,163.33) .. controls (132.1,163.33) and (133,164.23) .. 
(133,165.33) -- (133,183.33) .. controls (133,184.44) and (132.1,185.33) .. 
(131,185.33) -- (103,185.33) .. controls (101.9,185.33) and (101,184.44) .. 
(101,183.33) -- cycle  ;
\draw (117,174.33) node  [align=left] {ws{\scriptsize 3}};
\draw    (149,186.33) .. controls (149,185.23) and (149.9,184.33) .. 
(151,184.33) -- (179,184.33) .. controls (180.1,184.33) and (181,185.23) .. 
(181,186.33) -- (181,204.33) .. controls (181,205.44) and (180.1,206.33) .. 
(179,206.33) -- (151,206.33) .. controls (149.9,206.33) and (149,205.44) .. 
(149,204.33) -- cycle  ;
\draw (165,195.33) node  [align=left] {ws{\scriptsize 4}};
\draw (272,165.29) node  [align=left] {...};
\end{tikzpicture}
\label{fig:workflowfefintion}
\end{figure}
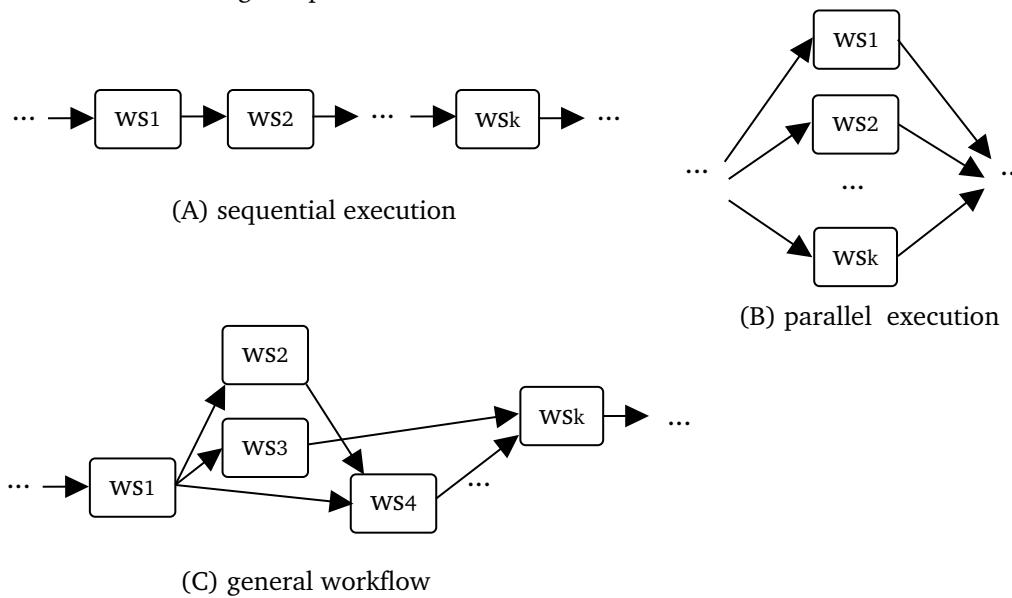
\end{center}

Regarding QoS metrics, it is enough to define how they are computed on (A) 
sequential execution and (B) parallel execution. The more generic workflow 
scenario (C) can be decomposed to these two cases, for example, $ws_2$ and $ws_3$ 
are executed in parallel, but $ws_4$ has to wait for $ws_2$ to finish. We will 
define the QoS for a composition of services; including the elementary metrics 
but also real QoS measures, described as in \cite{gabrel2018qos}.

\subsection{QoS Metrics}
\textbf{The total number of services} is the simplest metric considered, which was 
evaluated in the first edition of the Web Services Challenge in 2005. 
It is not a real QoS metric as it makes no sense to describe a single service, 
but for a composition of service is relevant. Clearly, the value is the same 
for parallel and sequential execution, as it simply sums up the total number of 
used services.

\textbf{Execution path} is also relevant only if the execution of services in 
parallel is possible, and it was used in the 2008 edition of the Web Services 
Challenge. Execution path is a more refined metric than the number of services, since 
at parallel execution, only one execution step is added; resulting in a better 
approximation of the execution time of the composition. For the workflow (C) in 
Figure \ref{fig:workflowfefintion}, the execution path is 4, which is the 
minimum number of steps required to execute all services, but counting only one, 
parallel, execution when multiple services become available. This metric is used 
especially in Section \ref{sec:complexity:algorithms} or in the evaluation 
criteria of the 2008 edition of the Services Challenge\cite{bansal2008wsc}.

\textbf{Execution time} is an important and more popular QoS metric. The 
execution time of services may be obtained from the service definitions provided 
by the service developers, as a WSLA value. In this case, the contract specifies 
an upper bound on the value. Alternatively, it can be observed and computed by a 
service monitor; and in this case, it is not known until the service is 
executed. 

The execution time for a composition is the sum of the execution times for each 
service in sequential execution (A) and the maximum execution time of any 
service in parallel execution (B).

\textbf{The cost} of execution refers to the amount that has to be paid to 
access a service, if services are priced. Clearly, it is the sum of the costs of all services 
for both parallel and sequential execution, and it can also be seen as a 
generalization of the \emph{number of services} metric, which is similar to a 
cost of \emph{1} for each service.

\textbf{The throughput} of a service refers to the number of service calls that 
can be invoked within a time frame, usually a second. It is expected that a 
service behaves with normal SLA (Service Level Agreement) if the declared 
throughput or bandwidth is not exceeded, and service provides can implement  
some throttling mechanism if this limitation is exceeded. For both sequential and 
parallel execution cases, the throughput of a composition is the minimum value 
of the throughput values of the component services.

\textbf{The reliability} measures the expected probability of a service to 
correctly return the response value (with no error messages, within time limit, 
etc). Usually, it is measured by a number between \emph{0} and \emph{1}, the 
ideal reliability is \emph{1}. For both parallel and sequential execution, the 
\emph{reliability} value of a composition is the product of the reliabilities of 
component services; because a successful execution depends on the success of all 
the services involved.

Other QoS measures exist but they are less popular and can be modeled as a combination of 
the above. Ideally, a composition model and a composition solution (algorithm) 
would consider them and try to find the optimal compositions. Moreover, it is 
not realistic to work with only one of the metrics above: most often, more of 
them are used together. In a very expressive model, one could include complex QoS 
constraints in the composition requests; for example, a user might prefer the 
fastest composition possible that safely allows at least 1000 calls per second 
and does not exceed a budget of 10\$ per minute. In the real world, another issue 
with designing QoS-aware systems is that not all service provides declare or 
follow their SLAs. If we imagine a composition framework that allows the 
registration of services from untrusted developers, extra validation must be implemented, e.g., monitoring the service's status, health or SLAs.

We leave a more detailed analysis of QoS models or algorithms as future work. As a 
final remark, the computational complexity of the composition problem with QoS is 
significantly increased. For most versions, the optimization problem is 
\emph{NP-Complete} even for the simplest versions of parameter matching. However, the algorithms that optimize the smallest by the number of services can be adapted to more complicated QoS metrics. 
}

\section{Stateful Compositions}
\label{sec:models:stateful}
\comment{Many papers also study states of services or states of the world. This 
is more complicated to model and computationally to solve, but a summary of what 
has been done is relevant.}

\iftoggle{fullThesis}{

Generally, Web Services can be defined as software components that publicly expose functionalities over the network. In particular, in most studies related to (automatic) composition, the concept of a service is simplified to a single functionality, without any loss of generality. But if more service examples are considered, we can observe that the functionality of a service can be described on two levels \cite{lecue2008towards}: the (proper) functional level, and the procedural level. By functional level we refer to the interaction between the service caller, and the service itself; through input and output parameters. This describes service invocation as a request-response communication process where all existing information is inside the parameter values. The service is an atomic endpoint that for some given values of input parameters always returns the same values as output. This type of services is also called \emph{information providing} \cite{zhao2012automatic}. An example is a service that sums up two numbers: for input \emph{a} and \emph{b}, the service will always return \emph{a+b} and there is nothing else that is changed within or outside of the service before and after a call to the service.

By the process level, we refer to anything that involves one of the following: description of service states, \emph{world} states (describing anything similar to states outside the service), and descriptions of more complex interactions between services and the service consumers. One of the simplest examples can be a weather providing service - that can return the temperature for a given input city. Clearly, even if no state is changing in a system that uses the weather service, just the fact that two service invocations are initiated at different times will lead to different return values.

In the rest of the thesis (outside this section), only the functional aspects of the services is analyzed. But stateful services are very frequently met in practice. A very popular cloud computing platform consists only on stateful services - the Amazon Web Services\footnote{Amazon AWS -- \url{https://aws.amazon.com/}}. To use this kind of services, one must initiate a session of communication - where some user identity is established, the user can then authenticate, use the service with a price plan, log out and so on. The service manages internally all the information requiring this functionality. The model required to define the automatic composition of stateful services is clearly more complex; and the arising problems are computationally harder as well, for example as proved in \cite{kila2011computational}. However, without considering states in the composition model, the applicability is drastically limited. Even the \emph{QoS}-aware model from Section \ref{sec:models:qos} cannot practically implement a \emph{cost} for service calls if the services do not have a method to authenticate users - as done for example in AWS services.

The \emph{Behavioral Description-based Web Service Description Model}, described for composition for example in \cite{kil2008computational}, or \cite{kil2013behavioural}, defines web services as complex structures, consisting of five parts:

\begin{enumerate}
 \item a set of \textbf{internal variables} that the service controls. A service state is described by a valuation of all its internal variables.
 \item a set of \textbf{input variables}, similar to input parameters in our previous models, through which the information is transmitted to the service at a service call.
 \item a set of \textbf{output variables}, used to communicate the result of a service call.
 \item an \textbf{initial predicate} over internal variables, used to describe the initial service state(s).
 \item a \textbf{transition predicate}, defined over triples of internal, input and \emph{primed} variable valuations. This describes the possible transitions: for a current service state and the input values with which the service is accessed; specifies through \emph{primed} $(\in$ internal) variables, possible successor states of the service. The \emph{transition predicate} can be deterministic (if there is only one successor state) or non-deterministic.
\end{enumerate}

Modeling service descriptions in such a manner enables the (formal) definition of automatic composition of stateful services. This summary description is intended to give a rough idea of how stateful services are modeled for composition, but for a complete definition, see the referenced bibliography.

Within this thesis, we do not propose contributions for stateful service composition model, its complexity or solution algorithms; as it falls out of our scope. Stateful composition models are too complex to generate approachable problems, and for this reason, they were not included in the Web Services Challenge series either.
}


\section{Online Composition Problem}
\label{sec:models:online}

\comment{ \textcolor{red}{[contribution]} Discuss less standard aspects of WSC, 
most important: \textbf{online} or \textbf{dynamic} problem. In this version, 
some \textbf{Composition Manager} unit handles user requests which are not 
transient as in the in classical model; but rather used over a period of time. 
This means that composition requests can be removed/dropped, added or updated. 
Moreover, the repository is dynamic: services can be added or deleted, and 
possibly, QoS metrics can get updated. As this is orthogonal with other 
dimensions of the problem, it can be analyzed in conjunction with other aspects 
like complexity, semantic expressiveness, etc. A separate sub-section of this 
will describe the model proposed in the paper published at ICSOFT 2019, and the 
algorithm in the following section.}

\iftoggle{fullThesis}{

In this section, we use the term \emph{online problem} to refer to problems that process input data continuously, therefore processing \emph{queries} that can either update some state or request some information about that state.

Web Service Composition can naturally be analyzed as an online problem. Obviously, web services themselves provide some functionality \emph{online}, and following SoA principles, their registration, updates, downtime, and eventual decommissioning; are dynamic events that can occur. Systems developed on top of services need to prepare for these events. This can be very important for automatic composition as well.

Currently, there is little research in this context for automatic composition. Some papers provide fault-tolerance for composite services, such as \cite{cardinale2011fault} and \cite{rao2007fault}; but they do not define the composition explicitly as an online problem. Also, they do not analyze the theoretical complexity of the operations involved but propose engineering solutions like replication of services.

\subsection{Maintenance of Compositions}

This online version of service composition is independent of any model described in Chapter \ref{sec:models}. This means that it can be used with any parameter matching model, semantic model, considering QoS metrics or not, etc. Regarding service states, the online composition model considers the service state just at some elementary level: the availability of services. More in-depth analysis of service states in the online model can be considered but is left for future work (for example, we can imagine a system that automatically changes compositions for user requests based on service response time updates).

The model definition below is presented in \cite{diac2019failover}, which proposes a failover mechanism for service composition. That is the main part of the proposed solution for the generic online problem.

In the Online Service Composition, we consider three types of actors: the service providers, the users requesting for compositions, and the composition generator. Usually generating the composition is a static method applied only once, generating  the composition and returning it to the user. 
But in real-world tough, most often, we expect that the user would run the resulting composition for a continuous period of time, over which the services  should remain available. Also, we can imagine that service providers can shut down services, or similarly, they can add services to registries, or update the service interfaces. All of these are actually frequently met in practice. Figure \ref{fig:online_composition_actors} displays this general scenario, with the composition engine positioned in the center - the component implementing the solution for the \emph{online composition problem}. The \emph{service monitor} component, part of the \emph{composition engine} verifies the service's health - and can detect if services break down.

\begin{center}
\begin{figure}
\centering
\caption[Actors of the \textbf{Online Composition}]{Actors involved in the \emph{Online Web Service Composition} problem.}
\label{fig:online_composition_actors}
\vspace{0.5cm}

\tikzset{every picture/.style={line width=0.75pt}} 

\begin{tikzpicture}[x=0.75pt,y=0.75pt,yscale=-1,xscale=1]

\draw (417.75,106.75) node  {\includegraphics[width=28.13pt,height=28.13pt]{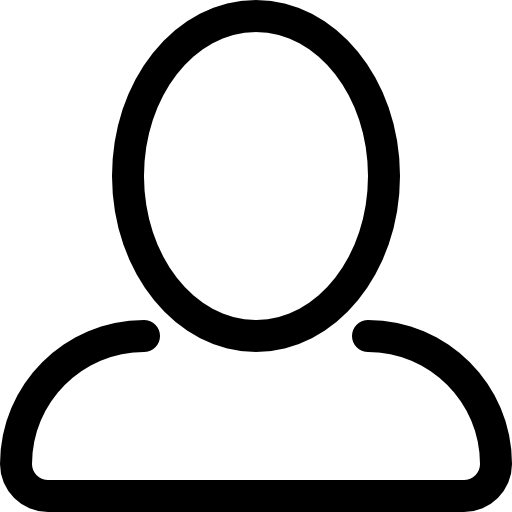}};
\draw (439.75,64.75) node  {\includegraphics[width=28.13pt,height=28.13pt]{images/actor.png}};
\draw  [line width=1.5]  (163,8.86) .. controls (163,7.75) and (163.9,6.86) .. (165,6.86) -- (326.5,6.86) .. controls (327.6,6.86) and (328.5,7.75) .. (328.5,8.86) -- (328.5,129) .. controls (328.5,130.1) and (327.6,131) .. (326.5,131) -- (165,131) .. controls (163.9,131) and (163,130.1) .. (163,129) -- cycle ;
\draw [line width=0.75]    (163,35) -- (329.5,35) ;

\draw  [line width=0.75]  (292.5,43.17) -- (292.5,62.37) .. controls (292.5,65.22) and (271.57,67.54) .. (245.75,67.54) .. controls (219.93,67.54) and (199,65.22) .. (199,62.37) -- (199,43.17)(292.5,43.17) .. controls (292.5,46.02) and (271.57,48.34) .. (245.75,48.34) .. controls (219.93,48.34) and (199,46.02) .. (199,43.17) .. controls (199,40.31) and (219.93,38) .. (245.75,38) .. controls (271.57,38) and (292.5,40.31) .. (292.5,43.17) -- cycle ;
\draw (75.75,76.75) node  {\includegraphics[width=28.13pt,height=28.13pt]{images/actor.png}};
\draw (96.75,121.75) node  {\includegraphics[width=28.13pt,height=28.13pt]{images/actor.png}};
\draw    (103.5,75) .. controls (139.95,86.82) and (121.09,64.68) .. (160.65,63.06) ;
\draw [shift={(162.5,63)}, rotate = 538.64] [fill={rgb, 255:red, 0; green, 0; blue, 0 }  ][line width=0.75]  [draw opacity=0] (8.93,-4.29) -- (0,0) -- (8.93,4.29) -- cycle    ;

\draw    (112.5,116) .. controls (131.22,118.96) and (124.7,93.77) .. (160.82,82.5) ;
\draw [shift={(162.5,82)}, rotate = 523.86] [fill={rgb, 255:red, 0; green, 0; blue, 0 }  ][line width=0.75]  [draw opacity=0] (8.93,-4.29) -- (0,0) -- (8.93,4.29) -- cycle    ;

\draw    (418.5,71) .. controls (380.88,65.06) and (381.48,110.09) .. (331.04,71.69) ;
\draw [shift={(329.5,70.5)}, rotate = 397.90999999999997] [fill={rgb, 255:red, 0; green, 0; blue, 0 }  ][line width=0.75]  [draw opacity=0] (8.93,-4.29) -- (0,0) -- (8.93,4.29) -- cycle    ;

\draw   (170,88) .. controls (170,87.45) and (170.45,87) .. (171,87) -- (218.79,87) .. controls (219.34,87) and (219.79,87.45) .. (219.79,88) -- (219.79,117.86) .. controls (219.79,118.41) and (219.34,118.86) .. (218.79,118.86) -- (171,118.86) .. controls (170.45,118.86) and (170,118.41) .. (170,117.86) -- cycle ;
\draw    (179.66,85.88) -- (207.34,67.12) ;
\draw [shift={(209,66)}, rotate = 505.89] [fill={rgb, 255:red, 0; green, 0; blue, 0 }  ][line width=0.75]  [draw opacity=0] (8.93,-4.29) -- (0,0) -- (8.93,4.29) -- cycle    ;
\draw [shift={(178,87)}, rotate = 325.89] [fill={rgb, 255:red, 0; green, 0; blue, 0 }  ][line width=0.75]  [draw opacity=0] (8.93,-4.29) -- (0,0) -- (8.93,4.29) -- cycle    ;
\draw    (202.25,85.44) -- (215.75,68.56) ;
\draw [shift={(217,67)}, rotate = 488.66] [fill={rgb, 255:red, 0; green, 0; blue, 0 }  ][line width=0.75]  [draw opacity=0] (8.93,-4.29) -- (0,0) -- (8.93,4.29) -- cycle    ;
\draw [shift={(201,87)}, rotate = 308.66] [fill={rgb, 255:red, 0; green, 0; blue, 0 }  ][line width=0.75]  [draw opacity=0] (8.93,-4.29) -- (0,0) -- (8.93,4.29) -- cycle    ;
\draw    (329.5,111.5) .. controls (362.65,135.39) and (374.89,108.42) .. (398.65,106.13) ;
\draw [shift={(400.5,106)}, rotate = 537.71] [fill={rgb, 255:red, 0; green, 0; blue, 0 }  ][line width=0.75]  [draw opacity=0] (8.93,-4.29) -- (0,0) -- (8.93,4.29) -- cycle    ;

\draw  [color={rgb, 255:red, 155; green, 155; blue, 155 }  ,draw opacity=1 ] (351.33,151.39) .. controls (351.15,148.22) and (353.57,145.49) .. (356.74,145.31) .. controls (359.91,145.12) and (362.63,147.54) .. (362.81,150.71) .. controls (363,153.88) and (360.58,156.6) .. (357.41,156.79) .. controls (354.24,156.97) and (351.52,154.56) .. (351.33,151.39) -- cycle ;
\draw  [color={rgb, 255:red, 155; green, 155; blue, 155 }  ,draw opacity=1 ] (375.7,139.93) .. controls (375.52,136.76) and (377.93,134.04) .. (381.1,133.85) .. controls (384.27,133.67) and (387,136.09) .. (387.18,139.26) .. controls (387.37,142.43) and (384.95,145.15) .. (381.78,145.33) .. controls (378.61,145.52) and (375.89,143.1) .. (375.7,139.93) -- cycle ;
\draw  [color={rgb, 255:red, 155; green, 155; blue, 155 }  ,draw opacity=1 ] (373.77,158.08) .. controls (373.58,154.91) and (376,152.19) .. (379.17,152) .. controls (382.34,151.81) and (385.06,154.23) .. (385.25,157.4) .. controls (385.43,160.57) and (383.01,163.29) .. (379.84,163.48) .. controls (376.67,163.67) and (373.95,161.25) .. (373.77,158.08) -- cycle ;
\draw  [color={rgb, 255:red, 155; green, 155; blue, 155 }  ,draw opacity=1 ] (392.38,150.97) .. controls (392.19,147.8) and (394.61,145.08) .. (397.78,144.89) .. controls (400.95,144.71) and (403.67,147.13) .. (403.86,150.3) .. controls (404.05,153.47) and (401.63,156.19) .. (398.46,156.37) .. controls (395.29,156.56) and (392.57,154.14) .. (392.38,150.97) -- cycle ;
\draw [color={rgb, 255:red, 155; green, 155; blue, 155 }  ,draw opacity=1 ]   (335.59,149.12) .. controls (346,140.58) and (332.17,123.61) .. (377.47,133.67) ;
\draw [shift={(378.86,133.98)}, rotate = 192.84] [fill={rgb, 255:red, 155; green, 155; blue, 155 }  ,fill opacity=1 ][line width=0.75]  [draw opacity=0] (8.93,-4.29) -- (0,0) -- (8.93,4.29) -- cycle    ;

\draw [color={rgb, 255:red, 155; green, 155; blue, 155 }  ,draw opacity=1 ]   (362.81,150.71) .. controls (369.11,146.84) and (368.25,146.65) .. (374.23,141.25) ;
\draw [shift={(375.7,139.93)}, rotate = 498.63] [fill={rgb, 255:red, 155; green, 155; blue, 155 }  ,fill opacity=1 ][line width=0.75]  [draw opacity=0] (8.93,-4.29) -- (0,0) -- (8.93,4.29) -- cycle    ;

\draw [color={rgb, 255:red, 155; green, 155; blue, 155 }  ,draw opacity=1 ]   (335.55,155.57) -- (349.4,151.9) ;
\draw [shift={(351.33,151.39)}, rotate = 525.15] [fill={rgb, 255:red, 155; green, 155; blue, 155 }  ,fill opacity=1 ][line width=0.75]  [draw opacity=0] (8.93,-4.29) -- (0,0) -- (8.93,4.29) -- cycle    ;

\draw [color={rgb, 255:red, 155; green, 155; blue, 155 }  ,draw opacity=1 ]   (356.42,156.88) .. controls (358.37,161.97) and (366.11,162.7) .. (376.93,163.44) ;
\draw [shift={(378.85,163.57)}, rotate = 183.91] [fill={rgb, 255:red, 155; green, 155; blue, 155 }  ,fill opacity=1 ][line width=0.75]  [draw opacity=0] (8.93,-4.29) -- (0,0) -- (8.93,4.29) -- cycle    ;

\draw [color={rgb, 255:red, 155; green, 155; blue, 155 }  ,draw opacity=1 ]   (387.18,139.26) .. controls (389.93,137.43) and (401.82,137.37) .. (413.85,139.31) ;
\draw [shift={(415.75,139.63)}, rotate = 190.13] [fill={rgb, 255:red, 155; green, 155; blue, 155 }  ,fill opacity=1 ][line width=0.75]  [draw opacity=0] (8.93,-4.29) -- (0,0) -- (8.93,4.29) -- cycle    ;

\draw [color={rgb, 255:red, 155; green, 155; blue, 155 }  ,draw opacity=1 ]   (384.39,160.58) .. controls (387.59,165.7) and (393.17,161.64) .. (397.61,157.53) ;
\draw [shift={(399.04,156.17)}, rotate = 495.93] [fill={rgb, 255:red, 155; green, 155; blue, 155 }  ,fill opacity=1 ][line width=0.75]  [draw opacity=0] (8.93,-4.29) -- (0,0) -- (8.93,4.29) -- cycle    ;

\draw [color={rgb, 255:red, 155; green, 155; blue, 155 }  ,draw opacity=1 ]   (403.86,150.3) -- (414.27,148.04) ;
\draw [shift={(416.22,147.61)}, rotate = 527.76] [fill={rgb, 255:red, 155; green, 155; blue, 155 }  ,fill opacity=1 ][line width=0.75]  [draw opacity=0] (8.93,-4.29) -- (0,0) -- (8.93,4.29) -- cycle    ;

\draw (402,25) node  [align=left] { \ \ \ \ \ \ \ \ \ users\\(service consumers)};
\draw (454,97) node  [align=left] {\textbf{. . .}};
\draw (245,22) node  [align=left] {\textbf{composition engine}};
\draw (247,56.96) node  [align=left] {{\small repository}};
\draw (85,25) node  [align=left] { \ service \\providers};
\draw (60.5,113) node  [align=left] {\textbf{. . .}};
\draw (130.31,60.95) node [rotate=-334.2] [align=left] {{\footnotesize service}};
\draw (124.96,48.7) node [rotate=-334.2] [align=left] {{\footnotesize register}};
\draw (140.23,122.93) node [rotate=-320.23] [align=left] {{\footnotesize service}};
\draw (133.34,114.09) node [rotate=-320.23] [align=left] {{\footnotesize remove}};
\draw (195,104) node  [align=left] {{\footnotesize service}\\};
\draw (195,116) node  [align=left] {{\footnotesize monitor}\\};
\draw (275,75) node [scale=0.8] [align=left] {{\footnotesize service registration}};
\draw (277,87) node [scale=0.8] [align=left] {{\footnotesize generate composition}};
\draw (275,99) node [scale=0.8] [align=left] {{\footnotesize delete service}};
\draw (277,111) node [scale=0.8] [align=left] {{\footnotesize drop compositon}};
\draw (272,121) node [scale=0.8] [align=left] { . . .};
\draw  [color={rgb, 255:red, 155; green, 155; blue, 155 }  ,draw opacity=1 ]  (317.55,145.73) .. controls (317.55,145.18) and (318,144.73) .. (318.55,144.73) -- (334.35,144.73) .. controls (334.9,144.73) and (335.35,145.18) .. (335.35,145.73) -- (335.35,162.47) .. controls (335.35,163.02) and (334.9,163.47) .. (334.35,163.47) -- (318.55,163.47) .. controls (318,163.47) and (317.55,163.02) .. (317.55,162.47) -- cycle  ;
\draw (326.45,154.1) node [scale=0.8,rotate=-356.63] [align=left] {in};
\draw  [color={rgb, 255:red, 155; green, 155; blue, 155 }  ,draw opacity=1 ]  (416.2,136.45) .. controls (416.2,135.9) and (416.65,135.45) .. (417.2,135.45) -- (440.99,135.45) .. controls (441.54,135.45) and (441.99,135.9) .. (441.99,136.45) -- (441.99,153.66) .. controls (441.99,154.21) and (441.54,154.66) .. (440.99,154.66) -- (417.2,154.66) .. controls (416.65,154.66) and (416.2,154.21) .. (416.2,153.66) -- cycle  ;
\draw (429.1,145.05) node [scale=0.8,rotate=-356.63] [align=left] {out};
\draw (365.49,56.78) node [rotate=-2.17] [align=left] {{\footnotesize composition}};
\draw (363.84,67.51) node [rotate=-2.17] [align=left] {{\footnotesize (in, out)}};
\draw (365.44,46.78) node [rotate=-2.17] [align=left] {{\footnotesize find}};
\draw (356.63,150.44) node [rotate=-354.59] [align=left] {{\tiny si}};
\draw (381.66,139.03) node [rotate=-354.59] [align=left] {{\tiny sj}};
\draw (379.38,157.33) node [rotate=-354.59] [align=left] {{\tiny sk}};
\draw (397.73,149.57) node [rotate=-354.59] [align=left] {{\tiny sl}};
\end{tikzpicture}

\end{figure}
\end{center}

In Section \ref{sec:models:online:example}, we will present an detailed example with a dynamic repository of services and two composition requests. Removal of a service affects one of the two composition responses.

\subsection{Formal Definition}
We formally summarize in the following the elements of the model and present its basic mode of operation. For simplicity, we will use the initial, simplest parameter matching by name. Also, we do not use QoS values, and only favor shorter compositions to longer ones, counting only the total number of services used.

\textbf{Parameters.} Let $\mathbb{P}$ be the set of all parameters, identified by names, that appear in any service definition or in the user request. $\mathbb{P}$ is not restricted in any way, and it can include any string.

\textbf{Web Services.} A service is identified by a name and contains two sets of parameters, input, and output: $ws_{i} = \langle name,In,Out\rangle$, where $In, Out \subseteq \mathbb{P}$.

\textbf{Service Repository.} The service repository, written as $\mathbb{R}$, is the set of all services, $\mathbb{R} = \bigcup\limits_{\forall i} ws_{i}$.

\textbf{Composition Query.} A \emph{user request} for a composition or a \emph{composition query}  can be thought of as a service that does not exist yet, but the user has an interest in it. It is defined over a set of input parameters that the user initially knows and a set of requested (output) parameters that the user needs. All active queries $q_{x} = \langle name,In,Out\rangle$;  $In, Out \subseteq \mathbb{P}$ make the set $\mathbb{Q}$.

\textbf{Valid Composition.} For query $q_{x}$, a valid service composition, is a sequence of services \emph{$\langle ws_{1}, ws_{2}, ... ws_{k}\rangle$}, where: each service input set must be included in the set of all previously called services outputs, or in the user initially known parameters, and the final user query output must be covered by services outputs.

\newcommand\bigforall{\mbox{\Large $\mathsurround7pt\forall$}}

\begin{equation*}
\begin{split}ws_{i}.In \subseteq \Bigg( q_{x}.In \cup \bigg(\bigcup_{j=1}^{i-1} ws_{j}.Out\bigg) \Bigg), \bigforall i = \overline{1..k} \ \ \
\vspace{-0.3cm} \\
and \ \ \ q_{x}.Out \subseteq \Bigg(\bigcup_{i=1}^{k} ws_{i}.Out\Bigg)
\end{split}
\end{equation*}

\textbf{Online Web Service Composition Problem}. Following the notations above, a solution for the online version of the problem has to dynamically resolve the series of operations:

\begin{itemize}
	\item \emph{registerService($ws_i$)} - a service developed adds a service to the registry. Ideally, such an operation can modify existing compositions as well, that were found at an earlier time. For example, if considering $ws_i$ would make some compositions shorter, or if some compositions requests were unsolvable and became solvable.
	\item \emph{removeService($ws_i$)} - a service developer stops providing a service. At least, all compositions that basted on $ws_i$ need to be re-computed (their solutions can increase in length or they can become unsolvable).
	\item \emph{detectServiceDown($ws_i$}) - the internal service monitor detects that a service broke. In this case, the composition engine can only implement the same strategy as for the \emph{removeService($ws_i$)} above, eventually notifying the service developer as well. If more elaborate QoS metrics are considered, the service monitor would detect such situations as well.
	\item \emph{findComposition($q_x$)} - a service consumers requests for a new composition. On top of providing a solution if it exists, the composition manager has to store the request, response, and the client internally, to be able to notify the respective user when relevant services become inaccessible and to provide alternate compositions.
	\item \emph{dropCompositionRequest($q_x$)} - as compositions are maintained, the users requesting compositions can also notify if some composition request is no longer used.
\end{itemize}

The operations above are the most important for the online version of the problem, but other useful operations can be imagined. For example, if a service is detected to exhibit problems (is unavailable or has delayed response times), in many cases other services developed by the same provider - are most probably affected as well. For example, this can happen because of services from one provider can be deployed on the same server, or have some other common dependency. In such a case, the service monitor can notify the composition engine of this exact situation and appropriate strategies can be implemented.

Similarly, in the case of particular QoS metrics, the composition engine has to implement intelligent strategies to respond to queries. For example, the 
\emph{throughput} of a web service is a particular metric if clients concurrently use the service. If more requests use the same service, the total number of service invocations should not exceed the service's throughput.

It is easy to imagine other aspects as the ones presented above. Therefore, the \emph{online} perspective of the automatic composition opens many new research directions. Our failover strategy presented in paper \cite{diac2019failover} and, more generally in the complexity Section \ref{sec:complexity:online} is just the first step in this direction.

\subsection{Updating Compositions Example}
\label{sec:models:online:example}

A more detailed example of the \emph{online} version of the service composition problem is described below. Name matching is used for the parameter model as defined in Section \ref{sec:models:names}, a dynamic repository of eight services is considered and two composition requests are maintained. Three steps or stages are shown in Figure \ref{fig:online_example} and described below:

\begin{enumerate}
  \item in this state the two composition requests are satisfied. First, \emph{getDrivingConditions} can be resolved by the following sequence of six services: \emph{locatePhone}, \emph{getWeather}, \emph{getLatLon}, \emph{getMap}, \emph{nearbyStreet}, and \emph{trafficInfo}. Secondly, the \emph{getCityMap} service can be solved by a composition of only three services: \emph{getCityDistrict}, \emph{getCityCenter}, and \emph{getMap}.
  \item then, suppose one of the services, \emph{getLatLon}, is removed or breaks, either of the two having the same effect on the composition engine.
  \item afterward, the composition solving the \emph{getCityMap} request is not affected as it does not include the inaccessible service. But the composition for the \emph{getDrivingConditions} request is affected. However, \emph{getLatLon} can be replaced in this context with service \emph{getCityCenter}. Even if the services are not identical they both satisfy parameter matching restrictions. The new composition for \emph{getDrivingConditions} request is: \emph{locatePhone}, \emph{getWeather}, \emph{getCityCenter}, \emph{getMap}, \emph{nearbyStreet}, and \emph{trafficInfo}.
\end{enumerate}

\begin{figure}[H]
  \caption[Failover compositions after \textbf{Service Remove}]{\textbf{query} and \textbf{update} operations interleaved on \textbf{repository} and \textbf{requests}.\\Composition for \textbf{getDrivingConditions} changes after \textbf{getLatLon} is deleted.\\More details are given in the above Section \ref{sec:models:online:example}.}
  \vspace{0.5cm}
  \includegraphics[width=1.1\linewidth]{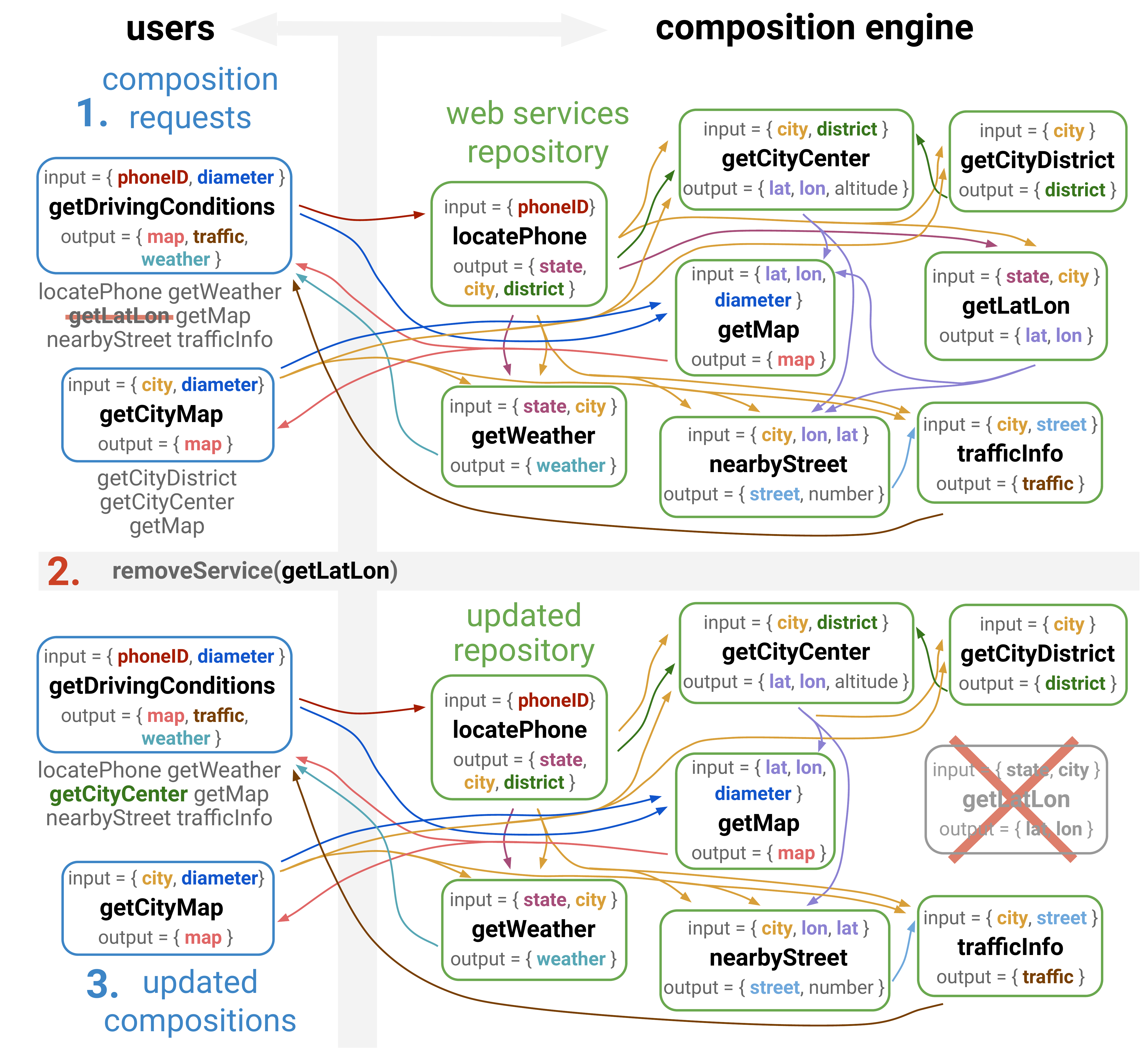}
  \label{fig:online_example}
\end{figure}

\section{Conclusion}

Apart from the models and all the elements described in Chapter \ref{sec:models}, many others can be considered. Automatic Service Composition has many aspects and all of them are potential interesting research topics. If we investigate practical applications as in Chapter \ref{sec:applications}, other situations arise as well, for example, many service definition languages allow \emph{optional} input parameters, without formally describing the semantics behind, which is a requirement for automation. Also, security can be an important topic in service composition, like in any distributed system that exposes data over the network. Out of all these, we have chosen the models to analyze based on two reasons: if they are computationally interesting (resulting in optimization or computationally hard problems) or if the existing models required some additions to be able to express different constraints, present in manual composition (when the composition is constructed by a human person) on cases that we believe are important.

Regarding potential future work, for the composition models, there are many possibilities of continuation. Another contribution to this topic is the API and data model designed for a modern composition framework. It is described in Section \ref{sec:applications:formalFramework} and uses elements from almost all of the presented models: the \emph{Relational} and \emph{Object-Oriented} models, with contextual knowledge of objects with types organized \emph{hierarchically}. Clearly, as future work, it can be extended to also include interesting \emph{QoS} properties, \emph{stateful} information about services or compositions, in particular: maintain compositions as in the \emph{online} version and also consider potential security weaknesses. All of these can be combined all together or just some of them; and the most practical contribution is to implement such a framework and make service developers use it.

In this chapter, we proposed several composition problem models and in the next chapter, we will propose solutions and algorithms for each of them.

}

%
\chapter{\underline{Complexity} \ of Composition Models}
\label{sec:complexity}
\comment{Unlike the previous chapter, all subsections present, in a greater amount personal contributions. \\ Iterate variants presented in the previous chapter and discuses the computational complexity of the arising problems, on each operation, phase. Algorithms are also presented here.}

\iftoggle{fullThesis}{
This chapter provides a detailed analysis of the \emph{computational} complexity of various versions and operations of the automatic composition problem. It also presents statistics about the optimization problem, in which the number of services used is reduced, which can be seen as an approximation for the running time or cost. Algorithms are first presented as proofs of the upper bound measure of complexity, and in Section \ref{sec:complexity:algorithms} they are evaluated relative to the solutions that won in the challenges organized in 2005 and 2008. Algorithms solving the newly proposed models are evaluated on benchmarks artificially created for these models. They cannot be compared with other algorithms since there are no other algorithms yet to solve the new models. For the synthetic benchmarks, we also describe how our generators simulate all functionality, even edge cases and increase the run-time for searching compositions, resulting in the intended worst-case scenarios.

The next two sections avoid the detailed description of algorithms and focus on the theoretical complexity of the underlying problems, on different models: Section \ref{sec:complexity:upper} for the upper bound measure of complexity, where a short description of the algorithms is provided; and Section \ref{sec:complexity:lower} for the more theoretical lower bound.
}

\section{Upper Bounds - Overview}
\label{sec:complexity:upper}
\comment{Briefly presents algorithms solving the defined problems, their complexity and some resulting conclusions.}

\iftoggle{fullThesis}{
Generally, the upper bound is a measure of the computational complexity of problems, that is the complexity of the fastest known algorithm that solves the given problem. We will analyze the run-time complexity of the best-performing algorithms (generally, i.e. the overall performance), that we propose in Section \ref{sec:complexity:algorithms}. For some given operation, for example, building a composition, this can potentially be the effective upper bound, but it also depends on the requirement that the composition should be: shortest possible, preferable shorter, or if no requirement is specified in this regard. The algorithms are designed to perform the best relative to the challenges evaluation criteria, that consider the composition length, and the choice is usually to reduce the composition size but without significantly increasing the theoretical complexity or the run-time. For most of the algorithms, we also analyze the pre-processing phase efficiency.

\subsection{Name Matching Model}
The simplest model for parameter matching leads to the computationally-easiest version of the problem - thus, the fastest algorithm, relative to the number of services. Also, at the time of designing the algorithm for this model, which is in 2017, in \cite{diac2017engineering}, our focus was on low-level optimizations and almost only on producing the fastest algorithm. This is beyond theoretical complexity as constant-factors were also considered, programming language, read operations and others.

The Algorithm \ref{algo:fornamematch} maintains a set of \emph{known parameters} that gradually increases; and for each web service, a set of remaining required input parameters that are yet unknown. Once such a set becomes empty, the service is marked as available to be called, i.e. \emph{callable}. The next service to call can be chosen without any criteria, or as in Algorithm \ref{algo:fornamematch}, guided by a heuristic score that estimates the usefulness of services. In our implementation, including this score was done with almost no drawback to efficiency. Shortly, the algorithm is similar to a breadth-first traversal of a graph with services as nodes and accessibility guided by input requirements. In BFS, one node can be selected when it is neighbor of a node visited already; like a service that can be called if all its inputs are known.

Following the notations in Section \ref{sec:models:names}: if $|\mathbb{R}| = $ the size of the repository, or the number of services; and $P = $ the total number of parameters, i.e. the sum of the number of parameters of each service, both input, and output, we were able to design an algorithm of \textbf{$O(|\mathbb{R}|+P)$} run-time complexity. This is the resulting complexity if: (1) we do not consider the representation of parameter names as strings, but consider them as numbers, or pre-compute their hash values; and (2) we do not try to reduce the composition size. If we do reduce the composition size, in a way that proved very effective on benchmarks, the complexity is increased only to $O(|\mathbb{R}| \cdot log|\mathbb{R}|+P)$, because of a priority queue operation. The first linear complexity is clearly optimal, as it is the problem input size.

We did not implement an approach to guarantee the shortest composition, because the optimization problem is $\texttt{NP-Hard}$ as shown in Section \ref{sec:complexity:lower}. However, besides the scoring mechanism, we also implemented a reduction phase, that essentially drops the unused services which may be added to the composition for various reasons. If applied repeatedly, this can increase the complexity, and we do not consider it in the theoretical complexity analysis. The number of reductions applied is limited by the number of services,  $|\mathbb{R}|$, but in practice, only a few reductions were applied on test cases (more details in the description of the algorithm). One reduction runs in amortized linear time.

In Sections \ref{sec:complexity:upper} and Section \ref{sec:complexity:lower}, the focus is on the complexities of different problem models. The proposed algorithms are summarized in Section \ref{sec:complexity:upper} and will be detailed in Section \ref{sec:complexity:algorithms}.

\subsection{Hierarchical Model}

The main contribution of the paper \cite{tucar2018semantic} is proving that the hierarchical model, at least as introduced in the 2008 challenge, does not increase the computational complexity. This was done by a pre-processing phase. Shortly, a Euler-Tour traversal of the hierarchy helps to reduce the execution time of finding sub/super-concept relations to $O(1)$. This, together with other more commonly used data structures and optimizations, made it possible to implement a very efficient algorithm, that again, uses a heuristic score to reduce the composition size.

Another metric introduced by the 2008 challenge is the \emph{execution path}, that counts the number of \emph{levels} that a composition includes, where at a \emph{level} multiple services are executed in parallel. This considers that the execution time does not add in parallel execution. However, since, hypothetically, the maximum execution time of parallel processes is relevant, this is a better approximation of the total execution time. Finding the composition with optimum \emph{execution path} is solvable in polynomial-time, and our proposed Algorithm \ref{algo:hierarchical} computes such a composition.

The overall complexity of the implemented algorithm, following the notations in Section \ref{sec:models:hierarchical} is $O(|\mathbb{C}| + |\mathbb{R}| \cdot log|\mathbb{R}| + P)$ if we ignore the optional reduction phase; where $|\mathbb{C}|$ is the number of concepts, $\mathbb{R}$ the repository and, again, $P$ the total number of parameters of services. We state this as an upper bound, as our algorithm has a much faster run time.

\subsection{Relational Model}

The relational model was designed to significantly increase the expressiveness of how input parameters are declared and matched, to allow more types of restrictions on service calls. A consequence of adding binary relations over the set of parameters in the model is the fact that now parameters with the same name can potentially appear in different relational contexts, thus adding the need to distinguish between \emph{instances} of the same concept. This has an important effect on the complexity as well. Even without it, the problem becomes $\texttt{NP-Complete}$ even for a single service parameter match (finding whether a service is callable in a given knowledge state); as it will be shown later. But if we consider that, now, there are multiple instances of a concept, the problem is even harder - as the number of generated instances is theoretically unlimited. In this case, it may be impossible to say that an instance has a solution or not, for specific problem instances. However, in practice, this issue was solved by limiting the number of instances of the same concept, according to their relational context.

The high-level description of the algorithm is broadly similar to previous algorithms solving the first two versions of composition models. It starts with the initially known \emph{knowledge state} and at each step in a loop tries to call a service with a new \emph{parameter match} - a new set of concept instances matched to inputs. The state contains the instances known of each concept type, with associated relations to other instances. The knowledge is now similar to a labeled directed graph unlike before where it was a simple set. To match parameters to a given service, and backtracking procedure is implemented. The procedure adds any instance matching the concepts of each input parameter at a time and checks the matching relations with parameters and instances already matched. To match a web service \emph{ws} to a knowledge state $\mathbb{K} = \langle \mathbb{O}, \mathbb{R} \rangle$ it takes a worst-case run-time of $O(|\mathbb{O}|^{|ws.I|} \cdot |ws.I|^{2})$ since all possible objects can be matched on any input parameter, and verifying relations can take $O(|ws.I|^{2})$. Obviously, in practice, this runs much faster because the relations are verified at each level and large portions of the execution tree are pruned. Also, type matches are checked as well: candidate instances matched to parameters must be sub-types of the type of the parameter. More details about the theoretical complexity of the relational model are less important, other than that the run time is exponential.

\subsection{Object-Oriented Model}

The Object-Oriented model was designed with two purposes: to translate the automatic service composition problem to modern service definition formats (like OpenAPI, schema.org, REST services); and to be able to model the composition on known examples where the previous model failed. It was designed during the same period as the Relational Model, and at that time they were somehow different solutions to the same problem: lack of expressiveness of traditional or hierarchical models. The so-called Object-Oriented approach was inspired by modern web standards, most notably the \textbf{schema.org} data model; but at the design time, the complexity was not a concern at all. The important part was designing a service composition model that can actually apply to real-world scenarios. The nice surprise about the Object-Oriented model was that it kept the polynomial complexity upper bound, even if no compromise was intended in this regard. The proposed algorithm (which will be detailed in Section \ref{sec:complexity:objects}) has a run time complexity of $O\big(\sum \{$number of properties of $c $ $|$ $ c \in \mathbb{C}\} + \sum \{|ws.I|+|ws.O| \big|$ $ ws \in \mathbb{R}\} \big)$, i.e. linear in the input size, if the composition length is not minimized, like in our implementation. The conclusion is that the complexity is not increased, more than the \emph{resolution} of the \emph{knowledge state} is increased: from single concepts to concept properties.

\subsection{Online Problem Operations}
The Online version of the Composition Problem defines new operations, other than the traditional and single \emph{findComposition} operation. In the online or dynamic problem, multiple composition requests can be created and removed, as well as services. The system or the algorithm maintains a list of composition requests with associated composition solutions if existent; and the dynamic repository of services. In our work published in paper \cite{diac2019failover}, the algorithm prepares for a fail-over solution for the scenario in which a service brakes that may be used in some compositions. If a backup is ready, the affected users are notified of the outage and provided with alternative composition. The parameter model used was the initial model, with trivial name-based matching. Thus, the complexity of fining the first composition for a query is the same, $O(|\mathbb{R}| + P)$, but once the  composition is returned to the user, a background thread searches for backups for the composition. If the composition length is $\texttt{L}$, then at most $\texttt{2$\cdot$L}$ new composition queries are processed. Therefore the backups search can take $O(|\mathbb{R}| \cdot (|\mathbb{R}| + P))$ but the execution is asynchronous. Creating the known and required parameters for these queries is implemented efficiently, in amortized linear complexity.

Once a web service breaks, the complexity of switching to backup composition - if existent - is also implemented efficiently, in $O(U)$, where $U$ is the number of compositions using that service. This is trivial since we maintain an index of all the usages of each service. Of course, a new search for backups is initiated in the background - that searches for backups for the new active composition, which was a backup itself previously. We did not yet found an efficient solution to the operation of adding a new service - other than iterating all existing composition requests and processing a new query for each. This is part of future work.

But the important improvement of the backup scheme is that when a used service breaks, for the queries for which backups exist, the user should notice no delay spent in searching new compositions, as they are provided with the pre-computed backups instantly. This is not measured by the theoretical complexity analysis but would be very important in practice.
}

\section{Lower Bounds}
\label{sec:complexity:lower}
\comment{Some reductions proving lower bounds of specific problems.
Reminder: also discuss the use of planning, answer set programming and why on most common composition definitions they increase the worst-case running time and are, in personal opinion, an inappropriate choice.}

\iftoggle{fullThesis}{
In the previous Section \ref{sec:complexity:upper} we shortly presented some theoretical complexity analysis of the algorithms designed for several composition models considered; and in the following section, we will present the algorithms themselves, together with some empirical evaluation. From the perspective of \emph{problem complexity} however, these provide measures of how efficient or fast can particular models be solved. The opposite measure of problem complexity is usually referred to as  \emph{lower bound}. It defines the limits for which there are proofs showing that problems cannot be solved faster than those limits. In this section, we present a few results in this direction. In theory, together with upper bounds, these give a view on how close the research community is to \emph{closing} a problem, i.e. proving that a problem cannot be solved faster than the run-time of an algorithm known to solve that problem. Our analysis is less precise but the following complexity results give some highlights, mainly proving that some (sub)problems are \texttt{NP-Complete}.

First of all, for the initial, raw version of composition: with the simplest \emph{name-matching} parameters and without consideration on the length of the resulting composition, we proposed an algorithm of \textbf{$O(|\mathbb{R}|+P)$} run-time complexity. Since $|\mathbb{R}|$ is the number of web services, that need to be read at least; and their configuration as well: $P$ - total number of parameters; this algorithm is clearly optimal. This is because the asymptotic complexity is equal to the size of the input. This result is similar to the one for the hierarchical version of the composition and the object-oriented model; we presented algorithms that are linear or close to linear run-time complexities relative to the problem instances size.

The polynomial-time complexity algorithm for the name match model was the initial starting point of the thesis. In paper \cite{zou2014dynamic}, a solution is proposed which reduces service composition instances to planning instances, i.e. A.I. planning in PDDL format, that are then solved by a generic planner. This is intriguing as we propose a linear algorithm for the same composition problem, and almost linear heuristic to reduce composition size with good results in practice; while planning is \texttt{PSPACE-complete} generally \cite{mcdermott1998pddl}. Clearly, reducing instances of an easier problem to a harder problem (computationally) does not seem optimal. However, planning solvers are studied a lot and are optimized with many techniques, so they were able to outperform in 2014 the solutions presented at the competition in 2005. But similarly, our proposed algorithm had much better results than the planning-based approaches. In general, it is not clear either if the plan solvers are trying to reduce the plan size, i.e. the number of actions used, which translates into the number of web services in the composition. Indirectly, this is probably true since they reduce the execution time, which is most often proportional to plan size. Other advantages for planning-based approaches are: they do not require the implementation of an algorithm, but only a "reduction" / conversion of inputs and outputs; and perhaps more importantly, they are more suitable to be adapted to stateful versions of composition (discussed in Section \ref{sec:models:stateful}).

\subsection{Shortest Composition is \texttt{NP-Hard}}
\label{sec:complexity:lower:shortest}
\textbf{Set Cover Problem:} For a universe (set of elements) $U = \{1, 2, ... n\}$; and a collection $S$ of $m$ sets whose union equals the universe, find the \textbf{smallest} set of sets from $S$ which \emph{covers} the universe: their union is $U$. The decision problem associated to \textbf{Set Cover} is one of the Karp's 21 \texttt{NP-Complete} problems, shown to be \texttt{NP-Complete} in 1972.

\textbf{Set Cover} reduction to Web Service Composition (the version where the shortest composition is required, namely \textbf{Min-WSC}). As exemplified in Figure \ref{fig:reduce_setcover}, each set \textbf{S}$_i$ is transformed into a web service \textbf{ws}$_i$ with empty input and a parameter at the output for each element from the set, with a corresponding name. The user request also has an empty set as the set of initially known parameters and requires finding all parameters with names from the elements of the universe. The reduction is linear so an algorithm that would guarantee the shortest composition in linear time, would also resolve the \textbf{Set Cover} in polynomial-time, which is possible only if \texttt{P = NP}.

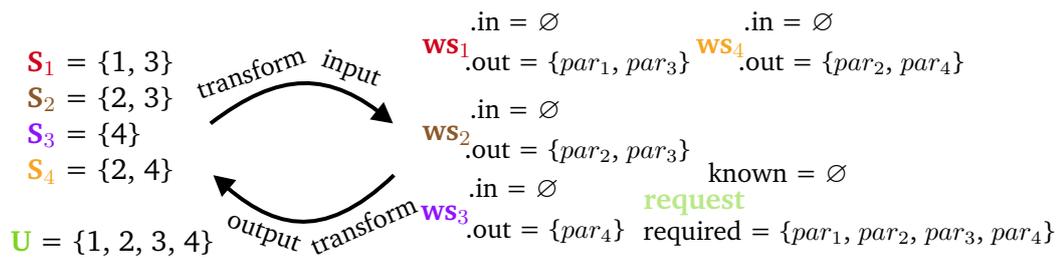
\begin{figure}[h]
\caption[Example of \textbf{SetCover} reduction to \textbf{Min-WSC}]{Example transformation of a \textbf{SetCover} instance to a \textbf{Min-WSC} instance. Shortest composition yields the smallest cover set.}
\label{fig:reduce_setcover}
\begin{center}

\tikzset{every picture/.style={line width=0.75pt}} 

\begin{tikzpicture}[x=0.67pt,y=0.75pt,yscale=-1,xscale=1]

\draw [line width=1.5]    (114.76,64) .. controls (153.25,37.54) and (181.44,37.01) .. (214.72,62.41) ;
\draw [shift={(216.76,64)}, rotate = 218.32] [fill={rgb, 255:red, 0; green, 0; blue, 0 }  ][line width=1.5]  [draw opacity=0] (11.61,-5.58) -- (0,0) -- (11.61,5.58) -- cycle    ;

\draw [line width=1.5]    (217.27,90) .. controls (176.5,125.1) and (148.33,116.47) .. (118.59,91.92) ;
\draw [shift={(116.29,90)}, rotate = 400.36] [fill={rgb, 255:red, 0; green, 0; blue, 0 }  ][line width=1.5]  [draw opacity=0] (11.61,-5.58) -- (0,0) -- (11.61,5.58) -- cycle    ;

\draw (60,79) node  [align=left] { \ \ \textbf{\textcolor[rgb]{0.82,0.01,0.11}{S$_1$}} = \{1, 3\}\\ \ \ \textbf{\textcolor[rgb]{0.55,0.34,0.16}{S$_2$}} = \{2, 3\}\\ \ \ \textbf{\textcolor[rgb]{0.56,0.07,1}{S$_3$}} = \{4\}\\ \ \ \textbf{\textcolor[rgb]{0.96,0.65,0.14}{S$_4$}} = \{2, 4\}\\\\\textbf{\textcolor[rgb]{0.49,0.83,0.13}{U}} = \{1, 2, 3, 4\}};
\draw (247,26) node  [align=left] {\textbf{\textcolor[rgb]{0.82,0.01,0.11}{ws$_1$}}};
\draw (285,13) node  [align=left] {{\small .in =} $\varnothing$};
\draw (320,34) node  [align=left] {{\small .out = \{$par_1$, $par_3$\}}};
\draw (247,70) node  [align=left] {\textbf{\textcolor[rgb]{0.55,0.34,0.16}{ws$_2$}}};
\draw (285,57) node  [align=left] {{\small .in =} $\varnothing$};
\draw (320,78) node  [align=left] {{\small .out = \{$par_2$, $par_3$\}}};
\draw (246,110) node  [align=left] {\textbf{\textcolor[rgb]{0.56,0.07,1}{ws$_3$}}};
\draw (284,97) node  [align=left] {{\small .in =} $\varnothing$};
\draw (302,118) node  [align=left] {{\small .out = \{$par_4$\}}};
\draw (400,26) node  [align=left] {\textbf{\textcolor[rgb]{0.96,0.65,0.14}{ws$_4$}}};
\draw (438,13) node  [align=left] {{\small .in =} $\varnothing$};
\draw (473,34) node  [align=left] {{\small .out = \{$par_2$, $par_4$\}}};
\draw (385,104) node  [align=left] {\textbf{\textcolor[rgb]{0.72,0.91,0.53}{request}}};
\draw (432,89) node  [align=left] {{\small known =} $\varnothing$};
\draw (474,119) node  [align=left] {{\small required = \{$par_1$, $par_2$, $par_3$, $par_4$\} }};
\draw (137.11,38) node [scale=0.9,rotate=-340.77] [align=left] {transform};
\draw (193.18,37) node [scale=0.9,rotate=-21.28] [align=left] {input};
\draw (200.19,115) node [scale=0.9,rotate=-336.71] [align=left] {transform};
\draw (144.11,120) node [scale=0.9,rotate=-20.13] [align=left] {output};
\end{tikzpicture}

\end{center}
\end{figure}

Therefore, finding the shortest composition is \texttt{NP-Hard} on the simple name matching model. This extends to most of our other models: for the optimization problem, requesting the shortest composition. This is true because, extending the models with a hierarchy of concepts, relations or concept properties can be particularized in such a way that any instance of the name-matching model is solvable through enhanced models. For example, a flat hierarchy of concepts, where no subsumption would be possible, can model any instance of the name-matching case. Likewise, the relational model can work without relations and the object-oriented model without properties.

\subsection{Relational Parameter Matching is \texttt{NP-Complete}}

In the relational model, presented in Section \ref{sec:models:relational}, parameters represented by concepts are enhanced with a set of binary relations. In the problem definition, multiple instances of the same concept are allowed, distinguished by their relational context. At some stage of the composition, the \textbf{knowledge state} is defined by the set of all instances of concepts, together with the relations defined over them. The input parameters of a Web Service are defined likewise: a set of types, with all relations between them required to call the service. To verify if a match exists: assigning known instances to each parameter, such that all relations in the service definition are satisfied i.e. matched; is a particular case of \emph{labeled subgraph isomorphism}. Subgraph isomorphism is \texttt{NP-Complete} \cite{cordella2004sub}. The instance-to-parameter matching problem is more general because: (1) the relations are \emph{directed} edges, and (2) an instance and parameters are typed, with types organized in the hierarchy of concepts, with the polymorphism implications. To prove that \texttt{NP-Complete} complexity holds, (2) is not an issue because: a dummy hierarchy of a \emph{single node/concept} simulates the non-hierarchical model. More precisely, since for the proof we are reducing \emph{labeled subgraph isomorphism} to \emph{relational parameter matching}, and the nodes of the graphs can match without any type restriction (any node to any node), a single-node tree of concepts, with all instances and parameters of that node's type, will represent the isomorphism problem case. The first difference (1), the directed relations, is also a generalization of the un-directed edges in graphs, as they can simply be replaced by two edges covering both directions, with the same label. Therefore, the problem of verifying just if a Web Service can be validly called in a knowledge state is \texttt{NP-Complete}. The general composition problem is obviously at least as hard since it requires applying the match at any stage of the composition, and moreover properly choosing the service to match as well.

The heuristic score-based approaches presented to shorten the composition, for example in Section \ref{sec:complexity:algorithms:name_heuristic}, give good results on challenge benchmarks, other tests generated by tools such as WS-Ben or proposed generator algorithms. However, it would be interesting to find if there is a limit on the error rate for the proposed heuristic or similar solutions. This is left for future work.

}

\section{Algorithms and Performance on Challenges and Self Assessment} 
\label{sec:complexity:algorithms}
\comment{Describe with more details; the work done in the two papers presented at KES 2017 and 2018, for the Web Services Challenges: 2005 and 2008 editions. Many things to write here ... for example, also talk about the test generators and why they are relevant, i.e. tests are much harder computationally compared with other generators.}

\iftoggle{fullThesis}{

The first two stages of our research are the implementations of efficient algorithms aiming to provide the best performing solutions to the 2005 \cite{blake2005eee} and 2008 \cite{bansal2008wsc} challenges. These are presented in detail in papers \cite{diac2017engineering} and \cite{tucar2018semantic} and both scored the highest relative to the solutions present at the challenges years before, on challenges benchmarks and with the challenges scoring system. The algorithms are described in the following, together with their empirical evaluation relative both to the old benchmarks and on newly generated tests.

\subsection{Name Matching Model - Algorithm}

The algorithm makes valid calls to available services from the repository until the goal is reached: all the user required parameters are found. To implement this efficiently, a set of known parameters \emph{K} is maintained that is first initialized with the user's initially known parameters. We define an \emph{accessible service}, a service from the repository that has all input parameters included in the currently known parameters. The algorithm chooses a new \emph{accessible} service repeatability, while it is possible, or the user request is already satisfied. The output of the chosen service is added to \emph{K} and the process continues. If at any point all goal parameters are completely included in the known parameter set, the search stops since a satisfying composition is found already. The search also stops if there are no \emph{accessible} services and in this case, the problem instance has no solution.

\begin{algorithm}[h]
\caption{Composition Search on Name Match model.}\label{algo:fornamematch}
\begin{algorithmic}[1]
\Function {FindComposition}{R, i, g}
\State $\textit{sol} \gets \text{empty list \small{\textcolor{gray}{// (of nodes)}}}$
\State $\textit{K} \gets i.O \text{ \small{\textcolor{gray}{ // set of known parameters}}}$
\State $ni \gets newAccessibleService(R, sol, K)$
\While {ni $\neq NULL$}
\State $sol.add(ni)$
\State $K \gets K \cup ni.O$
\If {$(g.I \subseteq K)$}
\State \Return \emph{sol}  \small{\textcolor{gray}{// and exit}}
\EndIf
\State $ni \gets \small{newAccessibleService(R, sol, K)}$
\EndWhile
\State \Return \emph{NULL}
\EndFunction
\vspace{0.5pc}

\Function {newAccessibleService}{R, sol, K}
\For {$n \in {R}$}
\If {($n.I \subseteq K) \land (n \not\in sol)$}
\Return {$n$}
\EndIf
\EndFor
\State \Return \emph{NULL}
\EndFunction
\end{algorithmic}
\end{algorithm}

In Algorithm \ref{algo:fornamematch} we have the input \emph{R} -- the repository as a set of web services, \emph{(i, g)} -- the initial and goal parameters structured as services (with \emph{i.I = $\emptyset$} and \emph{g.O = $\emptyset$}). \emph{sol} is the ordered list of services in the constructed composition solution and \emph{K} is the set of currently know parameters.

To implement \emph{newAccessibleService()} efficiently some other structures are implemented:
\begin{itemize}
  \vspace{-0.2cm}
  \item{\textbf{inputParameterOf}} -- a map from any parameter to the set of all services that have this parameter as input.
  \item{\textbf{unknownInputParameters}} -- a map from any service to the set of its input parameters that are currently unknown.
  \item{\textbf{accessibleServices}} -- services $ws_{i}$ with \emph{unknownInputParameters[$ws_{i}$]} = $\emptyset$.
  \item{\textbf{userUnknownParameters}} -- set of required parameters that are yet unknown.
  \vspace{-0.2cm}
\end{itemize}

\emph{newAccessibleService()} will return any element of \emph{accessibleServices} set if nonempty and \emph{NULL} otherwise. If the set is empty, and the \emph{userUnknownParameters} set is not empty, the problem instance has no solution. This is correct because we mark all parameters that are found on any invocation and once found a parameter is never forgotten. Moreover, if - for any reason - a service is invoked but it does not add any new parameter, that has no negative effect anywhere in the problem formulated as in Section \ref{sec:models:names}.

The data structures have to be updated after a service call. For any of the web service's output parameters, if that parameter was not in \emph{K} (known) set, then it is a new learned parameter. This will only happen once, which is relevant to the overall complexity. In this case, we iterate through all the services in \emph{inputParameterOf} that parameter and remove it from \emph{unknownInputParameters} for the service. This does not increase the time complexity since any parameter is found only once and then added to \emph{K}. When removing a parameter from the \emph{unknownInputParameters} of a web service, we check if there are still any unknown parameters for that service. If there are none, we add the service to \emph{accessibleServices}. This also can happen only one time.

Also, when a service is called it is permanently removed from \emph{accessibleServices}: it will never add any benefit to call it again. \emph{userUnknownParameters} is also updated when a new parameter is learned to enable constant time complexity of termination condition.

\subsection{Shortening the Compositions}\label{sec:complexity:algorithms:name_heuristic}

Finding the shortest composition is an \texttt{NP-Complete} problem, as it was shown in the previous lower bound Section \ref{sec:complexity:lower:shortest}. But in practice and on the challenge tests, adding a computationally-simple method for reducing the composition size had very good results in this regard. This method consists of a simple, score-based heuristic for reducing the solution length which does not significantly increase the running time.

Each distinct parameter and Web Service is associated with a positive floating-point score that approximates the expected benefit of finding that parameter or calling that Web Service. First, the scores are calculated for all parameters and services. Then the composition algorithm follows as described with the only difference that when one service is chosen from the currently accessible services the one with the highest score is chosen. The set is simply replaced by a set ordered decreasingly by the score values, and the time complexity is increased from \textbf{O(1)} to \textbf{O(logN)} if using Binary Search Tree or Heap structures.

The score-assigning algorithm follows two key principles:

\textbf{I. Service score cumulates the score of its output parameters.} This is because the score is a measure of the benefit of choosing that one service to call next, not the possibility of calling it, so the output parameters are favored to input parameters. Since the output parameters can be used further independently \textbf{sum} of scores is suitable.

\textbf{II. Parameter score cumulates the scores of all services for which they are inputs.} This is the other way around: if one parameter is found then any service that has it as input could possibly become accessible. This is a good estimate although services might have other harder-to-reach parameters as input and remain inaccessible.
Intuitively the score \emph{''flows''} from parameters to services and from services to parameters. The initial score is associated to the user goal parameters that can be assigned with any positive constant number (e.g. each goal parameter $\gets$ 1)

Working with the above score implications can result in cycles since one parameter can play the role of both input and output in different services. Some parameter scores might increase the score of services that increase the score of parameters and so on and it is possible to reach the initial parameters on this path. It is impossible to compute all the scores in this manner so a limitation has to be imposed.
Our chosen trade-off is to process any service only once.
The algorithm first assigns scores to parameters in the user request. Then for any service that provides any of these parameters, we process that service by calculating the score of that service. While doing this, some output parameters of that service might have no score assigned yet, and we handle them as having score 0. Next, the score of its input parameters is increased with the service score divided by the number of input parameters. Some new services might output these parameters so they are added to the processing queue. Then the next service from the queue is processed and so on until the queue is empty.

At the end of the iteration of all services that could be reached, the sum of output parameters score is computed again for each service. This is a simple but effective improvement since at the time of first processing of each service some parameters might had a lower score than at the end. This can be extended further in several ways, for example: if service score increases at the last step, we can increase the parameters score accordingly. For simplicity and performance concerns only the first step was implemented.

\iftoggle{fullThesis}{
\begin{figure}[h]
\caption[Heuristic \textbf{score} propagation reduces composition length]{Score propagation from user required parameters towards accessible services.}
\vspace{-0.3cm}
\label{fig:scorefig}
\begin{center}
\begin{tikzpicture}[node distance=1.7cm and 1cm, auto]
 \node[] (fdummy) {};
 \node[punkt, right = 4cm of fdummy ] (rin) {{\smaller in=$\emptyset$}\\\textbf{userReq\_In}\\{\smaller out=\{e, f\}}};
 \node[below right=0.1cm and -2cm of rin] (dummy) {};
 \node[punkt, left=2cm of dummy, text width=5.5em] (ws1) {{\smaller in=\{e\}}\\$\boldsymbol{ws_1}$\\{\smaller out=\{d, g\}}};
 \node[punkt, text width=5.7em, below right=1.5 and 0.7 cm of ws1.west] (ws2) {{\smaller in=\{b\}}\\$\boldsymbol{ws_2}$\\{\smaller out=\{a, c, e, h\}}};
 \node[punkt, below right=0.5cm and 1.4cm of dummy, text width=5.5em] (ws3) {{\smaller in=\{d, f\}}\\$\boldsymbol{ws_3}$\\{\smaller out=\{b\}}};
 \node[punkt, below right=1cm and 0.8cm of ws2.east, text width=7em] (rout) {{\smaller in=\{a, b, c\}}\\\textbf{userReq\_Out}\\{\smaller out=$\emptyset$}};
 \path ($(rout.north) + (-0.2, +0.1)$) edge [pil, bend right =0, dashed, sloped] node [above] {\textcolor{gray}{a, c}} (ws2);
 \path (rout.north) edge [pil, bend right =0, dashed, sloped] node [midway] {\textcolor{gray}{b}} (ws3);
 \path (ws2) edge [pil, bend right =0, dashed, sloped] node [midway] {\textcolor{gray}{b}} (ws3);
 \path (ws3) edge [pil, bend right =0, dashed, sloped] node [above] {\textcolor{gray}{d}} (ws1);
\end{tikzpicture}
\vspace{-0.5cm}
\end{center}
\end{figure}
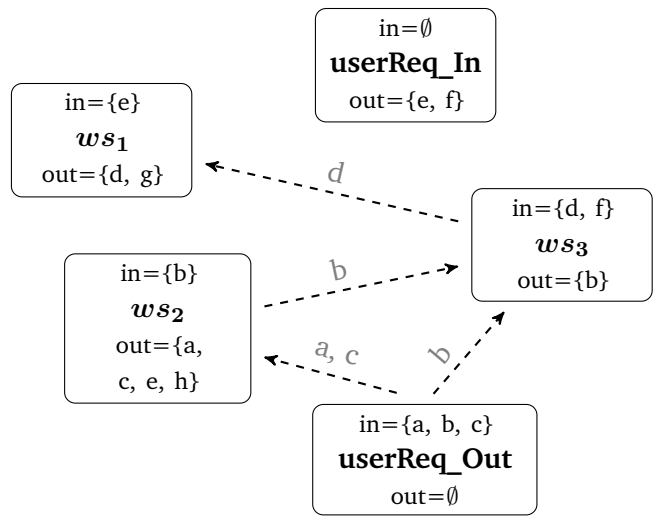
}

For the example, in Figure \ref{fig:scorefig}, the first scores will be assigned to parameters \emph{a, b, c} with the value of \textbf{1}. Services \emph{$ws_2$} and \emph{$ws_3$} are added to the queue since they produce at least one of these parameters. For service \emph{$ws_2$} the score is now \textbf{2} which is the sum of the scores of \emph{a}, \emph{c}, \emph{e} and \emph{h} that are \textbf{1}, \textbf{1}, \textbf{0} and \textbf{0}. Also, \emph{b}'s score is increased by \textbf{2} thus reaching \textbf{3}.

Service \emph{$ws_3$} is processed next and assigned the score of \emph{b} that is \textbf{3}. \emph{d} and \emph{f} score is assigned to \textbf{3/2 = 1.5}, and \emph{$ws_1$} is added to the queue and processed next. Its score is the sum of \emph{d} and \emph{g} that is \textbf{1.5+0=1.5}. \emph{e} score is \textbf{1.5/1 = 1.5} and the queue is empty.

It the end, all services scores are recalculated as the sum of their output parameters and in our case, this changes only the score of \emph{$ws_2$} that now includes the nonzero score of \emph{e} and reaches at \textbf{3.5}.

Algorithm \ref{algo:fornamematch} follows the score allocation algorithm with the accessible services set sorted decreasingly by these final scores. However, the provided solution can possibly be shortened further by several different tests. These have been proven important in practice.

One first observation is that any service that outputs no new parameter relative to previous services output in the chain is removed. Also, any output parameter that is not used by further services or the final user request is useless and thus services that output only parameters of this this type can also be removed. To include all the corner cases here, we iterate all services in the solution chain, keeping two information for all parameters: if there is any service already processed that provides the parameter as output (if it is already provided), and how many unprocessed-yet services require each parameter as their input (if it will be required further). To implement this efficiently, the solution chain has to be iterated twice: the first time all parameter's input appearances are counted. On the second iteration, this counter can be simply decreased on any new input appearance; the remaining counter will represent the appearances of a parameter in further services. This way, the time complexity of the algorithm is not increased, if proper data structures are used, like hash table containers. Finally, one service is useful if it outputs at least one parameter that is not provided by previous services but is required by further services in the chain, or in the final user required parameters. If one service is not useful by this definition, it can be dropped from the composition without invalidating it.
These final steps are effective on benchmarks reducing the solution length on tests from two different benchmarks.

\subsection{Results on the 2005 Challenge and WS-Ben}

The performance of our solution is compared with the planning based approach described in Zou et al \cite{zou2014dynamic} used with two different solvers: GraphPlan \cite{blum1997fast} and FastForward \cite{hoffmann2001ff}. For more details on planning-based solutions see our paper \cite{diac2017engineering}. Each of the three solutions is executed on tests from three sources: \emph{ICEBE'05} competition \cite{blake2005eee}, WS-Ben \cite{oh2006wsben}, a tool that creates instances for service composition problems; and tests from the generator in Section \ref{sec:complexity:algorithms:generatornamemodel}. Two metrics are considered: the \textbf{running time} to deliver a solution and the \textbf{solution length}, which is the number of web services in the composition. All solutions provided are validated with a separate program that verifies the accessibility of each service in the composition in order and the provisioning of all user required parameters at the end. Results on each benchmark are presented on separate tables.

\def\arraystretch{1.2}

\begin{table}[h]
\caption[\textbf{Name Matching} model: evaluation on \textbf{ICEBE 2005} tests]{Results on a selection of ICEBE'05 tests.}
\label{table1}
\begin{tabular*}{\hsize}{@{\extracolsep{\fill}}cccccccc@{}}
\toprule
\multirow{2}{*}{\large{$|R|$}} & \multirow{2}{*}{\large{file name}} & \multicolumn{2}{c}{\small{Algorithm \ref{algo:fornamematch}}} & \multicolumn{2}{c}{\small{GraphPlan}} & \multicolumn{2}{c}{\small{Fast-Fwd}}\\ \cline{3-4}  \cline{5-6} \cline{7-8}
& &\small{time ({{s}})}&\small{length}&\small{time ({{s}})}&\small{length}&\small{time ({{s}})}&\small{length}\\
\cline{1-8}
2656&composition1-50-16&0.46&2&0.3&2&2.99&2\\
4156&composition1-100-4&0.26&4&0.64&4&2.6&4\\
5356&composition2-50-32&2.1&7&6.6&7&14.3&7\\
8356&composition2-100-32&3.6&7&3.0&7&22.1&7\\
\bottomrule
\end{tabular*}
\end{table}

\emph{$|R|$} is the number of web services in the \emph{.WSDL} file or the repository size. Time is measured in seconds and \emph{len} column specifies the length of the provided composition in the number of web services.

The \emph{ICEBE'05} tests are structured in two sets: \emph{composition1} and \emph{composition2} where the second is computationally harder. Our first two tests are from \emph{composition1} and the last two are from \emph{composition2}, where we can see a significant increase in the running time of all solutions. Another important parameter of the tests is the number of parameters per service. The selected tests have  different choices for this parameter, specified in the last number of the file name (4, 16, 32). Also, each file in \emph{ICEBE'05} provides \emph{11} user requests from which one is selected randomly. The running times are compared on only this selected query that is the same for the three solutions.

The conclusion from \emph{ICEBE'05} results is that \emph{Fast-Forward} times are significantly higher especially on larger repositories and \emph{GraphPlan} has about the same performance as our polynomial solution. On all tests, the solution length is the same but interestingly, the solutions are not identical and each of the three solutions provides almost disjoint compositions, with only a few services in common.

\begin{table}[h]
\caption[\textbf{Name Matching} model: evaluation on \textbf{WS-Ben} tests]{Results on a selection of WSBen (Scale-Free option).}
\label{table:eval_name}
\begin{tabular*}{\hsize}{@{\extracolsep{\fill}}ccccccc@{}}
\toprule
\multirow{2}{*}{\large{$|R|$}} & \multicolumn{2}{c}{\small{Algorithm \ref{algo:fornamematch}}} & \multicolumn{2}{c}{\small{GraphPlan}} & \multicolumn{2}{c}{\small{Fast-Fwd}}\\ \cline{2-3} \cline{4-5} \cline{6-7}
& \small{time ({{s}})}&\small{length}&\small{time ({{s}})}&\small{length}&\small{time ({{s}})}&\small{length}\\
\cline{1-7}
300    & 0.03 & 9 & 0.08 & 9 & 0.07 & 9 \\
1000  & 0.1 & 7 & 0.3 & 11 & 1.4 & 6 \\
5000  & 0.7 & 8 & 2.7 & 13 & 4.6 & 6 \\
10000 & 1.7 & 9 & 3.6 & 15 & \textcolor{red}{error} & ? \\
\bottomrule
\end{tabular*}
\end{table}

The \emph{WSBen} tests are generated with the \emph{Scale-Free} option. Other options (\emph{Small World} or \emph{Random}) and other parameter changes did not reveal significantly different behavior. As seen in Table \ref{table:eval_name}, the running times are generally smaller relative to \emph{ICEBE'05 composition-2} on the same repository size, thus the computational complexity is not higher. \emph{Fast-Forward} is again the slowest and even fails on repositories with more than ten thousand services. To make \emph{Fast-Forward} and even \emph{GraphPlan} run on large repositories, several constants defining data-structure limits had to be increased in the code, like array limits and the maximum number of operators. But even after this change \emph{Fast-Forward} could not run on all tests in our experiments and failed with parsing errors. On \emph{WSBen}, \emph{GraphPlan} running times are closer to \emph{Fast-Forward} than to our polynomial solution which is again the fastest.

But more important and different from previous results, \emph{WSBen} tool revealed significant information about the provided solution length. \emph{GraphPlan} is faster than \emph{Fast-Forward} but it provides a much longer solution, our polynomial algorithm provides a solution almost as short as \emph{Fast-Forward} but much faster. This was not the initial case as without the heuristics in Section \ref{sec:complexity:algorithms:name_heuristic} and the final improvements added to reduce the solution length, the algorithm would reveal solutions that are roughly closer to the ones provided by \emph{GraphPlan}.

\subsection{Generating Complex Compositions}\label{sec:complexity:algorithms:generatornamemodel}

In paper \cite{zou2014dynamic} and other works, the \emph{ICEBE05} tests, described in \cite{blake2005eee}, are used for comparing WSC solutions. This benchmark was created for one of the first competitions on automatic WSC in 2005, and from our experiments, it is not computationally hard or relevant for the solution length. Later, the paper \cite{oh2006wsben} introduces \emph{WSBen} tool that was also largely used with better results, both for running time and for the solution length. However, to show the efficiency of the polynomial-time algorithm proposed in Section \ref{algo:fornamematch} over the planning methods, a new set of tests was created with a special generator that reveals a much higher running time variation of different solvers.

On both ICEBE05 and WSBen generated tests the solution found using either planning techniques or the polynomial algorithm is of at most 12 Web Services out of at most tens of thousands of web services within a repository. Since it is expected that the running time would be highly influenced by the solution length, this is not sufficient to differentiate between solutions, so the following generator is proposed.

The generator creates a large repository of Web Services with random parameters and then chooses an ordered list of distinct Web Services that is later changed to contain a solution. It creates the user request as two fictive Web Services in the solution chain: the first is a service with output parameters as the user's request initially known parameters and $\emptyset$ as input. Then for any service in the list in order, it changes the \textbf{input} parameter list. The number of elements stays the same but the parameters are replaced by a (uniform) random choice of distinct parameters from the union of the sets of \textbf{outputs} of previous services in the chain, including the user's request. The last extra "service" will have no output parameters but the input is constructed similarly, and that input set is the goal required parameters.

Algorithm \ref{solbased} is the summarized test generator algorithm. It uses four relevant input parameters that control the input and output size for the generated instance.

\begin{algorithm}[h]
\caption{Solution-based test generator.} \label{solbased}
\begin{algorithmic}[1]

\State $\text {\textcolor{gray}{ \small{/* \textbf{numWebServices} - number of Services,}}} {\textcolor{gray}{ \small{\hspace{0.4cm}\textbf{parsPerService} \text{ - max parameter set size}}}}$
\State $\text {\textcolor{gray}{ \small{\hspace{0.4cm}\textbf{numParameters} - total distinct pars.,}}} {\textcolor{gray}{ \small{\hspace{0.4cm}\textbf{numWSinSolution} \text{ - max solution length */}}}}$

\For {$i \gets 0, numWebServices$} \hspace{0.7cm} \text { \small{\textcolor{gray}{   // create web service $ws_i$\space:}}}
\State $ws_i.in \gets \text{\small{random parameter set}} $ \hspace{0.7cm} $\text{\textcolor{gray}{\small{// set size is random $\in$ [1, parsPerService]}}}$
\State $ws_i.out \gets \text{\small{random parameter set}} $ \hspace{0.7cm} $\text{\textcolor{gray}{\small{// with values $\in$ [1, numParameters]}}}$
\EndFor

\State $sol_0 \gets 0$ \text{\textcolor{gray}{\small{\space // $ws_0$.out are initially known}}}
\For {$i \gets 1, numWSinSolution+1$}
\If {$(i < numWSinSolution+1)$}
\State $sol_i \gets \text{random service index}$
\Else
\State $sol_i \gets 0 $\text{\small{\space \textcolor{gray}{// user request required}}} 
\EndIf
\State $ws_{sol_i}.in \gets $ random parameters from $\bigcup_{j=0}^{i-1} {\hspace{0.05cm}ws_{sol_j}}.out $
\EndFor
\State $\text{\small{\textcolor{gray}{// repository: $ws_1, ws_2, ... , ws_{numWebServices}$; (known, required) $\gets (sol_0.out, sol_0.in) $ }}} $

\end{algorithmic}
\end{algorithm}

\begin{figure}[h]
\caption[Building arbitrary-\textbf{long} compositions]{Inputs are constructed based on the output of previous services in the chain.}
\begin{center}
\label{solbasedfig}
\begin{tikzpicture} [->, >=latex]
   \graph[nodes={draw, rectangle, rounded corners, minimum width=60pt},
           branch down=1.4cm,
           grow right sep=2.6cm] {subgraph I_nm [
            V={{$ws_{sol_{0}}in$}, {$ws_{sol_{1}}in$},{$ws_{sol_{2}}in$},
                  {$ws_{sol_{len}}in$}, {$ws_{sol_{len+1}}in$}},
           W={{$ws_{sol_{0}}out$},{$ws_{sol_{1}}out$},{$ws_{sol_{2}}out$},
			 {$ws_{sol_{len}}out$}, {$ws_{sol_{len+1}}out$}
                 } ];
};

\draw[gray, rounded corners] ($({$ws_{sol_{0}}in$}.north west) + (-0.15, +0.15)$) rectangle ($({$ws_{sol_{0}}out$}.south east) + (+0.15, -0.15)$);
\draw[gray, rounded corners] ($({$ws_{sol_{1}}in$}.north west) + (-0.15, +0.15)$) rectangle ($({$ws_{sol_{1}}out$}.south east) + (+0.15, -0.15)$);
\draw[gray, rounded corners] ($({$ws_{sol_{2}}in$}.north west) + (-0.15, +0.15)$) rectangle ($({$ws_{sol_{2}}out$}.south east) + (+0.15, -0.15)$);
\draw[gray, rounded corners] ($({$ws_{sol_{len}}in$}.north west) + (-0.15, +0.15)$) rectangle ($({$ws_{sol_{len}}out$}.south east) + (+0.15, -0.15)$);
\draw[gray, rounded corners] ($({$ws_{sol_{len+1}}in$}.north west) + (-0.15, +0.15)$) rectangle ($({$ws_{sol_{len+1}}out$}.south east) + (+0.15, -0.15)$);

\draw[-latex] ({$ws_{sol_{1}}in$}.east)->($({$ws_{sol_{0}}out$}.west) + (0, +0.15)$);
\draw[-latex] ({$ws_{sol_{2}}in$}.east)->($({$ws_{sol_{0}}out$}.west) + (0, +0.05)$);
\draw[-latex] ({$ws_{sol_{2}}in$}.east)->($({$ws_{sol_{1}}out$}.west) + (0, +0.1)$);
\draw[-latex] ({$ws_{sol_{len}}in$}.east)->($({$ws_{sol_{0}}out$}.west) + (0, -0.07)$);
\draw[-latex] ({$ws_{sol_{len}}in$}.east)->($({$ws_{sol_{1}}out$}.west) + (0, +0.0)$);

\draw[-latex] ({$ws_{sol_{len}}in$}.east)->($({$ws_{sol_{2}}out$}.west) + (0, +0.05)$);
\draw[->] ({$ws_{sol_{len+1}}in$}.east) to ($({$ws_{sol_{0}}out$}.west) + (0, -0.23)$);
\draw[->] ({$ws_{sol_{len+1}}in$}.east) to ($({$ws_{sol_{1}}out$}.west) + (0, -0.12)$);

\draw[->] ({$ws_{sol_{len+1}}in$}.east) to ($({$ws_{sol_{2}}out$}.west) + (0, -0.05)$);
\draw[->] ({$ws_{sol_{len+1}}in$}.east) to ({$ws_{sol_{len}}out$}.west);

\node [label={[shift={(1,-3.9)}]\textbf{\vdots}}] {};
\node [label={[shift={(5.8,-3.9)}]\textbf{\vdots}}] {};

\end{tikzpicture}
\end{center}
\end{figure}
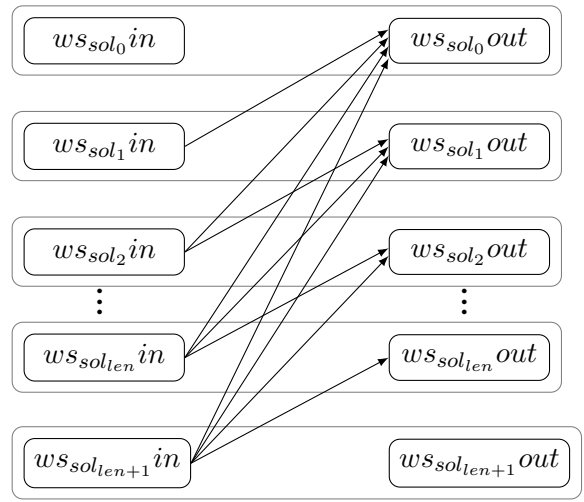

\subsection{Results on Large Compositions}

\begin{table}[h]
\caption[\textbf{Name Matching} model: evaluation on \textbf{generated} tests]{Results on tests generated by Algorithm \ref{solbased}.}
\label{table:eval_namegen}
\begin{tabular*}{\hsize}{@{\extracolsep{\fill}}ccccccccc@{}}
\toprule
    \multirow{2}{*}{\large{$|R|$}} & \multirow{2}{*}{\makecell{sol. \\ length}} & \multirow{2}{*}{\makecell{pars. per \\ service}} & \multicolumn{2}{c}{Algorithm \ref{algo:fornamematch}} & \multicolumn{2}{c}{GraphPlan} &\multicolumn{2}{c}{Fast-Fwd}\\ \cline{4-5} \cline{6-7} \cline{8-9}
    & & &time ({{s}})&len.&time ({{s}})&len.&time ({{s}})&len.\\
\cline{1-9}
300 & 100 & 15 & 0.07 & 50 & 1.5 & 51 & 3.3 & 50 \\
300 & 100 & 40 & 0.2 & 95 & 61 & 98 & 43 & 95 \\
200 & 150 & 70 & 0.3 & 141 & $>$3 hours & ? & 22 min & 141 \\
1000 & 200 & 50 & 1.6 & 130 & 39.9 & 133 & \textcolor{red}{error} & ? \\
1000 & 500 & 20 & 0.5 & 314 & $>$3 hours & ? & \textcolor{red}{error} & ? \\
1000 & 100 & 20 & 0.4 & 86 & 4.8 & 87 & \textcolor{red}{error} & ? \\
\bottomrule
\end{tabular*}
\end{table}

The newly generated tests revealed much more information as seen in Table \ref{table:eval_namegen}, for comparing not only our solution but even the relative behavior of the two planners. First, more test parameters are to consider than just the repository size, and the most important is the \emph{desired solution length} represented in the second column. This is the number of services chained to build a valid solution as in Section \ref{sec:complexity:algorithms:generatornamemodel}, but not all these services are strictly required. Often, shorter valid compositions are found, because one service's output parameters might not be chosen in the random subset of parameters; that constitutes the input of the next service. The easiest way to decrease the gap between the \emph{desired solution length} and length of discovered solutions is to increase the number of parameters per service, shown in the third column. Each service's input and output parameters sets will have a size of a uniform random between 1 and this number of elements. Increasing this eventually increases the probability of one service being strictly dependent on a previous service.

For example, the first two tests have \emph{300} repository sizes, and both have the \emph{desired solution length} set as \emph{100} but the first has \emph{15} maximum parameters per service while the second has \emph{40}. This produces a significant difference in the resulting compositions: from a solution of length of only \emph{50} for the first test the length is increased to \emph{95} on the second.
The last three tests are generated with different parameter values to provide long compositions both by increasing the \emph{desired solution length} and the maximum \emph{parameters per service}, and we can observe the solutions to be found from \emph{86} to at most \emph{314} web services.
The observed solution length has the highest influence in the running times of planning based solutions. For more than one or two hundred services required in the composition, both planners simply block for hours or provide errors while the polynomial solution runs in at most a few seconds.

The test with the repository size of \emph{200} was very useful to test the solution shortening improvements: initially, the heuristic provided a \emph{148} services-long solution that only after the improvements was reduced first to \emph{143} and finally to \emph{141}; that, interestingly, is exactly as short as the solution of \emph{Fast-Forward}. Also, this test reveals a huge difference in the running times: from \emph{0.3} seconds for the polynomial algorithm to \emph{22} minutes for \emph{Fast-Forward}. \emph{GraphPlan} was stopped after \emph{3} hours of running without providing any solution.

\subsection{Hierarchical Model - Algorithm}\label{algo:hierarchical}

In the following years the \emph{Web Services Challenge} evolved, and the first steps were to extend the semantics describing the parameter matching, as we also described in Section \ref{sec:models:hierarchical}. For the exact problem version of the 2008 challenge, we also implemented a solution algorithm that was later published in paper \cite{tucar2018semantic}. This was done as part of a bachelor's thesis of the main author of the paper. The essence of the paper, that was the bachelor's student idea was to pre-compute the Euler Traversal ordering of the concept-nodes of the hierarchy. In this manner, the sub-concept relations can be solved in $O(1)$ run-time, and together with appropriate data structures, it leads to an almost linear run time complexity, relative to the problem's input size. This Euler Traversal optimization can be used in many applications where tree-like structures are queried for node - child relations.

However, the 2008 challenge had other new elements as well. One is that more than just the number of services in a composition was measured. A new metric evaluated called \emph{execution path} measures the number of service calls if calls that can be executed in parallel are counted only as one. Obviously, this has implications in the algorithm design as well, that often uses the therm \emph{layer of services} - for a set of services that can be executed in parallel in a constructing composition.

The algorithm runs two processes in a loop. The first step, the \emph{construction phase}, consists of building a valid composition, attempting to use as few services as possible. The second step, the \emph{reduction phase}, removes useless services in the current composition based on some specific criteria. The loop ends when the second step provides no further improvement. At any time, the algorithm maintains a dynamic scoring of the services. Each service score estimates the amount of new information that can be obtained after adding the service to the composition. This score is used to minimize the size of the composition and is computed on the fly.

After selecting a service, its output parameters together with all instances that are more generic than these are learned.

More details of the algorithm are presented in the paper \cite{tucar2018semantic}, we will only describe the pre-processing that helped keep the complexity linear (or almost linear if the scores are considered).

First, an important preprocessing phase is presented, that is particular to our solution and that is crucial to the improved complexity if used together with several other suitable data structures, that are more common. This optimization can often be used when working with tree data structures for querying ancestor-descendant relations in constant time.

\vspace{-0.4cm}
\textbf{Hierarchy traversal: reduce run-time complexity of \emph{subsume} queries to $O(1)$}
The following Algorithm \ref{algo:eulertraversal} describes the method that computes the "\emph{entry time}" and "\emph{exit time}" of each node of a concept in the taxonomy, performing a linearization of the taxonomy tree. The \emph{time} variable is a global variable initialized with 0 that will keep the position in a modified Euler Tour of the tree.

\begin{algorithm}[h]
\caption{Euler-Traversal computing \emph{entry} and \emph{exit} time of each concept.} \label{algo:eulertraversal}
\begin{algorithmic}[1]

\Function{linearization}{$concept$}
    \State $time \gets time + 1$
    \State $entryTime[concept] \gets time$
    \For {$subconcept \in subconcepts[concept]$}
        \State $linearization(subconcept)$
    \EndFor
    \State $time \gets time + 1$
    \State $exitTime[concept] \gets time$
\EndFunction
\end{algorithmic}
\end{algorithm}

Algorithm \ref{algo:eulertraversal} builds a \textbf{Euler Tour} of the tree. It is useful based on the following observation: node $b$ is a successor of node $a$ iff $entryTime[a]<entryTime[b]<exitTime[a]$, thus the reduction to $O(1)$. This method is further used within the subsumes function throughout the algorithm, which returns true if a node is a successor, i.e. sub-concept of another.
For the rest of the algorithm stages see \cite{tucar2018semantic}.

\subsection{Results on the 2008 Challenge}
In the 2008 challenge, the solutions were evaluated based on three metrics: execution time, minimum number of services and minimum execution path. We evaluated our algorithm on three tests from the benchmark for which we found results for the other solutions from the challenge, summarized in paper \cite{bleul2009web}.

The Algorithm \ref{algo:hierarchical} was first validated for correctness with a specially created checker. Then, output parameters and execution times were compared with the results from the WSC-08 competition \cite{bansal2008wsc}, and with a later solution published in paper \cite{rodriguez2011automatic}. We will skip the later evaluation, which can be found in paper \cite{tucar2018semantic}. Original tests from the competition have been used. The algorithm performance would rank first in the competition based on the same scoring rules, and runs much faster than the algorithm in paper \cite{rodriguez2011automatic} for the second evaluation. The implementation was done in \texttt{Java} and ran on an \texttt{Intel(R) Core(TM) i5-3210M CPU @250 GHz} with 8 \texttt{GB RAM}.

The solutions have been evaluated over 3 tests, and on each of them, one solution could gain at most 18 points. 6 points were granted if the provided solution contains the minimum number of services, 6 for the minimum execution path. The first, second and third fastest running times would gain another 6, 4 and respectively 2 points, under the condition that they provide valid minimum path or number of services. Eight universities submitted solutions to the challenge, and the top three results are listed together with our proposed solution in Table \ref{table:evaltable_wsc2008}. The points for the top three fastest solutions are not updated for the solutions at the competition, even if our solution had the fastest running time on all tests (thus, this was done without decreasing their score).

\newcolumntype{L}[1]{>{\raggedright\let\newline\\\arraybackslash\hspace{0pt}}m{#1}}
\newcolumntype{C}[1]{>{\centering\arraybackslash\let\newline\\\arraybackslash\hspace{0pt}}m{#1}}
\newcolumntype{R}[1]{>{\raggedleft\let\newline\\\arraybackslash\hspace{0pt}}m{#1}}
\newcolumntype{P}[1]{>{\centering\arraybackslash}p{#1}}

\begin{table}[H]
\caption[\textbf{Hierarchical} model: evaluation on \textbf{WSC 2008} tests]{Results relative to WSC 08 challenge solutions.}
\label{table:evaltable_wsc2008}
\begin{tabular}{|P{0.3cm}|P{1.7cm}||P{0.85cm}|P{0.85cm}|P{1.1cm}|P{0.85cm}|P{1.1cm}|P{0.85cm}|P{0.85cm}|P{0.85cm}|}
	\cline{3-10}
	\multicolumn{2}{C{2.1cm}|}{} &\multicolumn{2}{C{2.1cm}|}{Tsinghua University }&
	\multicolumn{2}{C{2.1cm}|}{University of Groningen}&
	\multicolumn{2}{C{2.1cm}|}{Pennsylvania University} & 
	\multicolumn{2}{C{2.1cm}|}{\textbf{Proposed algorithm} \ref{algo:hierarchical}} \rule{0pt}{4ex}\\
     \cline{3-10}
	\multicolumn{2}{C{2.1cm}|}{} & result & points & result & points & result & points & result & points\\
	\hline
	\multirow{3}{*}{\rotatebox{270}{\hspace{-0.2cm}Test 4}}
	&min. serv. & 10 & +6 & 10 & +6 & 10 & +6 & \textbf{10} & \textbf{+6}\\
     &min. path & 5 & +6 & 5 & +6 & 5 & +6 & \textbf{5} & \textbf{+6}\\
	&time (ms) & 312 & +4 & 219 & +6 & 28078& &\textbf{34}& \textbf{+6}\\
	\hline
	\multirow{3}{*}{\rotatebox{270}{{\hspace{-0.2cm}Test 5}}}
	&min. serv. & 20 & +6 & 20 & +6 & 20 & +6 & \textbf{20} & \textbf{+6}\\
     &min. path & 8 & +6 & 10 &  & 8 & +6 & \textbf{8} & \textbf{+6}\\
	&time(ms) & 250 & +6 & 14734 & +4 & 726078& &\textbf{87}&\textbf{+6}\\
	\hline
	\multirow{3}{*}{\rotatebox{270}{{\hspace{-0.2cm}Test 6}}}
	&min. serv. & 46 &  & 37 & +6 & \multicolumn{2}{c|}{}& \textbf{45} &  \\
     &min. path & 7 & +6 & 17 & & \multicolumn{2}{c|}{no result} &\textbf{7} &\textbf{+6} \\
	&time(ms) & 406 & +6 & 241672 & +4 & \multicolumn{2}{c|}{} &\textbf{132} &\textbf{+6}\\
	\hline
	\multicolumn{2}{|C{2.5cm}|}{Score} & \multicolumn{2}{c|}{\underline{46 Points}} & \multicolumn{2}{c|}{\underline{38 Points}} & \multicolumn{2}{c|}{\underline{24 Points}} & \multicolumn{2}{c|}{\underline{\textbf{48 Points}}} \\
	\hline
\end{tabular}
\vspace{-0.5cm}
\end{table}


\section{Relational Model - Algorithm}
\label{sec:complexity:relational}
\comment{This section analyses the complexity of the relational model presented above. To increase the expressiveness, ended up with the first version of more than polynomial (subgraph isomorphism complexity) time needed for matching of service parameters to known objects.}

The proposed relational model highly increased the theoretical complexity of existing solutions for the composition problem, as proved in Section \ref{sec:complexity:lower}. However, our implemented solution includes some limitations to the number of generated instances and also reduces the branches of exponential execution if possible - for matching parameters. In this manner, we can show it applies to considerable-sized instances. The high-level of the algorithm is inspired by previous algorithms. The solution was published in \cite{diac2019relational}.

\subsection{Prerequisites}

During the construction of the composition, a \textbf{knowledge base} is kept in memory, which is constituted of \textbf{objects}. An object has a known type, which is a concept from the ontology, and also a set of \emph{relations} defined on the object. For example, the objects in Figure \ref{fig:knowledgeFig} represent a knowledge base. For each relation, the relation type and the pair object with which the relation is defined are kept in memory. Based on this information, the next service is chosen together with a matching between known objects and service inputs. The matching must satisfy the defined relations between service input parameters. After this fictive "call" (conceptually, the call is done by adding the service to the composition), the knowledge base is updated with the output of the service. New objects can be created after a call, and for known objects, new relations can be added.  After calling services, all inference rules are processed to see if they can be applied based on new objects or relations.

\begin{figure}[!h]
\centering

\caption[\textbf{Knowledge} structure example in the \textbf{Relational} model]{\textbf{Knowledge}: rectangles show types with their instance objects in circles and gray arrows represent relations.}
\label{fig:knowledgeFig}

\vspace{0.2cm}
\tikzset{every picture/.style={line width=0.75pt}} 

\begin{tikzpicture}[x=0.75pt,y=0.75pt,yscale=-1,xscale=1]

\draw  [line width=1.5]  (60.43,34.97) .. controls (60.43,30.02) and (64.44,26) .. (69.4,26) -- (136.39,26) .. controls (141.35,26) and (145.36,30.02) .. (145.36,34.97) -- (145.36,99.03) .. controls (145.36,103.98) and (141.35,108) .. (136.39,108) -- (69.4,108) .. controls (64.44,108) and (60.43,103.98) .. (60.43,99.03) -- cycle ;
\draw    (59.5,54) -- (146.5,53.67) ;

\draw  [line width=1.5]  (252,17.3) .. controls (252,12.16) and (256.16,8) .. (261.3,8) -- (327.73,8) .. controls (332.87,8) and (337.03,12.16) .. (337.03,17.3) -- (337.03,117.37) .. controls (337.03,122.5) and (332.87,126.67) .. (327.73,126.67) -- (261.3,126.67) .. controls (256.16,126.67) and (252,122.5) .. (252,117.37) -- cycle ;
\draw    (252,34.58) -- (337.5,34.58) ;

\draw   (269.85,61.16) .. controls (269.85,54.35) and (282.58,48.82) .. (298.27,48.82) .. controls (313.97,48.82) and (326.7,54.35) .. (326.7,61.16) .. controls (326.7,67.98) and (313.97,73.5) .. (298.27,73.5) .. controls (282.58,73.5) and (269.85,67.98) .. (269.85,61.16) -- cycle ;
\draw   (269.85,101.98) .. controls (269.85,95.17) and (282.58,89.64) .. (298.27,89.64) .. controls (313.97,89.64) and (326.7,95.17) .. (326.7,101.98) .. controls (326.7,108.8) and (313.97,114.33) .. (298.27,114.33) .. controls (282.58,114.33) and (269.85,108.8) .. (269.85,101.98) -- cycle ;
\draw  [line width=1.5]  (430.89,22.13) .. controls (430.89,16.53) and (435.43,12) .. (441.02,12) -- (513.37,12) .. controls (518.97,12) and (523.5,16.53) .. (523.5,22.13) -- (523.5,114.87) .. controls (523.5,120.47) and (518.97,125) .. (513.37,125) -- (441.02,125) .. controls (435.43,125) and (430.89,120.47) .. (430.89,114.87) -- cycle ;
\draw    (430,39) -- (523.5,39) ;

\draw   (450.79,65) .. controls (450.79,57.82) and (462.91,52) .. (477.86,52) .. controls (492.8,52) and (504.92,57.82) .. (504.92,65) .. controls (504.92,72.18) and (492.8,78) .. (477.86,78) .. controls (462.91,78) and (450.79,72.18) .. (450.79,65) -- cycle ;
\draw   (450.79,99) .. controls (450.79,91.82) and (462.91,86) .. (477.86,86) .. controls (492.8,86) and (504.92,91.82) .. (504.92,99) .. controls (504.92,106.18) and (492.8,112) .. (477.86,112) .. controls (462.91,112) and (450.79,106.18) .. (450.79,99) -- cycle ;
\draw [color={rgb, 255:red, 155; green, 155; blue, 155 }  ,draw opacity=1 ]   (326.7,61.16) .. controls (369.48,75.76) and (365.54,52.9) .. (449.52,64.82) ;
\draw [shift={(450.79,65)}, rotate = 188.23] [fill={rgb, 255:red, 155; green, 155; blue, 155 }  ,fill opacity=1 ][line width=0.75]  [draw opacity=0] (8.93,-4.29) -- (0,0) -- (8.93,4.29) -- cycle    ;

\draw [color={rgb, 255:red, 155; green, 155; blue, 155 }  ,draw opacity=1 ]   (326.7,98.98) .. controls (379.24,114.59) and (365.64,92.89) .. (449.52,98.91) ;
\draw [shift={(450.79,99)}, rotate = 184.25] [fill={rgb, 255:red, 155; green, 155; blue, 155 }  ,fill opacity=1 ][line width=0.75]  [draw opacity=0] (8.93,-4.29) -- (0,0) -- (8.93,4.29) -- cycle    ;

\draw [color={rgb, 255:red, 155; green, 155; blue, 155 }  ,draw opacity=1 ]   (129.58,82) .. controls (171.29,75.7) and (154.67,56.86) .. (268.13,61.1) ;
\draw [shift={(269.85,61.16)}, rotate = 182.23] [fill={rgb, 255:red, 155; green, 155; blue, 155 }  ,fill opacity=1 ][line width=0.75]  [draw opacity=0] (8.93,-4.29) -- (0,0) -- (8.93,4.29) -- cycle    ;

\draw [color={rgb, 255:red, 155; green, 155; blue, 155 }  ,draw opacity=1 ]   (129.58,82) .. controls (187.21,119.48) and (181.56,86) .. (268.53,101.74) ;
\draw [shift={(269.85,101.98)}, rotate = 190.46] [fill={rgb, 255:red, 155; green, 155; blue, 155 }  ,fill opacity=1 ][line width=0.75]  [draw opacity=0] (8.93,-4.29) -- (0,0) -- (8.93,4.29) -- cycle    ;

\draw   (73.42,82) .. controls (73.42,74.82) and (85.99,69) .. (101.5,69) .. controls (117.01,69) and (129.58,74.82) .. (129.58,82) .. controls (129.58,89.18) and (117.01,95) .. (101.5,95) .. controls (85.99,95) and (73.42,89.18) .. (73.42,82) -- cycle ;

\draw (106.84,41) node  [align=left] {\textbf{Person}};
\draw (101.5,82) node  [align=left] {pers1};
\draw (296.16,22.24) node  [align=left] {\textbf{University}};
\draw (298.27,61.16) node  [align=left] {univ1};
\draw (298.27,101.98) node  [align=left] {univ2};
\draw (476.53,27) node  [align=left] {\textbf{City}};
\draw (477.86,65) node  [align=left] {city1};
\draw (477.86,99) node  [align=left] {city2};
\draw (383.5,50.32) node [scale=0.7,color={rgb, 255:red, 155; green, 155; blue, 155 }  ,opacity=1 ,rotate=-354.18] [align=left] {isLocatedIn\\(univ1, city1)};
\draw (386,88) node [scale=0.7,color={rgb, 255:red, 155; green, 155; blue, 155 }  ,opacity=1 ,rotate=-352.81] [align=left] {isLocatedIn\\(univ2, city2)};
\draw (199,48) node [scale=0.7,color={rgb, 255:red, 155; green, 155; blue, 155 }  ,opacity=1 ,rotate=-355.2] [align=left] {isEmployeeOf\\ (pers1,univ1)};
\draw (202,86) node [scale=0.7,color={rgb, 255:red, 155; green, 155; blue, 155 }  ,opacity=1 ,rotate=-357.81] [align=left] {hasDestination\\ (pers1, univ2)};

\end{tikzpicture}
\end{figure}
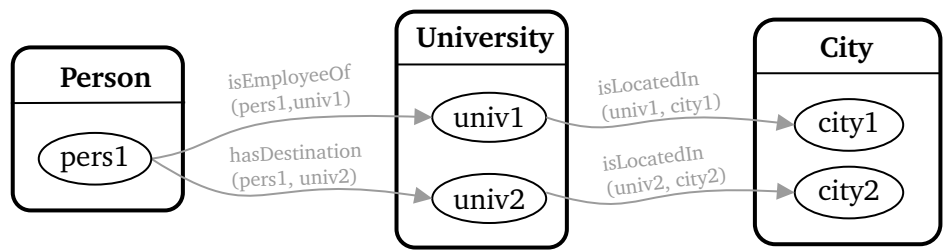 

To check if in a given \textbf{knowledge} state some service can be validly called, a backtracking procedure is implemented. For each service input parameter in order, all \emph{objects} of type or sub-types of the parameter type are iterated. The algorithm checks all the relations with objects that have already been matched with service parameters like shown from left to right in Figure \ref{fig:inputmatching}. Branches that do not satisfy any relation between completed levels are dropped. If finally, the last parameter could be matched with some object, then the match is complete, and the service can be called. Also, to avoid repeated loops, a history of calls is kept for every service, that includes matched objects (by using a hash value computed over the objects). Using the history of calls, calling services with parameters that have been used before is avoided.

For \emph{inference rules}, the only distinction is that variables do not have types, so any level of backtracking can work with any object. Rules with large premises and few relation restrictions would significantly slow down the search. This adds to the motivation of conceptually restricting service providers to create inference rules.

There are four main structures used in the algorithm: \textbf{ontology}, \textbf{repository}, \textbf{knowledge}, and \textbf{query}. First of all, the ontology is loaded, then the repository and the query that are verified to use the correct terms of the ontology. With the user query, the knowledge is instantiated with initially known objects and relations, and inference rules are applied for the first time.

\subsection{Find Match - Algorithm}

The bird's eye view of the main Algorithm \ref{algo:relationalmain} is simple: iterate all services and verify for each if the service can be called with the current knowledge, and do so if possible. After looping all services, the inference rules are applied, similarly. If at any time, the user required output is satisfied with all specified relations, the composition is complete, and it is then returned and the algorithm ends. If in some loop, no new service or rule could be applied, then the algorithm is blocked in the current state; thus the instance is unsolvable. The use of $query.out$ and $query.in$ as input and, respectively, output parameters of services is an implementation shortcut that can be done based on the structural similarity: $(\emptyset , query.in)$ is a fictive service that has $\emptyset$ as input and $query.in$ as output. \algoname{callService}($(\emptyset, query.in), \emptyset$) is just adding initially known objects to the knowledge, using the same method that adds service outputs later. The \algoname{applyInferenceRules()} is very similar to the core part of the composition search, Algorithm \ref{algo:relationalmain}. It iterates all rules and applies any applicable rule until no rule can be applied for objects it was not already applied on. Also, in the end, the algorithm checks if any services generating only unused parameters and relations and removes them.

\begin{figure*}[!htbp]
\centering

\caption[\textbf{Object} to input matching in \textbf{Relational} model]{Objects matching to input parameters, iterated from left to right. Parameter \textbf{p1} can match objects of \textbf{type1}, or subtypes \textbf{type4} or \textbf{type5}, but only \textbf{o1} and \textbf{o3} have the necessary relations, and similarly \textbf{o6} for \textbf{p2} on the current level.}
\label{fig:inputmatching}
\vspace{0.2cm}

\tikzset{every picture/.style={line width=0.75pt}} 
\begin{tikzpicture}[x=0.75pt,y=0.75pt,yscale=-1,xscale=1]
\draw [color={rgb, 255:red, 155; green, 155; blue, 155 }  ,draw opacity=0.3 ][line width=1.5]    (307.79,39.71) .. controls (309.46,41.38) and (309.46,43.04) .. (307.79,44.71) .. controls (306.12,46.38) and (306.12,48.04) .. (307.79,49.71) .. controls (309.46,51.38) and (309.46,53.04) .. (307.79,54.71) .. controls (306.12,56.38) and (306.12,58.04) .. (307.79,59.71) .. controls (309.46,61.38) and (309.46,63.04) .. (307.79,64.71) .. controls (306.12,66.38) and (306.12,68.04) .. (307.79,69.71) .. controls (309.46,71.38) and (309.46,73.04) .. (307.79,74.71) .. controls (306.12,76.38) and (306.12,78.04) .. (307.79,79.71) .. controls (309.46,81.38) and (309.46,83.04) .. (307.79,84.71) .. controls (306.12,86.38) and (306.12,88.04) .. (307.79,89.71) .. controls (309.46,91.38) and (309.46,93.04) .. (307.79,94.71) .. controls (306.12,96.38) and (306.12,98.04) .. (307.79,99.71) .. controls (309.46,101.38) and (309.46,103.04) .. (307.79,104.71) .. controls (306.12,106.38) and (306.12,108.04) .. (307.79,109.71) .. controls (309.46,111.38) and (309.46,113.04) .. (307.79,114.71) .. controls (306.12,116.38) and (306.12,118.04) .. (307.79,119.71) .. controls (309.46,121.38) and (309.46,123.04) .. (307.79,124.71) .. controls (306.12,126.38) and (306.12,128.04) .. (307.79,129.71) .. controls (309.46,131.38) and (309.46,133.04) .. (307.79,134.71) .. controls (306.12,136.38) and (306.12,138.04) .. (307.79,139.71) .. controls (309.46,141.38) and (309.46,143.04) .. (307.79,144.71) .. controls (306.12,146.38) and (306.12,148.04) .. (307.79,149.71) .. controls (309.46,151.38) and (309.46,153.04) .. (307.79,154.71) .. controls (306.12,156.38) and (306.12,158.04) .. (307.79,159.71) .. controls (309.46,161.38) and (309.46,163.04) .. (307.79,164.71) .. controls (306.12,166.38) and (306.12,168.04) .. (307.79,169.71) .. controls (309.46,171.38) and (309.46,173.04) .. (307.79,174.71) .. controls (306.12,176.38) and (306.12,178.04) .. (307.79,179.71) .. controls (309.46,181.38) and (309.46,183.04) .. (307.79,184.71) .. controls (306.12,186.38) and (306.12,188.04) .. (307.79,189.71) .. controls (309.46,191.38) and (309.46,193.04) .. (307.79,194.71) .. controls (306.12,196.38) and (306.12,198.04) .. (307.79,199.71) .. controls (309.46,201.38) and (309.46,203.04) .. (307.79,204.71) .. controls (306.12,206.38) and (306.12,208.04) .. (307.79,209.71) .. controls (309.46,211.38) and (309.46,213.04) .. (307.79,214.71) .. controls (306.12,216.38) and (306.12,218.04) .. (307.79,219.71) .. controls (309.46,221.38) and (309.46,223.04) .. (307.79,224.71) .. controls (306.12,226.38) and (306.12,228.04) .. (307.79,229.71) .. controls (309.46,231.38) and (309.46,233.04) .. (307.79,234.71) .. controls (306.12,236.38) and (306.12,238.04) .. (307.79,239.71) .. controls (309.46,241.38) and (309.46,243.04) .. (307.79,244.71) .. controls (306.12,246.38) and (306.12,248.04) .. (307.79,249.71) .. controls (309.46,251.38) and (309.46,253.04) .. (307.79,254.71) -- (307.79,255.71) -- (307.79,255.71) ;

\draw    (203,213) .. controls (227.01,243.87) and (261.58,235.56) .. (278.96,214.5) ;
\draw [shift={(280,213.2)}, rotate = 487.69] [fill={rgb, 255:red, 0; green, 0; blue, 0 }  ][line width=0.75]  [draw opacity=0] (8.93,-4.29) -- (0,0) -- (8.93,4.29) -- cycle    ;

\draw    (298,213.2) .. controls (305.43,257.92) and (452.02,258.98) .. (481.15,213.59) ;
\draw [shift={(482,212.2)}, rotate = 479.98] [fill={rgb, 255:red, 0; green, 0; blue, 0 }  ][line width=0.75]  [draw opacity=0] (8.93,-4.29) -- (0,0) -- (8.93,4.29) -- cycle    ;

\draw    (373,212.2) .. controls (356.34,231.93) and (332,233.16) .. (313.15,214.37) ;
\draw [shift={(312,213.2)}, rotate = 406.47] [fill={rgb, 255:red, 0; green, 0; blue, 0 }  ][line width=0.75]  [draw opacity=0] (8.93,-4.29) -- (0,0) -- (8.93,4.29) -- cycle    ;

\draw    (45,73) -- (57.73,57.54) ;
\draw [shift={(59,56)}, rotate = 489.47] [fill={rgb, 255:red, 0; green, 0; blue, 0 }  ][line width=0.75]  [draw opacity=0] (8.93,-4.29) -- (0,0) -- (8.93,4.29) -- cycle    ;

\draw    (85,73) -- (68.45,57.37) ;
\draw [shift={(67,56)}, rotate = 403.36] [fill={rgb, 255:red, 0; green, 0; blue, 0 }  ][line width=0.75]  [draw opacity=0] (8.93,-4.29) -- (0,0) -- (8.93,4.29) -- cycle    ;

\draw    (45,134) -- (45,118.67) ;
\draw [shift={(45,116.67)}, rotate = 450] [fill={rgb, 255:red, 0; green, 0; blue, 0 }  ][line width=0.75]  [draw opacity=0] (8.93,-4.29) -- (0,0) -- (8.93,4.29) -- cycle    ;

\draw [color={rgb, 255:red, 155; green, 155; blue, 155 }  ,draw opacity=1 ]   (239,151.2) .. controls (257.04,159.49) and (272.7,158.79) .. (288.76,142.53) ;
\draw [shift={(290,141.25)}, rotate = 493.32] [fill={rgb, 255:red, 155; green, 155; blue, 155 }  ,fill opacity=1 ][line width=0.75]  [draw opacity=0] (8.93,-4.29) -- (0,0) -- (8.93,4.29) -- cycle    ;

\draw [color={rgb, 255:red, 155; green, 155; blue, 155 }  ,draw opacity=1 ]   (239,91.2) -- (260.14,99.66) ;
\draw [shift={(262,100.4)}, rotate = 201.8] [fill={rgb, 255:red, 155; green, 155; blue, 155 }  ,fill opacity=1 ][line width=0.75]  [draw opacity=0] (8.93,-4.29) -- (0,0) -- (8.93,4.29) -- cycle    ;

\draw [color={rgb, 255:red, 155; green, 155; blue, 155 }  ,draw opacity=1 ]   (395,73) .. controls (348.21,60.39) and (328.59,67.29) .. (301.25,87.56) ;
\draw [shift={(300,88.5)}, rotate = 323.13] [fill={rgb, 255:red, 155; green, 155; blue, 155 }  ,fill opacity=1 ][line width=0.75]  [draw opacity=0] (8.93,-4.29) -- (0,0) -- (8.93,4.29) -- cycle    ;

\draw [color={rgb, 255:red, 155; green, 155; blue, 155 }  ,draw opacity=1 ]   (306,141.25) .. controls (351.54,182.83) and (462.75,174.38) .. (487.28,126.66) ;
\draw [shift={(488,125.2)}, rotate = 475.14] [fill={rgb, 255:red, 155; green, 155; blue, 155 }  ,fill opacity=1 ][line width=0.75]  [draw opacity=0] (8.93,-4.29) -- (0,0) -- (8.93,4.29) -- cycle    ;

\draw [color={rgb, 255:red, 155; green, 155; blue, 155 }  ,draw opacity=1 ]   (285,89) .. controls (314.85,40.91) and (430.83,28) .. (487.16,101.09) ;
\draw [shift={(488,102.2)}, rotate = 233.01] [fill={rgb, 255:red, 155; green, 155; blue, 155 }  ,fill opacity=1 ][line width=0.75]  [draw opacity=0] (8.93,-4.29) -- (0,0) -- (8.93,4.29) -- cycle    ;

\draw [color={rgb, 255:red, 155; green, 155; blue, 155 }  ,draw opacity=0.3 ][line width=1.5]    (276.5,33) -- (149,33) ;
\draw [shift={(146,33)}, rotate = 360] [color={rgb, 255:red, 155; green, 155; blue, 155 }  ,draw opacity=0.3 ][line width=1.5]    (14.21,-6.37) .. controls (9.04,-2.99) and (4.3,-0.87) .. (0,0) .. controls (4.3,0.87) and (9.04,2.99) .. (14.21,6.37)   ;

\draw [color={rgb, 255:red, 155; green, 155; blue, 155 }  ,draw opacity=0.3 ][line width=1.5]    (339.5,33) -- (440.5,33) ;
\draw [shift={(443.5,33)}, rotate = 180] [color={rgb, 255:red, 155; green, 155; blue, 155 }  ,draw opacity=0.3 ][line width=1.5]    (14.21,-6.37) .. controls (9.04,-2.99) and (4.3,-0.87) .. (0,0) .. controls (4.3,0.87) and (9.04,2.99) .. (14.21,6.37)   ;

\draw    (49.5,186) .. controls (49.5,184.9) and (50.4,184) .. (51.5,184) -- (148.5,184) .. controls (149.6,184) and (150.5,184.9) .. (150.5,186) -- (150.5,222) .. controls (150.5,223.1) and (149.6,224) .. (148.5,224) -- (51.5,224) .. controls (50.4,224) and (49.5,223.1) .. (49.5,222) -- cycle  ;
\draw (100,204) node  [align=left] { WebService1\\\textbf{Input} Params};
\draw    (165,193) .. controls (165,191.9) and (165.9,191) .. (167,191) -- (237,191) .. controls (238.1,191) and (239,191.9) .. (239,193) -- (239,211) .. controls (239,212.1) and (238.1,213) .. (237,213) -- (167,213) .. controls (165.9,213) and (165,212.1) .. (165,211) -- cycle  ;
\draw (202,202) node  [align=left] {p1 : type1};
\draw (157,202) node  [align=left] {{\LARGE :}};
\draw    (263,193) .. controls (263,191.9) and (263.9,191) .. (265,191) -- (335,191) .. controls (336.1,191) and (337,191.9) .. (337,193) -- (337,211) .. controls (337,212.1) and (336.1,213) .. (335,213) -- (265,213) .. controls (263.9,213) and (263,212.1) .. (263,211) -- cycle  ;
\draw (300,202) node  [align=left] {p2 : type2};
\draw    (355,192) .. controls (355,190.9) and (355.9,190) .. (357,190) -- (427,190) .. controls (428.1,190) and (429,190.9) .. (429,192) -- (429,210) .. controls (429,211.1) and (428.1,212) .. (427,212) -- (357,212) .. controls (355.9,212) and (355,211.1) .. (355,210) -- cycle  ;
\draw (392,201) node  [align=left] {p3 : type3};
\draw    (448,192) .. controls (448,190.9) and (448.9,190) .. (450,190) -- (520,190) .. controls (521.1,190) and (522,190.9) .. (522,192) -- (522,210) .. controls (522,211.1) and (521.1,212) .. (520,212) -- (450,212) .. controls (448.9,212) and (448,211.1) .. (448,210) -- cycle  ;
\draw (485,201) node  [align=left] {p4 : type4};
\draw  [color={rgb, 255:red, 155; green, 155; blue, 155 }  ,draw opacity=1 ][fill={rgb, 255:red, 155; green, 155; blue, 155 }  ,fill opacity=0.3 ]  (165,143) .. controls (165,141.9) and (165.9,141) .. (167,141) -- (237,141) .. controls (238.1,141) and (239,141.9) .. (239,143) -- (239,161) .. controls (239,162.1) and (238.1,163) .. (237,163) -- (167,163) .. controls (165.9,163) and (165,162.1) .. (165,161) -- cycle  ;
\draw (202,152) node [color={rgb, 255:red, 155; green, 155; blue, 155 }  ,opacity=1 ] [align=left] {o1 : type1};
\draw  [color={rgb, 255:red, 155; green, 155; blue, 155 }  ,draw opacity=1 ]  (165,112) .. controls (165,110.9) and (165.9,110) .. (167,110) -- (237,110) .. controls (238.1,110) and (239,110.9) .. (239,112) -- (239,130) .. controls (239,131.1) and (238.1,132) .. (237,132) -- (167,132) .. controls (165.9,132) and (165,131.1) .. (165,130) -- cycle  ;
\draw (202,121) node [color={rgb, 255:red, 155; green, 155; blue, 155 }  ,opacity=1 ] [align=left] {o2 : type4};
\draw  [color={rgb, 255:red, 155; green, 155; blue, 155 }  ,draw opacity=1 ][fill={rgb, 255:red, 155; green, 155; blue, 155 }  ,fill opacity=0.3 ]  (165,82) .. controls (165,80.9) and (165.9,80) .. (167,80) -- (237,80) .. controls (238.1,80) and (239,80.9) .. (239,82) -- (239,100) .. controls (239,101.1) and (238.1,102) .. (237,102) -- (167,102) .. controls (165.9,102) and (165,101.1) .. (165,100) -- cycle  ;
\draw (202,91) node [color={rgb, 255:red, 155; green, 155; blue, 155 }  ,opacity=1 ] [align=left] {o3 : type5};
\draw  [color={rgb, 255:red, 155; green, 155; blue, 155 }  ,draw opacity=1 ]  (165,51) .. controls (165,49.9) and (165.9,49) .. (167,49) -- (237,49) .. controls (238.1,49) and (239,49.9) .. (239,51) -- (239,69) .. controls (239,70.1) and (238.1,71) .. (237,71) -- (167,71) .. controls (165.9,71) and (165,70.1) .. (165,69) -- cycle  ;
\draw (202,60) node [color={rgb, 255:red, 155; green, 155; blue, 155 }  ,opacity=1 ] [align=left] {o4 : type1};
\draw  [color={rgb, 255:red, 155; green, 155; blue, 155 }  ,draw opacity=1 ]  (262,121) .. controls (262,119.9) and (262.9,119) .. (264,119) -- (334,119) .. controls (335.1,119) and (336,119.9) .. (336,121) -- (336,139) .. controls (336,140.1) and (335.1,141) .. (334,141) -- (264,141) .. controls (262.9,141) and (262,140.1) .. (262,139) -- cycle  ;
\draw (299,130) node [color={rgb, 255:red, 155; green, 155; blue, 155 }  ,opacity=1 ] [align=left] {o5 : type2};
\draw  [color={rgb, 255:red, 155; green, 155; blue, 155 }  ,draw opacity=1 ][fill={rgb, 255:red, 155; green, 155; blue, 155 }  ,fill opacity=0.3 ]  (262,91) .. controls (262,89.9) and (262.9,89) .. (264,89) -- (334,89) .. controls (335.1,89) and (336,89.9) .. (336,91) -- (336,109) .. controls (336,110.1) and (335.1,111) .. (334,111) -- (264,111) .. controls (262.9,111) and (262,110.1) .. (262,109) -- cycle  ;
\draw (299,100) node [color={rgb, 255:red, 155; green, 155; blue, 155 }  ,opacity=1 ] [align=left] {o6 : type2};
\draw  [color={rgb, 255:red, 155; green, 155; blue, 155 }  ,draw opacity=1 ]  (355,136) .. controls (355,134.9) and (355.9,134) .. (357,134) -- (427,134) .. controls (428.1,134) and (429,134.9) .. (429,136) -- (429,154) .. controls (429,155.1) and (428.1,156) .. (427,156) -- (357,156) .. controls (355.9,156) and (355,155.1) .. (355,154) -- cycle  ;
\draw (392,145) node [color={rgb, 255:red, 155; green, 155; blue, 155 }  ,opacity=1 ] [align=left] {o7 : type3};
\draw  [color={rgb, 255:red, 155; green, 155; blue, 155 }  ,draw opacity=1 ]  (355,105) .. controls (355,103.9) and (355.9,103) .. (357,103) -- (427,103) .. controls (428.1,103) and (429,103.9) .. (429,105) -- (429,123) .. controls (429,124.1) and (428.1,125) .. (427,125) -- (357,125) .. controls (355.9,125) and (355,124.1) .. (355,123) -- cycle  ;
\draw (392,114) node [color={rgb, 255:red, 155; green, 155; blue, 155 }  ,opacity=1 ] [align=left] {o8 : type6};
\draw  [color={rgb, 255:red, 155; green, 155; blue, 155 }  ,draw opacity=1 ]  (355,75) .. controls (355,73.9) and (355.9,73) .. (357,73) -- (427,73) .. controls (428.1,73) and (429,73.9) .. (429,75) -- (429,93) .. controls (429,94.1) and (428.1,95) .. (427,95) -- (357,95) .. controls (355.9,95) and (355,94.1) .. (355,93) -- cycle  ;
\draw (392,84) node [color={rgb, 255:red, 155; green, 155; blue, 155 }  ,opacity=1 ] [align=left] {o9 : type3};
\draw  [color={rgb, 255:red, 155; green, 155; blue, 155 }  ,draw opacity=1 ]  (448,105) .. controls (448,103.9) and (448.9,103) .. (450,103) -- (520,103) .. controls (521.1,103) and (522,103.9) .. (522,105) -- (522,123) .. controls (522,124.1) and (521.1,125) .. (520,125) -- (450,125) .. controls (448.9,125) and (448,124.1) .. (448,123) -- cycle  ;
\draw (485,114) node [color={rgb, 255:red, 155; green, 155; blue, 155 }  ,opacity=1 ] [align=left] {o2 : type4};
\draw (235,240) node  [align=left] {{\scriptsize rel1(p1,p2)}};
\draw (388,255.79) node [scale=0.7] [align=left] {{\small rel2(p2,p4)}};
\draw (364,230.79) node [scale=0.7,rotate=-351.77] [align=left] {{\small rel3(p3,p2)}};
\draw    (45.5,40) .. controls (45.5,38.9) and (46.4,38) .. (47.5,38) -- (80.5,38) .. controls (81.6,38) and (82.5,38.9) .. (82.5,40) -- (82.5,54) .. controls (82.5,55.1) and (81.6,56) .. (80.5,56) -- (47.5,56) .. controls (46.4,56) and (45.5,55.1) .. (45.5,54) -- cycle  ;
\draw (64,47) node [scale=0.8] [align=left] {type1};
\draw    (73.5,110) .. controls (73.5,108.9) and (74.4,108) .. (75.5,108) -- (108.5,108) .. controls (109.6,108) and (110.5,108.9) .. (110.5,110) -- (110.5,124) .. controls (110.5,125.1) and (109.6,126) .. (108.5,126) -- (75.5,126) .. controls (74.4,126) and (73.5,125.1) .. (73.5,124) -- cycle  ;
\draw (92,117) node [scale=0.8] [align=left] {type2};
\draw    (68.5,75) .. controls (68.5,73.9) and (69.4,73) .. (70.5,73) -- (103.5,73) .. controls (104.6,73) and (105.5,73.9) .. (105.5,75) -- (105.5,89) .. controls (105.5,90.1) and (104.6,91) .. (103.5,91) -- (70.5,91) .. controls (69.4,91) and (68.5,90.1) .. (68.5,89) -- cycle  ;
\draw (87,82) node [scale=0.8] [align=left] {type5};
\draw    (24.5,75) .. controls (24.5,73.9) and (25.4,73) .. (26.5,73) -- (59.5,73) .. controls (60.6,73) and (61.5,73.9) .. (61.5,75) -- (61.5,89) .. controls (61.5,90.1) and (60.6,91) .. (59.5,91) -- (26.5,91) .. controls (25.4,91) and (24.5,90.1) .. (24.5,89) -- cycle  ;
\draw (43,82) node [scale=0.8] [align=left] {type4};
\draw    (25.5,136) .. controls (25.5,134.9) and (26.4,134) .. (27.5,134) -- (60.5,134) .. controls (61.6,134) and (62.5,134.9) .. (62.5,136) -- (62.5,150) .. controls (62.5,151.1) and (61.6,152) .. (60.5,152) -- (27.5,152) .. controls (26.4,152) and (25.5,151.1) .. (25.5,150) -- cycle  ;
\draw (44,143) node [scale=0.8] [align=left] {type6};
\draw (65,15) node  [align=left] { \ \ \ types\\hierarchy};
\draw (264,164) node [color={rgb, 255:red, 155; green, 155; blue, 155 }  ,opacity=1 ,rotate=-354.33] [align=left] {{\scriptsize rel1(o1,o5)}};
\draw (263,81) node [color={rgb, 255:red, 155; green, 155; blue, 155 }  ,opacity=1 ,rotate=-10.23] [align=left] {{\scriptsize rel1(o3,o6)}};
\draw (383,41) node [color={rgb, 255:red, 155; green, 155; blue, 155 }  ,opacity=1 ,rotate=-1.4] [align=left] {{\scriptsize rel2(o6,o2)}};
\draw (363,62) node [color={rgb, 255:red, 155; green, 155; blue, 155 }  ,opacity=1 ,rotate=-1.4] [align=left] {{\scriptsize rel3(o9,o6)}};
\draw (380,174) node [color={rgb, 255:red, 155; green, 155; blue, 155 }  ,opacity=1 ,rotate=-1.4] [align=left] {{\scriptsize rel2(o5,o2)}};
\draw    (26.5,101) .. controls (26.5,99.9) and (27.4,99) .. (28.5,99) -- (61.5,99) .. controls (62.6,99) and (63.5,99.9) .. (63.5,101) -- (63.5,115) .. controls (63.5,116.1) and (62.6,117) .. (61.5,117) -- (28.5,117) .. controls (27.4,117) and (26.5,116.1) .. (26.5,115) -- cycle  ;
\draw (45,108) node [scale=0.8] [align=left] {type3};
\draw (309,22) node [scale=0.9,color={rgb, 255:red, 0; green, 0; blue, 0 }  ,opacity=0.7 ] [align=left] {current\\ \ level};
\draw (221,20) node [scale=0.9,color={rgb, 255:red, 155; green, 155; blue, 155 }  ,opacity=1 ] [align=left] {completed levels};
\draw (390,20) node [scale=0.9,color={rgb, 255:red, 155; green, 155; blue, 155 }  ,opacity=1 ] [align=left] {next levels};
\end{tikzpicture}

\end{figure*}
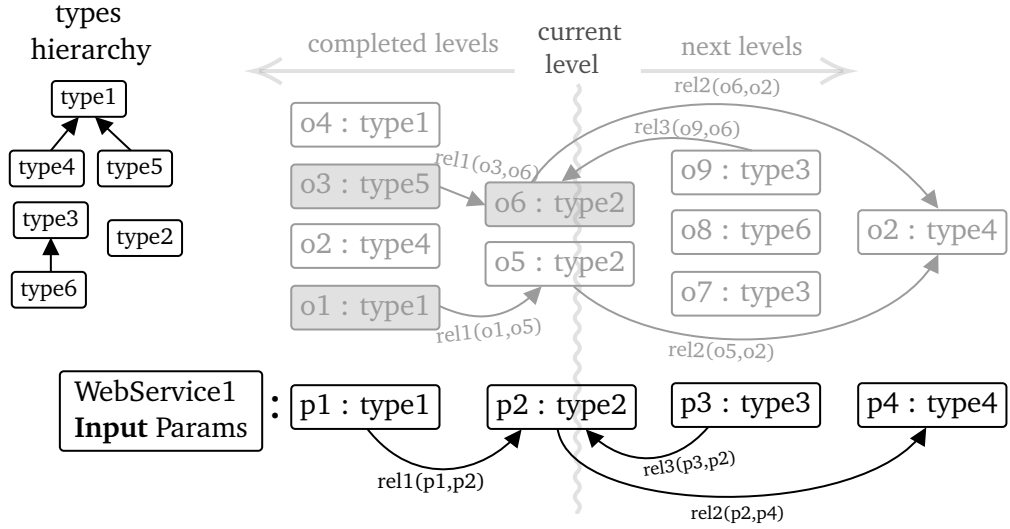

\begin{algorithm}
\caption{Composition search - main loop.}\label{algo:relationalmain}
\begin{algorithmic}[1]
\Function{searchCompositon}{$query$} \Comment{// $ontology$, $repository$ - are global}
\State \Call{callService}{$(\emptyset, query.in), \emptyset$};
\State \Call{applyInferenceRules}{$ $ $ $};
\State $newCall \gets true$;
\While{$\big(newCall $ $ \land $ $ \Call{findMatch}{(query.out, \emptyset)}=null$\big)}
  \State $newCall \gets false$;
  \ForAll {$service \in repository$}
    \State $matchObjects[$ $] \gets \Call{findMatch}{service}$;
  	\If{ $(matchObjects \neq null)$ }
  	  \State $newCall \gets true$;
  	  \State \Call{callService}{$service, matchObjects$};
    \EndIf
  \EndFor
  \State \Call{applyInferenceRules}{$ $ $ $};
\EndWhile
\If{\big(\Call{findMatch}{($query.out$,$\emptyset$)} $\neq$ $null$\big)} $ $ \Return{called services in order};
\State \Comment{// services for which \algoname{callService()} was invoked}
\State \Comment{// also, useless services are dropped}
\Else $ $ \Return{$null$};
\EndIf
\EndFunction
\end{algorithmic}
\end{algorithm}

\vspace{-0.2cm}
\algoname{callService()} - Algorithm \ref{algo:relationalcallservice} takes as parameters the $service$ to call and the already found matching objects. Output objects are created with their name generated from the joined service name with the parameter name and also the number of times the service was called in total. This can help for debugging and getting provenance \cite{ding2010vipen} information about objects later (though this is not yet implemented). After objects are created, all relations between input and output, or between output parameters, defined in the service are added to the matched or new objects according to the order known by their names.

\begin{algorithm}
\caption{Service call for known matched objects.}\label{algo:relationalcallservice}
\begin{algorithmic}[1]
\Function{callService}{$service, matchObjects$}
\State $hash \gets \Call{hashValue}{matchObjects}$;
\State $service.history \gets service.history \cup \{hash\}$; \hspace{0.3cm} $newObjects[$ $]\gets \emptyset$;
\ForAll {$parameter \in service.out$}
  \State $newObjects.add($new object of type $parameter.type)$;
\EndFor
\ForAll {$relation \in (service.relations \setminus service.in^{2})$}  
  \State $\Call{addRelation}{newObjects[x],newObjects[y]}$; \Comment{// for the objects matching parameters in $relation$}
\EndFor
\State $knowledge \gets knowledge \cup newObjects$;
\EndFunction
\end{algorithmic}
\end{algorithm}

\begin{algorithm}
\caption{Backtracking search for matching objects.}\label{algo:relationalfindmatch}
\begin{algorithmic}[1]
\Function{findMatch}{$service$}
  \State $matchFound \gets false$; \hspace{0.5cm} $matchObjects \gets new $ $Object[$ $]$;
  \State $\Call{bktParam}{0, service, matchObjects}$;
  \If {\big($\neg $ $ matchFound$\big)} \Return $null$;
  \Else $ $ \Return $matchObjects$;
  \EndIf
\EndFunction
\Function{bktParam}{$level, service, matchObjects[$ $]$}
\If {\big($level = service.in.size $\big)}
  \State $hash \gets \Call{hashValue}{matchObjects}$;
  \State $matchFound \gets (hash \notin service.history)$;
\Else
  \State $type \gets service.in[level].type$; 
  \State $subTypes \gets ontology.subTypes(type)$;
  \State $candidates \gets knowledge.objectsOfTypes(subTypes)$;
  \ForAll {$candidate \in candidates$}
    \If {($\neg$ $ $matchFound $ \land $ \Call{relationsMatch}{$0...level, candidate, ...$})}
      \State $matchObjects.add(candidate)$;
      \State $\Call{bktParam}{level+1, service, matchObjects}$;
    \EndIf
  \EndFor
\EndIf
\EndFunction
\end{algorithmic}
\end{algorithm}

The most runtime consuming method is \algoname{findMatch()} for services and its equivalent for rules that have exponential complexity. Searching for objects that match variables in rule premises is similar to services, and simpler to implement, as variables are not typed like parameters. The matching was defined in the problem model, exemplified in Figure \ref{fig:inputmatching} and is implemented in Algorithm \ref{algo:relationalfindmatch}. For each level of the backtracking, all possible objects of the type are tried, and object relations are checked to match any parameter relation from the current level to any previous level. Both possible orientations of relations are checked. Reaching one level higher than the number of input parameters indicates a complete match, for which we also lookup in service history. If the match is also new, $matchFound$ is set to $true$, and the search stops for all branches. There are a few intuitive special structures and methods used to get information faster. For example, $ontology.subTypes(...)$ returns all subtypes for a type including itself and $knowledge.objectsOfTypes(...)$ returns all objects known of specified types.

The algorithm presented is more a proof of concept implementation for the composition model. There are many possible enhancements and extra features that could be added.

\subsection{Synthetic Evaluation}
\vspace{-0.3cm}
To evaluate the Algorithm \ref{algo:relationalfindmatch}, we implemented a special test generator. The generator produces a problem instance containing the repository, the ontology and a user query that has a high probability of being solvable. First, it generates a set of services from which a composition can be extracted. To generate services with this property, we use this observation: as services are added in the composition, the knowledge is enhanced: more objects are obtained with their corresponding relations. So, to generate the repository, we start by incrementally generating the knowledge i.e. objects and relations, and saving all intermediate stages of this process. To generate the services we consider that every intermediate stage in the knowledge is the information gained during composition, thus, we need to create services that can generate these stages. Between each two stages $K_i$, $K_{i+1}$ we generate a "layer" of services with each service having as input parameters objects and relations between them from $K_i$ and output parameters in a similar way from $K_{i+1}$. The user query is generated from the input parameters and relations that are a subset of the first stage, while for output the parameters and relations which are a subset of the last stage of the knowledge. "Noise" is added in the repository and ontology, i.e. random concepts and services with to hide the composition. 

\begin{centering}
    \begin{table}
        \caption[\textbf{Relational} model: evaluation on \textbf{generated} tests]{Results on \textbf{generated tests}. Headers: number of services or repository size, composition length, number of applied rules, running time, and composition length if rules are ignored.}
        \label{table:relational_eval_generator}
        \centering
        {\renewcommand{\arraystretch}{1.1}
        \begin{tabular}{|c|c|c|c|c|}
             \hline
             \small{\makecell{\big| repository \big|}} & \small{\makecell{\big| solution \big|}} & \small{\makecell{number of \\ rules applied}} & \small{\makecell{run time \\ (seconds)}} & \small{\makecell{$|$ solution $|$ \\ (ignoring rules)}} \\
             \hline
             63 & 11 & 0 & 0.07 & 13 \\
             30 & 14 & 74 & 0.30 & 15 \\
             30 & 8 & 3 & 0.04 & 13 \\
             46 & 4 & 0 & 0.02 & 4 \\
            \hline
        \end{tabular}
        }
    \end{table}
\end{centering}

We evaluated its efficiency on tests generated as above, with and without inference rules (and also, ignoring them). Results in Table \ref{table:relational_eval_generator} show that the use of inference rules improves the composition, that becomes shorter on two tests; and that the algorithm is efficient. The slowest run is observed on a test-case where more rules are applied, which is expected. 

To compare our solution with others, we also tested the converted benchmark from the composition challenge \cite{bansal2008wsc}. These tests are obviously without any relations or rules, and these results are shown in Table \ref{table:relational_eval_2008}. 
Even if our algorithm is designed for the extended model, it finds a composition of a size comparable with the challenge winners (relative to the size of the repository) and in a shorter amount of time. The second and third columns in Table \ref{table:relational_eval_2008} show our algorithm performance. This experiment also proves that our algorithm implementation is compatible with previous models.

\begin{centering}
    \begin{table}[H]
        \caption[\textbf{Relational} model: evaluation on \textbf{converted WSC 2008} tests]{Results on \textbf{composition challenge tests} \cite{bansal2008wsc}. Headers: repository size; composition length and run time of Algorithm \ref{algo:relationalfindmatch}, composition length and run time of the winning solution of the composition challenge.}
        
        \vspace{-0.2cm}
        
        \label{table:relational_eval_2008}
        \centering
        {\renewcommand{\arraystretch}{1.1}
        \begin{tabular}{|c|c|c|c|c|c|}
             \hline
             \small{\makecell{\big| repository \big|}} & 
             \small{\makecell{\big| solution \big|}} & \small{\makecell{run time \\ (seconds)}} & \small{\makecell{\big| solution in \cite{bansal2008wsc} \big|}} & \small{\makecell{run time in \cite{bansal2008wsc} \\ (seconds)}} \\ 
             \hline
             1041 & 38 & 0.03 & 10 & 0.31 \\ 
             1090 & 62 & 0.04 & 20 & 0.25 \\  
             2198 & 112 & 0.07 & 46 & 0.40 \\
            \hline
        \end{tabular}
        }
    \end{table}
\end{centering}

}

\section{Object-Oriented Model - Algorithm}\label{sec:complexity:objects}
\comment{Back to a computationally simpler model (tough not sure yet ?), mainly presenting the paper submitted to WWW, ICE-B, and WI.}

\iftoggle{fullThesis}{
\subsection{Prerequisites}

The \emph{Object-Oriented Model} was designed specifically to include modern Semantic Web standards. Initially intended to follow the OpenAPI formats, and then more specifically the schema.org data model\footnote{Schema.org -- \url{https://schema.org/docs/datamodel.html}}: parameters (i.e. concepts) have associated properties, inherited through the hierarchy, as described in Section \ref{sec:models:hierarchical}. Surprisingly, the model does not significantly increase the computational complexity, as it is defined in Section \ref{sec:models:object-oriented}. In paper \cite{netedu2019openapi} the general method is analyzed from more perspectives: formats used for the semantic description of parameters, related works in the context of semantic web and the state of the art of Service-Oriented Architectures. We will focus here only on the algorithm and its empirical evaluation.

The proposed algorithm is intended to describe a generic solution that generates a valid composition. Tough it considers some basic optimizations, like special data structures and indexes, there are many ways in which it can be improved, so we shortly describe some of them after the basic algorithm description.

In a simplified form, the main entities have the following structure:
}

\comment{TODO: maybe replace these java-code-like listings with algorithmic style.}

\begin{minipage}{\textwidth}
\begin{lstlisting}[language=Java,basicstyle=\small]
class Concept { // used for full or partially defined
   String name; // considered as a label
   Concept parent; // the "isA" relation 	
   Set<Property> properties; // includes inherited properties
}
class Property {
   String name; 
   Concept type; 
}
class WebService {
   String name; 
   Set<Concept> in, out; // I/O parameters
}	
\end{lstlisting}
\end{minipage}

Global data structures that are most important and often used by the algorithm are presented below:

\vspace{-0.2em}
\begin{lstlisting}[basicstyle=\small,escapeinside={[}{]}\ttfamily]
Set<Concept> C; // knowledge: concepts, isA, and the properties
Set<WebService> webServices; // service repository
WebService Init, Goal; // user's query as two fictive services
Map<Concept, Set<Property>> known;
// partial concepts: known.get(c) = concept's known properties
Map <Concept, Map<Property, Set<WebService>>> required; 
// .get(c).get(p) services that have property p of C in input
Map <WebService, Map<Concept, Set<Property>>> remaining;
// .get(w).get(c) = properties of C necessary to call W
Set<WebService> callableServices;
// services with all input known in the current state
\end{lstlisting}
\vspace{-0.2em}

The algorithm described next uses the above structures, and is composed of three methods: initialization, the main composition search method which calls the last (utility) method, that updates the knowledge base with the new outputs learned from a service call. 

Several obvious instructions are skipped for simplicity, like parsing input data and initializing empty containers.

\vspace{-0.2em}
\begin{lstlisting}[basicstyle=\small,escapeinside={[}{]}\ttfamily]
void initialize() {
 // read the problem instance
 Init.in = Goal.out = [$\emptyset$]; webServices.add(Goal);
 for (WebService ws : webServices) {
  for (Concept c : ws.in) {
   for (Property p : c.properties) {
    // container creation skipped for simplicity
    required.get(c).get(p).add(ws);
    remaining.get(ws).get(c).add(p);       
   }
[\vspace{-20px}]
  }
[\vspace{-20px}]
 }
[\vspace{-20px}]
}    
\end{lstlisting}

\subsection{Property-Learner Composition Algorithm}
After reading the problem instance, the described data structures have to be \emph{loaded}. \textbf{Init} and \textbf{Goal} can be used as web services to reduce the implementation size, if they are initialized as above. Then, for each parameter in service's input, we add the corresponding concepts with their specified properties to the \emph{indexes} (maps) that efficiently get the services that have those properties at input and the properties that remain yet unknown but required to validly call a service.

\begin{lstlisting}[basicstyle=\small,escapeinside={[}{]}\ttfamily]
List<WebService> findComp(WebService Init, Goal) {
 List<WebService> composition; // the result
 callService(Init); // learn the initially known parameters
 
 while (!(required.get(Goal).isEmpty() || callableServices.isEmpty())) {
  WebService ws = callableServices.first();
      callableServices.remove(ws);
      composition.add(ws);
      callWebService(ws);
   }
   if (remaining.get(Goal).isEmpty()) { return composition; }
   } else { return null; } // no solution
[ \vspace{-15px}]
}
\end{lstlisting}

The main method that searches for a valid composition satisfying user's query is \emph{findComp()}. The result is simplified for now as an ordered list of services. As long as the \textbf{Goal} service is not yet callable, but we can call any other new service, we pick the \emph{first} service from the \emph{callableServices} set. Then we simulate its call, basically by learning all its output parameter information, with the help of \emph{callWebService()} method below. We add the selected service to the composition and remove it from \emph{callableServices} so it won't get called again. If \emph{callableServices} empties before reaching the \textbf{Goal}, then the query is unsolvable.

\begin{lstlisting}[basicstyle=\small,escapeinside={[}{]}\ttfamily]
void callWebService(WebService ws) {
 for (Concept c : ws.out) {
  Concept cp = c; // concept that goes up in tree
  boolean added = true; // if anything new was learned
  while (added && cp != null) {
   added = false;
   for (Property p : c.properties) {
    if (cp.properties.contains(p)
        &&!known.get(cp).contains(p)) {
     added = true; known.get(cp).add(p); // learn p at cp level
     for (WebService ws: required.get(cp).get(p)) {
      remaining.get(ws).get(cp).remove(p);
      if (remaining.get(ws).get(cp) .isEmpty()) {
       // all properties of cp in ws.in are known
       remaining.get(ws).remove(cp); }
      if (remaining.get(ws). isEmpty()) {
       // all concepts in ws.in known
       callableServices.add(ws); }
   }}}
   cp = cp.parent;
}}}  
\end{lstlisting}

\iftoggle{fullThesis}{

When calling a Web service, its output is learned and also expanded (we mark as learned properties for more generic concepts). This improves the algorithm's complexity, as it is better to prepare the detection of the newly callable services than to search for them by iteration after any call. This is possible by marking in the tree the known properties for each level and iterating only to levels that get any new information, as higher current service's output would be already known. 

The optimization also comes from the fact that all services with inputs with these properties are hence updated only once (for each learned concept and property). As it gets learned, information is removed from the \emph{remaining} data structure first at the property level and then at the concept level. When there's no property of any concept left to learn, the service gets callable. This can happen only once per service. The main loop might stop before reaching the tree root if at some generalization level, all current services' output was already known, and this is determined by the \emph{added} flag variable.

\subsection{Possible Improvements}

One important metric that the algorithm does not consider is the size of the produced composition. As can be seen from the overview description above, the solution is both deterministic and of polynomial-time complexity. This is possible because the length of the composition is not a necessary minimum. Finding the shortest composition is \texttt{NP-Hard} even for problem definitions that do not model semantics. The proposed model introduces \textit{properties}; this addition does not significantly increase the computational problem complexity. Even if the shortest composition is hard to find, there are at least two simple ways to favor finding shorter compositions with good results.

One is based on the observation that when a service is called, it is chosen from possibly multiple callable services. This choice is not guided by any criteria. It is possible to add at least a simple \textit{heuristic score} to each service that would estimate how useful is the information gained by that service call. Also, this score can be updated for the remaining services when information is learned.

Another improvement is based on the observation that services can be added to the composition even if they might produce no useful information -- there is no condition check that they add anything new, or the information produced could also be found from services added later to the composition. To mitigate this, another algorithm can be implemented that would search the resulting composition \textit{backward} and remove services that proved useless in the end. Both of the above improvements can have an impact on the running time as well, as the algorithm stops when the goal is reached.

\vspace{-0.2em}
\subsection{Empirical Evaluation}

To assess the performance on larger instances, a random test generator was built. The generator first creates a conceptual model with random concepts and properties and then a service repository based on the generated ontology. Service parameters are chosen from direct and inherited properties. In the last step, an ordered list of services is altered by rebuilding each service input. The first one gets its input from \textbf{Init}, which is randomly generated at the beginning. Each other service in order gets its input rebuilt from all previous services outputs, or valid properties of generalizations of those concepts. Finally, \textbf{Goal} has assigned a subset of the set of all outputs of the services in the list. The total number of services and the list size are input parameters for the generator. The intended dependency between services in the list is not guaranteed, so shorter compositions could potentially exist.

\setlength\extrarowheight{2pt}
\begin{table*}[ht]
  \centering
  \caption[\textbf{Object-Oriented} model: evaluation on \textbf{generated} tests]{Algorithm run-times and resulting composition size on random instances.}
  \label{table:oop_runtimes}
  \begin{tabular}{|c|c|c|c|c|}
    \hline
    ontology size & repository & run time & composition & dependencies\\
    \#classes + \#props. & \#services & seconds & \#services & \#services\\
    \hline
    10 (5 + 5) & 10 & 0.002 & \textbf{3} & 5 \\
    20 (10 + 10) & 20 & 0.003 & \textbf{4} & 10 \\
    50 (30 + 20) & 20 & 0.007 & \textbf{12} & 20 \\
    20 (10 + 10) & 50 & 0.011 & \textbf{6} & 20 \\
    \hline
\end{tabular}
\end{table*}

Table \ref{table:oop_runtimes} shows the algorithm performance. The first two columns briefly describe the input size, by the number of concepts and properties in the generated ontology and the total number of services in the repository. The column \emph{composition} specifies the length of the composition found by the algorithm. The last column, \emph{dependencies}, measures the length of the composition generated by the tests generator algorithm. The \emph{dependencies} represents the length of a  valid composition, hidden within the repository and that may contain useless services, as \emph{dependency} is not guaranteed.

}

\section{Online Composition - Algorithm}
\label{sec:complexity:online}
\comment{Presents the algorithm that computes backup compositions for composition requests, the tests generator described at the ICSOFT conference.}

\iftoggle{fullThesis}{

In the proposed definition for the \emph{online} composition, there are more operations to be resolved. The service repository is dynamic, and user requests for composition can be added or dropped as well. Presented in detail in paper \cite{diac2019failover} this version of the problem offers a new perspective on the composition scenario, one that is potentially more applicable. The implemented algorithm builds backup compositions that prepare for scenarios in which one service can fail/break or may be intentionally deleted. It is based on a traversal over the set of \emph{reachable} parameters, where services are used to find new parameters for building the composition. A heuristic score is assigned to services that is used when multiple services can be chosen. 

Our approach also handles dynamic removal of a service at the provider's request, which has similar implications with complete service failures at execution time. 

The composition algorithm is built upon an adjusted breadth-first search traversal over the set of parameters.
This starts from the user known parameters, aiming to obtain the user required parameters.
A loop tries to select the next services to call if any is available.
Learned output parameters are marked as such in a set, and for each service, all unknown input parameters are kept in a map.
When one such map becomes empty, the corresponding service becomes available to call.
The loop stops when all user required parameters are learned.
If multiple services become available at the same time, the one with the best score is selected.
This score approximates the usefulness of the service and is the length of the shortest possible distance from the service to any of the final user required parameters (distance is calculated in the number of services).
When selecting a service, there is no guarantee that the service will be useful in the final composition. Therefore, as a final step, we also execute a composition shortening algorithm.
This shortening algorithm navigates the services in the reverse order of the composition and deletes all services that do not output used parameters.

The design of the solution is to return the composition as soon as constructed to the user to provide a fast response time.
To provide the failover mechanism, the algorithm pre-computes alternative compositions for each user query.
These alternative compositions are searched in a background thread.
Each backup is stored internally at the time it completes.
If a service fails, the backup is provided to the user as the new active composition, and immediately takes an active role instead of the disrupted composition. 
Then, similarly, in the background, a new backup to the new active composition is re-computed, and the old one is deleted.
It is expected that these backup compositions are computed initially before any service failure.
However, in the other, unlikely and worst-case situation, it is easy to detect that the composition is broken without having a backup ready for replacement.
If this happens, the composition is re-built from scratch as for a new query, and the user will just notice a longer response time.

To compute the alternative backup compositions, the algorithm prepares for cases in which each (but only one) service goes down. If more services break at the same time, the recovery can take longer, as the solution is to process failed services sequentially.

The response composition for a query is represented as a simple ordered sequence of services. 
The order defines chronological execution of services and is determined according to how parameters matched between services. 
If one of the services fails, it might be replaceable by other services in the repository. 
To find a solution equivalent to the failed service, we search for a new backup composition.
This search is similar to solving a new user query, and we use again the same composition search algorithm.
However, we need to prepare new appropriate parameter sets. 
This is depicted in Figure \ref{fig:online_delete}, where $ws_{p}$ breaks.
All the services that are situated before the failed service provide a set of \emph{\{known\}} parameters, which includes the user's initially known parameters. 
These will be the input parameters of the new query.
We prepare the output parameters targeting to keep in the new composition the still valid services from the old suffix, after the failed service. 
All new required (output) parameters are collected in the \emph{\{required\} \textbackslash\  \{known\} \textbackslash\  \{gen\}} set, where \textbackslash\ is the set difference operator.
The \emph{\{required\}} set contains all input parameters of any service on the right of $ws_{p}$ and any initially user required parameter. 
Clearly, we can remove the \emph{\{known\}} parameters out of this new query output set.
Parameters in \emph{\{gen\}} are inputs of subsequent services that are also "generated" as outputs of other (previous) services on the right.
If we consider the old suffix after $ws_{p}$ still valid and try to keep it, \emph{\{gen\}} parameters are not actually necessary.
We can also remove them from the new required set. 
The reason for building the sets in this manner is that the more parameters are added to the known and the less are required, the shorter the composition will be.

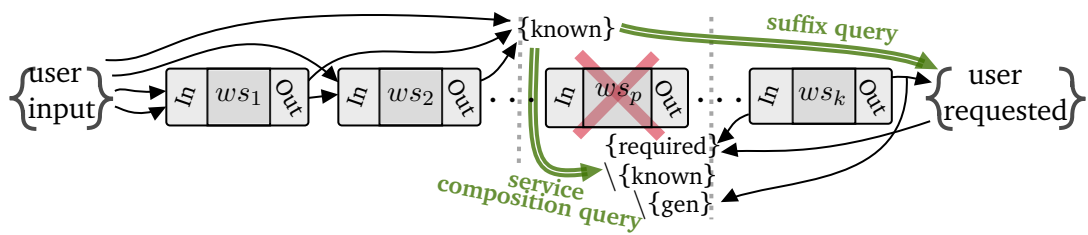
\begin{figure*}[!htbp]
\centering

\caption[Service backup scheme in \textbf{Online Composition} problem]{Service $ws_{p\leq k}$, part of composition \emph{$\langle ws_{1}, ws_{2}, ... , ws_{k}\rangle$} fails. $ws_{p}$ could be replaced by a new composition that uses all parameters learned from outputs of any previous service (on the left), and that generates all parameters that are at the input on the right. If it is not found, we can drop the services on the right of $ws_{p}$ and try an alternative composition replacing all the "suffix" that can potentially be affected by the failed service.}

\label{fig:online_delete}

\tikzset{every picture/.style={line width=0.75pt}}

\begin{tikzpicture}[x=0.70pt,y=0.65pt,yscale=-1,xscale=1]

\draw [color={rgb, 255:red, 208; green, 2; blue, 27 }  ,draw opacity=0.5 ][line width=4.5]    (309.5,29.5) -- (353.5,77.5) ;

\draw [color={rgb, 255:red, 208; green, 2; blue, 27 }  ,draw opacity=0.5 ][line width=4.5]    (350.5,30.33) -- (308.5,76.67) ;

\draw  [fill={rgb, 255:red, 155; green, 155; blue, 155 }  ,fill opacity=0.2 ] (89,40) .. controls (89,38.9) and (89.9,38) .. (91,38) -- (163,38) .. controls (164.1,38) and (165,38.9) .. (165,40) -- (165,69) .. controls (165,70.1) and (164.1,71) .. (163,71) -- (91,71) .. controls (89.9,71) and (89,70.1) .. (89,69) -- cycle ;
\draw  [fill={rgb, 255:red, 155; green, 155; blue, 155 }  ,fill opacity=0.2 ] (181,40) .. controls (181,38.9) and (181.9,38) .. (183,38) -- (255,38) .. controls (256.1,38) and (257,38.9) .. (257,40) -- (257,69) .. controls (257,70.1) and (256.1,71) .. (255,71) -- (183,71) .. controls (181.9,71) and (181,70.1) .. (181,69) -- cycle ;
\draw  [fill={rgb, 255:red, 155; green, 155; blue, 155 }  ,fill opacity=0.2 ] (292,40) .. controls (292,38.9) and (292.9,38) .. (294,38) -- (366,38) .. controls (367.1,38) and (368,38.9) .. (368,40) -- (368,70) .. controls (368,71.1) and (367.1,72) .. (366,72) -- (294,72) .. controls (292.9,72) and (292,71.1) .. (292,70) -- cycle ;
\draw  [fill={rgb, 255:red, 155; green, 155; blue, 155 }  ,fill opacity=0.2 ] (401,40) .. controls (401,38.9) and (401.9,38) .. (403,38) -- (475,38) .. controls (476.1,38) and (477,38.9) .. (477,40) -- (477,68) .. controls (477,69.1) and (476.1,70) .. (475,70) -- (403,70) .. controls (401.9,70) and (401,69.1) .. (401,68) -- cycle ;
\draw  [fill={rgb, 255:red, 155; green, 155; blue, 155 }  ,fill opacity=0.2 ] (312,38) -- (348.5,38) -- (348.5,72) -- (312,72) -- cycle ;
\draw  [color={rgb, 255:red, 74; green, 74; blue, 74 }  ,draw opacity=1 ][line width=1.5]  (19,38) .. controls (14.33,38.07) and (12.03,40.43) .. (12.1,45.1) -- (12.1,45.1) .. controls (12.19,51.77) and (9.91,55.13) .. (5.24,55.2) .. controls (9.91,55.13) and (12.29,58.43) .. (12.39,65.1)(12.34,62.1) -- (12.4,66.1) .. controls (12.47,70.77) and (14.83,73.07) .. (19.5,73) ;
\draw  [color={rgb, 255:red, 74; green, 74; blue, 74 }  ,draw opacity=1 ][line width=1.5]  (46.5,72.5) .. controls (51.17,72.5) and (53.5,70.17) .. (53.5,65.5) -- (53.5,65.5) .. controls (53.5,58.83) and (55.83,55.5) .. (60.5,55.5) .. controls (55.83,55.5) and (53.5,52.17) .. (53.5,45.5)(53.5,48.5) -- (53.5,45) .. controls (53.5,40.33) and (51.17,38) .. (46.5,38) ;
\draw  [color={rgb, 255:red, 74; green, 74; blue, 74 }  ,draw opacity=1 ][line width=1.5]  (509,38) .. controls (504.33,38.07) and (502.03,40.43) .. (502.1,45.1) -- (502.1,45.1) .. controls (502.19,51.77) and (499.91,55.13) .. (495.24,55.2) .. controls (499.91,55.13) and (502.29,58.43) .. (502.39,65.1)(502.34,62.1) -- (502.4,66.1) .. controls (502.47,70.77) and (504.83,73.07) .. (509.5,73) ;
\draw  [color={rgb, 255:red, 74; green, 74; blue, 74 }  ,draw opacity=1 ][line width=1.5]  (566.5,72.5) .. controls (571.17,72.5) and (573.5,70.17) .. (573.5,65.5) -- (573.5,65.5) .. controls (573.5,58.83) and (575.83,55.5) .. (580.5,55.5) .. controls (575.83,55.5) and (573.5,52.17) .. (573.5,45.5)(573.5,48.5) -- (573.5,45) .. controls (573.5,40.33) and (571.17,38) .. (566.5,38) ;
\draw    (61.5,51) .. controls (66.33,48.11) and (69.29,52.66) .. (87.01,51.18) ;
\draw [shift={(89,51)}, rotate = 534.14] [fill={rgb, 255:red, 0; green, 0; blue, 0 }  ][line width=0.75]  [draw opacity=0] (8.93,-4.29) -- (0,0) -- (8.93,4.29) -- cycle    ;

\draw    (61.5,60) .. controls (64.88,60.97) and (72.91,67.52) .. (87.4,60.79) ;
\draw [shift={(89,60)}, rotate = 512.7] [fill={rgb, 255:red, 0; green, 0; blue, 0 }  ][line width=0.75]  [draw opacity=0] (8.93,-4.29) -- (0,0) -- (8.93,4.29) -- cycle    ;

\draw    (60,42) .. controls (93.66,42.99) and (133.2,-1.1) .. (179.59,44.59) ;
\draw [shift={(181,46)}, rotate = 225.6] [fill={rgb, 255:red, 0; green, 0; blue, 0 }  ][line width=0.75]  [draw opacity=0] (8.93,-4.29) -- (0,0) -- (8.93,4.29) -- cycle    ;

\draw    (165,55) -- (179.02,53.25) ;
\draw [shift={(181,53)}, rotate = 532.87] [fill={rgb, 255:red, 0; green, 0; blue, 0 }  ][line width=0.75]  [draw opacity=0] (8.93,-4.29) -- (0,0) -- (8.93,4.29) -- cycle    ;

\draw [color={rgb, 255:red, 128; green, 128; blue, 128 }  ,draw opacity=0.7 ][line width=1.5]  [dash pattern={on 1.69pt off 2.76pt}]  (278.5,8) -- (277.5,94.33) ;

\draw [color={rgb, 255:red, 128; green, 128; blue, 128 }  ,draw opacity=0.7 ][line width=1.5]  [dash pattern={on 1.69pt off 2.76pt}]  (380.5,8) -- (380.5,125.5) ;

\draw    (56.5,33) .. controls (97,36) and (88,-4.8) .. (273,10.2) ;
\draw [shift={(273,10.2)}, rotate = 184.64] [fill={rgb, 255:red, 0; green, 0; blue, 0 }  ][line width=0.75]  [draw opacity=0] (8.93,-4.29) -- (0,0) -- (8.93,4.29) -- cycle    ;

\draw    (165,45) .. controls (194.21,10.35) and (233.21,44.5) .. (272.8,16.08) ;
\draw [shift={(274,15.2)}, rotate = 503.13] [fill={rgb, 255:red, 0; green, 0; blue, 0 }  ][line width=0.75]  [draw opacity=0] (8.93,-4.29) -- (0,0) -- (8.93,4.29) -- cycle    ;

\draw    (257,45) .. controls (266.03,41.68) and (269.18,34.29) .. (275.48,24.13) ;
\draw [shift={(276.5,22.5)}, rotate = 482.47] [fill={rgb, 255:red, 0; green, 0; blue, 0 }  ][line width=0.75]  [draw opacity=0] (8.93,-4.29) -- (0,0) -- (8.93,4.29) -- cycle    ;

\draw    (400.5,64.5) .. controls (390.22,68.24) and (391.29,73.73) .. (384.94,80.15) ;
\draw [shift={(383.5,81.5)}, rotate = 318.81] [fill={rgb, 255:red, 0; green, 0; blue, 0 }  ][line width=0.75]  [draw opacity=0] (8.93,-4.29) -- (0,0) -- (8.93,4.29) -- cycle    ;

\draw    (497.5,68.5) .. controls (434.45,91.78) and (407.15,84.01) .. (387.32,86.26) ;
\draw [shift={(385.5,86.5)}, rotate = 351.47] [fill={rgb, 255:red, 0; green, 0; blue, 0 }  ][line width=0.75]  [draw opacity=0] (8.93,-4.29) -- (0,0) -- (8.93,4.29) -- cycle    ;

\draw [color={rgb, 255:red, 65; green, 117; blue, 5 }  ,draw opacity=0.8 ][line width=1.5]    (287.5,26.07) .. controls (286.91,39.1) and (286.56,49.87) .. (286.56,58.76) .. controls (286.56,80.91) and (288.44,91.23) .. (294.83,95.08) .. controls (297.2,96.5) and (300.18,96.99) .. (303.82,96.99) .. controls (308.16,96.99) and (313.41,96.28) .. (313.82,96.19)(284.5,25.93) .. controls (283.91,39.01) and (283.56,49.83) .. (283.56,58.76) .. controls (283.56,83.12) and (286.44,93.53) .. (293.29,97.65) .. controls (296.07,99.32) and (299.53,99.99) .. (303.82,99.99) .. controls (308.28,99.99) and (313.67,99.28) .. (314.25,99.16) ;
\draw [shift={(322.5,96.5)}, rotate = 531.87] [fill={rgb, 255:red, 65; green, 117; blue, 5 }  ,fill opacity=0.8 ][line width=1.5]  [draw opacity=0] (11.61,-5.58) -- (0,0) -- (11.61,5.58) -- cycle    ;

\draw    (477,43) .. controls (485.74,42.08) and (489.82,43.7) .. (496.65,45.52) ;
\draw [shift={(498.5,46)}, rotate = 194.04] [fill={rgb, 255:red, 0; green, 0; blue, 0 }  ][line width=0.75]  [draw opacity=0] (8.93,-4.29) -- (0,0) -- (8.93,4.29) -- cycle    ;

\draw    (485,43) .. controls (487.46,108.5) and (417.64,100.75) .. (386.88,114.84) ;
\draw [shift={(385.5,115.5)}, rotate = 333.43] [fill={rgb, 255:red, 0; green, 0; blue, 0 }  ][line width=0.75]  [draw opacity=0] (8.93,-4.29) -- (0,0) -- (8.93,4.29) -- cycle    ;

\draw [color={rgb, 255:red, 65; green, 117; blue, 5 }  ,draw opacity=0.8 ][line width=1.5]    (332.48,12.58) .. controls (344.6,16.68) and (367.83,18.51) .. (393.74,20.37) .. controls (396.33,20.55) and (398.94,20.74) .. (401.57,20.93) .. controls (407.43,21.34) and (413.38,21.77) .. (419.33,22.24) .. controls (450.62,24.67) and (481.9,28.12) .. (494.56,33.87)(331.52,15.42) .. controls (343.79,19.58) and (367.29,21.48) .. (393.53,23.36) .. controls (396.11,23.55) and (398.73,23.73) .. (401.36,23.92) .. controls (407.21,24.34) and (413.15,24.76) .. (419.1,25.23) .. controls (450,27.63) and (480.92,30.96) .. (493.43,36.65) ;
\draw [shift={(501.79,38.43)}, rotate = 205.94] [fill={rgb, 255:red, 65; green, 117; blue, 5 }  ,fill opacity=0.8 ][line width=1.5]  [draw opacity=0] (11.61,-5.58) -- (0,0) -- (11.61,5.58) -- cycle    ;

\draw   (111,38) -- (144.5,38) -- (144.5,71) -- (111,71) -- cycle ;

\draw   (203,38) -- (236.5,38) -- (236.5,71) -- (203,71) -- cycle ;
\draw   (424,38) -- (457.5,38) -- (457.5,70) -- (424,70) -- cycle ;

\draw (33,54) node  [align=left] {user\\input};
\draw (538,55) node  [align=left] { \ \ \ user\\requested};
\draw  [color={rgb, 255:red, 155; green, 155; blue, 155 }  ,draw opacity=0.2 ][fill={rgb, 255:red, 155; green, 155; blue, 155 }  ,fill opacity=0.2 ]  (111.5,40) -- (144.5,40) -- (144.5,69) -- (111.5,69) -- cycle  ;
\draw (128,54.5) node  [align=left] {$ws_{1}$};
\draw (99,53.47) node [scale=0.8,rotate=-289.12] [align=left] {In};
\draw (155,54.5) node [scale=0.8,rotate=-60.07] [align=left] {Out};
\draw  [color={rgb, 255:red, 155; green, 155; blue, 155 }  ,draw opacity=0.2 ][fill={rgb, 255:red, 155; green, 155; blue, 155 }  ,fill opacity=0.2 ]  (203.5,40) -- (236.5,40) -- (236.5,69) -- (203.5,69) -- cycle  ;
\draw (220,54.5) node  [align=left] {$ws_{2}$};
\draw (191,53.47) node [scale=0.8,rotate=-289.12] [align=left] {In};
\draw (247,54.5) node [scale=0.8,rotate=-60.07] [align=left] {Out};
\draw (331,53) node  [align=left] {$ws_{p}$};
\draw (302,53.94) node [scale=0.8,rotate=-289.12] [align=left] {In};
\draw (358,55) node [scale=0.8,rotate=-60.07] [align=left] {Out};
\draw  [color={rgb, 255:red, 155; green, 155; blue, 155 }  ,draw opacity=0.2 ][fill={rgb, 255:red, 155; green, 155; blue, 155 }  ,fill opacity=0.2 ]  (424,39.5) -- (456,39.5) -- (456,68.5) -- (424,68.5) -- cycle  ;
\draw (440,54) node  [align=left] {$ws_{k}$};
\draw (411,53) node [scale=0.8,rotate=-289.12] [align=left] {In};
\draw (467,54) node [scale=0.8,rotate=-60.07] [align=left] {Out};
\draw (274,56) node  [align=left] {\textbf{. . .}};
\draw (385,56) node  [align=left] {\textbf{. . .}};
\draw (303,15) node  [align=left] {\{{\footnotesize known}\}};
\draw (354.5,92) node  [align=left] {\{{\footnotesize required}\}\\};
\draw (292.42,105.55) node [color={rgb, 255:red, 65; green, 117; blue, 5 }  ,opacity=0.8 ,rotate=-11.14] [align=left] {{\footnotesize \textbf{service}}};
\draw (360,117) node  [align=left] {\textbf{$\setminus$}\{{\footnotesize gen}\}};
\draw (353,101) node  [align=left] {\textbf{$\setminus$}\{{\footnotesize known}\}};
\draw (445.13,14.8) node [color={rgb, 255:red, 65; green, 117; blue, 5 }  ,opacity=0.8 ,rotate=-5.55] [align=left] {{\footnotesize \textbf{suffix query}}};
\draw (288.42,116.55) node [color={rgb, 255:red, 65; green, 117; blue, 5 }  ,opacity=0.8 ,rotate=-11.14] [align=left] {{\footnotesize \textbf{composition query}}};

\end{tikzpicture}
\end{figure*}

If no composition is found for the new ( \emph{\{known\}} $\rightarrow$ \emph{\{required\} \textbackslash\  \{known\} \textbackslash\  \{gen\}} ) query, it does not necessarily mean that no solution exists. 
There might be the case that services succeeding $ws_{p}$ use some parameters that are no longer accessible.
In such a situation, all the composition's suffix, i.e. services on the right of $ws_{p}$ must be reconstructed. 
This, again, can be done by a new composition query: ( \emph{\{known\}} $\rightarrow$ \emph{\{user requested\} \textbackslash\  \{known\}} ), displayed as the \emph{suffix query} in Figure \ref{fig:online_delete}.

If this does not succeed either, the user query is no longer solvable, because only keeping the services on the left of $ws_{p}$ can be at most useless, but not wrong.

Regardless of how the composition was re-constructed, the algorithm does not maintain the backups for the "partial" compositions themselves (i.e. the ones replacing just $ws_{p}$ or the suffix of the initial composition).
Backup solutions are built for the whole new composition.
The reduction algorithm is run on the new composition, eliminating useless or duplicate services that can appear.

Another action taken when a service fails is to rebuild all backup compositions that use it. 
These are backups for solutions of other user queries that are still working, i.e. that do not contain the failed service in the main composition but only in their backups.

\subsection{Data Structures}

We summarize in the following the data structures used in our proposed failover algorithms. 
We refer here only to the main global structures that keep high-level information like the services, the user queries and the solutions with their backups:

\vspace{0.2cm}

$\rightarrow$ Set$\langle$Service$\rangle$ \textbf{repository}: set $\mathbb{R}$ of all services;
\\ $\rightarrow$ Map$\langle$Parameter, Set$\langle$Service$\rangle$$\rangle$ \textbf{inputFor}: all services that have a specific parameter as input;
\\ $\rightarrow$ Set$\langle$Query$\rangle$ \textbf{requests}: set $\mathbb{Q}$ of user queries;
\\ $\rightarrow$ \textbf{Query} objects contain: \textbf{.In} and \textbf{.Out} : user known and requested parameters;
\\ $\rightarrow$ Map$\langle$Query, Solution$\rangle$ \textbf{compositions}: solutions;
\\ $\rightarrow$ \textbf{Solution} objects contain: Query \textbf{.query}, the resolved query; Array$\langle$Service$\rangle$ \textbf{.main}, services in solution; Map$\langle$Service, \text{Solution}$\rangle$ \textbf{.backup}, backup solution for each used service and Solution \textbf{.parent}, the composition it is backup to if $.main$ is a backup;
\\ $\rightarrow$ Map$\langle$Service, Set$\langle$Solution$\rangle\rangle$ \textbf{usages}: main or backup compositions using some service.

\subsection{Failover Algorithm}

The first functions used in Algorithm \ref{algo:online_findcomposition} create a composition for a user query. 
This integrates with computing the needed backups for failure situations, described in Algorithm \ref{algo:findbackup}. 
The functions of the algorithms use a series of temporary structures instantiated per query: \textbf{known}, \textbf{required}, \textbf{score}, and \textbf{ready}. 

\begin{algorithm}
\caption{Find composition and learning.}\label{algo:online_findcomposition}
\begin{algorithmic}[1]
\State Set$\langle$Parameter$\rangle$ \textbf{known}; \textcolor{gray}{// known parameters}
\State Map$\langle$Service, Set$\langle$Parameter$\rangle\rangle$ \textbf{required}; \textcolor{gray}{// input parameters of each service that are not yet \textbf{known}}
\State Map$\langle$Service, Int$\rangle$ \textbf{score}; \textcolor{gray}{// heuristic service score}
\State Set$\langle$Service$\rangle$ \textbf{ready}; \textcolor{gray}{// services with empty \textbf{required}$.get()$, so callable; ordered by scores.}
\State 
\Function{findCompositon}{$query$}
\State solution $\gets$ new Solution(); 
\State solution.query $\gets$ query; \ known, ready $\gets \emptyset;$
\State score $\gets $ \Call{computeServiceScores}{query};
\ForAll {service $\in$ repository}
    \State required.put(service, service.In);
\EndFor

\State \Call{learnParameters}{query.In};
\While{($\neg$\small{ }ready.empty() $\land $\small{ }query.Out$\small{ }\nsubseteq$ known\normalsize{})}
    \State nextService $\gets$ ready.first(); \textcolor{gray}{ // best scoring}
    \State solution.main.add(nextService);
    \State \Call{learnParameters}{nextService.Out};
\EndWhile
\If{(query.Out $\nsubseteq$ known)}
    \State \Return{$NULL;$} \textcolor{gray}{ // \textbf{query} is unsolvable}
\Else
    \State \Call{removeUselessServices}{solution, query};
\EndIf
\ForAll {(service $\in$ solution.main)}
    \State usages.get(service).add(solution);
\EndFor
\State \Return{solution};
\EndFunction
\State
\Function{learnParams}{Set$\langle$Parameter$\rangle $ pars}
\ForAll{(service $\in$ (repository $\setminus$ ready))}
    \State required.get(service).removeAll(pars);
    \If {(required.get(service).isEmpty())}
        \State ready.add(service); \textcolor{gray}{// just became callable}
    \EndIf
    
\EndFor
\State known $\gets$ known $\cup$ pars;
\EndFunction
\end{algorithmic}
\end{algorithm}

Their exact role is explained at the beginning of the listings. 
The \algoname{findComposition()} function is used to find the main, first solution for any composition query. 
This makes use of \algoname{learnParams()} helper function, which has the role to maintain the known parameters set up-to-date during the algorithm's execution. 
The function \algoname{findBackup()} is responsible for computing the backup for one component service in a composition.
The functions are wrapped in \algoname{backupComposition()}, which is essentially the function that the user should call for executing a composition with failover support.

\begin{algorithm}[H]
\caption{Finding backup alternatives.}\label{algo:findbackup}
\begin{algorithmic}[1]

\Function{backupComposition}{query}
\State solution $\gets$ \Call{findComposition}{query};

\ForAll{(service $\in$ solution.main)}
    \State bkp $\gets$ \Call{findBackup}{solution, service};
    \State solution.backup.put(service, bkp);
\EndFor
\EndFunction
\State 
\Function{findBackup}{sol, service}
\State p $\gets$ sol.main.indexOf(service);
\State $known_{bkp}$ $\gets$ sol.query.In;
\For{(i $\gets$ 0 to p - 1)}
    \State $known_{bkp}$ $\gets$ $known_{bkp}$ $\cup$ sol.main.get(i).Out;
\EndFor
\State $req_{bkp}$ $\gets$ sol.query.Out;
\For{(i $\gets$ sol.main.length down-to p + 1)}
    \State $req_{bkp}$ $\gets$ $req_{bkp}$ $\setminus$ sol.main.get(i).Out;
    \State \textcolor{gray}{// remove \{gen\}, as in Figure \ref{fig:online_delete}}
    \State $req_{bkp}$ $\gets$ $req_{bkp}$ $\cup$ sol.main.get(i).In;
\EndFor
\State s $\gets$ \Call{findComposition}{($known_{bkp}$, $req_{bkp}$)};
\If {(s == $NULL$)}
    \State query $\gets$ ($known_{bkp}$,sol.query.Out$\setminus$$known_{bkp}$);
    \State s $\gets$ \Call{findComposition}{query};
    \State \textcolor{gray} {// second type or ''suffix-query'' backup}
\EndIf
\If {(s $\neq$ $NULL$)} \ \ s.parent $\gets$ sol;
\EndIf

\State \Return{s};
\EndFunction
\end{algorithmic}
\end{algorithm}

For simplicity, we omitted the listing of some helper functions. 
\algoname{computeServiceScores()} computes scores indicating how good services are to be part of the composition.
The function performs a reversed order traversal of all services, starting from services that output any of the user's required parameters.
These get the best score which is \emph{1}. 
Following, services that output any input of services with score \emph{1} get a score of \emph{2}, and so on. 
Later on, line 14 of Algorithm \ref{algo:online_findcomposition}, the service with the best, lowest value score, is chosen out of all services currently available and added to the composition.
The \algoname{removeUselessServices()} is another helper function, that is a shortening algorithm and used for cleaning the composition of any services that are not strictly necessary.

Several optimizations can be applied to our algorithm construction. 
It is easy to observe that the calls to \algoname{findBackup()} for finding backups to each of the services in the composition can be executed in parallel. 
Otherwise, an improvement can also be obtained by building \emph{\{$known_{bkp}$\}} and \emph{\{$required_{bkp}$\}} sets only once instead of for each \algoname{findBackup()} call. 

In Algorithm \ref{algo:deleteService} we describe the delete service method.
This is called when a service provider specifically deletes a service or when a service fails. 
Both cases are treated similarly triggering replacement with a backup and a new backup re-computation.

\begin{algorithm}
\caption{Remove service method.}\label{algo:deleteService}
\begin{algorithmic}[1]
\Function{deleteService}{$ws$}
\State repository.remove($ws$);
\ForAll {(param $\in$ $ws$.In)}
    \State inputFor.get(param).remove($ws$);
\EndFor
\ForAll {(sol $\in$ usages.get($ws$))}
    \If{(sol.parent == $NULL$)} \textcolor{gray}{ // it is a main solution}
        \State bkpSol $\gets$ sol.backup.get($ws$);
        \State sol.backup.clear(); \textcolor{gray}{// also update \textbf{usages}}
        \State compositions.put(sol.query, bkpSol); \textcolor{gray}{// swap with backup, notify user}
        \If{(bkpSol $\neq$ $NULL$)}
            \ForAll {serv $\in$ bkpSol.main}
                \State newBkp $\gets$ \Call{findBackup}{bkpSol, serv};
                \State bkpSol.backup.put(serv, newBkp);
            \EndFor
        \EndIf
        \State \textcolor{gray}{// else \textbf{sol.query} becomes unsolvable, nothing to do}
    \Else \textcolor{gray}{$ $ // \textbf{sol} was a backup, try to find another}
        \State newBkp $\gets$ \Call{findBackup}{sol.parent, ws};
        \State sol.parent.backup.put(ws, newBkp);
    \EndIf
\EndFor
\State usages.get($ws$).clear();
\EndFunction
\end{algorithmic}
\end{algorithm}

\subsection{Empirical Evaluation}
\noindent
We evaluated our algorithms on synthetic generated data, created by a special tests generator that creates failover scenarios, with and without backups. First, several queries are generated, each from two disjoint sets of random parameters: input and output, i.e. initially known and requested parameters. 

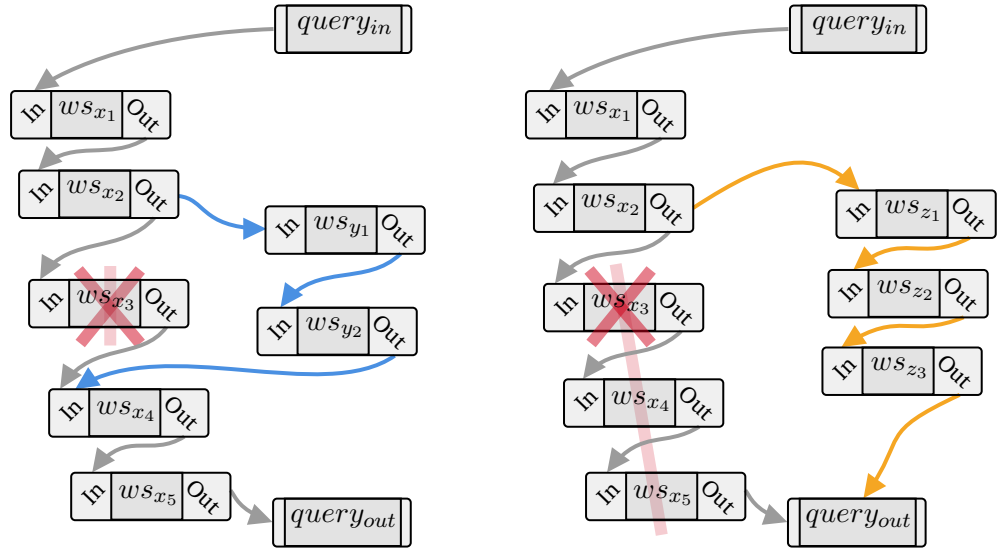
\begin{figure*}[!htbp]
\centering

\captionsetup{width=1\linewidth}
\caption[Two types of backups generated for \textbf{Online Composition}]{The two types of backups generated: replacing only one service on the left; and replacing all successive services or the "suffix", on the right.}
\label{fig:online_generator}
\vspace{0.2cm}

\tikzset{every picture/.style={line width=0.75pt}} 

\begin{tikzpicture}[x=0.75pt,y=0.75pt,yscale=-1,xscale=1]

\draw [color={rgb, 255:red, 208; green, 2; blue, 27 }  ,draw opacity=0.5 ][line width=4.5]    (44.5,140.2) -- (75.5,177.2) ;

\draw [color={rgb, 255:red, 208; green, 2; blue, 27 }  ,draw opacity=0.5 ][line width=4.5]    (75.5,140.2) -- (45.5,177.2) ;

\draw [color={rgb, 255:red, 208; green, 2; blue, 27 }  ,draw opacity=0.2 ][line width=4.5]    (60.5,138.2) -- (60.5,178.2) ;

\draw [color={rgb, 255:red, 208; green, 2; blue, 27 }  ,draw opacity=0.2 ][line width=4.5]    (312.77,137.28) -- (335.79,273.43) ;

\draw [color={rgb, 255:red, 74; green, 144; blue, 226 }  ,draw opacity=1 ][line width=1.5]    (94.5,103) .. controls (108.36,104.39) and (105.01,118.54) .. (135.57,119.45) ;
\draw [shift={(138.5,119.5)}, rotate = 180.12] [fill={rgb, 255:red, 74; green, 144; blue, 226 }  ,fill opacity=1 ][line width=1.5]  [draw opacity=0] (11.61,-5.58) -- (0,0) -- (11.61,5.58) -- cycle    ;

\draw [color={rgb, 255:red, 74; green, 144; blue, 226 }  ,draw opacity=1 ][line width=1.5]    (205.5,132) .. controls (188.13,148.73) and (164.24,136.27) .. (147.31,156.65) ;
\draw [shift={(145.5,159)}, rotate = 305.69] [fill={rgb, 255:red, 74; green, 144; blue, 226 }  ,fill opacity=1 ][line width=1.5]  [draw opacity=0] (11.61,-5.58) -- (0,0) -- (11.61,5.58) -- cycle    ;

\draw [color={rgb, 255:red, 74; green, 144; blue, 226 }  ,draw opacity=1 ][line width=1.5]    (202.5,183.33) .. controls (167.04,208.16) and (87.92,172.44) .. (44.94,198.95) ;
\draw [shift={(43,200.2)}, rotate = 325.81] [fill={rgb, 255:red, 74; green, 144; blue, 226 }  ,fill opacity=1 ][line width=1.5]  [draw opacity=0] (11.61,-5.58) -- (0,0) -- (11.61,5.58) -- cycle    ;

\draw [color={rgb, 255:red, 155; green, 155; blue, 155 }  ,draw opacity=1 ][line width=1.5]    (120.5,251) .. controls (123.34,254.78) and (124.38,260.35) .. (138.84,266.01) ;
\draw [shift={(141.5,267)}, rotate = 199.44] [fill={rgb, 255:red, 155; green, 155; blue, 155 }  ,fill opacity=1 ][line width=1.5]  [draw opacity=0] (11.61,-5.58) -- (0,0) -- (11.61,5.58) -- cycle    ;

\draw [color={rgb, 255:red, 155; green, 155; blue, 155 }  ,draw opacity=1 ][line width=1.5]    (97.5,224) .. controls (80.13,234.81) and (69.28,220.1) .. (53.26,239.73) ;
\draw [shift={(51.5,242)}, rotate = 306.47] [fill={rgb, 255:red, 155; green, 155; blue, 155 }  ,fill opacity=1 ][line width=1.5]  [draw opacity=0] (11.61,-5.58) -- (0,0) -- (11.61,5.58) -- cycle    ;

\draw [color={rgb, 255:red, 155; green, 155; blue, 155 }  ,draw opacity=1 ][line width=1.5]    (142,19) .. controls (88.44,20.93) and (46.53,34.03) .. (24.79,48.43) ;
\draw [shift={(22.5,50)}, rotate = 324.46000000000004] [fill={rgb, 255:red, 155; green, 155; blue, 155 }  ,fill opacity=1 ][line width=1.5]  [draw opacity=0] (11.61,-5.58) -- (0,0) -- (11.61,5.58) -- cycle    ;

\draw [color={rgb, 255:red, 155; green, 155; blue, 155 }  ,draw opacity=1 ][line width=1.5]    (78.5,74) .. controls (59.78,85.52) and (42.9,74) .. (26.54,88.1) ;
\draw [shift={(24.5,90)}, rotate = 315] [fill={rgb, 255:red, 155; green, 155; blue, 155 }  ,fill opacity=1 ][line width=1.5]  [draw opacity=0] (11.61,-5.58) -- (0,0) -- (11.61,5.58) -- cycle    ;

\draw [color={rgb, 255:red, 155; green, 155; blue, 155 }  ,draw opacity=1 ][line width=1.5]    (82.5,114) .. controls (66.09,132.34) and (40.38,122.73) .. (26.92,142.71) ;
\draw [shift={(25.5,145)}, rotate = 299.48] [fill={rgb, 255:red, 155; green, 155; blue, 155 }  ,fill opacity=1 ][line width=1.5]  [draw opacity=0] (11.61,-5.58) -- (0,0) -- (11.61,5.58) -- cycle    ;

\draw [color={rgb, 255:red, 155; green, 155; blue, 155 }  ,draw opacity=1 ][line width=1.5]    (88.5,169) .. controls (75.95,185.73) and (52.5,176.16) .. (37.14,197.55) ;
\draw [shift={(35.5,200)}, rotate = 301.89] [fill={rgb, 255:red, 155; green, 155; blue, 155 }  ,fill opacity=1 ][line width=1.5]  [draw opacity=0] (11.61,-5.58) -- (0,0) -- (11.61,5.58) -- cycle    ;

\draw  [fill={rgb, 255:red, 155; green, 155; blue, 155 }  ,fill opacity=0.15 ] (11,52.2) .. controls (11,51.1) and (11.9,50.2) .. (13,50.2) -- (88.5,50.2) .. controls (89.6,50.2) and (90.5,51.1) .. (90.5,52.2) -- (90.5,72.07) .. controls (90.5,73.17) and (89.6,74.07) .. (88.5,74.07) -- (13,74.07) .. controls (11.9,74.07) and (11,73.17) .. (11,72.07) -- cycle ;
\draw  [fill={rgb, 255:red, 155; green, 155; blue, 155 }  ,fill opacity=0.15 ] (31,50.2) -- (66.5,50.2) -- (66.5,73.79) -- (31,73.79) -- cycle ;

\draw  [fill={rgb, 255:red, 155; green, 155; blue, 155 }  ,fill opacity=0.15 ] (142.79,9.2) .. controls (142.79,8.1) and (143.68,7.2) .. (144.79,7.2) -- (208.79,7.2) .. controls (209.89,7.2) and (210.79,8.1) .. (210.79,9.2) -- (210.79,29.07) .. controls (210.79,30.17) and (209.89,31.07) .. (208.79,31.07) -- (144.79,31.07) .. controls (143.68,31.07) and (142.79,30.17) .. (142.79,29.07) -- cycle ;
\draw  [fill={rgb, 255:red, 155; green, 155; blue, 155 }  ,fill opacity=0.15 ] (148.59,7.2) -- (205.9,7.2) -- (205.9,30.79) -- (148.59,30.79) -- cycle ;

\draw  [fill={rgb, 255:red, 155; green, 155; blue, 155 }  ,fill opacity=0.15 ] (15,92.2) .. controls (15,91.1) and (15.9,90.2) .. (17,90.2) -- (92.5,90.2) .. controls (93.6,90.2) and (94.5,91.1) .. (94.5,92.2) -- (94.5,112.07) .. controls (94.5,113.17) and (93.6,114.07) .. (92.5,114.07) -- (17,114.07) .. controls (15.9,114.07) and (15,113.17) .. (15,112.07) -- cycle ;
\draw  [fill={rgb, 255:red, 155; green, 155; blue, 155 }  ,fill opacity=0.15 ] (35,90.2) -- (70.5,90.2) -- (70.5,113.79) -- (35,113.79) -- cycle ;

\draw  [fill={rgb, 255:red, 155; green, 155; blue, 155 }  ,fill opacity=0.15 ] (20,147.2) .. controls (20,146.1) and (20.9,145.2) .. (22,145.2) -- (97.5,145.2) .. controls (98.6,145.2) and (99.5,146.1) .. (99.5,147.2) -- (99.5,167.07) .. controls (99.5,168.17) and (98.6,169.07) .. (97.5,169.07) -- (22,169.07) .. controls (20.9,169.07) and (20,168.17) .. (20,167.07) -- cycle ;
\draw  [fill={rgb, 255:red, 155; green, 155; blue, 155 }  ,fill opacity=0.15 ] (40,145.2) -- (75.5,145.2) -- (75.5,168.79) -- (40,168.79) -- cycle ;

\draw  [fill={rgb, 255:red, 155; green, 155; blue, 155 }  ,fill opacity=0.15 ] (30,202.2) .. controls (30,201.1) and (30.9,200.2) .. (32,200.2) -- (107.5,200.2) .. controls (108.6,200.2) and (109.5,201.1) .. (109.5,202.2) -- (109.5,222.07) .. controls (109.5,223.17) and (108.6,224.07) .. (107.5,224.07) -- (32,224.07) .. controls (30.9,224.07) and (30,223.17) .. (30,222.07) -- cycle ;
\draw  [fill={rgb, 255:red, 155; green, 155; blue, 155 }  ,fill opacity=0.15 ] (50,200.2) -- (85.5,200.2) -- (85.5,223.79) -- (50,223.79) -- cycle ;

\draw  [fill={rgb, 255:red, 155; green, 155; blue, 155 }  ,fill opacity=0.15 ] (41,244.2) .. controls (41,243.1) and (41.9,242.2) .. (43,242.2) -- (118.5,242.2) .. controls (119.6,242.2) and (120.5,243.1) .. (120.5,244.2) -- (120.5,264.07) .. controls (120.5,265.17) and (119.6,266.07) .. (118.5,266.07) -- (43,266.07) .. controls (41.9,266.07) and (41,265.17) .. (41,264.07) -- cycle ;
\draw  [fill={rgb, 255:red, 155; green, 155; blue, 155 }  ,fill opacity=0.15 ] (61,242.2) -- (96.5,242.2) -- (96.5,265.79) -- (61,265.79) -- cycle ;

\draw  [fill={rgb, 255:red, 155; green, 155; blue, 155 }  ,fill opacity=0.15 ] (138,110.2) .. controls (138,109.1) and (138.9,108.2) .. (140,108.2) -- (215.5,108.2) .. controls (216.6,108.2) and (217.5,109.1) .. (217.5,110.2) -- (217.5,130.07) .. controls (217.5,131.17) and (216.6,132.07) .. (215.5,132.07) -- (140,132.07) .. controls (138.9,132.07) and (138,131.17) .. (138,130.07) -- cycle ;
\draw  [fill={rgb, 255:red, 155; green, 155; blue, 155 }  ,fill opacity=0.15 ] (158,108.2) -- (193.5,108.2) -- (193.5,131.79) -- (158,131.79) -- cycle ;

\draw  [fill={rgb, 255:red, 155; green, 155; blue, 155 }  ,fill opacity=0.15 ] (134,161.2) .. controls (134,160.1) and (134.9,159.2) .. (136,159.2) -- (211.5,159.2) .. controls (212.6,159.2) and (213.5,160.1) .. (213.5,161.2) -- (213.5,181.07) .. controls (213.5,182.17) and (212.6,183.07) .. (211.5,183.07) -- (136,183.07) .. controls (134.9,183.07) and (134,182.17) .. (134,181.07) -- cycle ;
\draw  [fill={rgb, 255:red, 155; green, 155; blue, 155 }  ,fill opacity=0.15 ] (154,159.2) -- (189.5,159.2) -- (189.5,182.79) -- (154,182.79) -- cycle ;

\draw  [fill={rgb, 255:red, 155; green, 155; blue, 155 }  ,fill opacity=0.15 ] (141.79,257.21) .. controls (141.79,256.11) and (142.68,255.21) .. (143.79,255.21) -- (208.5,255.21) .. controls (209.6,255.21) and (210.5,256.11) .. (210.5,257.21) -- (210.5,277.33) .. controls (210.5,278.43) and (209.6,279.33) .. (208.5,279.33) -- (143.79,279.33) .. controls (142.68,279.33) and (141.79,278.43) .. (141.79,277.33) -- cycle ;
\draw  [fill={rgb, 255:red, 155; green, 155; blue, 155 }  ,fill opacity=0.15 ] (147.65,255.21) -- (205.56,255.21) -- (205.56,279.04) -- (147.65,279.04) -- cycle ;

\draw [color={rgb, 255:red, 245; green, 166; blue, 35 }  ,draw opacity=1 ][line width=1.5]    (351.5,109) .. controls (392.24,82.16) and (411.61,79.57) .. (432.55,98.21) ;
\draw [shift={(434.5,100)}, rotate = 223.45] [fill={rgb, 255:red, 245; green, 166; blue, 35 }  ,fill opacity=1 ][line width=1.5]  [draw opacity=0] (11.61,-5.58) -- (0,0) -- (11.61,5.58) -- cycle    ;

\draw [color={rgb, 255:red, 155; green, 155; blue, 155 }  ,draw opacity=1 ][line width=1.5]    (377.5,251) .. controls (380.34,254.78) and (381.38,260.35) .. (395.84,266.01) ;
\draw [shift={(398.5,267)}, rotate = 199.44] [fill={rgb, 255:red, 155; green, 155; blue, 155 }  ,fill opacity=1 ][line width=1.5]  [draw opacity=0] (11.61,-5.58) -- (0,0) -- (11.61,5.58) -- cycle    ;

\draw [color={rgb, 255:red, 155; green, 155; blue, 155 }  ,draw opacity=1 ][line width=1.5]    (354.5,219) .. controls (337.13,229.81) and (326.28,219.76) .. (310.26,239.71) ;
\draw [shift={(308.5,242)}, rotate = 306.47] [fill={rgb, 255:red, 155; green, 155; blue, 155 }  ,fill opacity=1 ][line width=1.5]  [draw opacity=0] (11.61,-5.58) -- (0,0) -- (11.61,5.58) -- cycle    ;

\draw [color={rgb, 255:red, 155; green, 155; blue, 155 }  ,draw opacity=1 ][line width=1.5]    (399,19) .. controls (345.44,20.93) and (303.53,34.03) .. (281.79,48.43) ;
\draw [shift={(279.5,50)}, rotate = 324.46000000000004] [fill={rgb, 255:red, 155; green, 155; blue, 155 }  ,fill opacity=1 ][line width=1.5]  [draw opacity=0] (11.61,-5.58) -- (0,0) -- (11.61,5.58) -- cycle    ;

\draw [color={rgb, 255:red, 155; green, 155; blue, 155 }  ,draw opacity=1 ][line width=1.5]    (335.5,74) .. controls (316.78,85.52) and (299.9,80.45) .. (283.54,95.07) ;
\draw [shift={(281.5,97)}, rotate = 315] [fill={rgb, 255:red, 155; green, 155; blue, 155 }  ,fill opacity=1 ][line width=1.5]  [draw opacity=0] (11.61,-5.58) -- (0,0) -- (11.61,5.58) -- cycle    ;

\draw [color={rgb, 255:red, 155; green, 155; blue, 155 }  ,draw opacity=1 ][line width=1.5]    (339.5,121) .. controls (323.01,139.43) and (301.82,123.36) .. (284.13,144.89) ;
\draw [shift={(282.5,147)}, rotate = 306.12] [fill={rgb, 255:red, 155; green, 155; blue, 155 }  ,fill opacity=1 ][line width=1.5]  [draw opacity=0] (11.61,-5.58) -- (0,0) -- (11.61,5.58) -- cycle    ;

\draw [color={rgb, 255:red, 155; green, 155; blue, 155 }  ,draw opacity=1 ][line width=1.5]    (345.5,171) .. controls (329.18,186.68) and (311.94,177.17) .. (298.2,192.9) ;
\draw [shift={(296.5,195)}, rotate = 306.87] [fill={rgb, 255:red, 155; green, 155; blue, 155 }  ,fill opacity=1 ][line width=1.5]  [draw opacity=0] (11.61,-5.58) -- (0,0) -- (11.61,5.58) -- cycle    ;

\draw [color={rgb, 255:red, 245; green, 166; blue, 35 }  ,draw opacity=1 ][line width=1.5]    (489.5,124) .. controls (464.28,136.93) and (456.94,120.39) .. (433.7,138.24) ;
\draw [shift={(431.5,140)}, rotate = 320.41999999999996] [fill={rgb, 255:red, 245; green, 166; blue, 35 }  ,fill opacity=1 ][line width=1.5]  [draw opacity=0] (11.61,-5.58) -- (0,0) -- (11.61,5.58) -- cycle    ;

\draw [color={rgb, 255:red, 245; green, 166; blue, 35 }  ,draw opacity=1 ][line width=1.5]    (486.5,164) .. controls (459.2,175.05) and (455.67,159.99) .. (427.7,177.58) ;
\draw [shift={(425.5,179)}, rotate = 326.58000000000004] [fill={rgb, 255:red, 245; green, 166; blue, 35 }  ,fill opacity=1 ][line width=1.5]  [draw opacity=0] (11.61,-5.58) -- (0,0) -- (11.61,5.58) -- cycle    ;

\draw [color={rgb, 255:red, 245; green, 166; blue, 35 }  ,draw opacity=1 ][line width=1.5]    (484.5,203) .. controls (442.58,222.82) and (458.64,223.32) .. (438.15,252.68) ;
\draw [shift={(436.5,255)}, rotate = 305.99] [fill={rgb, 255:red, 245; green, 166; blue, 35 }  ,fill opacity=1 ][line width=1.5]  [draw opacity=0] (11.61,-5.58) -- (0,0) -- (11.61,5.58) -- cycle    ;

\draw  [fill={rgb, 255:red, 155; green, 155; blue, 155 }  ,fill opacity=0.15 ] (268,52.2) .. controls (268,51.1) and (268.9,50.2) .. (270,50.2) -- (345.5,50.2) .. controls (346.6,50.2) and (347.5,51.1) .. (347.5,52.2) -- (347.5,72.07) .. controls (347.5,73.17) and (346.6,74.07) .. (345.5,74.07) -- (270,74.07) .. controls (268.9,74.07) and (268,73.17) .. (268,72.07) -- cycle ;
\draw  [fill={rgb, 255:red, 155; green, 155; blue, 155 }  ,fill opacity=0.15 ] (288,50.2) -- (323.5,50.2) -- (323.5,73.79) -- (288,73.79) -- cycle ;

\draw  [fill={rgb, 255:red, 155; green, 155; blue, 155 }  ,fill opacity=0.15 ] (399.79,9.2) .. controls (399.79,8.1) and (400.68,7.2) .. (401.79,7.2) -- (465.79,7.2) .. controls (466.89,7.2) and (467.79,8.1) .. (467.79,9.2) -- (467.79,29.07) .. controls (467.79,30.17) and (466.89,31.07) .. (465.79,31.07) -- (401.79,31.07) .. controls (400.68,31.07) and (399.79,30.17) .. (399.79,29.07) -- cycle ;
\draw  [fill={rgb, 255:red, 155; green, 155; blue, 155 }  ,fill opacity=0.15 ] (405.59,7.2) -- (462.9,7.2) -- (462.9,30.79) -- (405.59,30.79) -- cycle ;

\draw  [fill={rgb, 255:red, 155; green, 155; blue, 155 }  ,fill opacity=0.15 ] (272,99.2) .. controls (272,98.1) and (272.9,97.2) .. (274,97.2) -- (349.5,97.2) .. controls (350.6,97.2) and (351.5,98.1) .. (351.5,99.2) -- (351.5,119.07) .. controls (351.5,120.17) and (350.6,121.07) .. (349.5,121.07) -- (274,121.07) .. controls (272.9,121.07) and (272,120.17) .. (272,119.07) -- cycle ;
\draw  [fill={rgb, 255:red, 155; green, 155; blue, 155 }  ,fill opacity=0.15 ] (292,97.2) -- (327.5,97.2) -- (327.5,120.79) -- (292,120.79) -- cycle ;

\draw  [fill={rgb, 255:red, 155; green, 155; blue, 155 }  ,fill opacity=0.15 ] (277,149.2) .. controls (277,148.1) and (277.9,147.2) .. (279,147.2) -- (354.5,147.2) .. controls (355.6,147.2) and (356.5,148.1) .. (356.5,149.2) -- (356.5,169.07) .. controls (356.5,170.17) and (355.6,171.07) .. (354.5,171.07) -- (279,171.07) .. controls (277.9,171.07) and (277,170.17) .. (277,169.07) -- cycle ;
\draw  [fill={rgb, 255:red, 155; green, 155; blue, 155 }  ,fill opacity=0.15 ] (297,147.2) -- (332.5,147.2) -- (332.5,170.79) -- (297,170.79) -- cycle ;

\draw  [fill={rgb, 255:red, 155; green, 155; blue, 155 }  ,fill opacity=0.15 ] (287,197.2) .. controls (287,196.1) and (287.9,195.2) .. (289,195.2) -- (364.5,195.2) .. controls (365.6,195.2) and (366.5,196.1) .. (366.5,197.2) -- (366.5,217.07) .. controls (366.5,218.17) and (365.6,219.07) .. (364.5,219.07) -- (289,219.07) .. controls (287.9,219.07) and (287,218.17) .. (287,217.07) -- cycle ;
\draw  [fill={rgb, 255:red, 155; green, 155; blue, 155 }  ,fill opacity=0.15 ] (307,195.2) -- (342.5,195.2) -- (342.5,218.79) -- (307,218.79) -- cycle ;

\draw  [fill={rgb, 255:red, 155; green, 155; blue, 155 }  ,fill opacity=0.15 ] (298,244.2) .. controls (298,243.1) and (298.9,242.2) .. (300,242.2) -- (375.5,242.2) .. controls (376.6,242.2) and (377.5,243.1) .. (377.5,244.2) -- (377.5,264.07) .. controls (377.5,265.17) and (376.6,266.07) .. (375.5,266.07) -- (300,266.07) .. controls (298.9,266.07) and (298,265.17) .. (298,264.07) -- cycle ;
\draw  [fill={rgb, 255:red, 155; green, 155; blue, 155 }  ,fill opacity=0.15 ] (318,242.2) -- (353.5,242.2) -- (353.5,265.79) -- (318,265.79) -- cycle ;

\draw  [fill={rgb, 255:red, 155; green, 155; blue, 155 }  ,fill opacity=0.15 ] (423,102.2) .. controls (423,101.1) and (423.9,100.2) .. (425,100.2) -- (500.5,100.2) .. controls (501.6,100.2) and (502.5,101.1) .. (502.5,102.2) -- (502.5,122.07) .. controls (502.5,123.17) and (501.6,124.07) .. (500.5,124.07) -- (425,124.07) .. controls (423.9,124.07) and (423,123.17) .. (423,122.07) -- cycle ;
\draw  [fill={rgb, 255:red, 155; green, 155; blue, 155 }  ,fill opacity=0.15 ] (443,100.2) -- (478.5,100.2) -- (478.5,123.79) -- (443,123.79) -- cycle ;

\draw  [fill={rgb, 255:red, 155; green, 155; blue, 155 }  ,fill opacity=0.15 ] (419,142.2) .. controls (419,141.1) and (419.9,140.2) .. (421,140.2) -- (496.5,140.2) .. controls (497.6,140.2) and (498.5,141.1) .. (498.5,142.2) -- (498.5,162.07) .. controls (498.5,163.17) and (497.6,164.07) .. (496.5,164.07) -- (421,164.07) .. controls (419.9,164.07) and (419,163.17) .. (419,162.07) -- cycle ;
\draw  [fill={rgb, 255:red, 155; green, 155; blue, 155 }  ,fill opacity=0.15 ] (439,140.2) -- (474.5,140.2) -- (474.5,163.79) -- (439,163.79) -- cycle ;

\draw  [fill={rgb, 255:red, 155; green, 155; blue, 155 }  ,fill opacity=0.15 ] (416,181.2) .. controls (416,180.1) and (416.9,179.2) .. (418,179.2) -- (493.5,179.2) .. controls (494.6,179.2) and (495.5,180.1) .. (495.5,181.2) -- (495.5,201.07) .. controls (495.5,202.17) and (494.6,203.07) .. (493.5,203.07) -- (418,203.07) .. controls (416.9,203.07) and (416,202.17) .. (416,201.07) -- cycle ;
\draw  [fill={rgb, 255:red, 155; green, 155; blue, 155 }  ,fill opacity=0.15 ] (436,179.2) -- (471.5,179.2) -- (471.5,202.79) -- (436,202.79) -- cycle ;

\draw  [fill={rgb, 255:red, 155; green, 155; blue, 155 }  ,fill opacity=0.15 ] (398.79,257.21) .. controls (398.79,256.11) and (399.68,255.21) .. (400.79,255.21) -- (465.5,255.21) .. controls (466.6,255.21) and (467.5,256.11) .. (467.5,257.21) -- (467.5,277.33) .. controls (467.5,278.43) and (466.6,279.33) .. (465.5,279.33) -- (400.79,279.33) .. controls (399.68,279.33) and (398.79,278.43) .. (398.79,277.33) -- cycle ;
\draw  [fill={rgb, 255:red, 155; green, 155; blue, 155 }  ,fill opacity=0.15 ] (404.65,255.21) -- (462.56,255.21) -- (462.56,279.04) -- (404.65,279.04) -- cycle ;

\draw [color={rgb, 255:red, 208; green, 2; blue, 27 }  ,draw opacity=0.5 ][line width=4.5]    (299.5,140.2) -- (330.5,177.2) ;

\draw [color={rgb, 255:red, 208; green, 2; blue, 27 }  ,draw opacity=0.5 ][line width=4.5]    (330.5,140.2) -- (300.5,177.2) ;

\draw (49.75,60.37) node  [align=left] {$ws_{x_{1}}$};
\draw (22,60.31) node [scale=0.8,rotate=-314.18] [align=left] {In};
\draw (77,62) node [scale=0.8,rotate=-43.44] [align=left] {Out};
\draw (177.75,17.37) node  [align=left] {$query_{in}$};
\draw (53.75,100.37) node  [align=left] {$ws_{x_{2}}$};
\draw (26,100.31) node [scale=0.8,rotate=-314.18] [align=left] {In};
\draw (81,102) node [scale=0.8,rotate=-43.44] [align=left] {Out};
\draw (58.75,155.37) node  [align=left] {$ws_{x_{3}}$};
\draw (31,155.31) node [scale=0.8,rotate=-314.18] [align=left] {In};
\draw (86,157) node [scale=0.8,rotate=-43.44] [align=left] {Out};
\draw (68.75,211.37) node  [align=left] {$ws_{x_{4}}$};
\draw (41,210.31) node [scale=0.8,rotate=-314.18] [align=left] {In};
\draw (96,212) node [scale=0.8,rotate=-43.44] [align=left] {Out};
\draw (79.75,253.37) node  [align=left] {$ws_{x_{5}}$};
\draw (52,252.31) node [scale=0.8,rotate=-314.18] [align=left] {In};
\draw (107,254) node [scale=0.8,rotate=-43.44] [align=left] {Out};
\draw (176.75,118.37) node  [align=left] {$ws_{y_{1}}$};
\draw (149,118.31) node [scale=0.8,rotate=-314.18] [align=left] {In};
\draw (204,120) node [scale=0.8,rotate=-43.44] [align=left] {Out};
\draw (172.75,169.37) node  [align=left] {$ws_{y_{2}}$};
\draw (145,169.31) node [scale=0.8,rotate=-314.18] [align=left] {In};
\draw (200,171) node [scale=0.8,rotate=-43.44] [align=left] {Out};
\draw (177.12,265.48) node  [align=left] {$query_{out}$};
\draw (434.12,265.48) node  [align=left] {$query_{out}$};
\draw (454.75,189.37) node  [align=left] {$ws_{z_{3}}$};
\draw (427,189.31) node [scale=0.8,rotate=-314.18] [align=left] {In};
\draw (482,191) node [scale=0.8,rotate=-43.44] [align=left] {Out};
\draw (457.75,150.37) node  [align=left] {$ws_{z_{2}}$};
\draw (430,150.31) node [scale=0.8,rotate=-314.18] [align=left] {In};
\draw (485,152) node [scale=0.8,rotate=-43.44] [align=left] {Out};
\draw (461.75,110.37) node  [align=left] {$ws_{z_{1}}$};
\draw (434,110.31) node [scale=0.8,rotate=-314.18] [align=left] {In};
\draw (489,112) node [scale=0.8,rotate=-43.44] [align=left] {Out};
\draw (336.75,253.37) node  [align=left] {$ws_{x_{5}}$};
\draw (364,254) node [scale=0.8,rotate=-43.44] [align=left] {Out};
\draw (309,252.31) node [scale=0.8,rotate=-314.18] [align=left] {In};
\draw (325.75,206.37) node  [align=left] {$ws_{x_{4}}$};
\draw (298,205.31) node [scale=0.8,rotate=-314.18] [align=left] {In};
\draw (353,207) node [scale=0.8,rotate=-43.44] [align=left] {Out};
\draw (315.75,157.37) node  [align=left] {$ws_{x_{3}}$};
\draw (288,157.31) node [scale=0.8,rotate=-314.18] [align=left] {In};
\draw (343,159) node [scale=0.8,rotate=-43.44] [align=left] {Out};
\draw (310.75,107.37) node  [align=left] {$ws_{x_{2}}$};
\draw (283,107.31) node [scale=0.8,rotate=-314.18] [align=left] {In};
\draw (338,109) node [scale=0.8,rotate=-43.44] [align=left] {Out};
\draw (434.75,17.37) node  [align=left] {$query_{in}$};
\draw (306.75,60.37) node  [align=left] {$ws_{x_{1}}$};
\draw (279,60.31) node [scale=0.8,rotate=-314.18] [align=left] {In};
\draw (334,62) node [scale=0.8,rotate=-43.44] [align=left] {Out};
\end{tikzpicture}
\end{figure*}

For each of these queries, we build the resolving composition as a sequence of services modified to have each input chosen randomly from the outputs of previous services or from the query's input parameters. 
To create backup possibilities, we build an alternative solution for a randomly chosen service of the main solution. 
The alternative solution can either replace just the chosen service - the \emph{first} type of backup or also all it's successive services - the \emph{second} type of backup. 
In Figure \ref{fig:online_generator}, gray edges show how parameters pass through a composition example $\mathcal{X} = \langle ws_{x_{1}}, ws_{x_{2}},  ws_{x_{3}} ws_{x_{4}}, ws_{x_{5}} \rangle$. 
Edges are displayed only between consecutive services of the composition, but parameters can be chosen randomly from all outputs of previous services. 
Blue edges track parameters passing through the sequence $\mathcal{Y} = \langle ws_{y_{1}}, ws_{y_{2}}\rangle$ that can replace $ws_{x_{3}}$ in $\mathcal{X}$ - the first type of backup. 
Orange edges follow the sequence $\mathcal{Z} = \langle ws_{z_{1}}, ws_{z_{2}},  ws_{z_{3}} \rangle$ that can replace $\langle ws_{x_{3}},  ws_{x_{4}} ws_{x_{5}} \rangle$ in $\mathcal{X}$ - the second type or "suffix" backup. 
If $ws_{x_{3}}$ fails, either $\mathcal{Y}$ or $\mathcal{Z}$ could replace it. 
In Algorithm \ref{algo:findbackup}, inside \algoname{findBackup()} function the first backup type is obtained at line 17, and the second type at line 20. 
In the solution we prioritize for finding the first type of backup since it is more likely to have a shorter length, therefore being more efficient.
The generator creates multiple of both of these backup types: first building paths of services and then assigning parameters accordingly. 

Table \ref{table:eval} presents our evaluation metrics. 
The first columns define the experiment instance size: the total number of distinct parameters, the total number of services in the initial repository and the total number of user queries. 
The following column specifies the total number of services used in all solutions found initially, for all queries. 
Out of these, we count how many have at least one possible backup: of the first or the second type. 
The last two columns include the running time consumed for solving all user requests and respectively for searching backups for each service used in compositions. 
Execution times are obtained for a Java implementation running on an $\texttt{Intel(R) Core(TM) i5 CPU @2.40 GHz}$ machine with $\texttt{8 GB RAM}$. 

For each query, one service is deleted, simulating its failure. 
This service can be either part of the initial composition found by the algorithm or of the backup built by the generator.
Our test setting ensures that the deletions are not limited to backup services, therefore covering all possible cases: either replacing a composition service with a backup (the branch at line 6 in Algorithm \ref{algo:deleteService}) or just computing a new backup if a previous backup service fails (the branch at line 15 in Algorithm \ref{algo:deleteService}).

Finally, we also measured the naive re-construction of all the compositions after the deletion of services. These produced similar run times as the initially built solutions. 
This is expected since the problem instance is almost as large as in the initial setting.

\begingroup
\setlength{\tabcolsep}{6pt} 
\renewcommand{\arraystretch}{1.2} 
\begin{table*}[h!]
\centering
\caption[\textbf{Online Composition} problem: evaluation on \textbf{generated} tests]{Experimental algorithm results. All run times are in seconds.}
\label{table:eval}
\begin{tabular}{|c c c|c|c c|c c|}
 \hline
 \makecell{$\mathbb{|P|}$ \\ \small{parameters}} & \makecell{$\mathbb{|R|}$ \\ \small{repository}} & \makecell{$\mathbb{|Q|}$ \\ \small{queries}} & \makecell{services \\ used} & \multicolumn{2}{c|}{\makecell{backups found \\ of each type}} & \makecell{build \\ solutions} & \makecell{search \\ backups}\\
 \hline
 1000 & 100 & 5 & 23 & \ \ \ 15 & 2 & 0.005 & 0.022 \\
 1000 & 500 & 20 & 171 & \ \ \ 63 & 12 & 0.07 & 0.35 \\
 10000 & 1000 & 100 & 464 & \ \ \ 232 & 29 & 0.56 & 2.02 \\
 10000 & 2500 & 20 & 905 & \ \ \ 273 & 132 & 0.41 & 9.05 \\
 \hline
\end{tabular}

\end{table*}
\endgroup

The measurements give an insight into the stress inflicted on the composition engine in practice.
Two choices would be possible in case of failure: re-compute a composition naively, which would have a similar cost with the initial computation of the composition, or using our pre-computed backups. 
The backup option is clearly preferable. 
This is because the switch to the alternative composition is done instantly eliminating the overhead on computing a new solution.
Even though the composition re-computation is not very high, it would still be desirable to avoid any disruption. 
Our solution practically eliminates any downtime caused by the failure.
Moreover, for situations when no pre-computed backups are available, the user or service administrator can be warned ahead of time of critical services in the composition that are unrecoverable, and for which other measures could be taken (i.e. adding a new equivalent service in the repository). 
Finally, we observe that the time for building backups is not very significant, and does not add delays for the user since this operation is performed asynchronously after the initial composition is retrieved. Also, because of the novelty of our approach, we could not find any similar work to compare the execution times.
}

\section{Conclusion}\label{sec:complexity:conclusion}
\comment{Conclusions on the complexity analysis and what it implies for defining models and designing algorithms for compositions.}

The complexities chapter presents some of the most important contributions of the thesis, other than the two newly proposed models of parameter matching. It starts with the chronologically-first algorithm designed for the initial name-match model, which has excellent results on the Services Challenge. Then we show how we can keep similar complexities for algorithms solving the first semantic model introduced by the hierarchy of concepts, and even for the newly proposed object-oriented model inspired by \emph{schema.org} data model.

In cases where such a low run-time complexity is provably not possible, we proposed some simple heuristics, that provide close to optimal results for the optimization version of the problem. The method can be generalized to more complex QoS metrics, and other variations on the problem definition. Even for the relational model, which is the hardest to solve efficiently, our algorithm provides results that prove that it can be applied in practice, where worst-case scenarios are very unlikely.

Other contributions that we consider important and, hopefully, relevant to the community in the future, are the methods proposed for generating insightful test cases. For the name-matching and hierarchical model, all tests and benchmarks already existing were solvable by compositions of at most tens of services. We generated tests that cannot be solved without hundreds of services, at least; and the run times on these tests are much higher and variate among solutions. Also, we described test generators for the relational model, capable to create tests where both relations and inference rules are used, even test cases where inference rules can be used to reduce the composition length even if not strictly necessary to solve the problem instance. Finally, the test generators for the online version of the composition simulate all the types of scenarios that can appear in a solution that builds backups for composition, part of the \emph{online composition problem}, as named.

For future work on the complexity analysis, there are several directions of continuation. One of them is to analyze the complexity of problems that use more then one of the models presented. In this direction, some steps are done, tough at design level only, in Section \ref{sec:applications:formalFramework} with the proposed framework design. It would also be interesting to compute the detailed complexity of problems and algorithms, which consider, for example, both the \emph{online} version and the semantic rich models of \emph{relational parameters} or \emph{object-oriented} types. Ideally, the complexity of a model including all of them could also be computed.

Another direction that should be extended is the \emph{lower bounds} Section \ref{sec:complexity:lower} where proving some proper lower-bound limits on optimization problems would be great (non-trivial polynomial-time complexities). Also, it would be interesting to know how close is the heuristic to the optimal result, in the number of services, in different models and we only discussed execution time complexity. On this topic, as the score based heuristic in Section \ref{sec:complexity:algorithms:name_heuristic} has good empirical results, it should be included in the enhanced model's algorithms, because a shorter, or more efficient composition is always preferable.

%
\chapter{\underline{Applications}}
\label{sec:applications}
\comment{A more practical approach on WSC, find as many examples of related work in industry, not necessarily defined precisely as service composition. The personal \textcolor{red}{[contribution]} here is lower.}


Although the number of research papers studying Web Service Composition is relatively large, there are yet only a few applications, tools or frameworks that use service composition models in practice, i.e. in the software industry. Most works are research-developed initiatives to provide proof of concept rather than real-world applications. We believe this is motivated by the fact that composition models require the definition of some standards, which are hard to follow in practice on such a large scale, and that the composition requires automation. Moreover, in the automatic composition context, naturally, service developers can create services independently, i.e. not knowing about each other. This makes it even harder to design a model that anybody would agree upon and follow over a long period. The developers and consumers of services use different design patterns and even the understanding of what a Web Service is can be surprisingly different. Often, research works over-simplify the view of Web Services, considering, for example: that services only provide information, are stateless, and can be implemented only based on parameter matching on a naive, trivial model (like the name matching model), omit execution, security issues and others that arise in practice.

Our proposed models: the \emph{relational parameter matching}, allowing the generation of \emph{multiple instances of the same type}, the \emph{object-oriented} model and the \emph{online}/dynamic view on the composition are trying to resolve some of these problems. In this section we show a few examples of practical implementations of the methods described. However, independently, they remain theoretical initiatives, more is to be done until reaching the maturity and level of details required to become easily applicable in practice. The examples that we present are inspired by the literature and some postings found on the web. We will analyze how these examples are related to our view of service composition. Our endeavor can be viewed as a step back from what is required in practice, most of the time without automation of composition, motivated by the need to get closer to the research results. Later, we will discuss some possible Natural Language Processing specific applications, based on some previous experience. Applicability of automatic composition models on specific (sub)domains goes beyond the trivial customization, as in the case of domain-specific ontologies for semantically-rich parameters. This is because specialized services can have different structures, for example in the NLP domain many services work with large fragments of texts annotated trough XML or other languages with extra information, which may play the role of parameters.

In the final section, we outline the design documentation of a composition framework, in this context. The implementation is currently left as future work.

\section{Composition In Software Development}
\label{sec:applications:industry}
\comment{For example Microsoft's tool, Yahoo Pipes, Altova Software, etc... describe each with insights on composition.}

We identified several software related to service composition. Even if not all of them also implement automation of composition; some design decisions and use-cases inspired by such applications are potentially useful in composition generally.

\subsection{Yahoo! Pipes}

\textbf{Yahoo! Pipes} \cite{pruett2007yahoo} is an old tool developed by Yahoo! that was shut down in 2015. It does not implement automation of composition as we presented it in the models of Chapter \ref{sec:models}. However, it is a visual tool that helps to build mashups of web services, among other things: web feeds, tools, resources, etc. Its naming suggests the similarity between \emph{matching}/aggregation of services/tools and \emph{piping} in a general sense. Yahoo! Pipes was one of the most popular software tools related to the composition domain, with more than 90000 developers (users of the application) and more than 5 million daily executions \cite{jones2009conversations}.

\begin{figure}[h]
  \includegraphics[width=\linewidth]{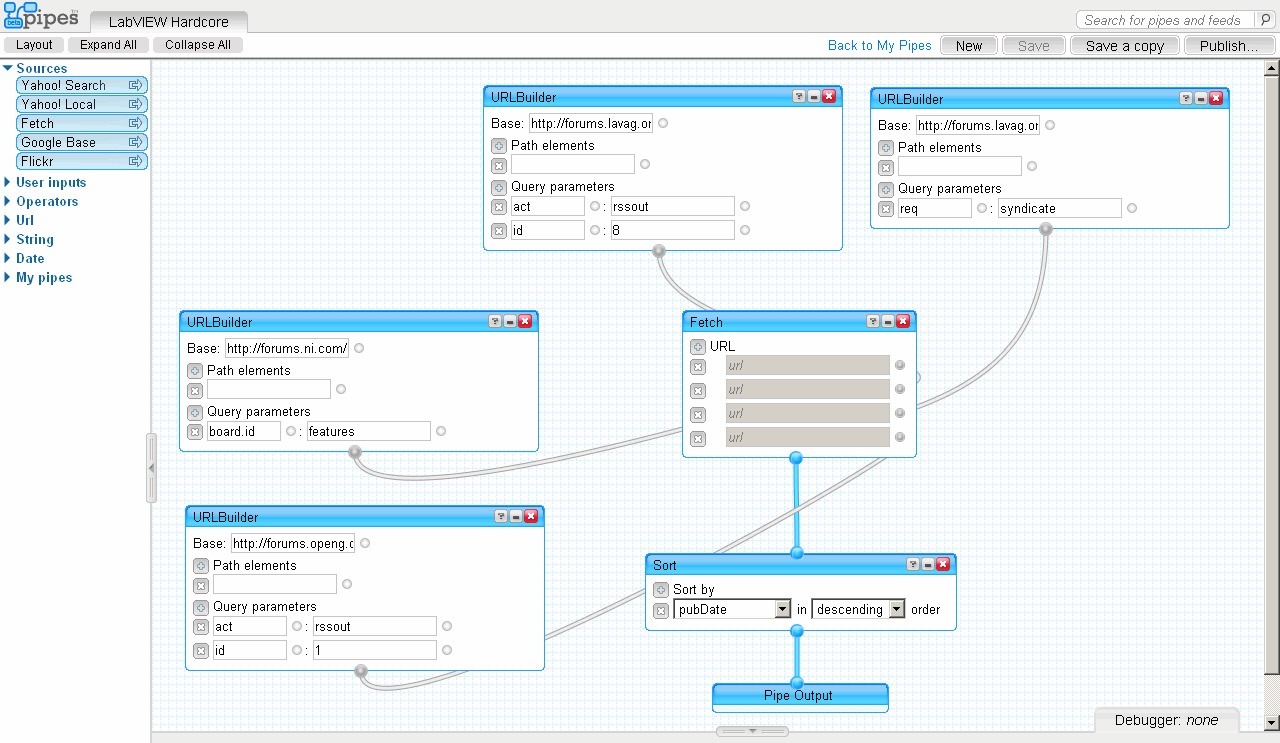}
  \caption[\textbf{Yahoo! Pipes} Mashup example]{Example of a Yahoo! Pipes Mashup with inputs and operators. Pipe's output is similar to the output of a composition in our model.}
  \label{fig:yahoopipes}
\end{figure}

Other large software companies developed similar tools, for example, Google's Mashup Editor which evolved into Google App Engine, Microsoft PopFly, Intel Mash Maker and IBM QEDWiki \cite{elmeleegy2008mashup}. These types of applications are still used and developed, even if many of them were discontinued, replaced/renamed or adapted.

There are several elements, part of Yahoo! Pipes design, as exemplified in Figure \ref{fig:yahoopipes}, which are equivalent or similar to the composition model. The most similar are the \emph{user inputs} \cite{pruett2007yahoo}, equivalent to the input parameters in automatic composition. The data types used are also similar to the model in Section \ref{sec:models:object-oriented}. Other functionalities, less related to automatic composition, are: execution of mash-ups, \emph{operators} that transform the data, resembling state-full composition and more. It is important to study \emph{Yahoo! Pipes} and other applications of its type before designing an (automatic) service composition framework. Also, it is interesting to apply the automation of composition in these types of applications or guide the manual composition process, by generating suggestions to the user based on some matching criteria.

\subsection{The Altova Software}

\begin{figure}[h]
  \includegraphics[width=\linewidth]{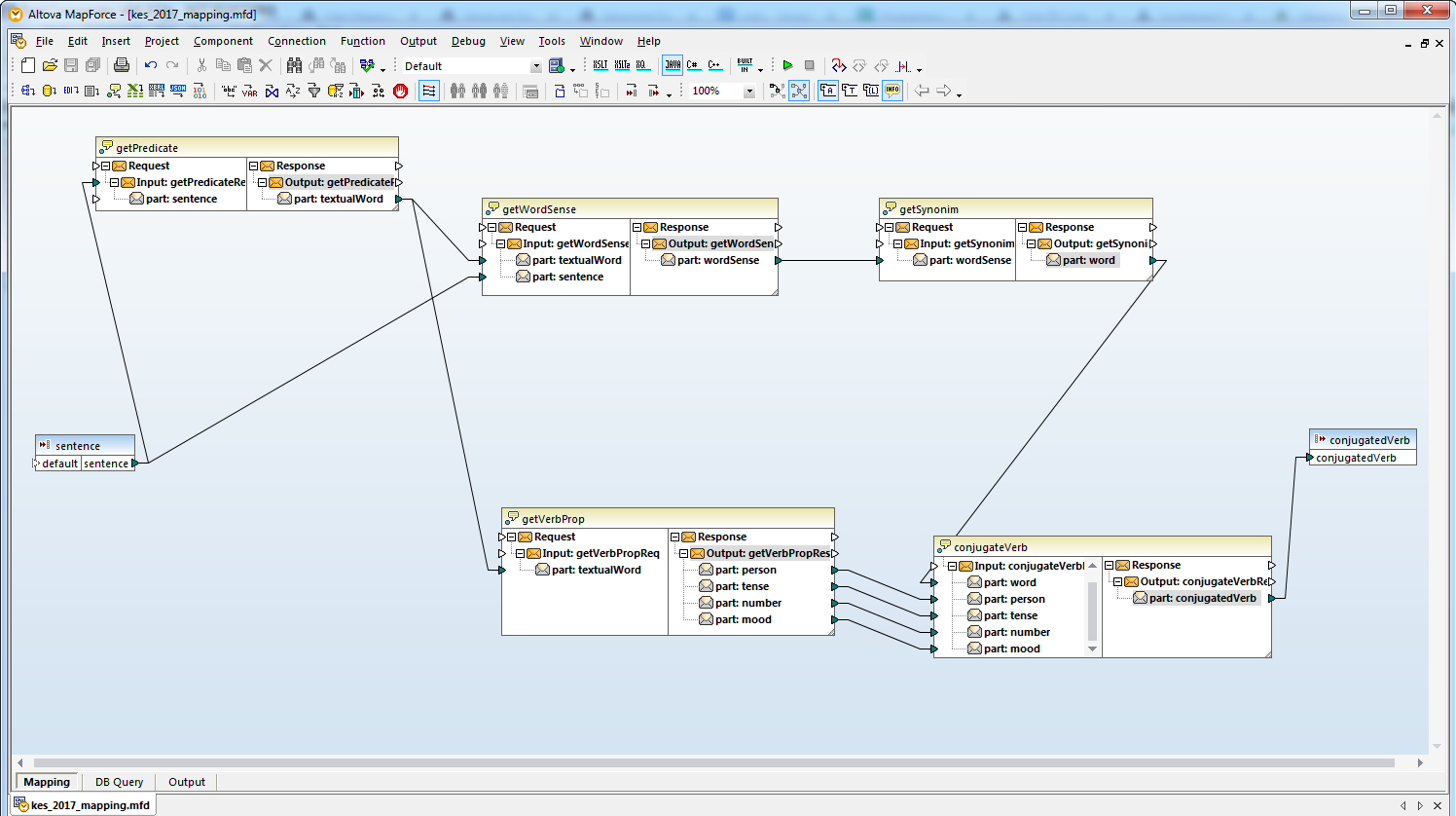}
  \caption[\textbf{Name Matching} example in \textbf{Altova MapForce}]{Modeling of our example in Section \ref{sec:models:names:example} with Altova MapForce 2020, \\ \url{https://www.altova.com/mapforce} using a 30-day evaluation license.}
  \label{fig:altovamapforce}
\end{figure}

For our analysis, Altova software is not fundamentally different than the tools previously presented. The reason why we mention it here though is that, unlike the above, this software is actively developed and upgraded. The last edition was launched on 9 October 2019, less than one year ago. Altova software consists of a suite of products, out of which MapForce is the most relevant to the composition domain\footnote{Altova MapForce -- \url{https://www.altova.com/mapforce}}. MapForce is described as a graphical data mapping tool used to integrate different data sources, data processing units, and to automate transformations. The \emph{connectors} displayed as edges in Figure \ref{fig:altovamapforce} correspond to parameter matching. There is an option to auto-connect matching children based on naming, but it has limited functionality (related to inheritance between data types). MapForce does not implement the execution of the work-flows itself, execution is done by a different tool: \emph{MapForce Server}. Finally, Altova's MapForce is a good candidate to integrate with automatic composition.

Also, some interesting features of Altova's MapForce that can be considered in service composition:

\begin{itemize}
  \setlength\itemsep{-1em}
  \item optional and required parameters: often inputs of services are declared as required, but in service composition this is omitted;
  \item case sensitiveness of name matching (generalized to the similarity of terms).
  \item possibility to combine both REST and SOAP services;
  \item integrate multiple standards, formats or technologies, i.e. web services, databases, Google protocol buffers, XBRL, and much more.
\end{itemize}



\section{Natural Language Processing Services}
\label{sec:applications:nlp}
\comment{In the discussed NLP-related application, service \emph{parameters} are defined differently, as annotated text in a knowledge graph. One service adds a new layer of information and depends on some previous annotations. The parameters are thus transformed into information we know about a certain text. The example should reach the NLP Tools example website, and also reference the QuoVadis project to exemplify - also enabling the reference of the paper.}

There are several ways in which NLP domain is related to Web Service Composition. Particularly related to service discovery is the idea of describing a required functionality in natural language. Service discovery is the phase preceding composition, but similarly, the NLP domain can potentially be used in the composition itself as well. For example, in \cite{bosca2006fly} restricted Natural Language is used to express requests, further used with multiple purposes: discovery, selection, and composition of web services. For this purpose, the request and service repository is processed to build a so-called \emph{Semantic Service Catalog}. This is generically similar to our proposed semantic extension for parameter description and services in Section \ref{sec:models:relational} and Section \ref{sec:models:object-oriented}. The distinction is that we do not explicitly use natural language, just some structured, extended semantic knowledge representation.

\subsection{Generalities}
Automatic Service Composition can also be considered an example of \emph{declarative programming}. In declarative programming, the \emph{programmer} expresses the computation it needs, in a structured language, without defining the control flow as in \emph{imperative programming}. For example, \emph{Logic Programming} is a form of declarative programming. \emph{Answer Set Programming} \cite{lifschitz2019answer} is a modern variant of logic programming. Interestingly, \cite{rainer2005web} applies \emph{answer set programming} to solve web service composition instances and in general, planning solvers are used to build compositions, as we described in the solutions sections of the complexity Chapter \ref{sec:complexity}. \comment{maybe it's better to move the Answer Set Programming reference to complexity chapter as well - it's kind of off-topic here.}

Following the \emph{declarative programming} general paradigm, in service composition, the user defines the composition request. The composer algorithm also takes as input the service repository. The composition is then automatically built by the algorithm, similarly with the \emph{solver} in declarative programming. The required functionality can be expressed or \emph{declared}, as specified, in a defined controlled \emph{language}. Such language may be the natural language or some intermediate language. The advantage is that, then, the programmer does not need to follow or even know any standardized language. Such a method is presented in \cite{sangers2013semantic} where a keyword-based method is used in conjunction with part-of-speech tagging, lemmatization, and word sense disambiguation. Also, it is possible to describe in the repository the service input and output parameters in natural language, and use that information to better guide the composition process. The semantics proposed in several sections of Chapter \ref{sec:models} can also be considered a step towards similar reasoning but trough \emph{knowledge representation}, and without explicit natural language.

We have yet described some ways to use NLP for Web Service Composition. But the other way is also possible: Web Service Composition can be used in NLP. The NLP domain like any other domain in computer science; is using web services in different applications. Particularities of language-related web services can be exploited by service composition. Even automation of the composition can be adapted to specific use-cases of NLP.

The vast majority of NLP web services deal with processing texts, generating some extra information about texts. One problem is representing both \emph{data} - the texts themselves, and \emph{metadata} - the extra information that we know about the texts. Metadata can include information about the documents like authors, title, editing houses, year of publication, etc, and annotations: different levels of linguistic expertise added to the text. Even written language, as a form of representing natural language, has extra information or \emph{metadata}, i.e. punctuation marks. These replace the relevant additional information naturally transmitted in verbal speech: gestures, intonation, even body language. NLP web services that work with texts usually require some information, for example, segmentation at the sentence level; and produce some new information, for example, part-of-speech tags or some other morphological information. Most of the research in NLP is based somehow on a processing workflow of this nature. One example developed by the local NLP Group\footnote{\url{http://nlptools.info.uaic.ro/}} is the construction of the Quo Vadis Corpus \cite{cristea2015quo}, containing a large number of entities, mostly references to characters in the novel, and different types of relations between them. Before the semi-supervised annotation of the corpus, a pre-processing phase was conducted. This phase consisted of a sequence of five web services that added information to the corpus: segmentation at the sequence level, tokenization, part-of-speech tagging, lemmatization, and noun phrase chunking. There are a few dependencies between these levels, for example, tokenization requires sentence segmentation. Therefore, similarly with the initial Definition \ref{chaindef} of \emph{Chained Matching} in the simplest matching model, the processors/services sequence must comply with a partial order defined by the dependence between services. Without thinking of standards, serialization or formats used; we can already see the similarity with our service composition. If all the meta-data involved in such a process can be organized in independent pieces and referenced in standard manner, that is adopted by service definitions and requests; then that leads exactly to service composition. If the number of web services is large, and there are many parameters expressed as meta-data, then automation is needed. Recently the industry started developing such repositories of services or APIs oriented towards NLP tasks: for example Amazon Comprehend\footnote{\url{https://aws.amazon.com/comprehend/}} or Google Cloud Natural Language API\footnote{\url{https://cloud.google.com/natural-language}}.

\subsection{NLP-Services Composition Systems}
Automating the composition of NLP web services is not the only problem in NLP workflows. There are still many issues in the area of standardizing the formats used to keep the meta-data about texts, usually done \emph{inline}, in the same files with the texts. There are many works in this direction, in general; and some of them are implemented by the NLP-Tools group in the \emph{Faculty of Computer Science} of \emph{"Alexandru Ioan Cuza" University of Iasi}. One of the most important is the ALPE system \cite{cristea2007alpe, pistolgraphical, cristea2008managing} that proposes a hierarchy of annotation schemas and processing tools that is also capable of automating the search of \emph{paths}, the equivalent of general work-flows.

ALPE (Automated Linguistic Processing Environment) is heavily focused on the representation, which is the first requirement in practice. At a high-level view, the ALPE model is based on a directed acyclic graph in which nodes represent annotation schemas, and (hyper)arcs are subsumption relations between schemas. Subsumption in an ALPE hierarchy is roughly similar to the subsumption in the semantic models in Chapter \ref{sec:models}, but it is defined at the annotation level rather than on the semantic interpretation of a sub/super-concept. The graph is augmented with three possible operations: simplification, pipeline and merge. The pipeline is conceptually similar to our parameter matching, tough it is again defined at annotation state/schema level, and not parameter level. In ALPE, the simplification operation reduces an annotated file to the format of another subsuming format in the hierarchy, and merging combines two or more annotations applied to the same hub document onto one \cite{pistol2009managing}. Merge and simplification operations in the view of service composition are implicit because of the model itself: parameters that are not needed are ignored, and all known parameters are kept together in the stateful \emph{knowledge} anyway, without the need for an explicit merge. In service composition, there is no \emph{support} document. In a way, elements corresponding to the (hyper)arcs in the ALPE schemas graph correspond to some form of (hyper)arcs - connecting multiple nodes (from general hypergraphs \cite{bretto2013hypergraph}). Another important feature in ALPE is the execution of workflows, while in this thesis the execution is tackled only abstractly in the online version of composition, in Section \ref{sec:models:online}. \comment{left here, introduce the 4th chapter of the thesis here as a conclusion or reference for more documentation.} For more details about the ALPE framework, see Chapter 4 - \emph{A hierarchy for annotation formats and processing modules} and Chapter 5 - \emph{Automated Linguistic Processing Environment} of the thesis \cite{pistolthesis}.

Another automated processing system for NLP is the CLARIN project\footnote{\url{https://clarin.dk/}}. Similar to ALPE (and the automatic service composition); CLARIN builds a \emph{chain} of tools that reach the goal of a user request, if no single service satisfies it \cite{jongejan2013workflow}. The search for such a workflow in CLARIN is similar to a general path-finding algorithm. The result may not be unique and the user can choose which one is considered the best fit using the user interface. Unlike ALPE, CLARIN data model is hierarchical. The \emph{composition} search algorithm is recursive, and a maximum depth of 20 recursive steps is set. This may be enough in practice, but small relative to our algorithms which compute much longer compositions: even up to tens of thousands of services on simple models. Another positive aspect of CLARIN is that it does implement execution.

\comment{maybe make a connection to TEPROLIN from ReTeRom and the integration with my framework.}

A step towards the applicability of Automatic Service Composition is the framework presented in the next section. While it is not specialized for the NLP domain, it can be easily adapted to the NLP domain - or other domains - based on configurable ontologies.


\newpage
\section{Design Document of an Abstract Composition Framework}
\label{sec:applications:formalFramework}
\comment{Similar to the HTN PlanGraph formal definition, e.g. \href{https://www.sciencedirect.com/science/article/pii/S0004370296000471}{here (link).} }

\comment{adapt the framework description to the context and motivate abut the missing implementation}

\subsection{Short description}
We propose an new abstract web service composition design, focusing on the data model and functionalities. The intention is to provide an example of what could be useful in practice, relative to previous works on service composition and the analysis on what is currently being developed, in the previous sections of this chapter. It defines semantic web service composition with modern-language, expressive service definition capabilities. 
The data model includes both: the ontologies that enable the semantic description of parameters and a dynamic repository of services. Most important, it can resolve composition requests. Internally, it can potentially also cache frequently used service constructs, such that future requests can be solved faster.

\subsection{Functionalities and API Description}
The framework's functionalities are organized into three categories, named levels. The Ontological Level is the new part, which didn't exist in previous models. It is motivated by the need for interoperability: the framework is intended to be used simultaneously for more than one domain, since a general all-in-one ontology does not seem feasible. Therefore, we allow multiple ontologies.

The \textbf{Ontological Level} is represented by the set of \emph{types - concepts}, defined under the data model described in the next section of this document. Adding, removing and updating ontologies are infrequent operations, so there is no need to implement them efficiently. Also, hypothetically, service providers should be restricted to update ontologies and should only be able to get information about them, by accessing getter methods: read but not write. Ontologies can potentially be referenced by a unique name or id. Most important methods:

\begin{itemize}
  \setlength\itemsep{-1em}
  \item \textit{getOntologies(...)} $\rightarrow$ returns a list with all ontologies registered
  \item \textit{addOntology(...)} $\rightarrow$ registers a new ontology
  \item \textit{removeOntology(...)} $\rightarrow$ removes an unused ontology
  \item \textit{updateOntology(...)} $\rightarrow$ a more complex method, used for example to add a type to an ontology
\end{itemize}

One problem with the methods \emph{add/remove/update ontology} is that they can invalidate not only the cached compositions but, potentially the services themselves.

The \textbf{Service Level} consists of the set or repository of web services. Each web service is also defined on the data model, as parameters are defined based on a known and existing ontology. Web services can also be read, added, removed. Service providers can use these methods. In a simplified version, there is no QoS or monitoring information associated with services. Methods:

\begin{itemize}
  \setlength\itemsep{-1em}
  \item \textit{addWebService(...)} $\rightarrow$ creates a new web service in the repository.
  \item \textit{removeWebService(...)} $\rightarrow$ removes an existing web service that gets deprecated \textit{removeWebService(...)} + \textit{addWebService(...)} can be used for a service update.
  \item \textit{getWebServiceRepository(...)} $\rightarrow$ get the list of all services in the repository defined over a given ontology.
  \item \textit{getWebServicesUsingTypes(...)} $\rightarrow$ get a list of services that work with the specified types (within an ontology). This is just an utility method used, for example, to get all services that take as input one certain type when it becomes available.
\end{itemize}

The \textbf{Composition Level} handles compositions requests and the (cached) responses. Users of any type can create a request for a composition (not only service developers). The request consists of the ontology and the known and required types, which are structurally similar to service parameters. The response is an ordered structure/workflow of services that validly satisfy the request. It may not necessarily be a simple list, as parallel execution is possible. Methods:

\begin{itemize}
  \setlength\itemsep{-1em}
  \item \textit{getComposition(...)} $\rightarrow$ given a set of types with some of their properties known, and a required list of types with properties; returns a workflow of services that can be validly called in that order to achieve the required parameters (types with specified properties). All are defined under the same ontology.
  \item \textit{getCompositionWithQoS(...)} is an enhanced method used to specify which QoS measures should be prioritized (for example preference over cheaper or faster compositions); or limit the total number of services in the result, total cost, etc. Even more interesting, a degree of how strictly the matching rules should be applied could be specified.
\end{itemize}

\subsection{Data Model}

\textbf{Types and Properties}. The data model was designed following the example of \emph{schema.org}\footnote{\url{https://schema.org/docs/datamodel.html}}, a collaborative initiative (proposed by several companies) for structured semantic data on the web, recommended for the inter-operable description of services. This model was also chosen because it provided the expressibility that solved some of the examples where previous composition language was inapplicable. The basic elements defining basic \textbf{concepts} of the ontology are named \textbf{types} and are organized in a hierarchy. \textbf{Multiple inheritance} is allowed, this is fundamentally different from the model in Section \ref{sec:models:object-oriented}. Types have \textbf{properties} that are structurally similar to members of classes in object-oriented programming. In \emph{composition}, after \emph{calling} some services, we might know that there is information only about a subset of a type's properties. For example, we might know a person's name but not the person's email address even if the ontology specifies that people generally have email addresses. Properties also have their \textbf{type}, which is defined over the same taxonomy of \textbf{types} (or concepts). Actually, as in schema.org, one property might be defined over a set of types. Inheritance transfers properties of types to all direct and indirect subtypes.

\textbf{Parameters}. Service's input and output parameters have the same structure: each parameter is composed of a type and a subset of its properties. They define what is required to call the service and what is returned after the call. Based on this, the framework knows what it can use further in composition. Information is modeled at the concept level, and not the instance level: one does not work with concrete values, but their (abstract) types and properties, knowing that, at execution, values for these types will be returned. The framework does not implement execution but, instead, it provides an executable structure of services. For service input, subtypes of the required type can be used because of the \emph{isA} interpretation of inheritance. At the output, properties of supertypes are learned, as \emph{knowledge} is represented by a set of types with associated properties, which can be obtained by running the composition. For example, if we got a \emph{student}’s \emph{name}, we also know a \emph{person}’s \emph{name} if the \emph{student isA person}.

In automatic composition, this can lead to some problems. If different properties of a type are learned from different (incompatible) subtypes, then they don't describe the same concept.

\begin{figure}[h]
 \centering
  \caption[\textbf{Type inheritance} in framework]{Types inheritance issue with inferred properties from incompatible types.}
  \vspace{-0.3cm}
  \includegraphics[width=\linewidth]{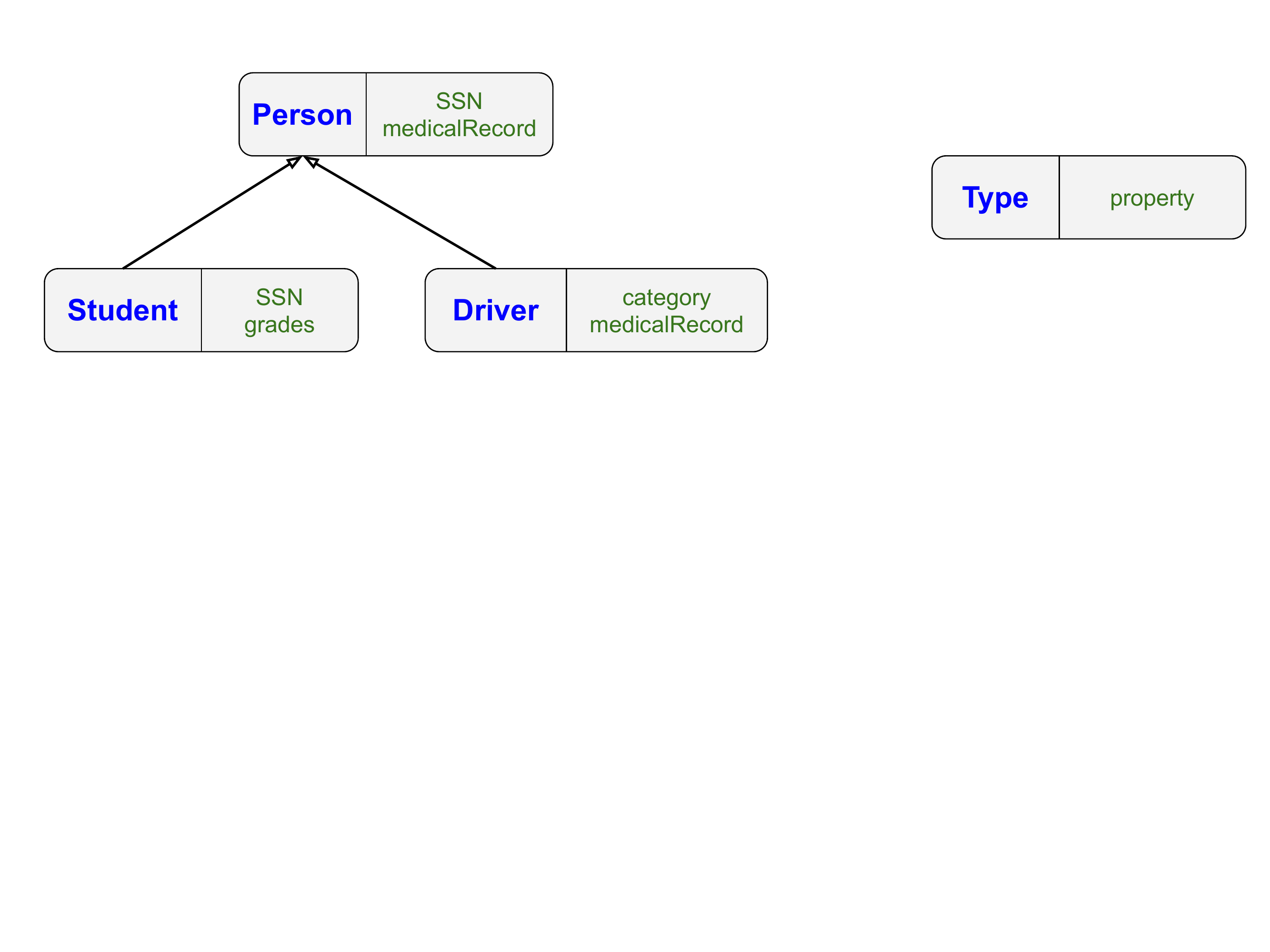}
  \vspace{-6.5cm}
  \label{fig:design_doc1}
\end{figure}

For example, as in Figure \ref{fig:design_doc1}, one service might return the \textbf{SSN} of \textbf{Students} from an institution and another service the \textbf{medicalRecord} of any \textbf{Driver} from whatever database. If we further need some \textbf{Person}’s both  \textbf{SSN} and the \textbf{medicalRecord}, even if currently both exist, they might not refer to the same \textbf{Person}.

To solve this issue, another element is introduced: the \textbf{object}.
An \textbf{object} is a \textbf{type} with a subset of its properties that are known. From this point of view, it is the same as the \textbf{object} in Section \ref{sec:models:object-oriented}. Even if the properties have been learned from different services or sources, we also know that they refer to the same \emph{object}, meaning that, at execution, it is expected they hold values for the same instance of that object.
In the same flavor, the web service definition is extended. If a service has at input and output two parameters of the same type (most probably with different properties), it can also specify if they refer to the same object or not. If yes, we can consider that in composition. Syntactically, this can be done using the names of parameters in service definitions. If one input and one output parameter of the same type also have the same \textbf{name}, then they refer to the same object. If names are different, then the objects are different even if they have the exact same type.

This further extends the representation of \textbf{knowledge} acquired during the construction of the composition. Knowledge is now a set of objects and we can have more objects of the same type, eventually even with the same subset of known properties. This is also useful for expressing service's requirements at input: one service can have more than one input of the same type and the same number of objects of that type must be known to call that service (together with corresponding properties). In this case, to determine if a service can be called, gets much more complicated on edge cases, but is still solvable. Again, even more complicated, it is now possible to call the same service multiple times, with different parameters. This increases the expressiveness, since, as far as we know, it was never applied before in stateless service composition, although it naturally occurs in manual composition scenarios.

\textbf{super-properties} and \textbf{sub-properties}. Like in schema.org, some properties can be in a similar inheritance relation. \emph{Knowing} a property implies the knowledge of its super-properties on the same types, just like in \textbf{type}s inheritance. For example, \url{https://schema.org/identifier} has subproperties \url{https://schema.org/flightNumber} and \url{https://schema.org/isbn}, which are different types of identifiers.

\begin{figure}[h]
  \centering
  \caption[\textbf{Inheritance of properties} in framework]{Inheritance is possible for types but also for properties.}
  \includegraphics[width=\linewidth]{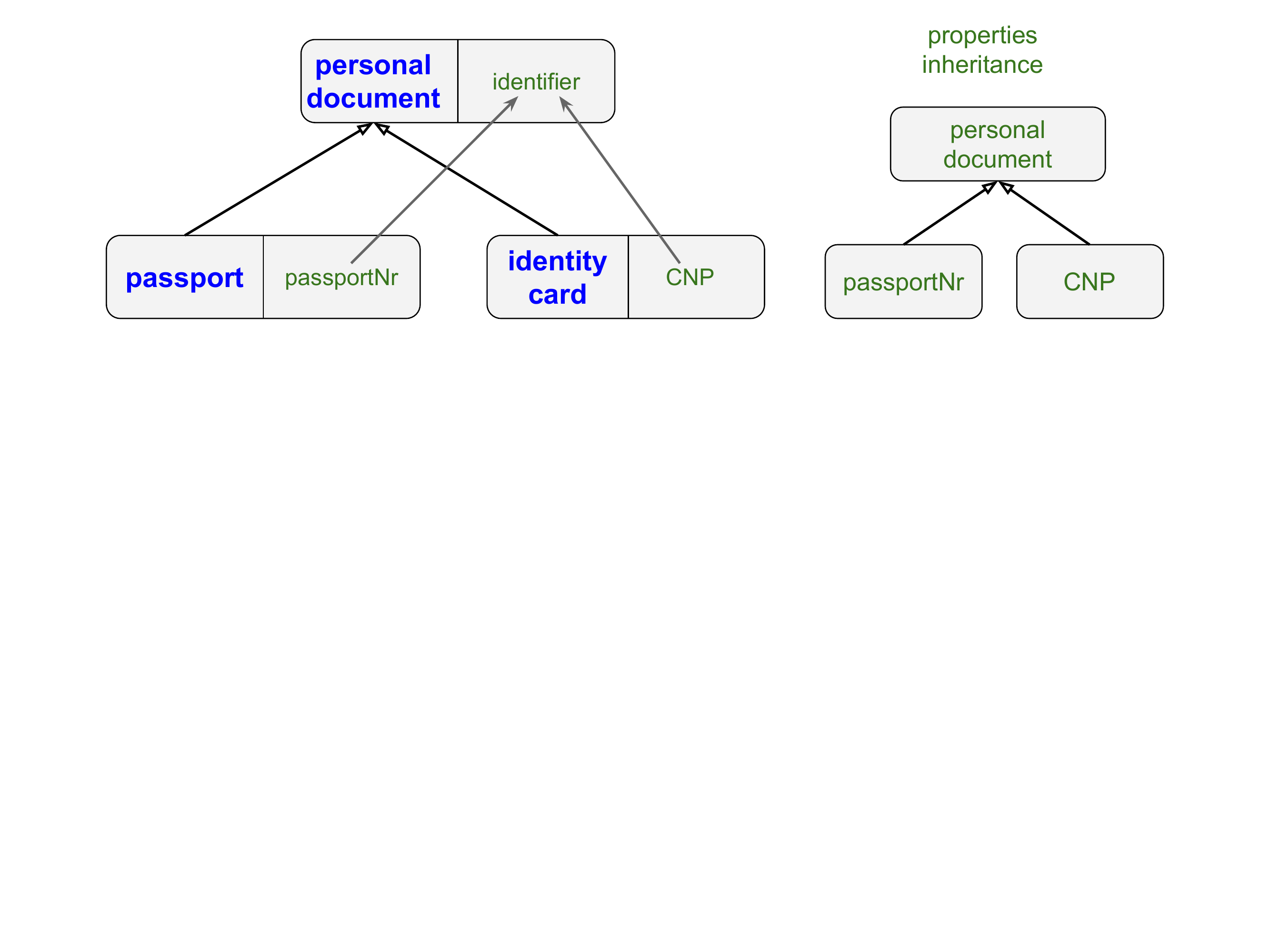}
  \vspace{-7cm}
  \label{fig:design_doc2}
\end{figure}

A more specific \textbf{type} cannot have a less specific \textbf{property} if the more general \textbf{type} doesn't (i.e, the gray slim arrows in Figure \ref{fig:design_doc2} cannot go down). This is motivated by the way property inheritance works with sub-properties: the most specific version of the property is inherited to subtypes. This also holds for \textbf{objects} and their known \textbf{properties} versions. Currently, multiple inheritance is not allowed on properties, and there is no such case in schema.org.

\textbf{Inverse properties}. Some properties can have an \textbf{inverseOf} relation with another specified property. If any of these properties are learned within a composition, the reserve is inferred automatically. To understand how this can work, the \textit{unboxing} defined below is required.

\begin{figure}[h]
 \centering
  \caption[Properties \textbf{inverse-of} relation in framework]{Properties in the \emph{inverseOf} relation: \textbf{alumni} and \textbf{alumniOf}. \\ Types of properties must match.}
  \includegraphics[width=\linewidth]{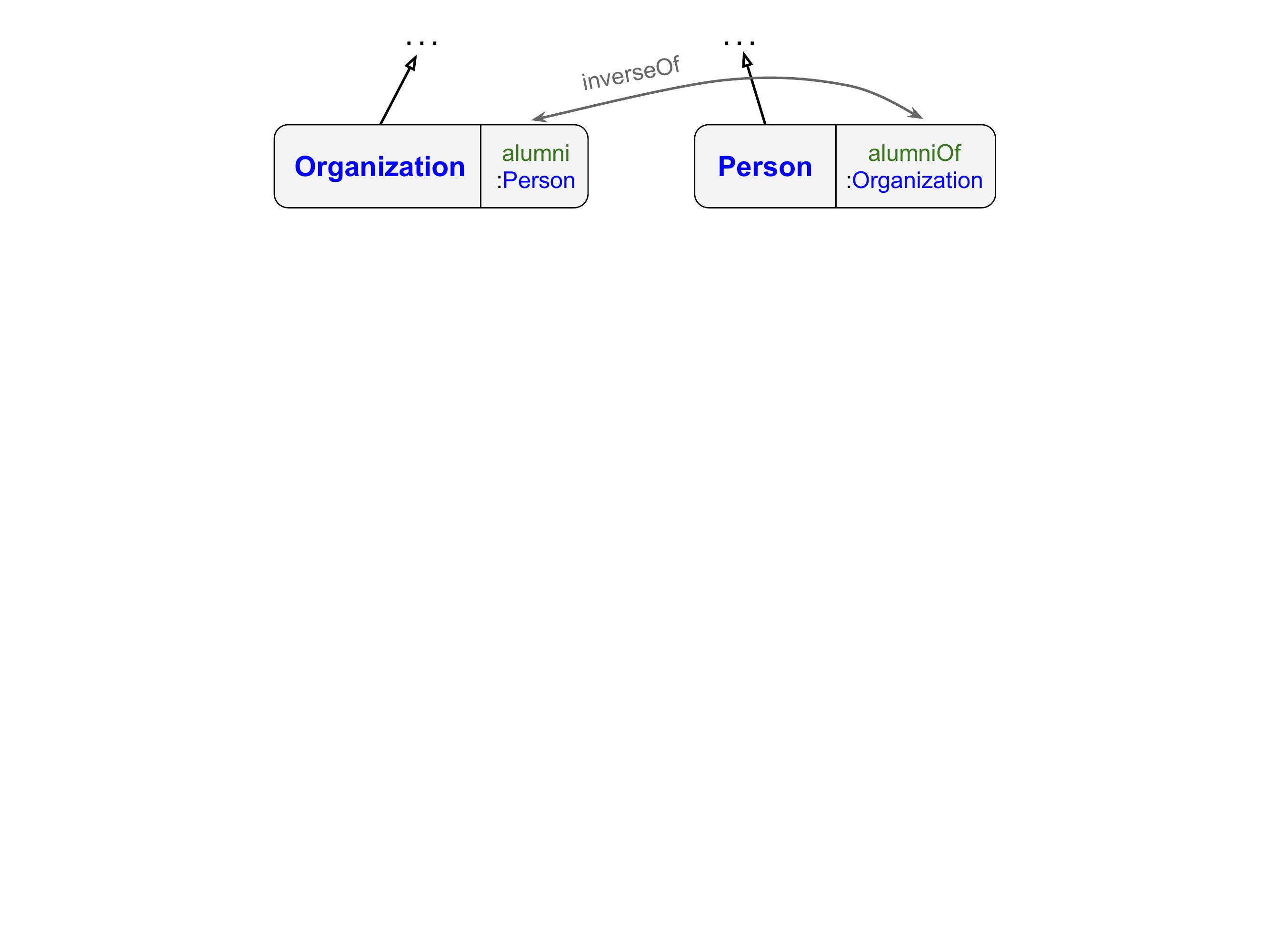}
  \vspace{-8cm}
  \label{fig:design_doc3}
\end{figure}

\textbf{Unboxing}. It is possible to \emph{convert} one property to an object, as properties themselves have the same domain of types as objects. For example, if one service has as an input parameter an \textbf{address} and we only know one address which is a property of a student's \emph{residency:address}, then it is valid to call the service with the student's residency address.

Finally, it is possible to have an object without knowing any of its properties, and it is different from not knowing the object at all.

\subsection{Open Questions}
\label{sec:applications:open_questions}

There are other several design decisions to take. They are left as future work, since are either dependent of the context in which the framework will be implemented, or open questions for now. We will present them here as they can provide interesting ideas and considerations.


\begin{itemize}
  \setlength\itemsep{-1em}
  \item The framework can be implemented as a library, providing code for the main functionalities; or, more complicated, as a deployable application that handles all the requests at a given endpoint.
  \item It would be useful to add the possibility to model \emph{"containers"}, i.e. Lists, Arrays, Sets; for example, by using schema's \emph{ItemList} type. The model would allow a service that takes an input of a type to be used for a container of elements of that type. The framework would call the service for each element and would return the output for each. It is easy to find motivating examples for this feature.
  \item It would be useful to model relations between objects, similarly with the relational model in Section \ref{sec:models:relational}.
  \item It would be useful to include the possibility to specify \emph{required} an \emph{optional} input parameters, as they are often met in practice, in service definitions.
  \item Similarly, for any object, some properties could be defined as \emph{required}, with the meaning that if absent, the objects could not be defined.
\end{itemize}

\section{Conclusion}\label{sec:applications:conclusion}

\comment{TODO: Conclusions on the Applications Chapter.}

From our analysis in the first section of this chapter and the analysis of the applied research initiatives, it results that there is still a big gap between research and industry in automatic service composition. This is because composition is a very complex area, and the software industry requires that all phases generally known from Service Oriented Architecture are actually implemented: design, discovery, compute the composition, up to execution, maintenance, failure mitigation and many more.

We described some software tools, both older and deprecated, and more recent ones, which are being actively developed nowadays. Unfortunately, the automation of composition is rarely included effectively. Natural Language Processing services, in general, require a slightly different paradigm on service composition. But, with some adjustments and transformation, NLP provides an excellent context for applying automatic composition, because it already has many standardization initiatives: schemas, standards or systems. The model proposed for the composition framework in Section \ref{sec:applications:formalFramework} can be potentially used for this case as well.

The application part of the thesis is probably the best source for future work ideas. First, the extension of the framework model design (e.g., open questions in Section \ref{sec:applications:open_questions}) and, most of all, the implementation of such a framework are big milestones.

%
\chapter{Conclusion}
\label{sec:conclusion}

Web Service Composition is an active research domain, important for the modern service-oriented programming styles, applicable in many particular domains or use-cases. It can be seen as a tool for programming, or meta-programming, accelerating integration, development, and favoring re-usability. Automating the process of building compositions is critical in more applications every day, not only because many independent services are created everyday. Moreover and more recently, institutions or companies provide large web service-based platforms. Most of these are domain-specific, therefore use the same standards and technologies for up to thousands of services or more. Some specifications like the OpenAPI are proposed by groups of companies and academics, cooperating for the benefit of all. These trends give great opportunities to apply the earlier research on automatic composition.

The thesis discusses the most frequent aspects of composition: semantics, computational complexity, service metrics and quality of service, algorithm performance, stateful services, comparative evaluation, test generation, a fail-over mechanism of the dynamic version of the composition problem, and highlights real-world applications.

\section{Contributions}
\label{sec:conclusion:contributions}
\comment{Outline personal contributions in the field of Composition: fast (optimal?) algorithms, new design, framework functionality, reduction of NLP-style services to classic web services, complexity analysis. Some high-level conclusions...}

Our initial efforts were in designing the fastest possible composition algorithm, solving the simplest parameter matching model. We succeeded not only to find the algorithm performing relatively the fastest on the known benchmarks, but also to prove its complexity is optimal for the composition problem that does not require the shortest composition. Also, adding some heuristic scores, the composition size is reduced very close to the best known results, without a significant increase in performance. To emphasize the execution time spent searching compositions, we implemented generators that create the need for arbitrary-long compositions. For long compositions the difference with respect to the run time between our algorithms and others from literature becomes huge.

Further, we adapted the initial algorithm to the first semantic matching model that is the model of the organized composition challenge. To keep the best performance, a Euler traversal of the hierarchy of concepts is efficiently used.

However, we are also aware that the first semantic match model has a number of limitations. Manual compositions are with consideration to many other elements not included in a simple subsumption hierarchy. Therefore, we extend the semantic models while keeping the complete formal definition of automation. In the first extension we add binary relations between concepts. Perhaps more important, we lowered the abstraction level of parameters, with the addition of objects. Objects are more concrete than previous concepts, as multiple objects of the same concept are manipulated, but they do not hold actual values. Human reasoning in manual composition makes use of objects of very similar meaning.

Another contribution brings recent semantic web specifications like \emph{OpenAPI} and the \emph{schema.org} ontology to automatic composition. To do so, concepts are extended with inheritable properties similar to members in object-oriented programming or, more specifically, \emph{schema.org}'s types.

We believe that, especially in the automatic composition domain, research should be closely connected to application development. Therefore, we analyze older and current software development tools that open opportunities for automation. Particularly, we consider \emph{Natural Language Processing} services, which we have used in the past. Considering the proposed matching extensions and practical applications analyzed, we also wrote a design document with the data model and functionalities for a modern composition framework.

Altogether, the contributions of the thesis range from purely theoretical to applied, from algorithm implementation to semantic knowledge representation, from participating in composition competitions to creating new and insightful benchmarks.

\section{Future Work}
\label{sec:conclusion:future}
\comment{Many possibilities here ... for example framework extension, real user requirements, stateful services, NLP future work directions. }
\comment{idea1: if there is no composition, what would be the so-called minimum number of services or functionalities to be added?}
\comment{idea2: Cezar Andrici's email with methods signature}
\comment{idea3: Other reminders: addapt algorithms to the WSC 2009 and 2010 that have QoS.}

Most of the contributions of the thesis can be extended in several ways. In this section we highlight the most relevant or interesting possibilities of continuation, in the same order of the thesis structure.

When the first algorithm was implemented to solve the name matching model, the heuristic score used to shorten the compositions was thought to be dynamically updated when services are added or parameters are learned. Due to time restrictions and because the results were good enough on tests, this was not implemented. This feature can be done while building the composition, without a significant increase in run-time complexity; therefore it can give a neat improvement. Moreover, for the heuristic in general, proving some rate of approximation or, at least, finding the maximum number of loops of the final reduction phase of the algorithm would be revealing. Since the score has promising results, another idea is to adapt it for Quality of Service metrics. QoS was included in the last editions of the services challenge, the only one not addressed explicitly by our algorithms. This is a direction with promising results, especially if modern QoS specifications are used.

Returning, for simplicity, to name matching, there are many other interesting problems, inspired by graph theory. Imagine the automatic composition as a special path search problem. Similarly, other related graph problems can translate into new composition problems. For example, if there is no composition to satisfy a request, one can compute the minimum number of missing services that, if added, can resolve the request, together with the existing services. This is non-trivial but still realistic if the missing services are limited by some constraints, like the number of parameters or, for example, are imported from a foreign repository. Even flow problems can be converted to the composition variants. If the throughput of services is included, the user request may be associated with a minimum throughput requirement. One workflow of services from the repository may satisfy the throughput partially and there may be complementary compositions that add up to satisfy the request both functionally and quantitatively. And these are just to show a few examples of graph problems that may be translated to their composition corespondents, but many more can be imagined.

There are a lot of ways to extend the semantics of matching models. In the framework design, we combine the relational and object-oriented models. This is just a step forward. Many concepts of knowledge representation or reasoning can be added as well. The question is which of them is most useful and perhaps lightweight enough to be easily adopted by developers.

The applications section gives the most straightforward future work possibilities. The implementation of the framework is the most persuasive. Some feedback from communities of developers can help to make the right design decisions and answer open questions. A more feasible and smaller effort approach would customize the framework for a given domain or use-case. Natural Language Processing is an excellent candidate, providing large repositories of services defined with the same standards and working with the same concepts.

Considering all of the above, we believe that automatic (service) composition is an expansive, reliant and complex research area with many unexplored opportunities, that will reveal a boosting development in the near future.


{
\setstretch{1.1}
\renewcommand{\bibfont}{\normalfont\small}
\setlength{\biblabelsep}{0pt}
\setlength{\bibitemsep}{0.5\baselineskip plus 0.5\baselineskip}
\printbibliography[nottype=online,notkeyword=excluded]
\printbibliography[heading=subbibliography,title={Webseiten},type=online,prefixnumbers={@},notkeyword=excluded]

@article{zou2014dynamic,
  title={Dynamic Composition of Web Services Using Efficient Planners in Large-Scale Service Repository},
  author={Zou, Guobing and Gan, Yanglan and Chen, Yixin and Zhang, Bofeng},
  journal={Knowledge-Based Systems},
  volume={62},
  pages={98--112},
  year={2014},
  publisher={Elsevier}
}

@misc{mcdermott1998pddl,
  title={PDDL - the Planning Domain Definition Language},
  author={McDermott, Drew and Ghallab, Malik and Howe, Adele and Knoblock, Craig and Ram, Ashwin and Veloso, Manuela and Weld, Daniel and Wilkins, David},
  year={1998},
}

@inproceedings{blake2005eee,
  title={The EEE-05 Challenge: A New Web Service Discovery and Composition Competition},
  author={Blake, M Brian and Tsui, Kwok Ching and Wombacher, Andreas},
  booktitle={IEEE International Conference on e-Technology, e-Commerce and e-Service},
  pages={780--783},
  year={2005},
  organization={IEEE}
}

@inproceedings{diac2017engineering,
  title={Engineering Polynomial-Time Solutions for Automatic Web Service Composition},
  author={Diac, Paul},
  booktitle={International Conference on Knowledge-Based and Intelligent Information and Engineering Systems (KES)},
  volume={112},
  pages={643--652},
  year={2017},
  publisher={Elsevier}
}

@inproceedings{bansal2008wsc,
  title={WSC-08: Continuing the Web Services Challenge},
  author={Bansal, Ajay and Blake, M Brian and Kona, Srividya and Bleul, Steffen and Weise, Thomas and Jaeger, Michael C},
  booktitle={10th IEEE Conference on E-Commerce Technology and the Fifth IEEE Conference on Enterprise Computing, E-Commerce and E-Services},
  pages={351--354},
  year={2008},
  organization={IEEE}
}

@inproceedings{tucar2018semantic,
  title={Semantic Web Service Composition Based on Graph Search},
  author={{\c{T}}uc{\u{a}}r, Liana and Diac, Paul},
  booktitle={International Conference on Knowledge-Based and Intelligent Information and Engineering Systems (KES)},
  volume={126},
  pages={116--125},
  year={2018},
  publisher={Elsevier}
}

@inproceedings{diac2019relational,
  title={Relational Model for Parameter Description in Automatic Semantic Web Service Composition},
  author={Diac, Paul and {\c{T}}uc{\u{a}}r, Liana and Netedu, Andrei},
  booktitle={International Conference on Knowledge-Based and Intelligent Information and Engineering Systems (KES)},
  year={2019},
  publisher={Elsevier}
}

@inproceedings{diac2019formalism,
  title={Extending the Service Composition Formalism with Relational Parameters},
  author={Diac, Paul and {\c{T}}uc{\u{a}}r, Liana and Mereu{\c{t}}{\u{a}}, Radu},
  booktitle={Working Formal Methods Symposium (FROM)},
  year={2019}
}

@inproceedings{berbner2006heuristics,
  title={Heuristics for QoS-Aware Web Service Composition},
  author={Berbner, Rainer and Spahn, Michael and Repp, Nicolas and Heckmann, Oliver and Steinmetz, Ralf},
  booktitle={International Conference on Web Services (ICWS)},
  pages={72--82},
  year={2006},
  organization={IEEE}
}

@inproceedings{kona2009wsc,
  title={WSC-2009: A Quality of Service-Oriented Web Services Challenge},
  author={Kona, Srividya and Bansal, Ajay and Blake, M Brian and Bleul, Steffen and Weise, Thomas},
  booktitle={Conference on Commerce and Enterprise Computing},
  pages={487--490},
  year={2009},
  organization={IEEE}
}

@article{gabrel2018qos,
  title={QoS-Aware Automatic Syntactic Service Composition Problem: Complexity and Resolution},
  author={Gabrel, Virginie and Manouvrier, Maude and Moreau, Kamil and Murat, Cecile},
  journal={Future Generation Computer Systems},
  volume={80},
  pages={311--321},
  year={2018},
  publisher={Elsevier}
}

@inproceedings{sferruzza2018extending,
  title={Extending OpenAPI 3.0 to Build Web Services from their Specification},
  author={Sferruzza, David and Rocheteau, J{\'e}r{\^o}me and Attiogb{\'e}, Christian and Lanoix, Arnaud},
  year={2018}
}

@article{guha2016schema,
  title={Schema.org: Evolution of Structured Data on the Web},
  author={Guha, Ramanathan V and Brickley, Dan and Macbeth, Steve},
  journal={Communications of the ACM},
  volume={59},
  number={2},
  pages={44--51},
  year={2016},
  publisher={ACM}
}

@inproceedings{netedu2019openapi,
  title={A Web Service Composition Method Based on OpenAPI Semantic Annotations},
  author={Netedu, Andrei and Buraga, Sabin and Diac, Paul and {\c{T}}uc{\u{a}}r, Liana},
  booktitle={International Conference on e-Business Engineering (ICEBE)},
  year={2019},
}

@book{allemang2011semantic,
  title={{Semantic Web for the Working Ontologist: Effective Modeling in RDFS and OWL}},
  author={Allemang, Dean and Hendler, James},
  year={2011},
  publisher={Elsevier}
}

@inproceedings{cardinale2011fault,
  title={Fault Tolerant Execution of Transactional CompositeWeb Services: An Approach},
  author={Cardinale, Yudith and Rukoz, Marta},
  booktitle={Fifth International Conference on Mobile Ubiquitous Computing, Systems, Services and Technologies (UBICOMM)},
  year={2011},
  organization={IARIA}
}

@inproceedings{rao2007fault,
  title={Fault Tolerant Web Services Composition as Planning},
  author={Rao, Dongning and Jiang, Zhihua and Jiang, Yunfei},
  year={2007},
  booktitle={International Conference on Intelligent Systems and Knowledge Engineering},
  publisher={Atlantis Press}
}

@inproceedings{diac2019failover,
  title={Towards Integrated Failure Recovery for Web Service Composition},
  author={Diac, Paul and Onica, Emanuel},
  booktitle={International Conference on Software Technologies (ICSOFT)},
  year={2019},
}

@inproceedings{lecue2008towards,
  title={Towards the Composition of Stateful and Independent Semantic Web Services},
  author={L{\'e}cu{\'e}, Freddy and Delteil, Alexandre and L{\'e}ger, Alain},
  booktitle={ACM Symposium on Applied Computing},
  pages={2279--2285},
  year={2008},
  organization={ACM}
}

@article{kila2011computational,
  title={On the Computational Complexity of Behavioral Description-Based Web Service Composition},
  author={Kila, Hyunyoung and Namb, Wonhong and Leea, Dongwon},
  year={2011},
  publisher={Citeseer}
}

@article{zhao2012automatic,
  title={Automatic Composition of Information-Providing Web Services Based on Query Rewriting},
  author={Zhao, WenFeng and Liu, ChuanChang and Chen, JunLiang},
  journal={Science China Information Sciences},
  volume={55},
  number={11},
  pages={2428--2444},
  year={2012},
  publisher={Springer}
}

@inproceedings{kil2008computational,
  title={Computational Complexity of Web Service Composition Based on Behavioral Descriptions},
  author={Kil, Hyunyoung and Nam, Wonhong and Lee, Dongwon},
  booktitle={20th International Conference on Tools with Artificial Intelligence},
  volume={1},
  pages={359--363},
  year={2008},
  organization={IEEE}
}

@article{kil2013behavioural,
  title={Behavioural Description Based Web Service Composition Using Abstraction and Refinement},
  author={Kil, Hyunyoung and Nam, Wonhong and Lee, Dongwon},
  journal={International Journal of Web and Grid Services},
  volume={9},
  number={1},
  pages={54--81},
  year={2013},
  publisher={Inderscience Publishers Ltd.}
}

@article{hoffmann2001ff,
  title={FF: The Fast-forward Planning System},
  author={Hoffmann, J{\"o}rg},
  journal={AI magazine},
  volume={22},
  number={3},
  pages={57},
  year={2001}
}

@article{blum1997fast,
  title={Fast Planning Through Planning Graph Analysis},
  author={Blum, Avrim L and Furst, Merrick L},
  journal={Artificial Intelligence},
  volume={90},
  number={1},
  pages={281--300},
  year={1997},
  publisher={Elsevier}
}

@inproceedings{oh2006wsben,
  title={WSBen: A Web Services Discovery and Composition Benchmark},
  author={Oh, Seog-Chan and Kil, Hyunyoung and Lee, Dongwon and Kumara, Soundar R.T.},
  booktitle={International Conference on Web Services},
  pages={239--248},
  year={2006},
  organization={IEEE}
}

@article{bleul2009web,
  title={The Web Service Challenge - A Review on Semantic Web Service Composition},
  author={Bleul, Steffen and Weise, Thomas and Geihs, Kurt},
  journal={Electronic Communications of the EASST},
  volume={17},
  year={2009}
}

@article{cordella2004sub,
  title={A (Sub) Graph Isomorphism Algorithm for Matching Large Graphs},
  author={Cordella, Luigi P and Foggia, Pasquale and Sansone, Carlo and Vento, Mario},
  journal={Transactions on Pattern Analysis and Machine Intelligence},
  volume={26},
  number={10},
  year={2004},
  publisher={IEEE}
}

@inproceedings{rodriguez2011automatic,
  title={Automatic Web Service Composition with a Heuristic-based Search Algorithm},
  author={Rodriguez-Mier, Pablo and Mucientes, Manuel and Lama, Manuel},
  booktitle={International Conference on Web Services},
  pages={81--88},
  year={2011},
  organization={IEEE}
}

@inproceedings{ding2010vipen,
  title={ViPen: A Model Supporting Knowledge Provenance for Exploratory Service Composition},
  author={Ding, Weilong and Wang, Jing and Han, Yanbo},
  booktitle={International Conference on Services Computing},
  pages={265--272},
  year={2010},
  organization={IEEE}
}

@book{pruett2007yahoo,
  title={Yahoo Pipes},
  author={Pruett, Mark},
  year={2007},
  publisher={O'Reilly}
}

@inproceedings{jones2009conversations,
  title={Conversations in Developer Communities: A Preliminary Analysis of the Yahoo! Pipes Community},
  author={Jones, M. Cameron and Churchill, Elizabeth F.},
  booktitle={Fourth International Conference on Communities and Technologies},
  pages={195--204},
  year={2009}
}

@inproceedings{elmeleegy2008mashup,
  title={Mashup Advisor: A Recommendation Tool for Mashup Development},
  author={Elmeleegy, Hazem and Ivan, Anca and Akkiraju, Rama and Goodwin, Richard},
  booktitle={International Conference on Web Services},
  pages={337--344},
  year={2008},
  organization={IEEE}
}

@article{sangers2013semantic,
  title={Semantic Web Service Discovery Using Natural Language Processing Techniques},
  author={Sangers, Jordy and Frasincar, Flavius and Hogenboom, Frederik and Chepegin, Vadim},
  journal={Expert Systems with Applications},
  volume={40},
  number={11},
  pages={4660--4671},
  year={2013},
  publisher={Elsevier}
}

@book{lifschitz2019answer,
  title={Answer Set Programming},
  author={Lifschitz, Vladimir},
  year={2019},
  publisher={Springer International Publishing}
}

@inproceedings{rainer2005web,
  title={Web Service Composition Using Answer Set Programming},
  author={Rainer, Albert},
  booktitle={Workshop ”Planen, Scheduling und Konfigurieren, Entwerfen” Koblenz},
  year={2005}
}

@incollection{cristea2015quo,
  title={Quo Vadis: A Corpus of Entities and Relations},
  author={Cristea, Dan and G{\^\i}fu, Daniela and Colhon, Mihaela and Diac, Paul and Bibiri, Anca-Diana and M{\u{a}}r{\u{a}}nduc, C{\u{a}}t{\u{a}}lina and Scutelnicu, Liviu-Andrei},
  booktitle={Language Production, Cognition, and the Lexicon},
  pages={505--543},
  year={2015},
  publisher={Springer}
}

@article{cristea2007alpe,
  title={ALPE as LT4eL Processing Chain Environment},
  author={Cristea, Dan and Pistol, Ionut Cristian and For{\u{a}}scu, Corina},
  journal={Natural Language Processing and Knowledge Representation for eLearning Environments},
  pages={3},
  year={2007}
}

@article{pistolgraphical,
  title={A Graphical Interface for Computing and Distributing NLP Flows},
  journal={Web Services and Processing Pipelines in HLT: Tool Evaluation, LR Production and Validation},
  author={Pistol, Ionuț Cristian and Arusoaie, Andrei and Vasiliu, Andrei and Iftene, Adrian},
}

@article{cristea2008managing,
  title={Managing Language Resources and Tools using a Hierarchy of Annotation Schemas},
  author={Cristea, Dan and Pistol, Ionut Cristian},
  journal={Sustainability of Language Resources and Tools for Natural Language Processing},
  pages={1},
  year={2008}
}

@inproceedings{pistol2009managing,
  title={Managing Metadata Variability within a Hierarchy of Annotation Schemas},
  author={Pistol, Ionut and Cristea, Dan},
  booktitle={NLPCS},
  pages={111--116},
  year={2009}
}

@article{bosca2006fly,
  title={On-the-Fly Construction of Web Services Compositions from Natural Language Requests},
  author={Bosca, Alessio and Corno, Fulvio and Valetto, Giuseppe and Maglione, Roberta},
  journal={Journal of Software},
  volume={1},
  number={1},
  pages={40--50},
  year={2006}
}

@article{bretto2013hypergraph,
  title={Hypergraph Theory},
  author={Bretto, Alain},
  journal={An Introduction. Mathematical Engineering},
  year={2013},
  publisher={Springer}
}

@phdthesis{pistolthesis,
  title={The Automated Processing of Natural Language},
  author={Pistol, Ionut Cristian},
  school = {Alexandru Ioan Cuza University of Iasi},
  year={2011},
  type = {phdthesis},
}

@inproceedings{jongejan2013workflow,
  title={Workflow Management in CLARIN-DK},
  author={Jongejan, Bart},
  booktitle={Workshop on Nordic Language Research Infrastructure at NODALIDA},
  number={089},
  pages={11--20},
  year={2013},
  organization={Link{\"o}ping University Electronic Press}
}

@inproceedings{diac2017warp,
  title={WARP: Efficient Automatic Web Service Composition},
  author={Diac, Paul},
  booktitle={19th International Symposium on Symbolic and Numeric Algorithms for Scientific Computing (SYNASC)},
  pages={284--285},
  year={2017},
  organization={IEEE}
}

@inproceedings{amariei2018cell,
  title={Cell Grid Architecture for Maritime Route Prediction on AIS Data Streams},
  author={Amariei, Ciprian and Diac, Paul and Onica, Emanuel and Ro{\c{s}}ca, Valentin},
  booktitle={12th ACM International Conference on Distributed and Event-based Systems (DEBS)},
  pages={202--204},
  year={2018},
}

@inproceedings{colhon2014quovadis,
  title={Quo Vadis Research Areas--Text Analysis},
  author={Colhon, Mihaela and Diac, Paul and M{\u{a}}r{\u{a}}nduc, C{\u{a}}t{\u{a}}lina and Perez, Augusto},
  booktitle={Linguistic Resources And Tools For Processing The Romanian Language (LREC)},
  pages={45},
  year={2014}
}

@inproceedings{bibiri2014statistics,
  title={Statistics Over a Corpus of Semantic Links -- Quo Vadis},
  author={Bibiri, Anca-Diana and Colhon, Mihaela and Diac, Paul and Cristea, Dan},
  booktitle={Linguistic Resources And Tools For Processing The Romanian Language (LREC)},
  pages={33},
  year={2014}
}

@inproceedings{amariei2017optimized,
  title={Optimized Stage Processing for Anomaly Detection on Numerical Data Streams},
  author={Amariei, Ciprian and Diac, Paul and Onica, Emanuel},
  booktitle={11th ACM International Conference on Distributed and Event-based Systems (DEBS)},
  pages={286--291},
  year={2017}
}

@inproceedings{rocsca2018predicting,
  title={Predicting Destinations by Nearest Neighbor Search on Training Vessel Routes},
  author={Ro{\c{s}}ca, Valentin and Onica, Emanuel and Diac, Paul and Amariei, Ciprian},
  booktitle={12th ACM International Conference on Distributed and Event-based Systems (DEBS)},
  pages={224--225},
  year={2018}
}

@inproceedings{pascaru2018vehicle,
  title={Vehicle Routing and Scheduling for Regular Mobile Healthcare Services},
  author={Pascaru, Cosmin and Diac, Paul},
  booktitle={IEEE 30th International Conference on Tools with Artificial Intelligence (ICTAI)},
  pages={480--487},
  year={2018},
  organization={IEEE}
}

@inproceedings{diac2018relationships,
  title={Relationships and Sentiment Analysis of Fictional or Real Characters},
  author={Diac, Paul and Colhon, Mihaela and M{\u{a}}r{\u{a}}nduc, C{\u{a}}t{\u{a}}lina},
  booktitle={International Conference on Computational Linguistics and Intelligent Text Processing (CICLing)},
  year={2018}
}

@inproceedings{perrey2003service,
  title={Service-Oriented Architecture},
  author={Perrey, Randall and Lycett, Mark},
  booktitle={Symposium on Applications and the Internet Workshops},
  year={2003},
  organization={IEEE}
}

@misc{christensen2001web,
  title={Web Services Description Language WSDL 1.1},
  author={Christensen, Erik and Curbera, Francisco and Meredith, Greg and Weerawarana, Sanjiva and others},
  year={2001},
  publisher={Citeseer}
}

@article{levesque1986knowledge,
  title={Knowledge Representation and Reasoning},
  author={Levesque, Hector J},
  journal={Annual Review of Computer Science},
  volume={1},
  number={1},
  pages={255--287},
  year={1986},
}

@incollection{alonso2004web,
  title={Web Services},
  author={Alonso, Gustavo and Casati, Fabio and Kuno, Harumi and Machiraju, Vijay},
  booktitle={Web Services},
  pages={123--149},
  year={2004},
  publisher={Springer}
}

@book{daigneau2012service,
  title={Service Design Patterns: Fundamental Design Solutions for SOAP/WSDL and Restful Web Services},
  author={Daigneau, Robert},
  year={2012},
}

@book{newman2015building,
  title={Building Microservices: Designing Fine-Grained Systems},
  author={Newman, Sam},
  year={2015},
  publisher={" O'Reilly Media, Inc"}
}

@article{albers2003online,
  title={Online algorithms: a survey},
  author={Albers, Susanne},
  journal={Mathematical Programming},
  volume={97},
  number={1-2},
  pages={3--26},
  year={2003},
  publisher={Springer}
}

@inproceedings{blake2010wsc,
  title={WSC-2010: Web Services Composition and Evaluation},
  author={Blake, M Brian and Weise, Thomas and Bleul, Steffen},
  booktitle={International Conference on Service-Oriented Computing and Applications (SOCA)},
  pages={1--4},
  year={2010},
  organization={IEEE}
}

@inproceedings{blake2006wsc,
  title={WSC-06: The Web Service Challenge},
  author={Blake, M Brian and Cheung, William and Jaeger, Michael C and Wombacher, Andreas},
  booktitle={International Conference on E-Commerce Technology and International Conference on Enterprise Computing, E-Commerce, and E-Services (CEC/EEE)},
  pages={62--62},
  year={2006},
  organization={IEEE}
}

@inproceedings{blake2007wsc,
  title={WSC-07: Evolving the Web Services Challenge},
  author={Blake, M Brian and Cheung, William KW and Jaeger, Michael C and Wombacher, Andreas},
  booktitle={International Conference on E-Commerce Technology and International Conference on Enterprise Computing, E-Commerce and E-Services (CEC/EEE)},
  pages={505--508},
  year={2007},
  organization={IEEE}
}

@article{mcguinness2004owl,
  title={OWL Web Ontology Language Overview},
  author={McGuinness, Deborah L and Van Harmelen, Frank and others},
  journal={W3C Recommendation},
  pages={2004},
  year={2004}
}

@article{gudgin2003soap,
  title={SOAP Version 1.2},
  author={Gudgin, Martin and Hadley, Marc and Mendelsohn, Noah and Moreau, Jean-Jacques and Nielsen, Henrik Frystyk and Karmarkar, Anish and Lafon, Yves},
  journal={W3C Recommendation},
  volume={24},
  pages={12},
  year={2003}
}

@book{weerawarana2005web,
  title={Web Services Platform Architecture: SOAP, WSDL, WS-policy, WS-addressing, WS-BPEL, WS-reliable Messaging and More},
  author={Weerawarana, Sanjiva and Curbera, Francisco and Leymann, Frank and Storey, Tony and Ferguson, Donald F},
  year={2005},
  publisher={Prentice Hall PTR}
}

@article{ludwig2003web,
  title={Web Service Level Agreement (WSLA) Language Specification},
  author={Ludwig, Heiko and Keller, Alexander and Dan, Asit and King, Richard P and Franck, Richard},
  journal={IMB Corporation},
  pages={815--824},
  year={2003}
}

@article{venkatachalam2016comprehensive,
  title={Comprehensive Survey on Semantic Web Service Discovery and Composition},
  author={Venkatachalam, K and Karthikeyan, NK and Kannimuthu, S},
  journal={Advances in Natural and Applied Sciences},
  volume={10},
  number={5},
  pages={32--41},
  year={2016},
  publisher={American-Eurasian Network for Scientific Information}
}

@inproceedings{syu2012survey,
  title={A Survey on Automated Service Composition Methods and Related Techniques},
  author={Syu, Yang and Ma, Shang-Pin and Kuo, Jong-Yih and FanJiang, Yong-Yi},
  booktitle={International Conference on Services Computing},
  year={2012},
  organization={IEEE}
}

@inproceedings{rao2004survey,
  title={A Survey of Automated Web Service Composition Methods},
  author={Rao, Jinghai and Su, Xiaomeng},
  booktitle={International Workshop on Semantic Web Services and Web Process Composition},
  pages={43--54},
  year={2004},
  organization={Springer}
}

@inproceedings{kourtesis2008combining,
  title={Combining SAWSDL, OWL-DL and UDDI for Semantically Enhanced Web Service Discovery},
  author={Kourtesis, Dimitrios and Paraskakis, Iraklis},
  booktitle={European Semantic Web Conference},
  pages={614--628},
  year={2008},
  organization={Springer}
}

@article{van2000universal,
  title={Universal Description, Discovery and Integration},
  author={van Steenderen, Margaret},
  journal={SA Journal of Information Management},
  volume={2},
  number={4},
  year={2000}
}

@inproceedings{chao2004analysis,
  title={Analysis of Grid Service Composition With BPEL4WS},
  author={Chao, Kuo-Ming and Younas, Muhammad and Griffiths, Nathan and Awan, Irfan and Anane, Rachid and Tsai, Chen-Fang},
  booktitle={International Conference on Advanced Information Networking and Applications},
  pages={284--289},
  year={2004},
}

@inproceedings{wu2003automating,
  title={Automating DAML-S Web Services Composition using SHOP2},
  author={Wu, Dan and Parsia, Bijan and Sirin, Evren and Hendler, James and Nau, Dana},
  booktitle={International Semantic Web Conference},
  pages={195--210},
  year={2003},
  organization={Springer}
}

@article{nematzadeh2014qos,
  title={QoS Measurement of Workflow-based Web Service Compositions Using Colored Petri Net},
  author={Nematzadeh, Hossein and Motameni, Homayun and Mohamad, Radziah and Nematzadeh, Zahra},
  journal={The Scientific World Journal},
  year={2014},
  publisher={Hindawi}
}

@article{bekkouche2017qos,
  title={QoS-aware Optimal and Automated Semantic Web Service Composition with User’s Constraints},
  author={Bekkouche, Amina and Benslimane, Sidi Mohammed and Huchard, Marianne and Tibermacine, Chouki and Hadjila, Fethallah and Merzoug, Mohammed},
  journal={Service Oriented Computing and Applications},
  volume={11},
  number={2},
  pages={183--201},
  year={2017},
  publisher={Springer}
}

@article{baker2017energy,
  title={An Energy-aware Service Composition Algorithm for Multiple Cloud-based IoT Applications},
  author={Baker, Thar and Asim, Muhammad and Tawfik, Hissam and Aldawsari, Bandar and Buyya, Rajkumar},
  journal={Journal of Network and Computer Applications},
  volume={89},
  pages={96--108},
  year={2017},
  publisher={Elsevier}
}

@inproceedings{berrani2018towards,
  title={Towards a New Framework for Service Composition in the Internet of Things},
  author={Berrani, Samir and Yachir, Ali and Aissani, Mohamed},
  booktitle={International Conference on Computer Science and its Applications},
  pages={57--66},
  year={2018},
}

@inproceedings{ke2019implementation,
  title={An Implementation of Service Composition for Enterprise Business Processes},
  author={Ke, Jian and Xu, Jian Bo and Feng, Shu},
  booktitle={IOP Conference Series: Earth and Environmental Science},
  volume={234},
  number={1},
  year={2019},
  organization={IOP Publishing}
}

@inproceedings{wang2018client,
  title={A Client MicroServices Automatic Collaboration Framework Based on Fine-Grained APP},
  author={Wang, Ru and Chen, Shizhan and Feng, Zhiyong and Huang, Keman},
  booktitle={International Conference on Services Computing (SCC)},
  pages={25--32},
  year={2018},
  organization={IEEE}
}

@article{garriga2016restful,
  title={RESTful Service Composition at a Glance: A Survey},
  author={Garriga, Martin and Mateos, Cristian and Flores, Andres and Cechich, Alejandra and Zunino, Alejandro},
  journal={Journal of Network and Computer Applications},
  volume={60},
  pages={32--53},
  year={2016},
  publisher={Elsevier}
}

@inproceedings{kim2016ontology,
  title={Ontology-based Open API Composition Method for Automatic Mashup Service Generation},
  author={Kim, Sang Il and Kim, Hwa Sung},
  booktitle={International Conference on Information Networking (ICOIN)},
  pages={351--356},
  year={2016},
  organization={IEEE}
}

@book{richardson2008restful,
  title={RESTful Web Services},
  author={Richardson, Leonard and Ruby, Sam},
  year={2008},
  publisher={" O'Reilly Media, Inc."}
}

@article{sirin2004htn,
  title={HTN Planning for Web Service Composition Using SHOP2},
  author={Sirin, Evren and Parsia, Bijan and Wu, Dan and Hendler, James and Nau, Dana},
  journal={Journal of Web Semantics},
  volume={1},
  number={4},
  pages={377--396},
  year={2004},
  publisher={Elsevier}
}

@article{omid2017context,
  title={Context-Aware Web Service Composition based on AI Planning},
  author={Omid, Maryam},
  journal={Applied Artificial Intelligence},
  volume={31},
  number={1},
  pages={23--43},
  year={2017},
}

@article{jatoth2015computational,
  title={Computational Intelligence Based QoS-aware Web Service Composition: A Systematic Literature Review},
  author={Jatoth, Chandrashekar and Gangadharan, G.R. and Buyya, Rajkumar},
  journal={Transactions on Services Computing},
  volume={10},
  number={3},
  pages={475--492},
  year={2015},
  publisher={IEEE}
}

@inproceedings{pop2010ant,
  title={Ant-inspired Technique for Automatic Web Service Composition and Selection},
  author={Pop, Cristina Bianca and Chifu, Viorica Rozina and Salomie, Ioan and Dinsoreanu, Mihaela and David, Tudor and Acretoaie, Vlad},
  booktitle={International Symposium on Symbolic and Numeric Algorithms for Scientific Computing},
  pages={449--455},
  year={2010},
  organization={IEEE}
}

@inproceedings{boussalia2014optimizing,
  title={Optimizing QoS-based Web Services Composition by Using Quantum Inspired Cuckoo Search Algorithm},
  author={Boussalia, Serial Rayene and Chaoui, Allaoua},
  booktitle={International Conference on Mobile Web and Information Systems},
  pages={41--55},
  year={2014},
  organization={Springer}
}

@inproceedings{oh2005bf,
  title={BF*: Web Services Discovery and Composition as Graph Search Problem},
  author={Oh, Seog-Chan and On, Byung-Wo and Larson, Eric J and Lee, Dongwon},
  booktitle={International Conference on e-Technology, e-Commerce and e-Service},
  pages={784--786},
  year={2005},
  organization={IEEE}
}
}
\cleardoublepage

\phantomsection
\addcontentsline{toc}{chapter}{\listfigurename}
\listoffigures
\cleardoublepage

\phantomsection
\addcontentsline{toc}{chapter}{\listtablename}
\listoftables
\cleardoublepage


\cleardoublepage
%
\appendix
\addcontentsline{toc}{chapter}{Appendix}

\chapter{Name-Match Algorithm \ref{algo:fornamematch} Poster}
\label{sec:appendix:poster}

\comment{will add here: SYNASC poster - which was not included in the proceedings, with a short description, and descriptions and references to other resources created.}

Algorithm \ref{algo:fornamematch}, solving the composition for the initial name-matching model, was presented as a poster \cite{diac2017warp} at the \emph{19th International Symposium on Symbolic and Numeric Algorithms for Scientific Computing (SYNASC)}. The poster is a visual and concise representation of the work in paper \cite{diac2017engineering}, and as it was not included in the proceedings, therefore we added in on the next page. In this context, the algorithm was named \textbf{WARP}.

\textbf{WARP}, or Algorithm \ref{algo:fornamematch}, is a new algorithm designed to achieve the highest performance in Automatic Web Service Composition. In its first version, it solves the simplest form of composition, providing very low running times but relatively long compositions. Further, a heuristic score is associated to services that promote the most relevant services when multiple are accessible. After this improvement, compositions are short and computed almost as fast as before. Two well-known benchmarks presented in \cite{blake2005eee} and \cite{oh2006wsben} are used to compare WARP with previous planning based approaches from \cite{zou2014dynamic}. More revealing instances are generated by a special designed tests generator, presented in the poster.

\newpage
\pagestyle{empty}
\includepdf{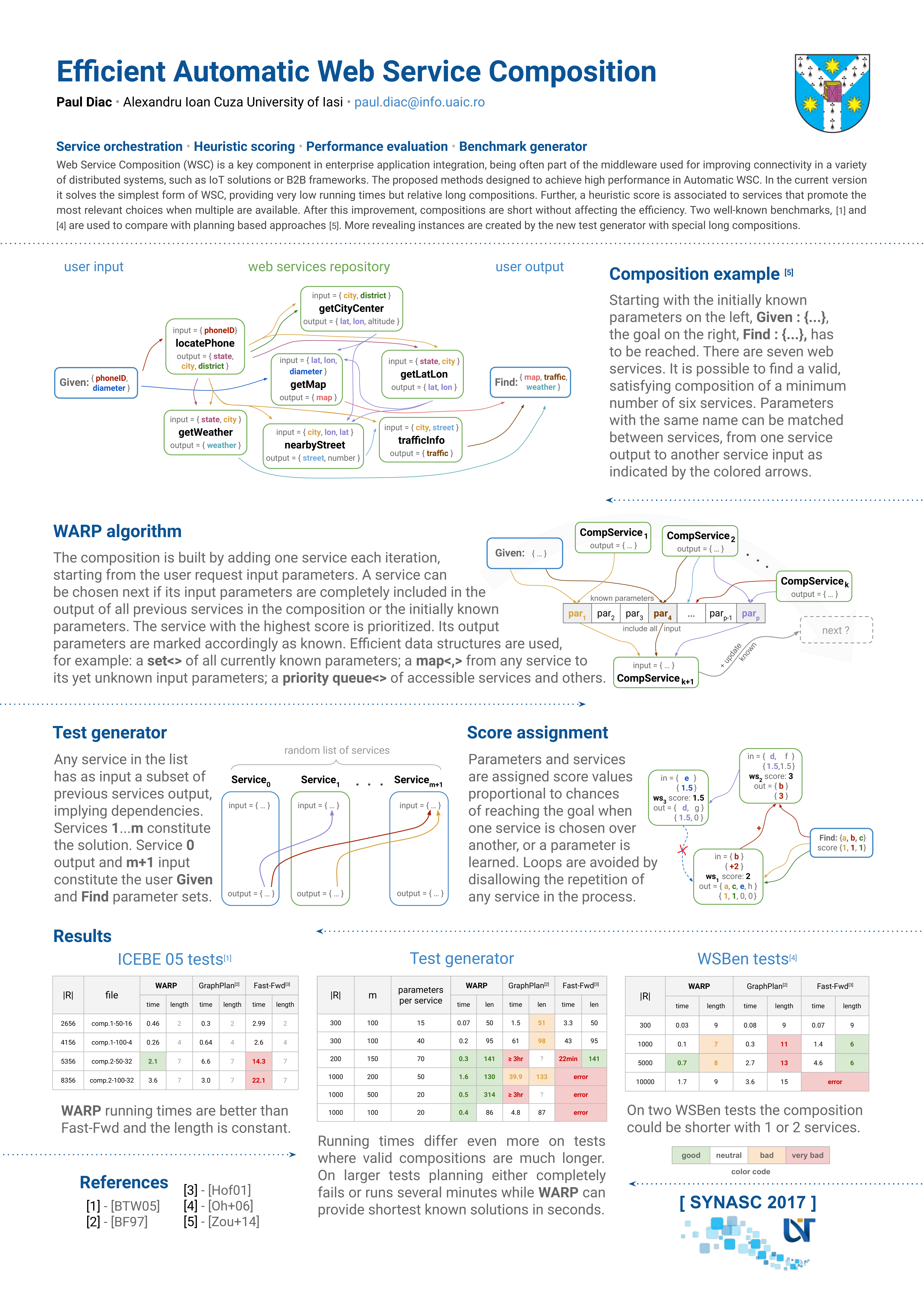}

\chapter{Resources: Code, Benchmarks and Corpus}
\label{sec:appendix:resources}
\comment{add some references to: code for the various algorithms on some GitHub account, benchmarks, QuoVadis Corpus, and maybe others.}

\vspace {-1.3cm}
The resources developed during the PhD are published on \href{https://github.com/pauldiac/}{\textcolor{blue}{\underline{GitHub}}}\footnote{repositories \textbf{WebServiceComposition} and \textbf{QuoVadis} -- 
\url{https://github.com/pauldiac/}} in \textbf{WebServiceComposition} and \textbf{QuoVadis} repositories. The structure is presented below, with some extra details on the right side. The \emph{README.md} files provide more structured information about content and usage. Except for some public tools (\emph{planners}, \emph{WSBen}), all the code and the benchmarks are personal contributions. \\ All code is under \emph{MIT License}\footnote{MIT License -- \url{https://choosealicense.com/licenses/mit/}}.
\vspace {0.6cm}

\begin{lstlisting}[basicstyle=\small, keywordstyle=\bfseries, morekeywords={name, match, hierarchical, relational, object, oriented, online, WebServiceComposition, QuoVadis}]
+---WebServiceComposition
|   +---name match
|   |   |   2017 KES.pdf
|   |   |   2017 SYNASC.pdf       papers
|   |   +---benchmark             generated and converted tests
|   |   \---projects
|   |       +---ConvertWSC        generator and converter
|   |       +---planning          planning solvers
|   |       +---psolver           polynomial algorithm
|   |       \---wsben             another generator tool
|   +---hierarchical
|   |       2018 KES.pdf          paper
|   |       wsc2008.zip           tests
|   +---relational
|   |   |   2019 FROM.pdf
|   |   |   2019 KES.pdf          papers
|   |   |   benchmark.zip         generated tests
|   |   \---RelationalWSC         algorithm implementation
|   +---object oriented
|   |   |   2019 ICEBE.pdf        paper
|   |   \---CompositionGenerator  test generator
|   \---online
|       |   2019 ICSOFT.pdf       paper
|       \---OnlineWSC             solver and test generator
\---QuoVadis
    |   2014 LPCL.pdf             paper
    |   Quo Vadis.zip             corpus
    +---Aggregator                aggregator tool
    \---CorefGraph                co-referential chains tool

\end{lstlisting}

\clearpage
%

\addcontentsline{toc}{chapter}{Declaration}
\pdfbookmark[0]{Declaration}{Declaration}

\titleformat{\chapter}{}{}{0pt}{\huge}

\chapter*{Declaration Regarding the Originality of the Doctoral Dissertation Content}
\label{sec:declaration}
\thispagestyle{empty}

I hereby declare on my own risk and acknowledge liability as warranted by the law that this doctoral dissertation obeys the quality and professional ethics standards and that the originality of its contents is ensured, according to the provisions of article 143 paragraph (4) and article 170 of The Law of National Education no. 1/2011, including subsequent amendments and additions, and article 65 paragraphs (5)-(7) of the Code for Doctoral Studies approved by Government Resolution no. 681/2011.

I hereby declare that the dissertation represents the result of my work, based on my own research and information obtained from sources that are accordingly quoted and indicated in the text, figures, tables and references according to the citation rules for academic writing and relevant copyright legislation. 

I declare that this dissertation has never been submitted to another superior education institution in order to obtain a scientific or didactic title.

In the case of subsequent discovery of forgery, I shall obey the provisions of the laws in force.

\bigskip

\noindent\textit{\thesisUniversity, \\ \thesisDate}

\smallskip

\begin{flushright}
	\begin{minipage}{5cm}
		\rule{\textwidth}{1pt}
		\centering\thesisName
	\end{minipage}
\end{flushright}




\end{document}